\pdfoutput=1
\documentclass[a4paper,fleqn,usenatbib]{mnras}

\usepackage[T1]{fontenc}
\usepackage{ae,aecompl}

\usepackage{mathrsfs} 
\usepackage{amsmath}  
\usepackage{amssymb}  
\usepackage{txfonts} 
\usepackage{mathtools} 
\usepackage{nccmath}  
\usepackage[ruled,lined,linesnumbered]{algorithm2e}
\usepackage{url}    
\usepackage{pifont} 
\usepackage{wasysym} 
\usepackage{xspace} 
\usepackage{ulem} 
\usepackage{multirow} 
\usepackage{tabularx} 
\usepackage{bigdelim} 
\usepackage{arydshln} 
\usepackage{enumitem} 
\usepackage{fancybox}
\usepackage{microtype} 
\usepackage[pdftex]{color}
\usepackage[pdftex,hiresbb]{graphicx}
\newcommand{\diff}{\mathrm{d}}

\newcommand{\Msolar}{\mathrm{M_{\odot}}}

\newcommand{\Zsolar}{\mathrm{Z_{\odot}}}
\newcommand{\Angstrom}{\mathrm{\AA}}

\newcommand{\pc}{\mathrm{pc}}

\newcommand{\yr}{\mathrm{yr}}
\newcommand{\kyr}{\mathrm{kyr}}

\newcommand{\eV}{\mathrm{eV}}
\newcommand{\keV}{\mathrm{keV}}

\newcommand{\kms}{\mathrm{km\; s^{-1}}}
\newcommand{\nden}{\mathrm{cm^{-3}}}
\newcommand{\mden}{\mathrm{g\; cm^{-3}}}

\newcommand{\cden}{\mathrm{cm^{-2}}}

\newcommand{\electron}{\mathrm{e^{-}}}
\newcommand{\proton}{\mathrm{H^{+}}}
\newcommand{\HI}{\mathrm{H^{0}}}
\newcommand{\Hminus}{\mathrm{H^{-}}}
\newcommand{\Hmol}{\mathrm{H_{2}}}
\newcommand{\Hmolplus}{\mathrm{H^{+}_{2}}}
\newcommand{\HeI}{\mathrm{He^{0}}}
\newcommand{\HeIMS}{\mathrm{He(2\;^{3}S)}} 
\newcommand{\HeII}{\mathrm{He^{+}}}
\newcommand{\HeIII}{\mathrm{He^{++}}}
\newcommand{\dust}{\mathrm{dust}}
\newcommand{\kR}[1]{k_{\mathrm{R#1}}}
\newcommand{\dex}{\mathrm{dex}}
\newcommand{\sigmaHmolPItot}{\sigma_{\rm H_{2},PI,tot}(E)}
\newcommand{\sigmaHmolPAtot}{\sigma_{\rm H_{2},PA,tot}(E)}
\newcommand{\BRC}{\mathrm{BR_{C93}(E)}}
\newcommand{\BRD}{\mathrm{BR_{DPI}}(E)}
\newcommand{\nele}{n(\mathrm{e^{-}})}

\newcommand{\ngr}{n_{\mathrm{gr}}}
\newcommand{\nH}{n({\mathrm{H}})}

\newcommand{\nHeq}{n_{\mathrm{H}}}
\newcommand{\nHmid}{n_{\mathrm{mid}}(\mathrm{H})}
\newcommand{\kBoltz}{k_{\mathrm{B}}}
\newcommand{\MHydro}{m_{\mathrm{H}}}
\newcommand{\Tgas}{T_{\mathrm{gas}}}
\newcommand{\Tgr}{T_{\mathrm{gr}}}
\newcommand{\Tsub}{T_{\mathrm{sub}}}
\newcommand{\Tcomp}{T_{\mathrm{comp}}}
\newcommand{\Tgasini}{T_{\mathrm{gas,0}}}
\newcommand{\Tgrini}{T_{\mathrm{gr,0}}}
\newcommand{\agr}{a_{\mathrm{gr}}}
\newcommand{\Qabs}{\mathcal{Q}_{\mathrm{abs}}(\nu,\agr)}
\newcommand{\Qsca}{\mathcal{Q}_{\mathrm{sca}}(\nu,\agr)}

\newcommand{\sigmagr}{\sigma_{\mathrm{gr}}}
\newcommand{\MBH}{M_{\mathrm{BH}}}

\newcommand{\Lcorona}{L_{\mathrm{corona}}}
\newcommand{\Lbol}{L_{\mathrm{bol}}}
\newcommand{\LX}{L_{\mathrm{X}}}

\newcommand{\hnuion}{h\nu_{\mathrm{ion}}}

\newcommand{\fX}{f_{\mathrm{X}}}

\newcommand{\Rsubiso}{R^{\mathrm{iso}}_{\mathrm{sub}}}
\newcommand{\pth}{p_{\mathrm{th}}}
\newcommand{\ethermal}{e_{\mathrm{th}}}
\newcommand{\gameff}{\gamma_{\mathrm{eff}}}
\newcommand{\LambdaMTL}{\Lambda_{\mathrm{metal}}}
\newcommand{\mH}{m_{\mathrm{H}}}

\newcommand{\bBH}{b_{\mathrm{BH}}}
\newcommand{\MNSC}{M_{\mathrm{NSC}}}
\newcommand{\bNSC}{b_{\mathrm{NSC}}}
\newcommand{\MNSD}{M_{\mathrm{NSD}}}
\newcommand{\bNSD}{b_{\mathrm{NSD}}}

\title[RHD simulations of AGN tori]{Subparsec-scale dynamics of a dusty gas disk exposed to anisotropic AGN radiation with frequency-dependent radiative transfer}
\author[D. Namekata and M. Umemura]{
Daisuke Namekata,$^{1}$\thanks{E-mail: namektds@gmail.com, namekata@ccs.tsukuba.ac.jp}, and
Masayuki Umemura$^{1}$ \thanks{E-mail: umemura@ccs.tsukuba.ac.jp}\\
$^{1}$Center for Computational Sciences, University of Tsukuba, 1-1-1 Tennodai, Tsukuba 305-8577 Ibaraki, Japan}

\date{Accepted XXX. Received YYY; in original form ZZZ}

\pubyear{2015}

\begin{document}
\label{firstpage}
\pagerange{\pageref{firstpage}--\pageref{lastpage}}
\maketitle

\begin{abstract}
We explore the gas dynamics near the dust sublimation radius of active galactic nucleus (AGN). For the purpose, we perform axisymmetric radiation hydrodynamic simulations of a dusty gas disk of radius $\approx 1\,\pc$ around a supermassive black hole of mass $10^{7}\,\Msolar$ taking into account (1) anisotropic radiation of accretion disk, (2) X-ray heating by corona, (3) radiative transfer of infrared (IR) photons reemitted by dust, (4) frequency dependency of direct and IR radiations, and (5) separate temperatures for gas and dust. As a result, we find that for Eddington ratio $\approx 0.77$, a nearly neutral, dense ($\approx 10^{6\operatorname{-}8}\;\nden$), geometrically-thin ($h/r<0.06$) disk forms with a high velocity ($\approx 200 \sim 3000\;\kms$) dusty outflow launched from the disk surface. The disk temperature is determined by the balance between X-ray heating and various cooling, and the disk is almost supported by thermal pressure. Contrary to \citet{krolik07:_agn}, the radiation pressure by IR photons is not effective to thicken the disk, but rather compresses it. Thus, it seems difficult for a radiation-supported, geometrically-thick, obscuring torus to form near the dust sublimation radius as far as the Eddington ratio is high ($\sim 1$). The mass outflow rate is $0.05\operatorname{-}0.1\;\Msolar/\yr$ and the column density of the outflow is $N_{\mathrm{H}}\lesssim 10^{21}\;\mathrm{cm^{-2}}$. To explain observed type-II AGN fraction, it is required that outflow gas is extended to larger radii ($r\gtrsim 10\;\pc$) or that a denser dusty wind is launched from smaller radii ($r\sim 10^{4}\;R_{g}$).
\end{abstract}


\section{Introduction} \label{sec:intro}
The activities of active galactic nuclei (AGNs) manifest in the form of intense radiation and relativistic winds or jets, which dynamically interact with the surrounding medium and could change significantly its physical state. These so-called AGN feedbacks are generally thought to play important roles in shaping various properties of galaxies. Thus, a detailed understanding of AGN activities is crucial to elucidate the formation and evolution of galaxies. There are a number of questions for AGN activities that need to be answered. Among them, the most important are (i) when and how does AGN phenomenon occurs?, (ii) how are the mode (i.e., radiative-mode or jet-mode; \citealt{heckman14:_coevol_galax_super_black_holes}) and the strength of AGN phenomenon determined?, (iii) how long does AGN phenomenon continue? In order to answer these questions, we must understand gas supply process to the accretion disk (AD) around a supermassive black hole (SMBH) and clarify how the rate and the duration of gas supply are determined.

According to the unified model of AGNs (e.g., \citealt{antonucci93:_unified_model_activ_galac_nuclei_quasar,urry95:_unified_schem_radio_loud_activ_galac_nuclei}), there is a dusty (molecular) torus around an AD. This torus is thought of as a gas reservoir for the AD. Hence, what we need to do is to examine gas supply process(es) from AGN tori to ADs. An important first step toward this is to explore the structures of density, temperature, and gas flow near the dust sublimation radius ($\Rsubiso$)\footnote{For a typical AGN spectrum, the dust sublimation radius is given by
\begin{eqnarray}
\Rsubiso & = &  0.121\;\pc \left(\dfrac{\Lbol}{10^{45}\;\mathrm{erg\;s^{-1}}}\right)^{0.5} \nonumber\\
&& \qquad \times \left(\dfrac{\Tsub}{1800\;\mathrm{K}}\right)^{-2.804} \left(\frac{\agr}{0.1\;\micron}\right)^{-0.510}, \label{eq:Rsubiso}
\end{eqnarray}
where $\Lbol$ is the bolometric luminosity of AGN, $\Tsub$ is the sublimation temperature of dust grain, and $\agr$ is the grain radius. In the derivation, we assume (i) that AGN radiates isotropically at all the frequency, (ii) the fiducial SED model described in \S~\ref{subsec:AGN_SED}, and (iii) the dust model described in \S~\ref{subsec:ISM_dust_model}. Note that the accuracy of the fit gets worse for $\agr \gtrsim 0.3\;\micron$.}, and outflow rate from there, because (i) these quantities are closely related to the efficiency of angular momentum transfer\footnote{Turbulent and self-gravitational torques depend on density and velocity fields.} and the net gas supply rate to AD, and (ii) regions near the dust sublimation radius are exposed to powerful ionizing radiation from AD and corona and are where gas inflow is most strongly hampered by the AGN radiation.

There have been many observational efforts to elucidate the density structure of AGN tori (for recent review, see \citealt{netzer15:_revis_unified_model_activ_galac_nuclei}). The scale height of AGN tori can be estimated from the type-II AGN fraction and observations show that the type-II fraction is $\approx 0.3\operatorname{-}0.6$ for AGNs with modest X-ray luminosities of $10^{43}\operatorname{-}10^{44}\;\mathrm{erg\;s^{-1}}$ (e.g., \citealt{hasinger08:_absor}). This means that the half opening angles $\theta_{\mathrm{OA}}$ of tori are $\approx 50^{\circ}\operatorname{-}70^{\circ}$. The luminosity dependence of the type-II fraction is still under debate (e.g., \citealt{hasinger08:_absor,toba13:_microm_lumin_funct_various_types_galax_akari,toba14:_lumin_soloan_digit_sky_survey,merloni14}). Fitting observed infrared (IR) spectral energy distributions (SEDs) by SEDs of phenomenological torus models shows that the internal density structures of tori need to be clumpy to some extent (e.g., \citealt{nenkova02:_dust_emiss_activ_galac_nuclei,dullemond05:_clump,onig06:_radiat_agn_ngc,nenkova08a:_agn_dusty_tori,nenkova08b:_agn_dusty_tori,stalevski12,feltre12:_smoot_agn,hatzminaoglou15}). The detailed modeling of IR SEDs of broad- and narrow-line AGNs suggests that hot pure-graphite dust clouds commonly exist near the dust sublimation radius (\citealt{mor12:_hot}). Direct observations of the spatial distribution of warm dust in the central several parsecs of a nearby AGN has become possible by recent near infrared (NIR)/MIR interferometers (e.g., \citealt{jaffe04:_ngc} for NGC 1068; \citealt{onig12:_parsec_scale_dust_emiss_from} for NGC 424; \citealt{tristram07:_resol_circin,tristram12:_agn,tristram14:_circin} for Circinus galaxy). \citet{tristram14:_circin} showed the dust emission in the central a few parsec in the Circinus galaxy comes from two components: a disk-like component and an component extending in polar direction. A similar result was obtained for NGC 424 (\citealt{onig12:_parsec_scale_dust_emiss_from}). The radii of the inner edges of tori are being probed by long-term IR reverberation mapping (RM) observations for a number of nearby AGNs. \citet{koshida14:_rever_seyfer} compared the K-band reverberation radii with the reverberation radii of broad Balmer emission lines obtained by \citet{bentz09} and the radii of hypothetical hot dust clouds obtained by \citet{mor12:_hot} and showed that the K-band reverberation radii are $\approx 4\operatorname{-}5$ times of the radii of broad Balmer line emission regions and that the hot pure-graphite dust clouds are located between the other two radii, indicating that the outer parts of ADs and the inner edges of tori are smoothly connected. However, even with state-of-the-art observational instruments, more detailed distributions of gas and dust and the structure of gas flow at sub-parsec scales cannot be probed due to the performance limitation. Thus, it is difficult to constrain gas supply process(es) from AGN tori to AD by observational studies alone.

There are also a number of theoretical and numerical modelings of AGN tori. Here, we briefly summarize the recent progress and the problems of radiation hydrodynamic (RHD) modeling of AGN tori because we are interested in the gas structures and the gas flow in regions exposed to powerful AGN radiation\footnote{Note that past efforts on modeling AGN tori and their difficulties are discussed in detail in \citet{krolik07:_agn} and \citet{chan15:_radiat}.}. \citet{krolik07:_agn} and \citet{shi08:_radiat_x} analytically showed that a radiation-supported, geometrically-thick, hydrostatic structure can be formed near an AGN. In their models, vertical support is provided by IR dust reemission. However, their models are based on idealized assumptions such as (i) AGN radiates isotropically at all the wavelengths, (ii) all the radiation emitted by AGN is turned into IR photons at the inner (vertical) boundary, and (iii) sub-Keplerian rotation. In reality, AD radiates anisotropically; a more stronger radiation is emitted for the direction parallel to the symmetric axis of AD (\citealt{netzer87:_quasar}; see also Eq.(\ref{eq:AD_flux})). Also, it is doubtful whether sub-Kepler rotation is reconciled with Kepler rotation suggested by observations of a $\mathrm{H_{2}O}$ mega-mesar disk in NGC 4258 (e.g., \citealt{greenhill95:_detec_ngc,herrnstein05:_ngc}). Thus, it is not clear whether such a structure forms if these simplifications are removed. \citet{wada12:_radiat} performed three-dimensional (3D) RHD simulations of a circumnuclear dusty gas disk in the central 60 parsecs of an AGN hosting galaxy taking into account anisotropic radiation from AD, X-ray heating\footnote{By X-ray heating, we mean photoelectric heating by hard X-ray photons and Compton heating.} from AD corona, and self-gravity of gas and showed that a geometrically-thick structure is formed by circulation flow driven by failed winds\footnote{The resultant SED is calculated by \citet{schartmann14:_tiem} and it agrees with typical AGN SEDs except for spectral features at shorter wavelengths. The luminosity dependence of gas structure is investigated by \citet{wada15:_obscur_fract_activ_galac_nuclei}.}. The gas supply rate measured at the distance of one parsec is $\approx 2\times 10^{-4} \operatorname{-} 10^{-3}\;\Msolar\;\yr^{-1}$. However, their simulations do not spatially resolve regions near the dust sublimation radius and hence the gas supply rate to AD is not clear. Gas dynamics at smaller scale was investigated by \citet{dorodnitsyn12:_activ_galac_nucleus_obscur_winds}, in which they performed axisymmetric RHD simulations of a dusty gas disk of radius $\approx 2\;\pc$ around an AGN for different values of Eddington ratios taking into account X-ray heating and the transfer of IR photons and showed that an IR-supported structure is formed, but its thickness is thinner than the predictions by \citet{krolik07:_agn} and \citet{shi08:_radiat_x}. An averaged outflow rate including failed winds is found to be $0.1\operatorname{-}0.2\;\Msolar\;\yr^{-1}$ for Eddington ratio of 0.6. However, there are several problems in this study. First, they consider an extremely high density case. The number density of the densest part of the disk is larger than $10^{24}\;\nden$ (see their Fig.1), which is twelve order magnitude larger than typical number densities of broad line region (BLR), $n_{\mathrm{BLR}}\sim 10^{10}\operatorname{-}10^{12}\;\nden$ (e.g., \citealt{netzer13:_physic_evolut_activ_galac_nuclei}). This could make it difficult for such a high density gas to heat enough that the disk inflates by dust reemission and a more realistic gas disk should be examined. Second, they also assumed that AGN radiates isotropically. Third, they assumed a large X-ray luminosity fraction ($0.5$). According to Fig.1 in \citet{ishibashi09:_agn_uv_x}, typical X-ray luminosity fraction is $\sim 0.1$ for UV luminosity $10^{45}\;\mathrm{erg\;s^{-1}}$. Fourth, their mid-plane boundary condition is not realistic; they change the mid-plane density manually in the course of the simulations. More recently, \citet{chan15:_radiat} investigated the gas structure and the gas flow near the dust sublimation radius by performing 3D RHD simulations and showed that a geometrically-thick, obscuring structure can be formed near the dust sublimation radius. However, they also assumed isotropic AGN radiation and did not take into account the effects of X-ray heating. Moreover, they assumed that the temperatures of gas and dust is always the same. This is not the case in general. \citet{dorodnitsyn15:_parsec} examined time evolution of a dusty torus located just outside the dust sublimation radius by performing axisymmetric RHD simulations taking into account the so-called $\alpha$-viscosity. In their study, they focused on low Eddington ratios ($0.01\operatorname{-}0.3$) and showed that a dusty torus keeps its structure during $400\;\kyr$. However, as the same as their previous study, isotropic AGN radiation and a large X-ray luminosity fraction are assumed in their study. In order to obtain realistic distributions of gas and dust and flow structure, we need to perform RHD simulations taking into account the effects of \ding{202} anisotropy of AD radiation, \ding{203} X-ray heating with typical X-ray luminosity fraction, and \ding{204} dust reemission.

In this study, we perform chemo-RHD simulations of a dusty gas disk at regions near the dust sublimation radius taking into account all of the three effects discussed above, as well as \ding{205} frequency dependency of direct and IR radiations and \ding{206} separate temperatures for gas and dust ($\Tgas$, $\Tgr$), to investigate the density and temperature structures, the characteristics of gas flow, and outflow rate realized near the dust sublimation radius of an AGN. In the following, we will show that it is very difficult to form a geometrically-thick, obscuring structure near the dust sublimation radius by radiation pressure of dust reemission \textit{alone} and that X-ray heating plays a role in confining of dense gas to the vicinity of the mid-plane. This paper is organized as follows. In Section \ref{sec:model}, we explain our models and basic assumptions. Then, we describe the detail of the numerical methods in Section \ref{sec:numerical_methods}. Next, in Section \ref{sec:numerical_results}, we show our numerical results. In Section~\ref{sec:discussions}, we discuss uncertainties of our numerical results and the relation between our study and previous studies and give some implications for AGN tori. Finally, in Section~\ref{sec:summary}, we summarize the present study.

\section{Model} \label{sec:model}
\subsection{Basic assumptions, model parameters, and our approach}\label{subsec:parameters}
\begin{figure}
\centering
\includegraphics[clip,width=\linewidth]{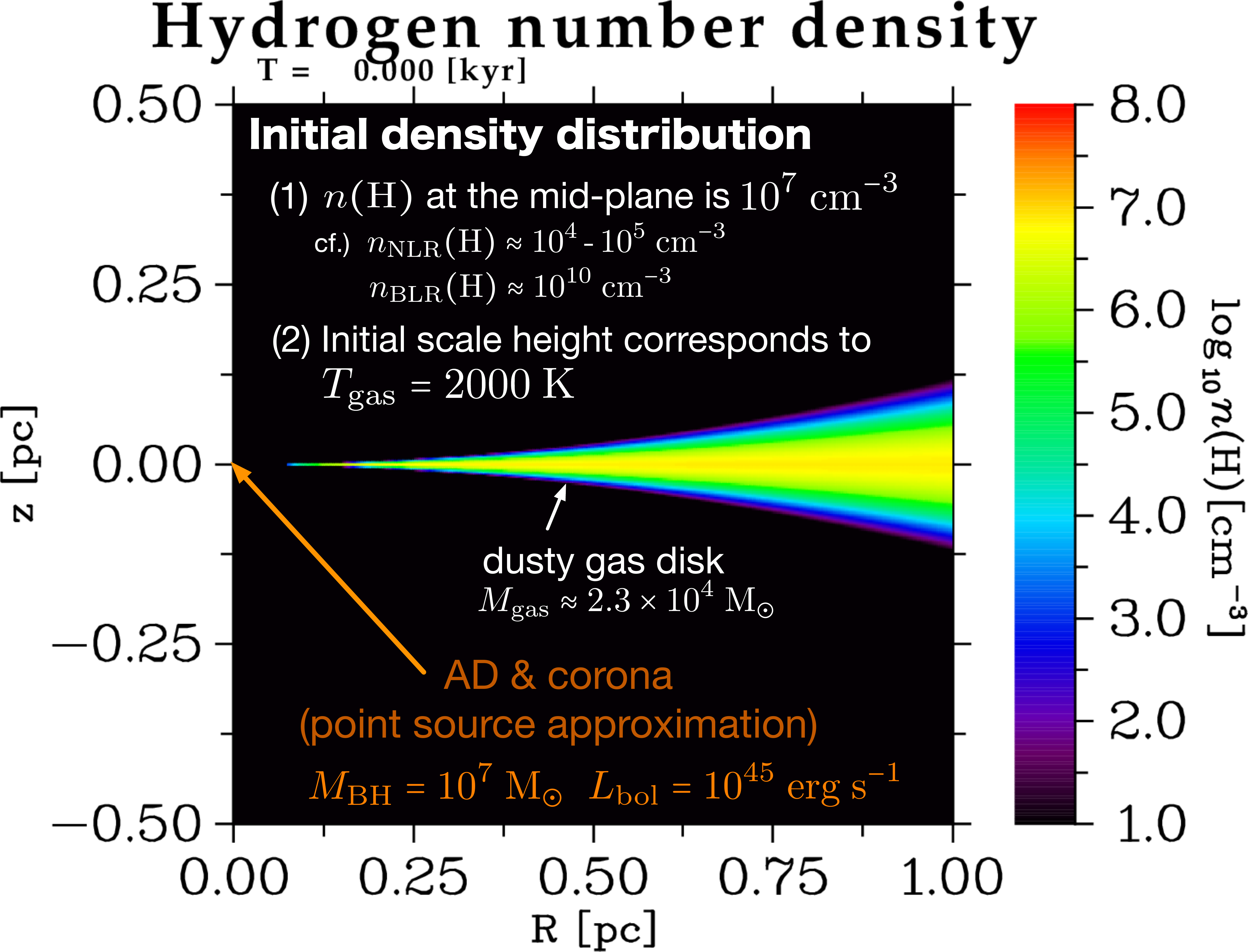}
\caption{The distribution of hydrogen number density of a dusty gas disk at the initial state. In this study, we use a cylindrical coordinate and $z=0$ corresponds to the mid-plane of galactic disk of host galaxy as well as that of the dusty gas disk. The mid-plane number density is $10^{7}\;\nden$, which is intermediate between number densities of narrow- and broad-line regions. The scale height of the disk corresponds to the gas temperature $2000\;\mathrm{K}$. The disk is surrounded by a warm ($\Tgas=3\times 10^{4}\;\mathrm{K}$), rarefied, dust-free medium.}
\label{fig:initial_model}
\end{figure}

As shown in Fig.~\ref{fig:initial_model}, we consider a dusty gas disk of radius $\approx 1\;\pc$ around a SMBH of mass $10^{7}\;\Msolar$. We assume that the gas disk is axisymmetric and its symmetric axis is parallel with that of the accretion disk (AD). The inner and outer radii of the disk are assumed to be $0.075\;\pc(\approx \Rsubiso)$ and $1\;\pc(\approx 8\Rsubiso)$, respectively. We assume that the disk is initially in the hydrostatic equilibrium in the vertical direction (i.e., $z$-direction in Fig.~\ref{fig:initial_model}) with external gravity due to the SMBH and host galaxy and it rotates with velocities so that the centrifugal force balances with the external gravity. The hydrogen number density at the mid-plane, $\nHmid$, is constant over galactic radii and is assumed to be $\nHmid = 10^{7}\;\nden$, which is intermediate between number densities of narrow-line region (NLR) and BLR. Thus, our choice of gas density is reasonable. In this study, we consider the RHD evolution of this gas disk when irradiated by an AGN with bolometric luminosity $\Lbol=10^{45}\;\mathrm{erg\;s^{-1}}$. The corresponding Eddington ratio is $\approx 0.77$ and it is higher than typical Eddington ratio of radiative-mode AGNs ($\sim 0.1$). The reason for this choice is because studying AGNs with high accretion rates is more important than those with low accretion rates in terms of AGN feedback.

We assume that radiation from AGN is emitted by the AD and the corona (e.g., \citealt{kawaguchi01:_broad_spect_energ_distr_of}), which are located at the origin in Fig.~\ref{fig:initial_model}. Following previous studies (\citealt{netzer87:_quasar,kawaguchi10:_orien_effec_on_the_inner,kawaguchi11:_near_infrar_rever_by_dusty,liu11:_dusty}), we assume that the AD radiates anisotropically due to the change in the apparent surface area with the line-of-sight (LOS) and the limb darkening effect. In this case, the monochromatic radiation flux of the AD is given by
\begin{equation}
F^{\mathrm{AD}}_{\nu}(\theta) = \frac{6}{7}\dfrac{L^{\mathrm{AD}}_{\nu}\exp(-\tau_{\nu})}{4\pi r^{2}}\cos\theta(1+2\cos\theta), \label{eq:AD_flux}
\end{equation}
where $\theta$ is the angle measured from the symmetric axis of the AD, $L^{\mathrm{AD}}_{\nu}$ is the monochromatic luminosity of the AD, and $\tau_{\nu}$ is the optical depth. On the other hand, we assume that the corona emits radiation isotropically and its monochromatic radiation flux is simply given by $F^{\mathrm{corona}}_{\nu}=L^{\mathrm{corona}}_{\nu}\exp(-\tau_{\nu})/(4\pi r^{2})$. The fraction of luminosity of the corona is controlled by the parameter $\fX \equiv \Lcorona/\Lbol$ and the fiducial value in this study is $0.08$. A detailed description of SEDs of both components is given in \S~\ref{subsec:AGN_SED}.

We assume that the external gravitational potential consists of a SMBH, a nuclear star cluster (NSC), and a nuclear stellar disk (NSD). All of these are modeled by the Plummer potential $\Phi=-GM/\sqrt{r^{2}+b^{2}}$ (\citealt{plummer11}). The parameters adopted are as follows: $\MBH=10^{7}\;\Msolar$, $\bBH=0.1\;\pc$, $\MNSC=10^{8}\;\Msolar$, $\bNSC=10\;\pc$, $\MNSD=10^{9}\;\Msolar$, and $\bNSD=100\;\pc$. Note that a large value for $\bBH$ is intended to avoid a very small timestep near the origin and hence numerical results near the origin are not reliable. 

The numerical simulation is started when the AGN is turned on and we numerically follow the subsequent evolution of the disk. In order to take into account the five important effects described in \S~\ref{sec:intro} (\ding{202}-\ding{206}), we perform hydrodynamic simulations taking into account (1) photoionization and photodissociation of gas, (2) radiative transfer (RT) of IR photons reemitted by heated dust grains, (3) radiation pressure on gas and dust, and (4) self-gravity of gas. The rates of photoionization, photodissociation, photoelectric heating and so on are evaluated by performing RT calculations of direct radiation from AD and corona. The resultant rates are then used to calculate non-equilibrium chemical reactions of the following chemical species: $\electron$, $\proton$, $\HI$, $\Hminus$, $\Hmol$, $\Hmolplus$, $\HeI$, $\HeII$, $\HeIII$, and dust. The effects of metal is considered in an approximate way. We take into account Compton heating/cooling in a simplest way as well as standard cooling processes pertaining to hydrogen, helium, and dust (see \S~\ref{subsec:ISM_dust_model} and \S~\ref{subsec:radtr_chemistry}). As shown later, a dusty outflow is launched from the disk surface heated strongly by X-ray/UV/optical photons emitted by the AD and the corona. We measure this outflow rate. X-ray photons also heat the whole of the disk almost uniformly and determine the gas pressure in the interior of the disk. We investigate how the structures of density and temperature of the disk are determined by X-ray heating and radiation pressure of IR photons. Thus, various physical processes should play roles in shaping the gas structure and the gas flow. In this study, we devote ourself clarifying the effects of each process. To this end, we perform a number of simulations with switching some process(es) on and off and with fixing the parameters that characterize the system such as $\MBH$, $\Lbol$, and $\nHmid$. The dependency on these parameters will be examined elsewhere.

Table~\ref{tbl:our_models} summarize the simulation runs and their names show which physical processes are switched on or off in the simulations.  Comparing different runs enables us to clarify the effects of each physical process on the structures of density, temperature, and gas flow and the outflow rate. For instance, we can discriminate the effects of the radiation pressure of IR photons by the comparison between  \texttt{gra01\_wo\_sca} and \texttt{gra01\_wo\_sca\_noRP-IR}. More detail for each run is described when we explain our numerical results (\S~\ref{sec:numerical_results}).

\begingroup
\renewcommand{\arraystretch}{1.0}
\begin{table*}
\centering
\begin{minipage}{\hsize}
\caption{Simulation runs. Here, we use the following abbreviations: SG --- self-gravity, RP --- radiation pressure, and LDA --- local dust absorption.}
\label{tbl:our_models}
\begin{tabular}{@{}llllllllp{13em}l@{}}
\hline
Model name$^{\dagger}$ & Resolution & SG & $\agr$      & RP         & Dust       & Metal & LDA & Notes & Result \\
                    &            &    & ($\micron$) &            & scattering &       & approx.    &       &  \\
\hline
\uline{\texttt{gra01\_wo\_sca\_SG}}$^{\ddagger}$      & $512^{2}$ & Yes & $0.1$ & Yes & No  & No  & No  & --- & \S~\ref{subsec:overview} \\
\uline{\texttt{gra01\_wo\_sca\_MTL\_SG}}$^{\ddagger}$ & $512^{2}$ & Yes & $0.1$ & Yes & No  & Yes & No  & --- & \S~\ref{subsec:overview} \\
\texttt{gra01\_wo\_sca}                    & $512^{2}$ & No  & $0.1$ & Yes & No  & No  & No  & --- & \S~\ref{subsec:overview} \\
\texttt{gra01\_w\_sca}                     & $512^{2}$ & No  & $0.1$ & Yes & Yes & No  & No  & --- & \S~\ref{subsec:overview} \\
\texttt{gra01\_wo\_sca\_noRP-IR}           & $512^{2}$ & No  & $0.1$ & $\triangle$ & No  & No  & No  & Only RP due to direct photons is taken into account. & \S~\ref{subsec:overview} \\
\texttt{gra01\_wo\_sca\_noRP-drct}         & $512^{2}$ & No  & $0.1$ & $\triangle$ & No  & No  & No  & Only RP due to IR photons is taken into account. & \S~\ref{subsec:overview} \\
\texttt{gra01\_wo\_sca\_noRP-all}          & $512^{2}$ & No  & $0.1$ & No  & No  & No  & No  & --- & \S~\ref{subsec:overview} \\
\texttt{gra01\_wo\_sca\_fX016}             & $512^{2}$ & No  & $0.1$ & Yes & No  & No  & No  & $\fX=0.16$ & \S~\ref{subsec:overview} \\
\texttt{gra01\_wo\_sca\_S05}               & $512^{2}$ & No  & $0.1$ & Yes & No  & No  & No  & \citet{schartmann05:_towar_activ_galac_nuclei}'s SED & \S~\ref{subsec:overview} \\
\texttt{gra01\_wo\_sca\_S05\_fX016}        & $512^{2}$ & No  & $0.1$ & Yes & No  & No  & No  & \citet{schartmann05:_towar_activ_galac_nuclei}'s SED, $\fX=0.16$ & \S~\ref{subsec:overview}  \\
\hdashline[0.4pt/1pt]
\texttt{gra01\_wo\_sca\_LDA1\_SG}          & $512^{2}$ & Yes & $0.1$ & Yes & No  & No  & Yes & --- & \S~\ref{subsec:cooling_photon_effects} \\
\texttt{gra01\_wo\_sca\_MTLLDA1\_SG}       & $512^{2}$ & Yes & $0.1$ & Yes & No  & Yes & Yes & --- & \S~\ref{subsec:cooling_photon_effects} \\
\hdashline[0.4pt/1pt]
\texttt{gra01\_wo\_sca\_LDA1\_SG\_hr}      & $1024^{2}$ & Yes & $0.1$ & Yes & No  & Yes & Yes & High spatial resolution & \S~\ref{subsec:resolution_effects}  \\
\texttt{gra01\_wo\_sca\_MTLLDA1\_SG\_hr}   & $1024^{2}$ & Yes & $0.1$ & Yes & No  & Yes & Yes & High spatial resolution & \S~\ref{subsec:resolution_effects}  \\
\hdashline[0.4pt/1pt]
\texttt{gra1\_wo\_sca}                     & $512^{2}$ & No & $1$   & Yes & No  & No  & No  & --- & \S~\ref{subsec:grain_size_effects}  \\
\hdashline[0.4pt/1pt]
\texttt{gra01\_wo\_sca\_Tts10\textasciicircum 6K\_SG} & $512^{2}$ & Yes & $0.1$ & Yes & No  & No  & No  & instantaneous thermal sputtering if $\Tgas>10^{6}\;\mathrm{K}$ & \S~\ref{subsec:thermal_sputtering_effects} \\
\texttt{gra01\_wo\_sca\_MTLLDA1\_ts\_SG\_hr} & $1024^{2}$ & Yes & $0.1$ & Yes & No  & Yes  & Yes  & instantaneous thermal sputtering if $\Tgas>10^{6}\;\mathrm{K}$ & \S~\ref{subsec:thermal_sputtering_effects} \\
\hdashline[0.4pt/1pt]
\texttt{gra01\_wo\_sca\_noCompton\_SG} & $512^{2}$ & Yes & $0.1$ & Yes & No  & No  & No  & no Compton heating/cooling & \S~\ref{subsec:Compton_heatcool_effects} \\
\hline
\end{tabular}
\end{minipage}
\begin{flushleft}
$^{\dagger}$ The model name shows which physical processes are switched on or off. A string `\texttt{gra}` shows that the composition of dust grain is graphite. Numbers following `\texttt{gra}` indicate the radius of dust grains in $\micron$. Strings `\texttt{\_w\_sca}` and `\texttt{\_wo\_sca}` indicate whether the simulation takes into account the dust scattering of IR photons or not. Strings `\texttt{\_SG}` and `\texttt{\_MTL}` indicate that self-gravity and metal cooling are taken into account, respectively. If a model name contains a string `\texttt{LDA1}`, we assume that all the cooling photons emitted from low temperature dense gas are absorbed by local dust (for details, see \S~\ref{subsec:cooling_photon_effects}). We call this local dust absorption (LDA) approximation. A string `\texttt{\_noRP-IR}` indicates that the radiation pressure of IR photons is ignored in the simulation although we do calculate the transfer of IR photons (hence, heating of dust grains due to absorption of IR photons is taken into account). A string `\texttt{\_noRP-all}` indicates that radiation pressure is completely neglected in the simulation. In the model with `\texttt{\_Tts10\textasciicircum 6K}`, dust destruction due to thermal sputtering is taken into account in a pretty simple manner. We ignore Compton heating/cooling processes in the model with `\texttt{\_noCompton}`. For the other keywords in model name, see the entry `Notes`. \\
$^{\ddagger}$ Reference models in this study, with which an other model is compared to discriminate the effects of a particular physical process.
\end{flushleft}
\end{table*}
\endgroup

\subsection{AGN SED} \label{subsec:AGN_SED}
The SED of AD is modeled by the SED model given by \citet{nenkova08a:_agn_dusty_tori} except for models whose name contain the word \texttt{\_S05}, for which we use the SED model given by \citet{schartmann05:_towar_activ_galac_nuclei} to check the effects of SED shape (for the detail of SED shape, see the original papers or \S~2.2 in \citealt{namekata14:_agn}).

\begin{figure}
\centering
\includegraphics[clip,width=\linewidth]{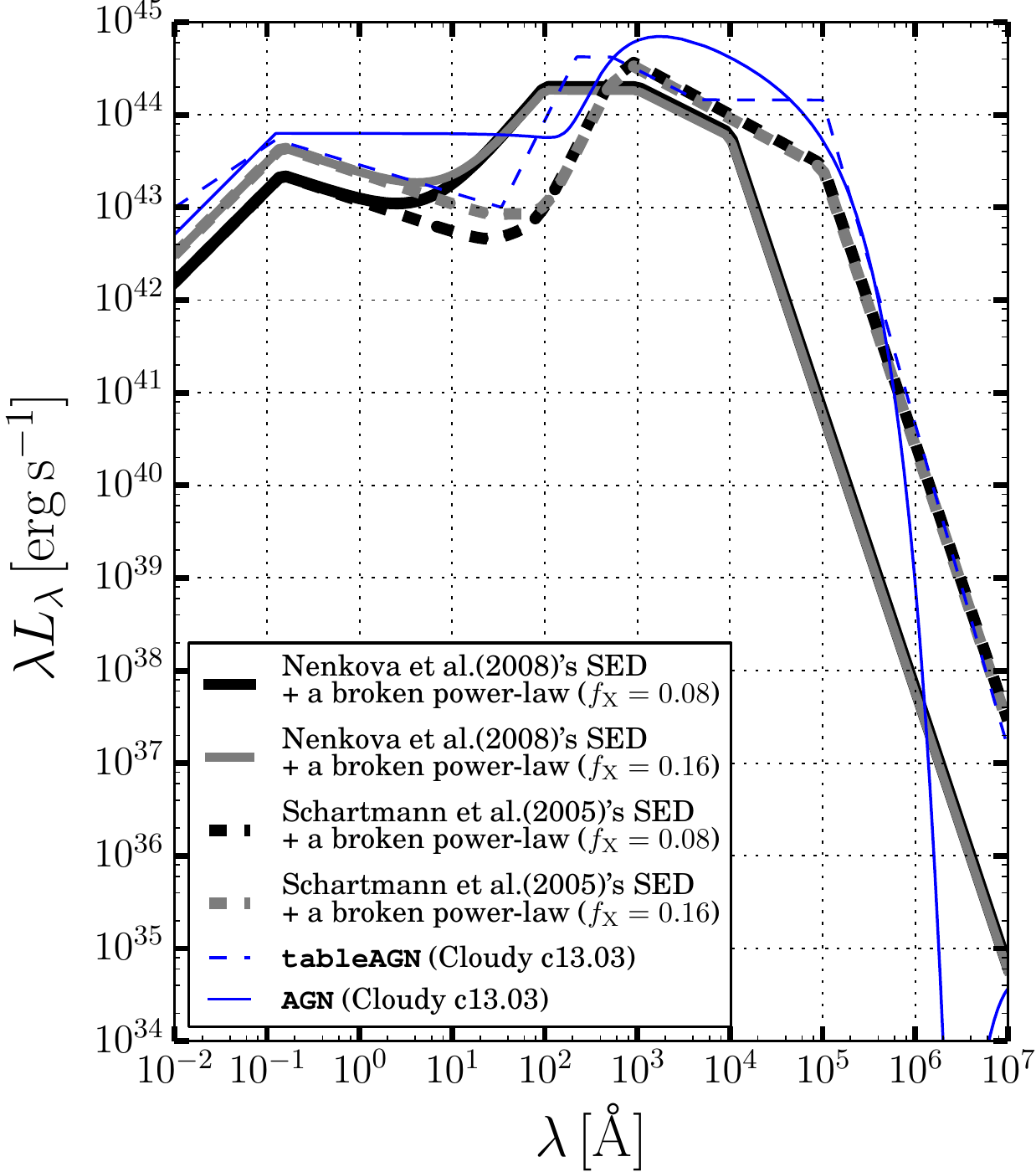}
\caption{AGN SED models for $\Lbol=10^{45}\;\mathrm{erg\;s^{-1}}$. For comparison, two SED models implemented in the \textsc{Cloudy} (\texttt{tableAGN} and \texttt{AGN}) are also shown by the blue lines. For the parameters of \texttt{AGN}, we assume T =1.5e5 k, a(ox) = -1.4, a(uv)=-0.5, and a(x)=-1.}
\label{fig:AGN_model_SEDs}
\end{figure}

The SED of corona is modeled by a broken power-law (BPL):
\begin{equation}
F^{\mathrm{corona}}_{\lambda} \propto 
\begin{cases}
1, & \lambda \leq \lambda_{\mathrm{b}}, \\
(\lambda_{\mathrm{b}}/\lambda)^{\frac{4}{3}}, & \lambda_{\mathrm{b}} < \lambda \leq \lambda_{\mathrm{h}}, \\
(\lambda_{\mathrm{b}}/\lambda_{\mathrm{h}})^{\frac{4}{3}}\exp\left(-\frac{\lambda - \lambda_{\mathrm{h}}}{\Delta\lambda_{\mathrm{co}}}\right), & \lambda > \lambda_{\mathrm{h}},
\end{cases}
\end{equation}
where $\lambda_{\mathrm{b}}=0.144762\;\Angstrom$, $\lambda_{\mathrm{h}}=30\;\Angstrom$, and $\Delta\lambda_{\mathrm{co}}=30\;\Angstrom$. The parameter values are determined so that the the BPL resembles in shape to \texttt{tableAGN}, which is a SED model for AGN implemented in the \textsc{Cloudy} (\citealt{ferland13:_releas_cloud}). The shapes and properties of the AGN SEDs adopted in this study are shown in Fig.~\ref{fig:AGN_model_SEDs} and Table~\ref{tbl:AGN_model_SED_properties}, respectively. The range of wavelength considered in this study is $\lambda=0.01 \operatorname{-} 10^{7}\Angstrom$, which are denoted by $[\lambda_{\min},\lambda_{\max}]$ (the corresponding frequency range is $[\nu_{\min},\nu_{\max}]$).

\begingroup
\renewcommand{\arraystretch}{1.0}
\begin{table*}
\centering
\begin{minipage}{\hsize}
\caption{The properties of the SED models}
\label{tbl:AGN_model_SED_properties}
\begin{tabular}{@{}llllll@{}}
\hline
SED & $L(>13.6\;\eV)/\Lbol$ & $L(>1\;\keV)/\Lbol$ & $L(2\operatorname{-}10\;\keV)/\Lbol$& $\hnuion$ (keV)$^{\dag}$ & $\Tcomp$ ($10^{7}\;\mathrm{K}$)$^{\ddag}$ \\
\hline
\citet{nenkova08a:_agn_dusty_tori} + BPL ($\fX=0.08$)$^{\clubsuit}$ & $0.6877$ & $0.08597$ & $0.01861$ & $6.659$ & $1.932$ \\
\citet{nenkova08a:_agn_dusty_tori} + BPL ($\fX=0.16$) & $0.7148$ & $0.15716$ & $0.03172$ & $13.18$ & $3.823$ \\
\citet{schartmann05:_towar_activ_galac_nuclei} + BPL ($\fX=0.08$) & $0.3674$ & $0.07244$ & $0.01357$ & $6.543$ & $1.898$ \\
\citet{schartmann05:_towar_activ_galac_nuclei} + BPL ($\fX=0.16$) & $0.4224$ & $0.14481$ & $0.02712$ & $13.07$ & $3.793$ \\
\hline
\end{tabular}
\begin{flushleft}
$^{\dag}$ Mean-energy of ionizing photon $\hnuion$, which is defined as $h\int^{\nu_{\max}}_{\nu_{\mathrm{L}}}\nu' L_{\nu'}\diff \nu'/\int^{\nu_{\max}}_{\nu_{\mathrm{L}}} L_{\nu'} \diff \nu'$, where $\nu_{\mathrm{L}}$ is the frequency at the Lyman limit. \\
$^{\ddag}$ Compton temperature $\Tcomp$, which is defined as $\langle h\nu \rangle/4\kBoltz$, where $\langle h\nu \rangle = h \int^{\nu_{\max}}_{\nu_{\min}}\nu' L_{\nu'} \diff \nu'/\Lbol$. \\
$^{\clubsuit}$ The fiducial SED model in this study.
\end{flushleft}
\end{minipage}
\end{table*}
\endgroup

\subsection{ISM and dust model} \label{subsec:ISM_dust_model}
We assume that an ISM consists of chemical species of $\electron$, $\proton$, $\HI$, $\Hminus$, $\Hmol$, $\Hmolplus$, $\HeI$, $\HeII$, $\HeIII$, dust, and metals and assume the tight dynamical coupling between gas and dust. We solve chemical reactions for all the species except for metal, whose effects are taken into account only through a cooling function (described later). The chemical reactions and radiative and thermal processes adopted in this study are taken from previous studies (e.g., \citealt{shapiro87:_hydrog_molec_radiat_coolin_pregal_shock,abel97:_model,glover03:_radiat_x,yoshida06:_format_of_primor_stars_in,glover08:_uncer_hd}) and are summarized in Tables~\ref{tbl:chemical_reactions} and \ref{tbl:radiative_processes}, respectively. The reaction coefficients for a few reactions are updated (see Appendix~\ref{appendix:sec:chemical_reactions}).

As in \citet{namekata14:_agn}, we compute the internal energy density of gas, $\ethermal$, by the equation
\begin{equation}
\ethermal = \frac{\pth}{\gameff -1},
\end{equation}
where $\pth$ is the thermal pressure and $\gameff$ is the effective specific heat ratio of gas, which is defined as
\begin{equation}
\dfrac{1}{\gameff -1} \equiv \sum_{k}\frac{X_{k}}{\gamma_{k}-1},
\end{equation}
where $X_{k}$ and $\gamma_{k}$ are the number fraction and the specific heat ratio of species $k$, respectively. We assume $\gamma_{k}=5/3$ except for molecular hydrogen for which we use the formula given by \citet{landau80:_statis_physic} (see also \citealt{yoshida06:_format_of_primor_stars_in}).

We assume that dust consists of graphite whose mass density is $2.26\;\mden$ (\citealt{laor93:_spect_const_on_the_proper}), because SED modelings of AGNs suggest the presence of hot graphite dust near the dust sublimation radius (\citealt{mor12:_hot}). We do not consider size distribution and assume that the grain radius is $\agr=0.1\;\micron$ for most of our models. The dust-to-gas mass ratio is assumed to be $0.01$. The optical constants such as the absorption efficiency $\Qabs$ and the scattering efficiency $\Qsca$ are calculated by the photoionization code \textsc{Cloudy} (version C13.03, last described by \citealt{ferland13:_releas_cloud}) and are essentially the same as those in \citet{laor93:_spect_const_on_the_proper}. The temperature of dust grain, $\Tgr$, is determined assuming thermal equilibrium:
\begin{eqnarray}
&& \int^{\nu_{\max}}_{\nu_{\min}}\diff\nu\; \dfrac{L_{\nu}(\theta)}{4\pi r^{2}}\exp(-\tau_{\nu})\Qabs \pi \agr^{2} \ngr \nonumber\\
&& +\int^{\nu_{\max}}_{\nu_{\min}}\diff\nu\; \int^{4\pi}_{0}\diff\Omega\; I^{\mathrm{IR}}_{\nu}(\Omega) \Qabs \pi \agr^{2} \ngr \nonumber\\
&& -\int^{\nu_{\max}}_{\nu_{\min}}\diff\nu\; 4\pi \agr^{2} \pi B_{\nu}(\Tgr) \Qabs \ngr - \Lambda_{\mathrm{g\operatorname{-}gr}} = 0, \label{eq:EQ_determining_Tgr}
\end{eqnarray}
where $r$ is the distance from AGN, $L_{\nu}(\theta)$ is the monochromatic luminosity of AGN, $\tau_{\nu}$ is the optical depth, $\ngr$ is the number density of grain, $\Omega$ is the solid angle, $I^{\mathrm{IR}}_{\nu}(\Omega)$ is the intensity of IR photons, $B_{\nu}$ is the Planck function, and $\Lambda_{\mathrm{g\operatorname{-}gr}}$ is the rate of energy transfer due to gas-dust collision (\citealt{burke83:_the_gas_grain_inter_in}; we assume the average accommodation coefficient $\overline{\alpha}_{T}=0.4$ [see \citealt{namekata14:_agn}]). In this study, we assume that the sublimation temperature of dust grain is $\Tsub=1800\;\mathrm{K}$. Sublimation and solidification of dust are treated as follows:
\begin{itemize}[leftmargin=*]
\item dust grain evaporates instantaneously if $\Tgr>\Tsub$.
\item the vapor of dust grain becomes solidified instantly if (i) $\Tgas < \Tsub$ and (ii) $\Tgr$ is smaller than $\Tsub$ after the solidification.
\end{itemize}
The destruction of dust grains due to thermal sputtering is considered in model \texttt{gra01\_wo\_sca\_Tts10\textasciicircum 6K\_SG} only.

\begin{figure}
\centering
\includegraphics[clip,width=\linewidth]{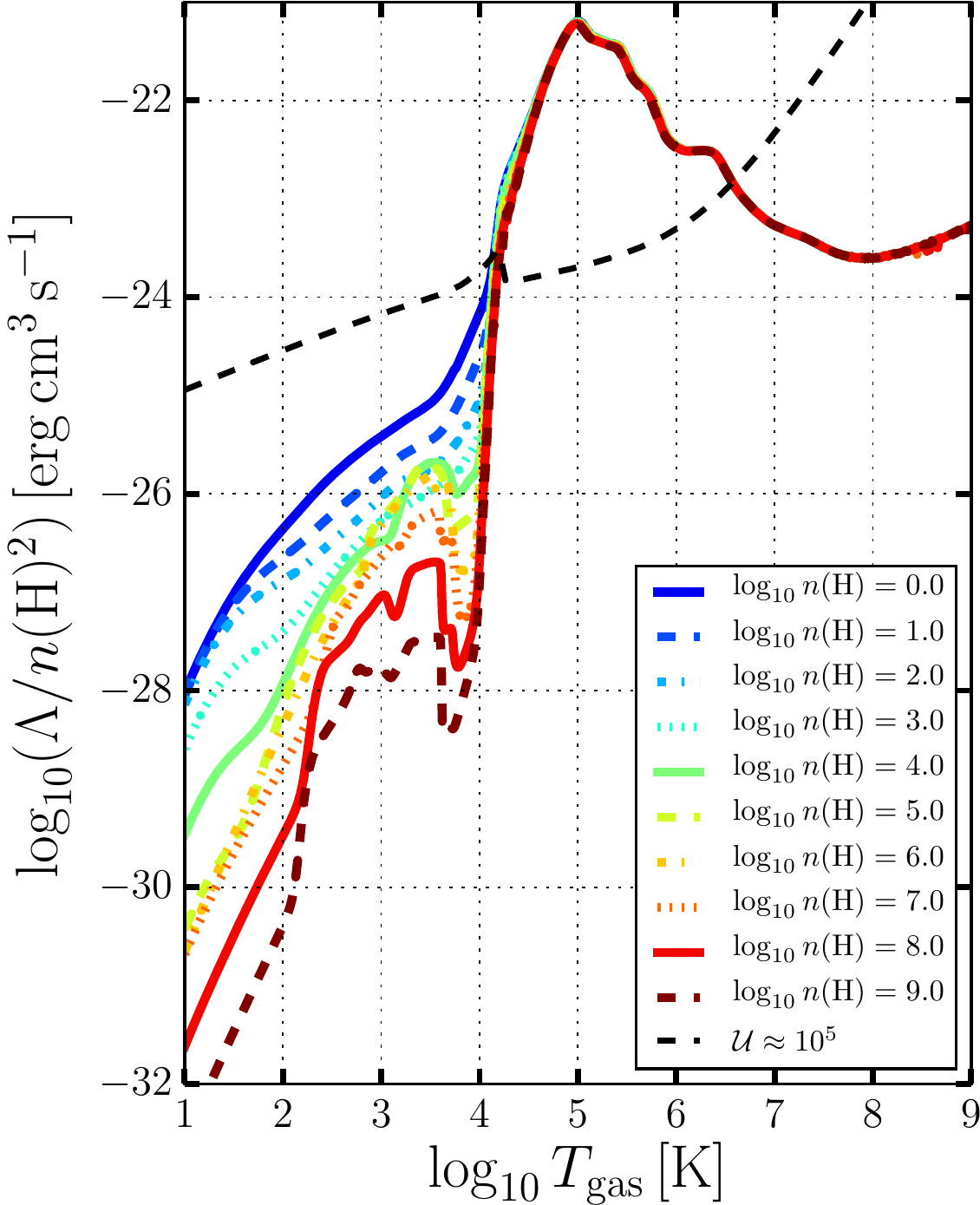}
\caption{Cooling function computed by the \textsc{Cloudy} (version C13.03, \citealt{ferland13:_releas_cloud}) for $Z=1\;\Zsolar$ (colored lines). In the calculation of the cooling function, we assume (i) elemental abundance similar to the \textsc{Cloudy}'s abundance set \texttt{ISM}, (ii) no dust grains, and (iii) no photoionization are assumed. At $\Tgas<10^{4}\;\mathrm{K}$, the cooling rate depends on hydrogen number density since most of cooling processes are spontaneous emission. The black dashed line shows an example of actual cooling curve for photoionized gas ($\nH\approx 1\;\nden$ and the ionization parameter $\mathcal{U}\approx 10^{5}$). When calculating it, we assume the equilibrium abundance at a given gas temperature and we do not take into account the destruction of dust grains due to thermal sputtering. The enhancement of the cooling rate at $\Tgas>10^{6}\;\mathrm{K}$ is due to Compton cooling and dust cooling.}
\label{fig:cooling_function}
\end{figure}

The effects of metals are taken into account in the form of cooling function, because it is numerically too costly to perform long-term RHD simulations with taking all of important chemical reactions relevant to metals into consideration even for a modern super computer. The cooling function is calculated by the \textsc{Cloudy} assuming the metallicity of ISM is solar. Figure~\ref{fig:cooling_function} shows the resultant cooling function. We approximate metal cooling, $\LambdaMTL$, by this cooling function, although the simultaneous use of $\LambdaMTL$ and the radiative processes listed in Table~\ref{tbl:radiative_processes} results in double counts of some of cooling processes such as $\Hmol$ cooling. This approximation is justified by the following additional measures:
\begin{enumerate}[leftmargin=*]
\item The cooling function is applied only for low temperature ($\Tgas\lesssim 10^{4}\;\mathrm{K}$) gas where gas density is generally high and the ionization parameter is low (see \S~\ref{sec:numerical_results}). In such regions, metal cooling (e.g., [\ion{O}{i}]$63\micron$, [\ion{C}{ii}]$157\micron$, emission lines from CO and $\mathrm{H_{2}O}$ molecules) is expected to dominate other coolings due to hydrogen and helium. Hence, the above approximation can be acceptable as a zeroth order approximation. To restrict the metal cooling to low temperature gas, $\LambdaMTL$ is computed by
\begin{equation}
\LambdaMTL(\nH,\Tgas) = \Lambda(\nH,\Tgas)\exp\left[-\left(\dfrac{\Tgas}{15000}\right)^{8}\right],
\end{equation}
where $\Lambda(\nH,\Tgas)$ is the cooling function computed by the \textsc{Cloudy}. As noted above, the use of $\LambdaMTL$ causes the double counts of some cooling processes. This property is, however, useful to place a \textit{lower limit} on the scale height of gas disk.
\item For regions with $\Tgas \gtrsim 10^{4}\;\mathrm{K}$, metal cooling is not applied. This is justified because high temperature is realized in the outflow region (see \S~\ref{sec:numerical_results}) where the ionization parameter is extremely high and recombination cooling by metal will be suppressed due to overionization\footnote{\citet{thoul96:_hydrod} have shown that a strong UV background radiation smooths out the peaks of the cooling function at $\Tgas=10^{4}\operatorname{-}10^{5}\;\mathrm{K}$, whose origin are recombination coolings due to hydrogen and helium (see their Fig.1). Similarly, it is expected that recombination coolings due to metals disappear in regions with large ionization parameters.}. Actually, a photoionization calculation by the \textsc{Cloudy} predicts that the gas temperature in the outflow region is completely determined by the balance between Compton heating and Compton cooling and recombination coolings due to metals are negligible.
\end{enumerate}

\begin{figure*}
\centering
\includegraphics[clip,width=0.666\linewidth]{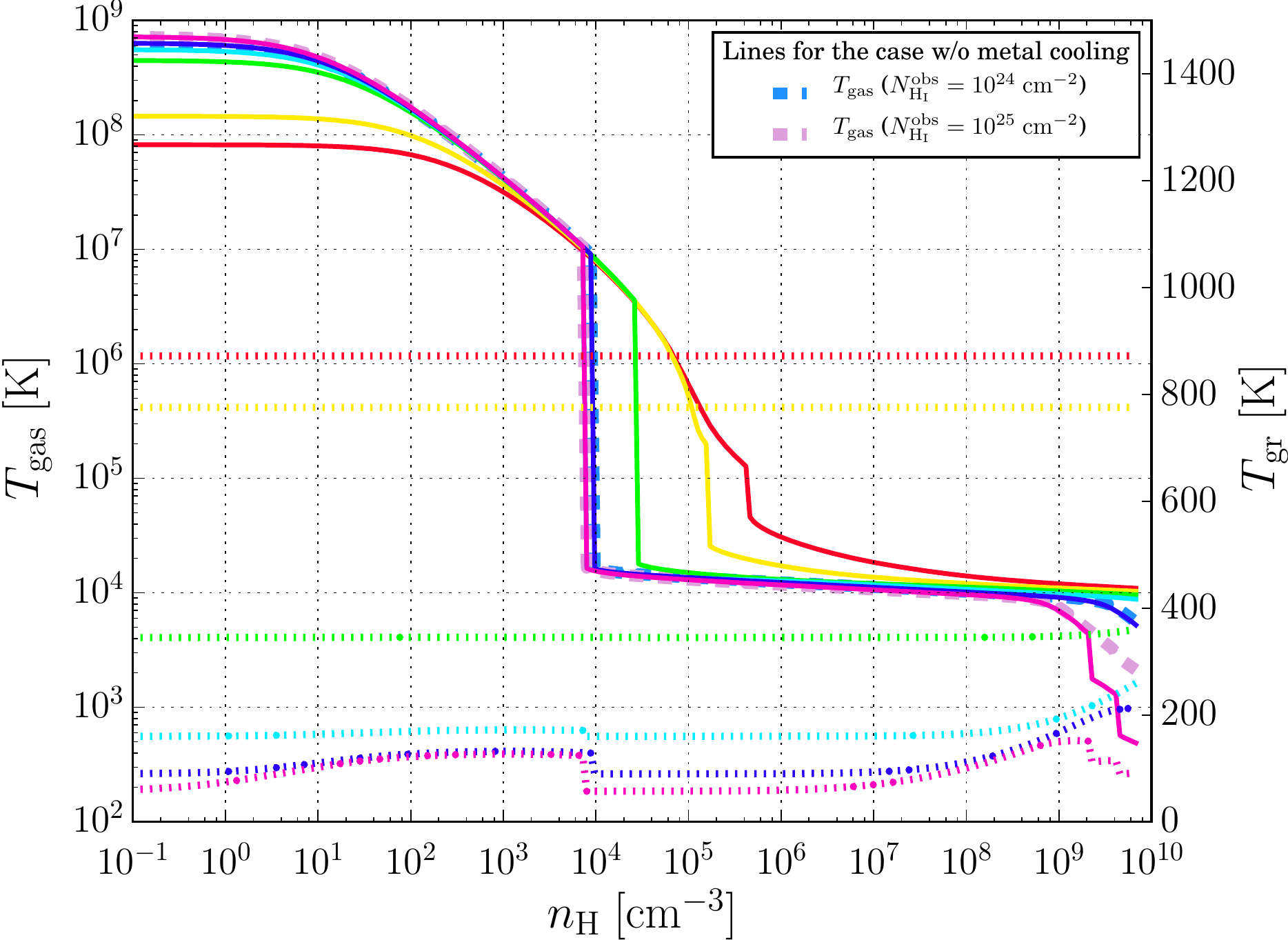}
\caption{Equilibrium temperatures of gas (solid, the left-$y$ axis, logarithmic scale) and dust (dotted, the right-$y$ axis, linear scale) for ISM irradiated by an AGN are plotted as a function of $\nHeq$ for various $N^{\mathrm{obs}}_{\ion{H}{i}}$, where $\nHeq \equiv \rho/m_{\mathrm{H}}$ and $N^{\mathrm{obs}}_{\ion{H}{i}}$ is the column density of obscuring $\ion{H}{i}$ gas. This obscuring gas contains dust with the dust-to-gas mass ratio $0.01$ and is used only to attenuate the incident radiation field. The radiation flux incident on the ISM is computed assuming (i) the fiducial AGN SED (see Fig.~\ref{fig:AGN_model_SEDs} and Table~\ref{tbl:AGN_model_SED_properties}), (ii) $\Lbol=5\times 10^{44}\;\mathrm{erg\;s^{-1}}$, (iii) $r=0.5\;\pc$, and (iv) that the AGN radiates isotropically at all wavelengths. The figure shows the case with metal cooling ($Z=1\;\Zsolar$). For comparison, we also plot two curves of equilibrium temperature of gas for the case without metal cooling (dash-dotted lines). The line colors, in the order of red to magenta, correspond to $N^{\mathrm{obs}}_{\ion{H}{i}}=10^{20}$, $10^{21}$, $10^{22}$, $10^{23}$, $10^{24}$, and $10^{25}\;\cden$. In the case shown here, the gas temperature in low density gas ($\nHeq<10^{2}\;\nden$) is determined by the balance between Compton heating and Compton cooling. Due to spectral hardening, the Compton temperature is larger for higher $N^{\mathrm{obs}}_{\ion{H}{i}}$. In the case without metal cooling, the gas temperature of heavily-obscured gas ($N^{\mathrm{obs}}_{\ion{H}{i}}=10^{25}\;\cden$) starts to go down for $\nHeq>10^{9}\;\nden$ since dust cooling becomes efficient in high density regime ($98\%$ of the gas cooling rate is due to the dust cooling at $\nHeq=10^{9}\;\nden$). $\Hmol$ fraction increases with this sharp decline of gas temperature and gas becomes fully molecular at $\nHeq=10^{9.8}\;\nden$. At this point, the fractions of dust cooling and $\Hmol$ cooling are $90\%$ and $10\%$, respectively. The metal cooling reinforces further declines in the gas temperature of heavily-obscured gas (see $\nHeq>10^{9.5}\;\nden$). The difference between with and without metal cooling is much clearer in lower luminosity cases, although not shown here. As for the dust temperature, we find a large jump between $N^{\mathrm{obs}}_{\ion{H}{i}}=10^{21}\;\cden$ and $10^{22}\;\cden$. This is because the photoheating rate of dust decreases rapidly for $N^{\mathrm{obs}}_{\ion{H}{i}}\gtrsim 10^{22}\;\cden$ ($N^{\mathrm{obs}}_{\ion{H}{i}}=10^{22}\;\cden$ roughly corresponds to the optical depth of unity). In this obscured regime, the dust temperature is mainly determined by collisional energy transfer between gas and dust. Hence, the dust temperature is basically larger for higher $\nHeq$. On the other hand, the dust temperature in unobscured regime is principally determined by the balance between photoheating and thermal emission. We note that the equilibrium temperatures depend somewhat on initial condition (i.e., starting point from which we seek a equilibrium point) and the above results are obtained with the initial temperatures: $\Tgasini =30000\;\mathrm{K}$ and $\Tgrini=10\;\mathrm{K}$.}
\label{fig:equilibrium_temperature_curve}
\end{figure*}

For the ISM model described above, we compute the equilibrium temperatures of gas and dust at a distance of $r=0.5\;\pc$ from an AGN with $\Lbol=5\times 10^{44}\;\mathrm{erg\;s^{-1}}$ and the result is shown in Fig.~\ref{fig:equilibrium_temperature_curve}. It is clear from the figure that ISM can become \textit{four-phases}: very hot ($\Tgas\approx \Tcomp$), hot ($\Tgas\approx (1\operatorname{-}2)\times 10^{4}\;\mathrm{K}$), warm ($10^{3}\;\mathrm{K} < \Tgas < 10^{4}\;\mathrm{K}$), and cold ($\Tgas < 10^{3}\;\mathrm{K}$) media.

\begin{table}
\centering
\begin{minipage}{\hsize}
\caption{Radiative and thermal processes}
\label{tbl:radiative_processes}
\begin{tabular}{@{}p{6cm}l@{}}
\hline
Process & References \\
\hline
\ding{110} \textbf{Cooling} & \\
Collisional ionization cooling: & \\
\qquad $\HI$    & (5),(7) \\
\qquad $\HeI$   & (5),(7) \\
\qquad $\HeII$  & (5),(7) \\
\qquad $\HeIMS$ & (5),(7)$^{\dag}$ \\
Recombination cooling: & \\
\qquad $\proton$ & (8) \\
\qquad $\HeII$   & (8) \\
\qquad $\HeIII$  & (8) \\
Dielectric recombination cooling: & \\
\qquad $\HeII$   & (1),(8) \\
Collisional excitation cooling: & \\
\qquad $\HI$ (all $n$)             & (5),(7) \\
\qquad $\HeII$ ($n=2$)             & (5),(7) \\
\qquad $\HeI$ ($n=2,3,4$ triplets) & (5),(7)$^{\dag}$ \\
Bremsstrahlung cooling & \\
\qquad all ions & (4) \\
Rovibrational line cooling: & \\
\qquad $\Hmol$  & (2),(9) \\
Compton cooling by direct photons from AGN & \S~\ref{subsec:radtr_chemistry} \\
Metal cooling  & \S~\ref{subsec:ISM_dust_model} \\
Other cooling: & \\
\qquad collisional dissociation of $\Hmol$ (R3,R14,R15,R37) & (4) \\
\qquad collisional electron detachment of $\Hminus$ (R7,R8) & (4) \\
\qquad $\Hminus$ formation (R5) & (4) \\
\qquad $\Hmolplus$ formation (R23) & (4) \smallskip\\
\hdashline[0.4pt/1pt]
\ding{110} \textbf{Heating} & \\
Heating by $\Hmol$ formation & \\
\qquad On grain surface & (2) \\
\qquad In gas-phase (R6,R17,R18,R24,R43)  & (2),(4) \\
Photoionization/photodissociation heating & \\
\qquad all species other than $\electron$, $\proton$, and dust & Table~\ref{tbl:chemical_reactions}, \S~\ref{subsec:radtr_chemistry} \\
Compton heating by direct photons from AGN & \S~\ref{subsec:radtr_chemistry} \smallskip\\
\hdashline[0.4pt/1pt]
\ding{110} \textbf{Others} & \\
Collisional gas-grain energy transfer & (4) \\
Compton cooling/heating with the cosmic wave background (CMB) photons & (6) \\
\hline
\end{tabular}
\end{minipage}
\begin{flushleft}
{\small REFERENCES.}---
(1) \citet{aldrovandi73:_radiat_dielec_recom_coeff_compl_ions};
(2) \citet{hollenbach79:_molec_format_and_infrar_emiss};
(3) \citet{burke83:_the_gas_grain_inter_in};
(4) \citet{shapiro87:_hydrog_molec_radiat_coolin_pregal_shock}
(5) \citet{cen92:_hydrod_approac_to_cosmol};
(6) \citet{peebles93:_princ_physic_cosmol}
(7) \citet{fukugita94:_reion};
(8) \citet{hui97:_equat};
(9) \citet{galli98:_univer};. \\
$^{\dag}$ We multiply the original formulae by $\exp\left[-\left(\frac{2\times 10^{4}\;\mathrm{K}}{\Tgas}\right)^{8}\right]$ to turn off these cooling at low temperature regime. This prescription is introduced to prevent a spurious phenomenon: in a strongly irradiated high density ($>10^{7}\;\nden$) gas, cooling rates due to these processes become extremely large since the term $n(\electron)^{2} n(\HeII)$ in the formulae becomes very large and, as a result, the gas temperature becomes unrealistically low in such regions, which is in disagreement with a prediction obtained by using the photoionization code \textsc{Cloudy} (version C13.03, last described by \citealt{ferland13:_releas_cloud}).
\end{flushleft}
\end{table}

\section{Numerical Methods} \label{sec:numerical_methods}
We numerically solve the following RHD equations:
\begin{eqnarray}
&& \frac{\partial \rho}{\partial t} + \nabla\cdot(\rho\bmath{v}) = 0, \label{eq:EOC} \\
&& \frac{\partial \rho^{(k)}}{\partial t} + \nabla\cdot(\rho^{(k)}\bmath{v}) = 0, \label{eq:EOC_species}
\end{eqnarray}
\begin{eqnarray}
&& \frac{\partial \rho v_{R}}{\partial t} + \nabla\cdot (\rho v_{R}\bmath{v}) = -\frac{\partial \pth}{\partial R} + \frac{\rho v^{2}_{\phi}}{R} -\rho\frac{\partial \Phi}{\partial R} + \rho a^{\mathrm{rad}}_{R}, \label{eq:EOM_R} \\
&& \frac{\partial L_{z}}{\partial t} + \nabla\cdot(L_{z}\bmath{v}) = 0, \label{eq:EOAMz} \\
&& \frac{\partial \rho v_{z}}{\partial t} + \nabla\cdot (\rho v_{z}\bmath{v}) = -\frac{\partial \pth}{\partial z} -\rho\frac{\partial \Phi}{\partial z} + \rho a^{\mathrm{rad}}_{z}, \label{eq:EOM_z}
\end{eqnarray}
\begin{eqnarray}
&& \frac{\partial E_{g}}{\partial t} + \nabla\cdot \left\{(E_{g}+p)\bmath{v}\right\} = - \rho\bmath{v} \cdot \nabla\Phi + \rho\bmath{v}\cdot\bmath{a}^{\mathrm{rad}} + \Gamma - \Lambda, \label{eq:EE} \\
&& \pth = \frac{\rho \kBoltz \Tgas}{\mu \MHydro}, \label{eq:EOS}
\end{eqnarray}
\begin{eqnarray}
&& \Phi = \Phi_{\mathrm{ext}} + \Phi_{\mathrm{sg}} \\
&& \Delta \Phi_{\mathrm{sg}} = 4\pi G \rho, \label{eq:Poisson_Eq}
\end{eqnarray}
\begin{eqnarray}
&& \frac{1}{c}\frac{\partial I_{\nu}}{\partial t} + \nabla\cdot (I_{\nu}\bmath{n}) = -\chi_{\nu} I_{\nu} + j_{\nu} + S_{\nu}, \label{eq:RTE}
\end{eqnarray}
where $\rho$ is the gas density, $\rho^{(k)}$ is the gas density of species $k$, $\bmath{v}=(v_{R},v_{\phi},v_{z})$ the velocity, $L_{z}\equiv \rho v_{\phi} R$ the $z$-component of the angular momentum, $E_{g}=\frac{1}{2}\rho\bmath{v}^{2}+\ethermal$ the total energy density, $\bmath{a}^{\mathrm{rad}}=(a^{\mathrm{rad}}_{R},0,a^{\mathrm{rad}}_{z})$ the radiative acceleration, $\Gamma$ the heating function, $\Lambda$ the cooling function, $\Tgas$ the gas temperature, $\mu$ the mean molecular weight relative to the mass of hydrogen atom, $\kBoltz$ the Boltzmann constant, $\Phi_{\mathrm{ext}}$ ($\Phi_{\mathrm{sg}}$) the potential of external (self-) gravity, $G$ the gravitational constant, $I_{\nu}$ the intensity of IR photons, $\chi_{\nu}$ the extinction coefficient, $j_{\nu}$ the emissivity, $S_{\nu}$ the scattering term.

The system of equations (\ref{eq:EOC}-\ref{eq:RTE}) are integrated with operator splitting technique. The original system is divided into four parts: \fbox{i} hydrodynamics-gravity part, \fbox{ii} RT part, \fbox{iii} non-equilibrium chemistry part, and \fbox{iv} radiation force part. In each step, we first solve the hydrodynamics-gravity part \fbox{i} using the method described in \S\ref{subsec:hydro_gravity}. Next, we evaluate the photoionization and photoheating rates by performing a RT calculation using the previously-updated density and temperature distributions (part \fbox{ii}). After that, we calculate chemical reactions using these rates (part \fbox{iii}). Finally, the momentum and kinetic energy are updated by the radiative acceleration obtained by performing a RT calculation again (part \fbox{iv}). Note that, in \citet{namekata14:_agn}, the parts \fbox{ii} and \fbox{iii} are iterated until relative errors of physical variables satisfy a specified convergence criterion, while we do not perform such an iteration in this study because numerical results do not depend on the presence or absence of the iteration (this is probably because the global timestep of our simulations is sufficiently small [$\Delta t\approx 0.01\operatorname{-}0.02\;\yr$]). In the following, we describe the details of numerical methods used for each part.

\subsection{Hydrodynamics and self-gravity} \label{subsec:hydro_gravity}
We solve the hydrodynamic part \fbox{i} of the governing equations using the M-AUSMPW$^{+}$ scheme (\citealt{kim05:_accur_effic_and_monot_numer_1}) and the multidimensional limiting process with fifth order interpolation (MLP5, \citealt{kim05:_accur_effic_and_monot_numer_2}). The M-AUSMPW$^{+}$ scheme, as the name implies, is one of the advection upstream splitting method (AUSM) type schemes (\citealt{liou93:_new_flux_split_schem}) and is more robust and less dissipative in multi-dimensional flows than the previous AUSM type schemes such as AUSM$^{+}$ (\citealt{liou96:_sequel_to_ausm}) and AUSMPW$^{+}$ (\citealt{kim01:_method_for_accur_comput_of_1}), owing to (i) discontinuity detection using Mach number and pressure and (ii) addition of correction step of the primitive variables at the cell interfaces based on the smoothness of the flow. In this study, we do not employ the modification of the pressure splitting function suggested by \citet{kim05:_accur_effic_and_monot_numer_1} (see \S 2.3.2 in their paper), because it results in negative thermal energy in some cases. Combined with MLP5, the M-AUSMPW$^{+}$ scheme achieves fifth-order accurate in space in smooth regions of the flow. For accurate advection of chemical species, we employ the consistent multifluid advection (CMA) method (\citealt{plewa99})\footnote{When combining the CMA method with the M-AUSMPW$^{+}$ scheme, we need to calculate $\bmath{Y}_{\mathrm{L[R],superbee}}$ which is the chemical abundances at the left and right sides of a cell interface and are obtained by the Monotonic Upstream-Centered Scheme for Conservation Laws (MUSCL) interpolation with the superbee limiter. In a certain situation, abundance of a particular element or all the elements can become zero and $\rho_{\mathrm{L[R],superbee}} = \sum_{k}\rho^{(k)}_{\mathrm{L[R],superbee}}$ does not hold in this case, where $\rho^{(k)}$ is density of species $k$. This happens for a cell interface separating a fully-ionized region and a fully neutral region, for example. In such cases, we simply use the chemical abundance of the nearest cell center as $\bmath{Y}_{\mathrm{L[R],superbee}}$.}. This combination is also adopted and tested in \citet{namekata11:_evolut_nuclear_disk_gas_galac_center}.

The gravitational force and its power are treated as source terms of the hydrodynamic equations. The self-gravity is computed by the tree method (\citealt{barnes86:_nlog_n}) with modification to treat an axisymmetric system (see Appendix \ref{appendix:sec:self_gravity}). The opening angle criterion, $\theta^{\mathrm{grv}}_{\mathrm{crit}}$, is assumed to be 0.35. We take into account the monopole components only and use a quadtree to construct group interaction lists (\citealt{barnes90:_modif_tree_code,makino91:_treec_special_purpos_proces}). 

The code is parallelized with the Message Passing Interface (MPI). Time integration is performed by the second order total variation diminishing (TVD) Runge-Kutta method (\citealt{shu88:_total_variat_dimin_time_discr}).

\subsection{Radiative transfer and non-equilibrium chemistry} \label{subsec:radtr_chemistry}
We solve the RT parts (\fbox{ii},\fbox{iv}) of the governing equations using a hybrid method, in which the RT of direct radiation from AD and corona is solved by the (photon-conservative) long-characteristics method (LCM) while we solve that of IR photons by the finite-volume method (FVM). In the RT calculations, the following assumptions are made:
\begin{enumerate}[label=(A\arabic*), topsep=0pt,leftmargin=*]
\item The speed of light is infinity.
\item Gas is assumed to be at the rest. Hence, the Poynting-Robertson effect and special relativistic effects such as the Doppler-shift and the radiation drag are neglected.
\item We assume that direct photons singly-scattered by electrons escape from the system without any interactions. In other words, we do not consider multiple electron (Compton) scattering of direct photons.
\item We ignore dust scattering of direct photons.
\item We also neglect electron (Compton) scattering of IR photons.
\item We assume that dust scattering of IR photons is isotropic.
\item We do not perform an exact RT calculation of cooling photons arisen by the cooling processes described in \S~\ref{subsec:ISM_dust_model}. Instead, we assume that these cooling photons are optically-thin in most of the runs, but, in the runs with `\texttt{LDA1}`, their RT effects are approximately taken into account by using the LDA approximation (see \S~\ref{subsec:parameters} and \S~\ref{subsec:cooling_photon_effects}).
\item We employ the so-called on-the-spot (OTS) approximation (\citealt{osterbrock06:_astrop_gaseous_nebul_activ_galac_nuclei}) for recombination photons of the reactions R2, R31, and R32 (see Table~\ref{tbl:chemical_reactions}).
\item We use an approximate formula for the photodissociation rate of $\mathrm{H_{2}}$ due to the two-step Solomon process (\citealt{draine96:_struc_of_station_photod_front}).
\end{enumerate}

\subsubsection{RT of direct radiation}
\begin{figure}
\centering
\includegraphics[clip,width=\linewidth]{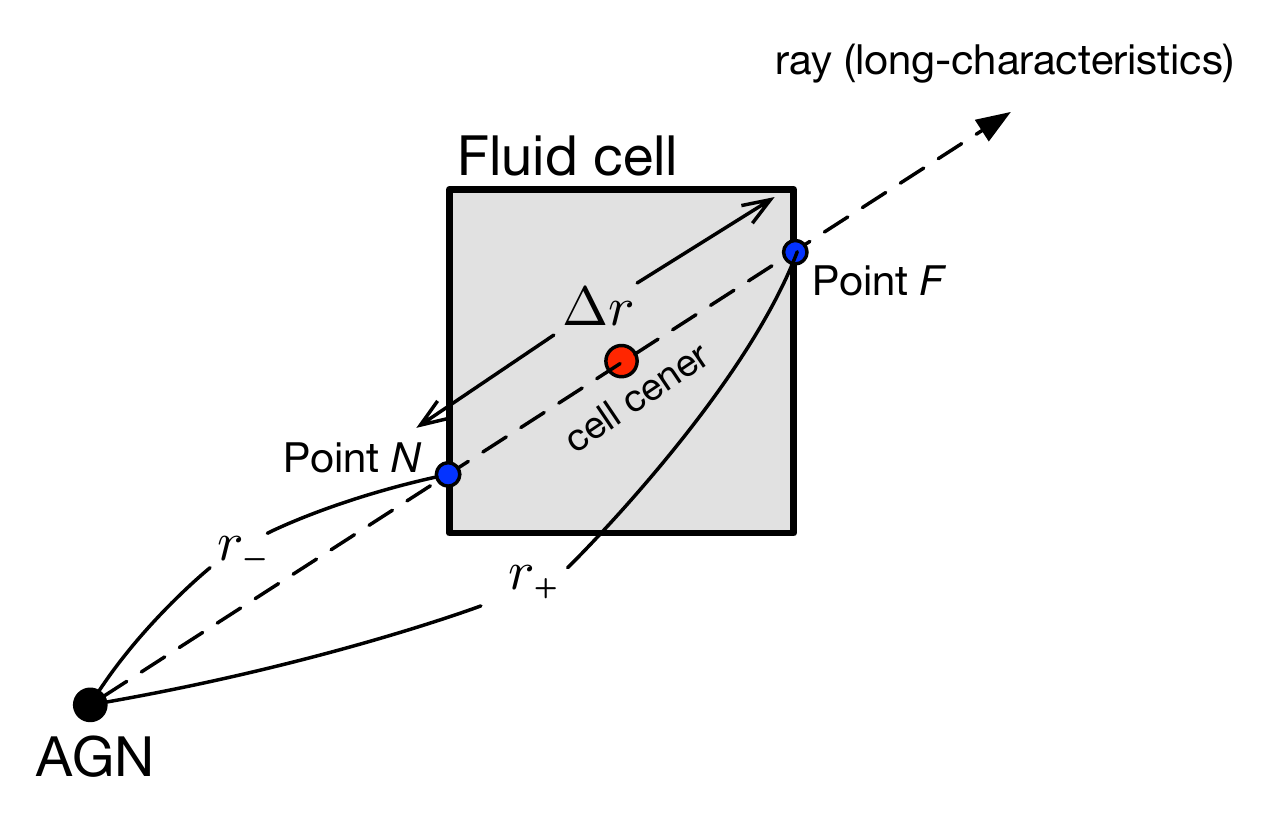}
\caption{A schematic illustration of the photon-conservative LCM.}
\label{fig:LCM}
\end{figure}

The photoionization or photodissociation rate $k^{\mathrm{rad}}_{\mathfrak{r}}$, the photoheating rate $\Gamma^{\mathrm{rad}}_{\mathfrak{r}}$, and the radiative acceleration $\bmath{a}^{\mathrm{rad}}_{\mathfrak{r}}$ due to reaction $\mathfrak{r}$ ($\mathfrak{r}$ = R4, R13, R19, R20, R21, R22, R27, R28, R35, and R36) are computed in a photon-conservative manner (e.g., \citealt{whalen06:_multis_algor_for_the_radiat,susa06:_smoot_partic_hydrod_coupl_radiat_trans}):
\begin{eqnarray}
k^{\mathrm{rad}}_{\mathfrak{r}}(\bmath{r}) & = & n(\mathfrak{r})\widehat{k}^{\mathrm{rad}}_{\mathfrak{r}}(\bmath{r}) = \int^{\nu_{\max}}_{\nu_{\min}} \frac{n(\mathfrak{r})\sigma_{\mathfrak{r}}(\nu)}{(n\sigma)_{\mathrm{tot}}}  \frac{\mathfrak{F}}{h\nu} \diff\nu, \label{eq:photoionization_rates}\\
\Gamma^{\mathrm{rad}}_{\mathfrak{r}}(\bmath{r}) & = & n(\mathfrak{r})\widehat{\Gamma}^{\mathrm{rad}}_{\mathfrak{r}}(\bmath{r}) = \int^{\nu_{\max}}_{\nu_{\min}} \frac{n(\mathfrak{r})\sigma_{\mathfrak{r}}(\nu)}{(n\sigma)_{\mathrm{tot}}}  \frac{\mathfrak{F}}{h\nu} (h\nu-h\nu_{\mathrm{th}}) \diff\nu, \label{eq:photoheating_rates} \\
\bmath{a}^{\mathrm{rad}}_{\mathfrak{r}}(\bmath{r}) & = & \frac{\hat{\bmath{e}}_{r}}{c\rho} \int^{\nu_{\max}}_{\nu_{\min}} \frac{n(\mathfrak{r})\sigma_{\mathfrak{r}}(\nu)}{(n\sigma)_{\mathrm{tot}}} \mathfrak{F}  \diff\nu, \label{eq:radiative_accels}, \\
\mathfrak{F} & = & \frac{L^{\mathrm{AGN}}_{\nu}(\theta)\exp(-\tau_{\nu})[1-\exp(-\Delta\tau_{\nu})]}{\Delta V}
\end{eqnarray}
where $n(\mathfrak{r})$ is the number density of ionized or dissociated species in reaction $\mathfrak{r}$, $\sigma_{\mathfrak{r}}(\nu)$ is the cross section for reaction $\mathfrak{r}$, $(n\sigma)_{\mathrm{tot}}\equiv \sum_{\mathfrak{r}'}n(\mathfrak{r}')\sigma_{\mathfrak{r}'}(\nu)$ ($\mathfrak{r}'$ takes all the photon-matter interactions including Compton scattering and dust absorption), $L^{\mathrm{AGN}}_{\nu}(\theta) = L^{\mathrm{AD}}_{\nu}(\theta) + L^{\mathrm{corona}}_{\nu}$, $\Delta V \equiv \frac{4\pi}{3}(r^{3}_{+}-r^{3}_{-})$ is the differential volume, $r_{\pm}$ are the distances between the AGN and the points of intersection between the ray passing through the center of a target cell and the cell faces (see Fig.~\ref{fig:LCM}), $\Delta r \equiv r_{+}-r_{-}$, $\tau_{\nu}\equiv \int^{r_{-}}_{0} (n\sigma)_{\mathrm{tot}}\diff s$ is the optical depth to the intersection point closest to the AGN, $\Delta \tau_{\nu}=(n\sigma)_{\mathrm{tot}} \Delta r$, and $h\nu_{\mathrm{th}}$ is the threshold energy for reaction $\mathfrak{r}$. The photoheating rate of dust is similarly calculated as
\begin{equation}
\Gamma^{\mathrm{rad}}_{\mathrm{gr}}(\bmath{r}) = \ngr\widehat{\Gamma}^{\mathrm{rad}}_{\mathrm{gr}}(\bmath{r}) = \int^{\nu_{\max}}_{\nu_{\min}} \frac{\ngr \sigma^{\mathrm{abs}}_{\mathrm{gr}}(\nu)}{(n\sigma)_{\mathrm{tot}}}\mathfrak{F} \diff\nu, \label{eq:dust_photoheating_rate}
\end{equation}
where $\sigma^{\mathrm{abs}}_{\mathrm{gr}}$ is the dust absorption cross section. The radiative acceleration due to dust absorption is calculated using the same formula as Eq.(\ref{eq:radiative_accels}). Compton heating/cooling rate and the radiative acceleration due to electron scattering are calculated as
\begin{eqnarray}
\Gamma^{\mathrm{rad}}_{\electron}(\bmath{r}) & = & \int^{\nu_{\max}}_{\nu_{\min}} \frac{\nele \sigma_{\mathrm{T}}}{(n\sigma)_{\mathrm{tot}}}\frac{\mathfrak{F}}{m_{e}c^{2}}(h\nu - 4\kBoltz\Tgas) \diff \nu, \label{eq:Compton_heating_cooling} \\
\bmath{a}^{\mathrm{rad}}_{\electron}(\bmath{r}) & = & \frac{\hat{\bmath{e}}_{r}}{c\rho} \int^{\nu_{\max}}_{\nu_{\min}} \frac{\nele \sigma_{\mathrm{T}}}{(n\sigma)_{\mathrm{tot}}} \mathfrak{F} \diff\nu, \label{eq:radiative_accel_ES}
\end{eqnarray}
where $\sigma_{\mathrm{T}}$ is the Thomson scattering cross section. 

In the part \fbox{ii}, we evaluate $\widehat{k}^{\mathrm{rad}}_{\mathfrak{r}}$, $\widehat{\Gamma}^{\mathrm{rad}}_{\mathfrak{r}}$, $\widehat{\Gamma}^{\mathrm{rad}}_{\mathrm{gr}}$, and the following quantities
\begin{eqnarray}
\widehat{\Gamma}^{\mathrm{rad}}_{\electron,\mathrm{1st}}(\bmath{r}) & = & \int^{\nu_{\max}}_{\nu_{\min}} \frac{\sigma_{\mathrm{T}}}{(n\sigma)_{\mathrm{tot}}}\frac{\mathfrak{F}}{m_{e}c^{2}} h\nu \diff \nu,  \\
\widehat{\Gamma}^{\mathrm{rad}}_{\electron,\mathrm{2nd}}(\bmath{r}) & = & \int^{\nu_{\max}}_{\nu_{\min}} \frac{\sigma_{\mathrm{T}}}{(n\sigma)_{\mathrm{tot}}}\frac{\mathfrak{F}}{m_{e}c^{2}} \diff \nu.
\end{eqnarray}
Then, we use them in the calculation of chemical reactions (i.e., part \fbox{iii}). In order to evaluate the above quantities, we first calculate the column density between the AGN and the point $N$ in Fig.~\ref{fig:LCM} using the LCM to calculate $\tau_{\nu}$. Then, we perform the frequency integrations. In practice, we integrate by wavelength with the trapezoidal rule and we use $\approx 800$ grid points for $\lambda < 912\;\Angstrom$ and $256$ grid points for $\lambda \geq 912\;\Angstrom$ (both grids are non-uniform). In the part \fbox{iv}, we evaluate the radiative accelerations (Eqs.\ref{eq:radiative_accels},\ref{eq:radiative_accel_ES}).

\subsubsection{RT of IR photons}
\begin{figure}
\centering
\includegraphics[clip,width=\linewidth]{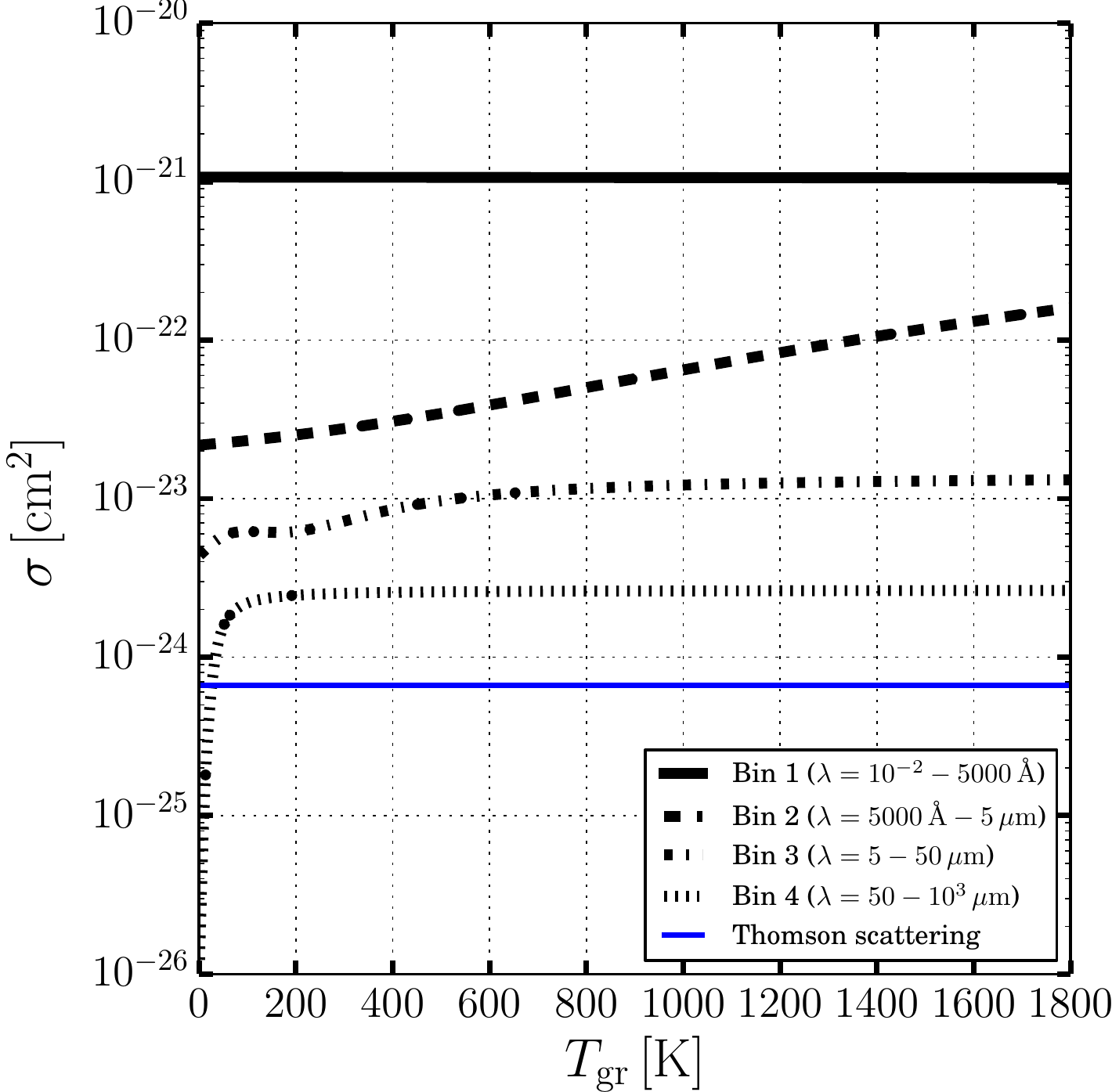}
\caption{Planck-mean absorption cross section per hydrogen nuclei for the dust-to-gas mass ratio $0.01$ and $\agr=0.1\;\micron$. For comparison, the Thomson scattering cross section is also shown by the blue line.}
\label{fig:Planck_mean_cross_section}
\end{figure}

Because solving the frequency-dependent RT equation (\ref{eq:RTE}) directly is numerically expensive, we solve a multi-group RT equation instead:
\begin{equation}
\frac{1}{c}\frac{\partial \overline{I}_{\mathfrak{v}}}{\partial t} + \nabla\cdot (\overline{I}_{\mathfrak{v}}\bmath{n}) = - \overline{\chi}_{P,\mathfrak{v}}\overline{I}_{\mathfrak{v}} + \overline{j}_{\mathfrak{v}} + \overline{S}_{\mathfrak{v}}, \label{eq:multigroup_RTE}
\end{equation}
where $\mathfrak{v}$ denotes a frequency range $\Delta\nu_{\mathfrak{v}}$,
\begin{eqnarray}
\overline{I}_{\mathfrak{v}} & = & \int_{\Delta\nu_{\mathfrak{v}}} I_{\nu} \diff\nu, \\
\overline{j}_{\mathfrak{v}} & = & \int_{\Delta\nu_{\mathfrak{v}}} j_{\nu} \diff\nu = \ngr \overline{\sigma}^{\mathrm{abs}}_{P,\mathfrak{v}}\overline{B}_{\mathfrak{v}}(\Tgr), \\
\overline{S}_{\mathfrak{v}} & = & \int_{\Delta\nu_{\mathfrak{v}}} S_{\nu} \diff\nu \approx \frac{\ngr\overline{\sigma}^{\mathrm{sca}}_{P,\mathfrak{v}}}{4\pi} \int\diff\Omega' \overline{I}_{\mathfrak{v}}(\Omega'), \\
\overline{B}_{\mathfrak{v}}(\Tgr) & = & \int_{\Delta\nu_{\mathfrak{v}}} B_{\nu}(\Tgr) \diff\nu.
\end{eqnarray}
$\overline{\chi}_{P,\mathfrak{v}} \equiv \ngr(\overline{\sigma}^{\mathrm{abs}}_{P,\mathfrak{v}}+\overline{\sigma}^{\mathrm{sca}}_{P,\mathfrak{v}})$ is the Planck-mean extinction coefficient, $\overline{\sigma}^{\mathrm{abs(sca)}}_{P,\mathfrak{v}}$ is calculated as
\begin{equation}
\overline{\sigma}^{\mathrm{abs(sca)}}_{P,\mathfrak{v}}(\Tgr) = \frac{\int_{\Delta\nu_{\mathfrak{v}}}\mathcal{Q}_{\mathrm{abs(sca)}}(\nu,\agr)\pi \agr^{2} B_{\nu}(\Tgr)\diff\nu}{\int_{\Delta\nu_{\mathfrak{v}}}B_{\nu}(\Tgr)\diff\nu}.
\end{equation}
In the above, we used the assumptions (A2), (A5), and (A6). In this study, we divide the entire range of wavelength into four bins (\texttt{bin1}: $10^{-2}\operatorname{-}5000\;\Angstrom$, \texttt{bin2}: $5000\;\Angstrom \operatorname{-} 5\;\micron$, \texttt{bin3}: $5\operatorname{-}50\;\micron$, \texttt{bin4}: $50-10^{3}\;\micron$). Figure~\ref{fig:Planck_mean_cross_section} shows the Planck-mean dust absorption cross section per hydrogen nuclei for the dust-to-gas mass ratio $0.01$ and $\agr=0.1\;\micron$. Equation (\ref{eq:multigroup_RTE}) is solved by the FVM further assuming (A1). The detail of the method is described in Appendix~\ref{appendix:sec:FVM_RT}.

The photoheating rate of dust and the radiative acceleration due to IR photons are calculated as follows
\begin{eqnarray}
\widehat{\Gamma}^{\mathrm{IR}}_{\mathrm{gr}}(\bmath{r}) & = & \sum_{\mathfrak{v}} \int\diff\Omega \overline{I}_{\mathfrak{v}}(\Omega) \overline{\sigma}^{\mathrm{abs}}_{P,\mathfrak{v}}, \\
\bmath{a}^{\mathrm{IR}}_{\mathrm{gr}}(\bmath{r}) & = & \frac{1}{c\rho}\sum_{\mathfrak{v}} \int\diff\bmath{\Omega} \overline{I}_{\mathfrak{v}}(\Omega) \overline{\chi}_{P,\mathfrak{v}}.
\end{eqnarray}
$\widehat{\Gamma}^{\mathrm{IR}}_{\mathrm{gr}}$ and $\bmath{a}^{\mathrm{IR}}_{\mathrm{gr}}$ are used in parts \fbox{iii} and \fbox{iv}, respectively.

\subsubsection{Non-equilibrium chemistry}
The chemical reactions and the equation of internal energy are solved in the same manner as \citet{namekata14:_agn}, in which we use subcycle technique and the $\alpha$-QSS method (\citealt{mott00:_quasi_stead_state_solver_stiff}) is used as an implicit equation solver. For details, please refer to \S~3.3 in \citet{namekata14:_agn}.

\subsection{Global timestep} \label{subsec:timestep}
Global timestep $\Delta t$ is determined as follows
\begin{eqnarray}
\Delta t & = & \min_{ij}\left(\Delta t_{\mathrm{hyd},ij}, \Delta t_{\mathrm{accl},ij}\right), \\
\Delta t_{\mathrm{hyd},ij} & = & C_{\mathrm{hyd}} \min\left(\frac{\Delta R_{i}}{|v_{R,ij}|+c_{s,ij}}, \frac{\Delta z_{j}}{|v_{z,ij}|+c_{s,ij}} \right), \\
\Delta t_{\mathrm{accl},ij} & = & \frac{C_{\mathrm{accl}}\max(|\bmath{v}_{ij}|, c_{s,ij})}{|\bmath{a}^{\mathrm{tot,eff}}_{ij}|},
\end{eqnarray}
where we use $(i,j)$ as an identifier of a fluid cell as in the usual manner, $c_{s,ij}$ is the adiabatic sound speed of cell $(i,j)$, and $\bmath{a}^{\mathrm{tot,eff}}_{ij}$ is the sum of gravitational, centrifugal, and radiative accelerations. In this study, we assume $C_{\mathrm{hyd}}=C_{\mathrm{accl}}=0.15$.

\subsection{Numerical setup} \label{subsec:numerical_setup}
An uniform spatial grid is used to cover a computational box with a radius $1.2\;\pc$ and a height $1.2\;\pc$. In most of models, we use $512^{2}$ grid points (the corresponding spatial resolution is $2.34\times 10^{-3}\;\pc$), but, some models are performed with $1024^{2}$ grid points to check the effects of spatial resolution. As for the discretization of solid angle, we assume $N_{\phi}=N_{\theta}=12$ (see Appendix~\ref{appendix:sec:FVM_RT}).

The initial density distribution of a dusty gas disk is obtained by solving the isothermal hydrostatic equation
\begin{equation}
\frac{\diff \ln\rho}{\diff z} = \frac{a^{\mathrm{grv,ext}}_{z}}{c^{2}_{s}},
\end{equation}
where $a^{\mathrm{grv,ext}}_{z}$ is the $z$ component of the acceleration of external gravity, $c_{s}$ is the adiabatic sound speed of gas. The outer edge of the disk $R_{\mathrm{out}}$ is smoothed by multiplying the density at $R>R_{\mathrm{out}}$ by a factor of $\exp\left[-20\frac{R-R_{\mathrm{out}}}{R_{\mathrm{out}}}\right]$, where we assume $R_{\mathrm{out}}=1\;\pc$ as described in \S~\ref{subsec:parameters}. We assume that the disk initially consists of a neutral atomic gas ($Y(\HI)=0.748125$, $Y(\HeI)=0.251875$, $Y(k)=0$; where $Y(\cdot)$ is mass abundance and $k$ denotes chemical species other than $\HI$ and $\HeI$) and the gas and dust temperatures are uniform in the disk with values of $\Tgas=2000\;\mathrm{K}$ and $\Tgr=100\;\mathrm{K}$. The initial velocity field is $\bmath{v}=\left(0, \sqrt{R|a^{\mathrm{grv,ext}}_{R}|}, 0\right)$, where $a^{\mathrm{grv,ext}}_{R}$ is the $R$ component of the acceleration of external gravity.

All the boundary conditions are fixed boundaries. Around the symmetric axis, we set up a boundary zone where deviations of primitive variables from the initial values and pressure fluctuation are artificially damped. This boundary zone is introduced to prevent negative density or negative internal energy at extremely low density regions that are formed near the symmetric axis due to strong radiation pressure from AGN. For details, see Appendix~\ref{appendix:sec:boundary_zone}.

\section{Numerical Results} \label{sec:numerical_results}
Firstly, in \S~\ref{subsec:overview}, we present the numerical results of the reference models (\texttt{gra01\_wo\_sca\_SG} and \texttt{gra01\_wo\_sca\_MTL\_SG}) and show typical density and temperature structures in regions near the dust sublimation radius and the effects of metal cooling on them. The effects of IR photons, dust scattering of IR photons, self-gravity, and $\fX$ are also examined there. Next, the numerical results of models \texttt{gra01\_wo\_sca\_LDA1\_SG} and \texttt{gra01\_wo\_sca\_MTLLDA1\_SG} are presented in \S~\ref{subsec:cooling_photon_effects} to show the effects of self-absorption of cooling photons. Then, the effects of grain size and dust destruction due to thermal sputtering are examined by using models \texttt{gra1\_wo\_sca} and \texttt{gra1\_wo\_sca\_Tts10\textasciicircum 6\_SG} in \S~\ref{subsec:grain_size_effects} and \S~\ref{subsec:thermal_sputtering_effects}, respectively. Finally, we investigate the roles of Compton heating/cooling using model \texttt{gra01\_wo\_sca\_noCompton\_SG} in \S~\ref{subsec:Compton_heatcool_effects}.

\subsection{Overview of reference models and the effects of metal cooling and IR photons} \label{subsec:overview}
\begin{figure*}
\centering
\includegraphics[clip,width=\linewidth]{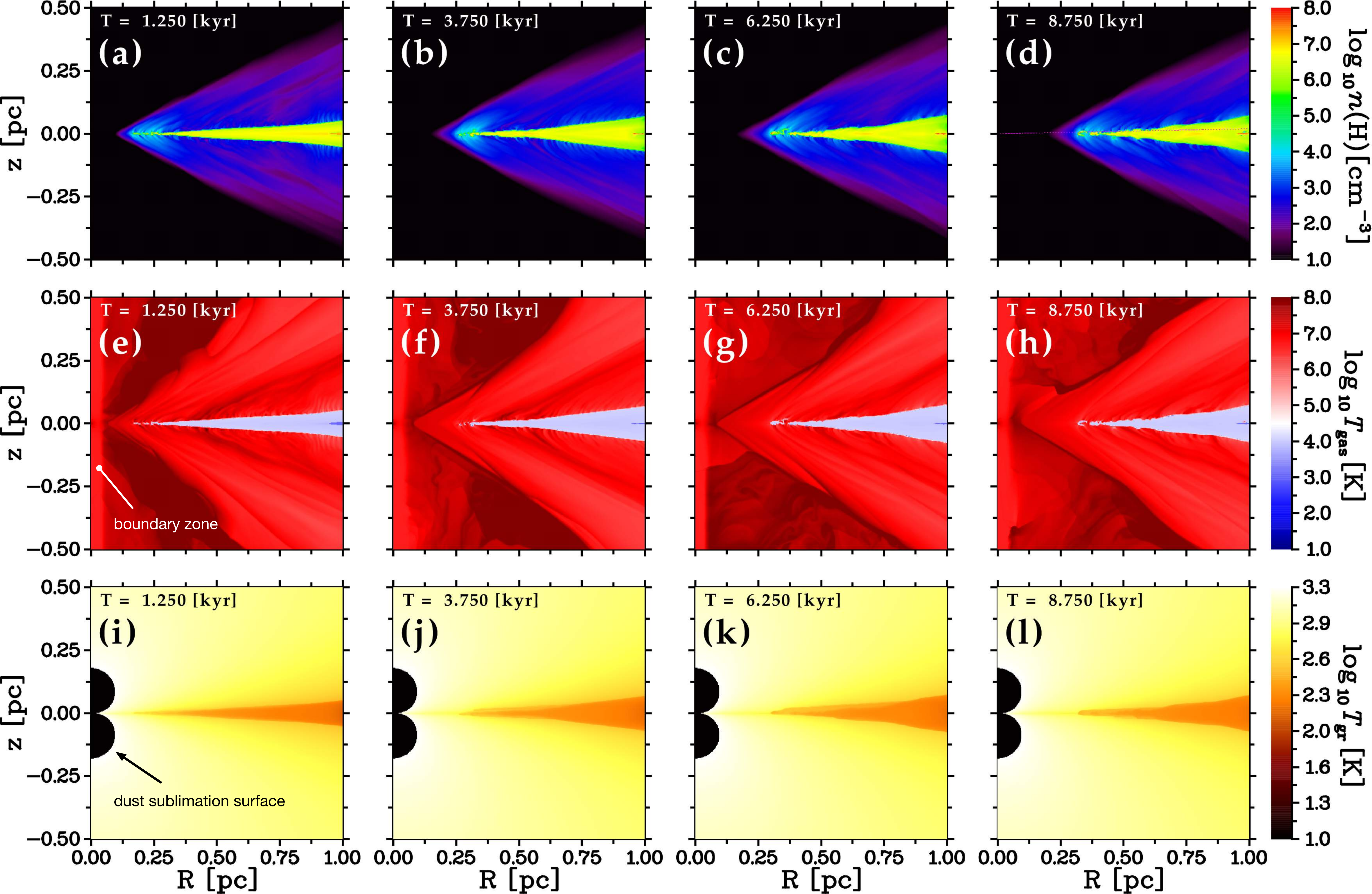}
\caption{Time evolution of the distributions of $\nH$, $\Tgas$, and $\Tgr$ in model \texttt{gra01\_wo\_sca\_SG}, in which metal cooling is switched off. In each panel, we show only a $(1\pc)^{2}$ region around the AGN that is located at the origin. The calculation time is shown at the upper left corner of each panel. The density profile along the magenta dotted line in panel (d) is used in the photoionization calculations in \S~\ref{subsec:uncertainties}.}
\label{fig:dists_gra01_wo_sca_SG}
\end{figure*}

\begin{figure*}
\centering
\includegraphics[clip,width=\linewidth]{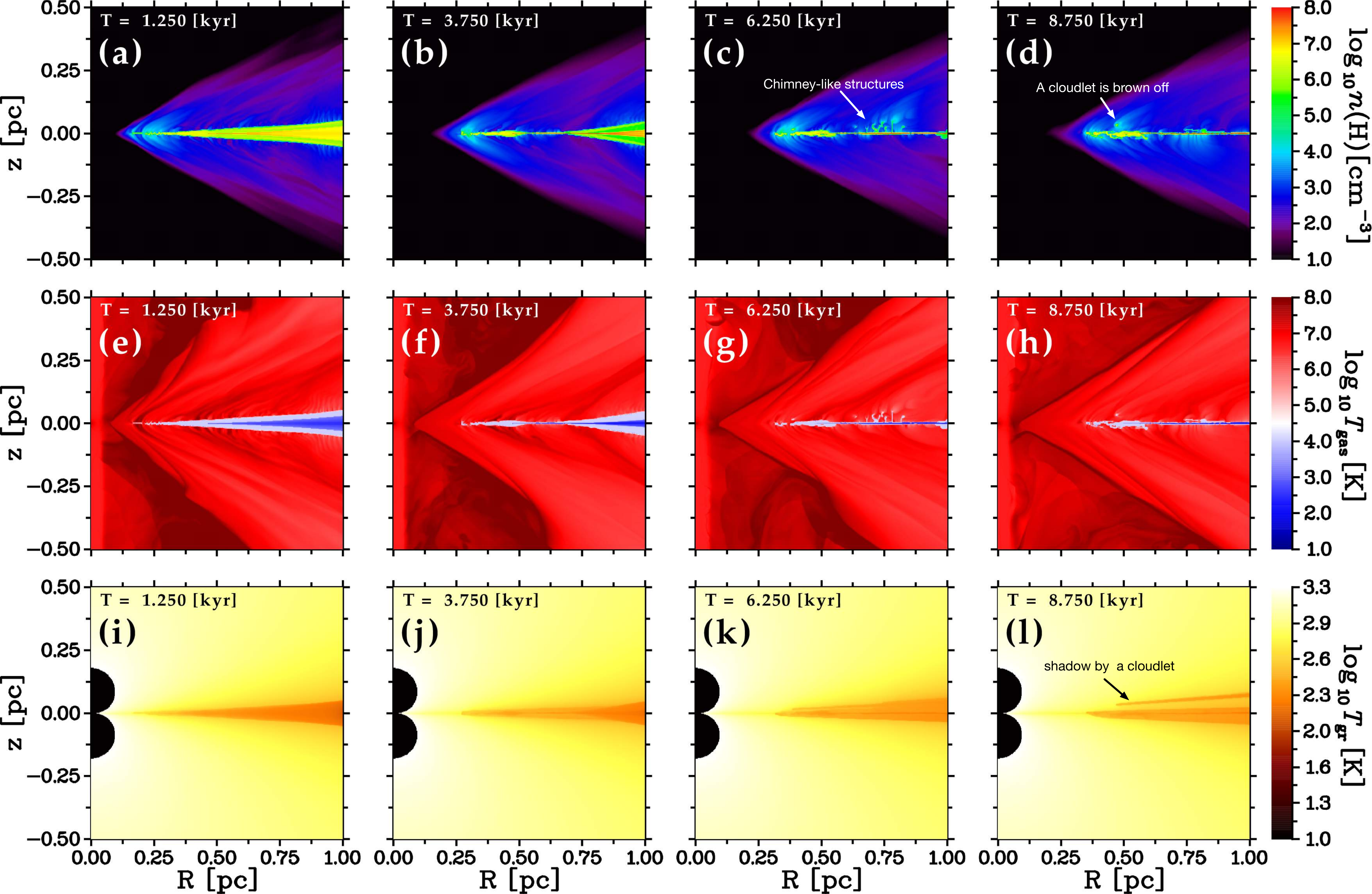}
\caption{The same as Fig.~\ref{fig:dists_gra01_wo_sca_SG}, but for model \texttt{gra01\_wo\_sca\_MTL\_SG}, in which metal cooling is switched on.}
\label{fig:dists_gra01_wo_sca_MTL_SG}
\end{figure*}

\begin{figure*}
\centering
\begin{tabular}{cc}
\begin{minipage}{0.5\hsize}
\includegraphics[clip,width=\linewidth]{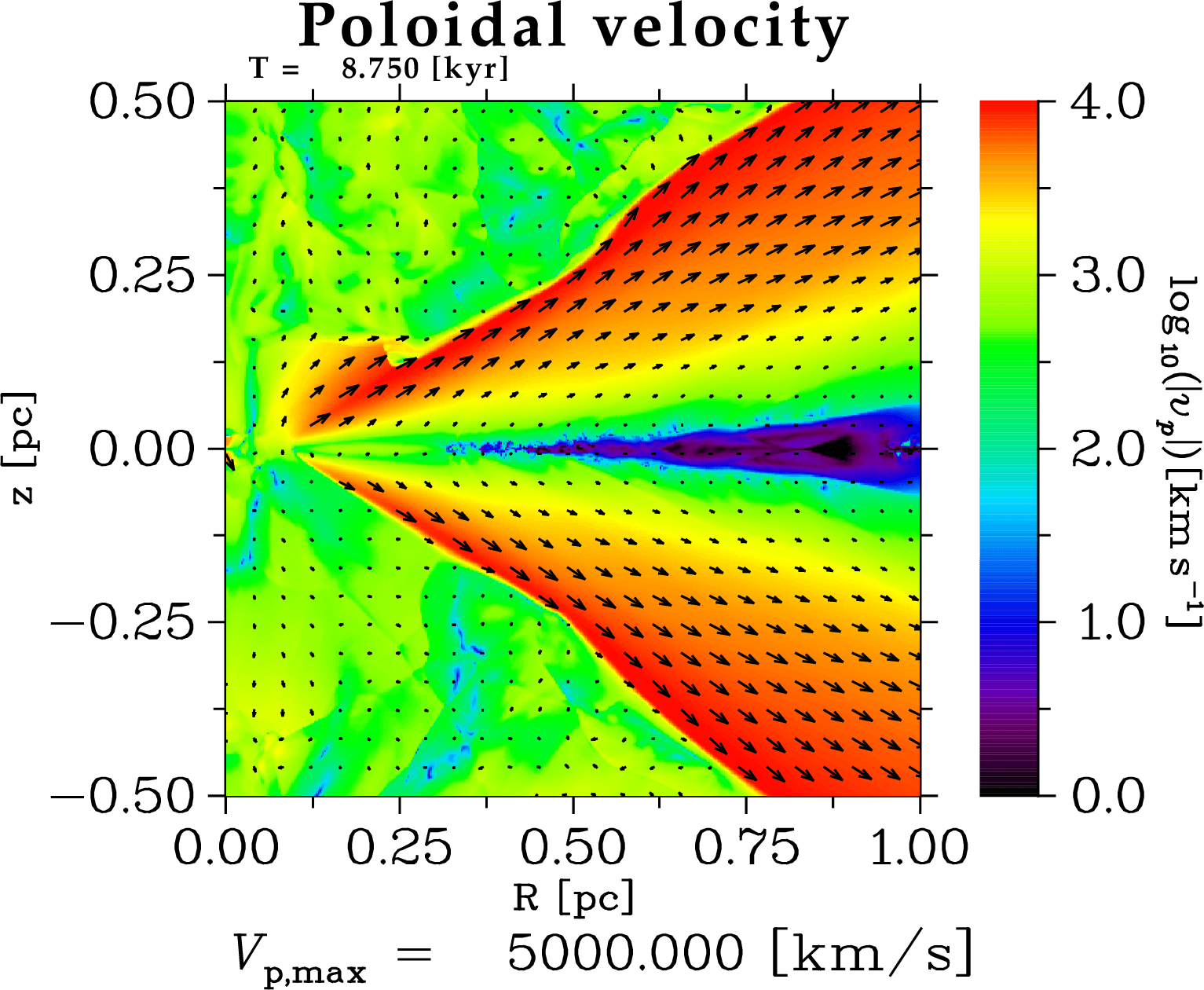}
\end{minipage}
\begin{minipage}{0.5\hsize}
\includegraphics[clip,width=\linewidth]{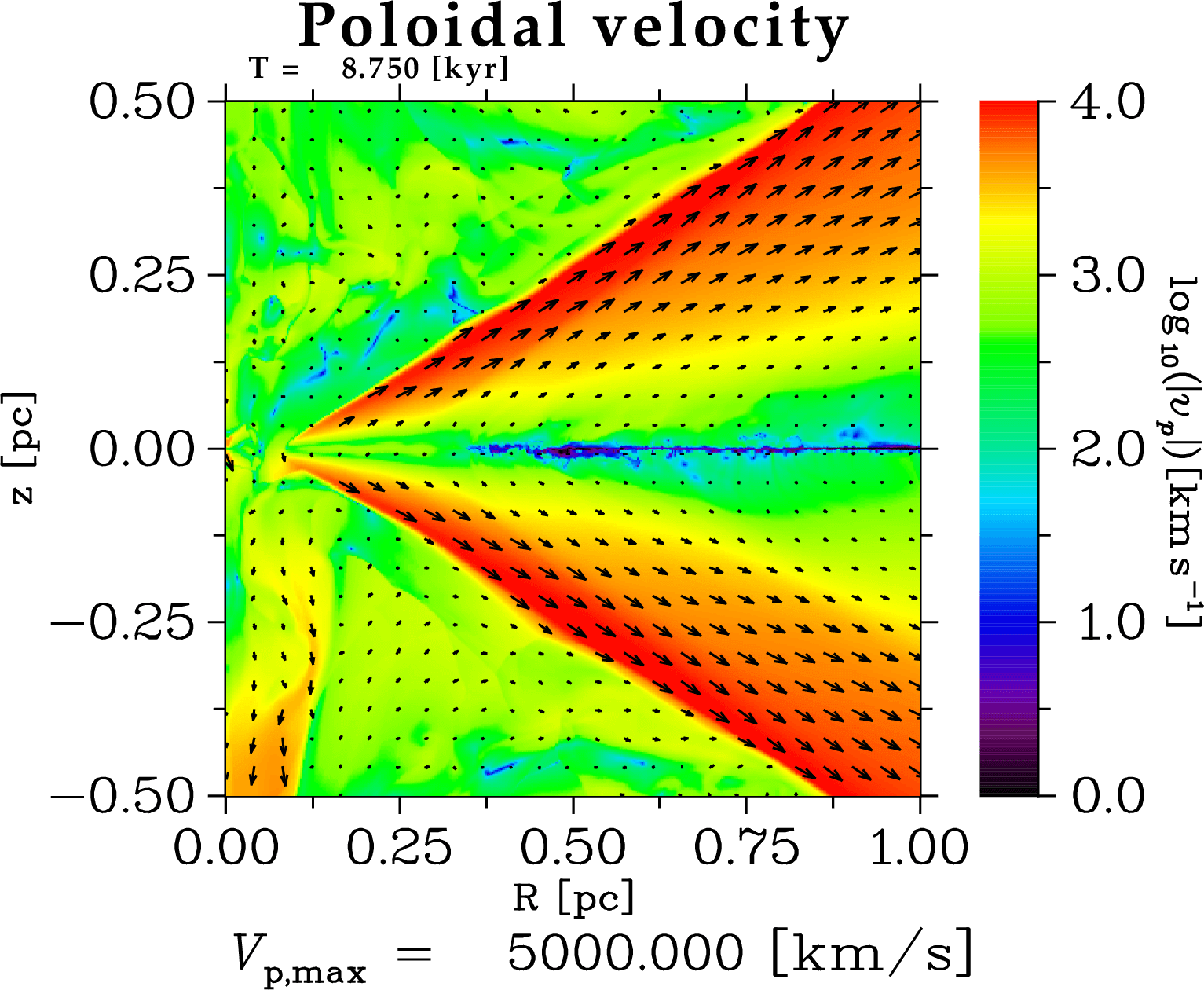}
\end{minipage}
\end{tabular}
\caption{Poloidal velocity fields in the reference models at $t=8.75\;\kyr$ (\textit{left}: \texttt{gra01\_wo\_sca\_SG}, \textit{right}: \texttt{gra01\_wo\_sca\_MTL\_SG}). The colors and the black arrows show the absolute values (i.e., $v_{p}\equiv \sqrt{v^{2}_{R}+v^{2}_{z}}$) and the directions of poloidal velocity vectors, respectively. The length of the arrow is proportional to the poloidal velocity, but, for visibility of the figure, we set up a maximum length: the lengths of all of the velocity vectors with $v_{p}>v_{p,\max}$ are set to the length corresponding to $v_{p,\max}$, which is described on the bottom of panel.}
\label{fig:velcfld_reference_models}
\end{figure*}

\begin{figure*}
\centering
\begin{tabular}{cc}
\begin{minipage}{0.5\linewidth}
\includegraphics[clip,width=\linewidth]{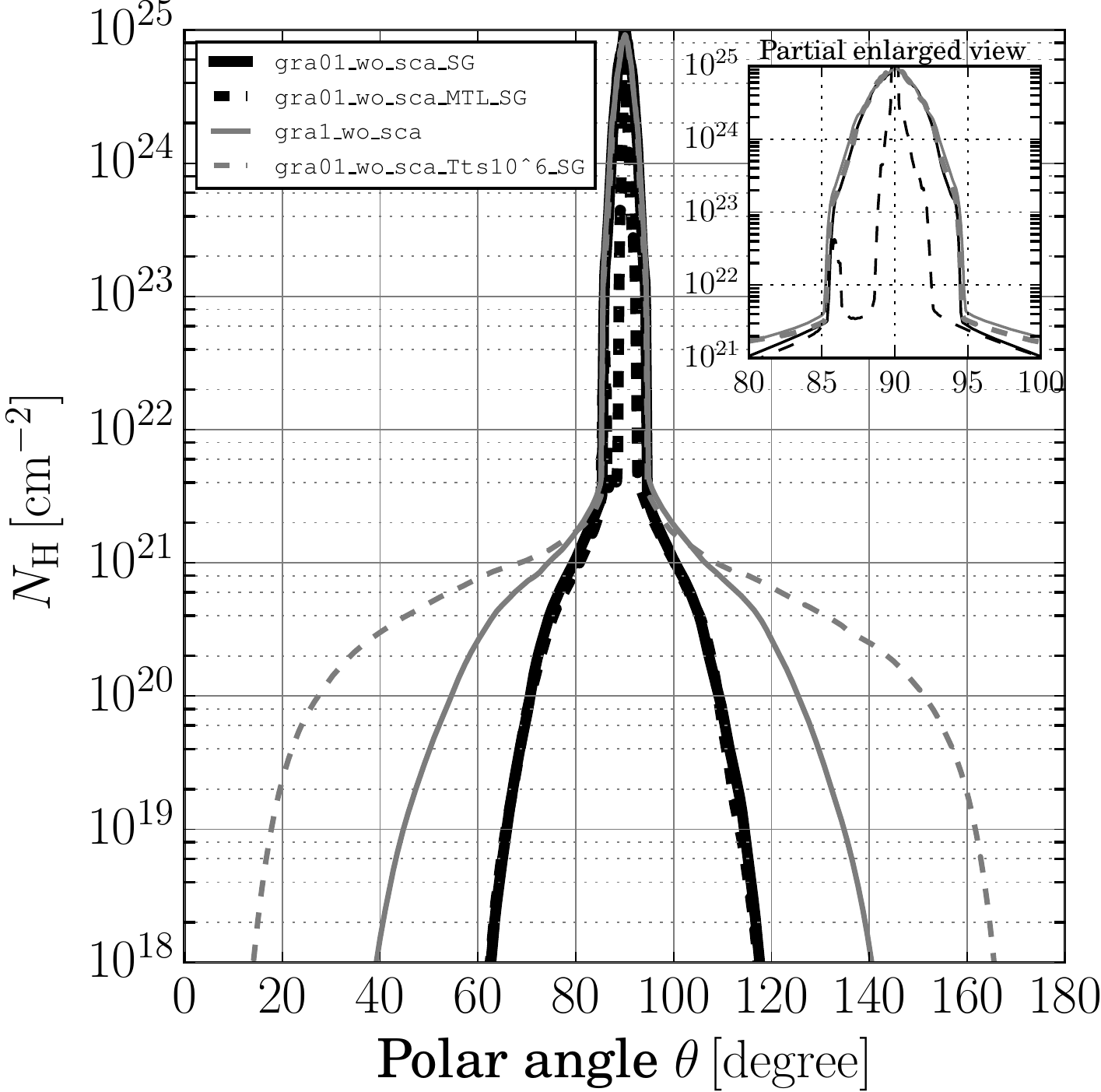}
\end{minipage}
\begin{minipage}{0.5\linewidth}
\includegraphics[clip,width=\linewidth]{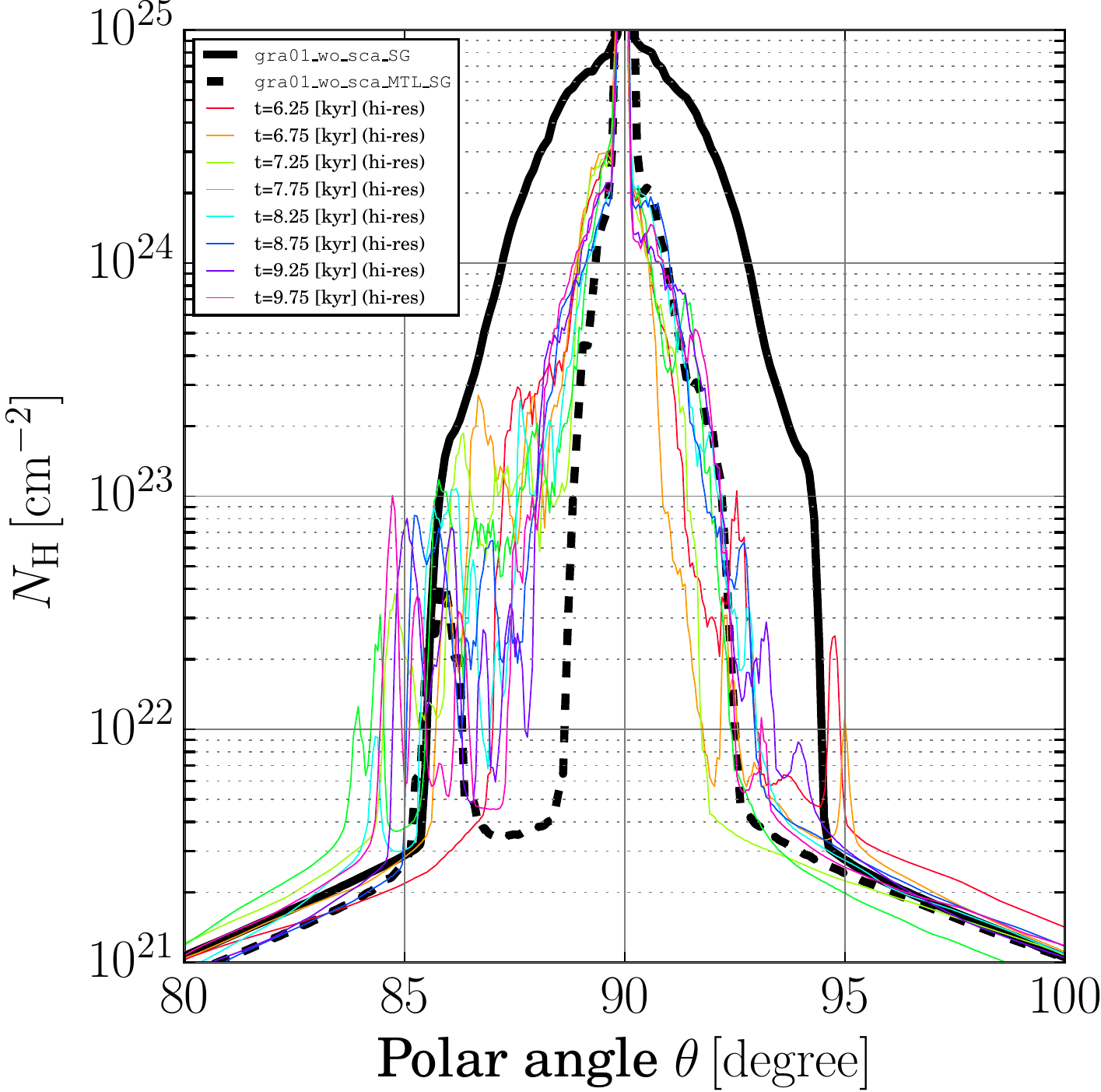}
\end{minipage}
\end{tabular}
\caption{Angular distribution of hydrogen column density $N_{\mathrm{H}}(\theta)$ in various models, where $\theta$ is the polar angle (the angle measured from the symmetric axis) and $N_{\mathrm{H}}(\theta)$ is the column density of hydrogen nuclei measured from the AGN to the boundary of the computational box along the given $\theta$. Left: $N_{\mathrm{H}}(\theta)$ at $t=8.75\;\kyr$ in the reference models (\texttt{gra01\_wo\_sca\_SG} and \texttt{gra01\_wo\_sca\_MTL\_SG}), models \texttt{gra1\_wo\_sca}, and \texttt{gra01\_wo\_sca\_Tts10\textasciicircum 6\_SG} are shown by the black solid, the black dashed, the gray solid, and the gray dashed lines, respectively. Right: $N_{\mathrm{H}}(\theta)$ in model \texttt{gra01\_wo\_sca\_MTLLDA1\_SG\_hr} at different calculation times are compared with those at $t=8.75\;\kyr$ in the reference models (see the legend for details). In both panels, a sharp increase in $N_{\mathrm{H}}(\theta)$ around $\theta=90^{\circ}$ corresponds to the obscuration due to the neutral disk.}
\label{fig:NH_dist}
\end{figure*}

Figures \ref{fig:dists_gra01_wo_sca_SG} and \ref{fig:dists_gra01_wo_sca_MTL_SG} show the time evolution of density, gas and dust temperatures in the reference models. Immediately after the simulations are started, the surface layer of the initial gas disk is strongly heated by absorption of ionizing photons from AGN and an dusty outflow is launched from the disk surface. Hard X-ray photons from corona quickly rise the gas temperature of the outflowing gas and it eventually settles at Compton temperature ($\approx 10^{7}\;\mathrm{K}$). Thus, the gas temperature is determined by the balance between Compton heating and Compton cooling (readers may have noticed that the gas temperature is large enough that small dust grains are rapidly destroyed by thermal sputtering. The effects of thermal sputtering is examined in \S~\ref{subsec:thermal_sputtering_effects}). Thermal pressure in the outflowing gas also increases and, as a result, the outflowing gas thermally expands in a vertical direction (its sound speed is comparable to circular velocity). At the same time, the outflow is strongly accelerated in the radial direction by dust absorption and electron scattering. This competition determines a scale height of the outflowing gas. The hydrogen number density of the outflow is $\nH\approx 10^{2\operatorname{-}3}\;\nden$ and the outflow velocity increases from $\sim 100\;\kms$ directly above the disk surface to $(2\operatorname{-}3)\times 10^{3}\;\kms$ at the outer layer of the outflow (Fig.~\ref{fig:velcfld_reference_models}). The hydrogen column density of the outflow region is $N_{\mathrm{H}}\lesssim 2\times 10^{21}\;\cden$ (Fig.~\ref{fig:NH_dist}), and hence, the outflow within $R<1\;\pc$ cannot be regard as an obscuring torus ($N_{\mathrm{H}}>3\times 10^{21}\;\cden$ is required to suppress broad UV lines according to \citealt{hasinger08:_absor}). The dust temperature in the outflow region is mainly determined by the balance between photoheating and cooling due to thermal emission and is $\Tgr\approx 500\operatorname{-}10^{3}\;\mathrm{K}$ (Figs.~\ref{fig:dists_gra01_wo_sca_SG}i-l and \ref{fig:dists_gra01_wo_sca_MTL_SG}i-l).

In the absence of metal cooling, the system reaches a quasi-steady state by $t=2\;\kyr$, which consists of a nearly neutral, dense ($\nH\approx 10^{6\operatorname{-}8}\;\nden$), thin ($h/r\approx 0.06$) disk and the dusty wind described above (Fig.~\ref{fig:dists_gra01_wo_sca_SG}b-d). X-ray heating determines the thermal state of the neutral thin disk and the gas and dust temperatures are $\approx (1\operatorname{-}3)\times 10^{4}\;\mathrm{K}$ and $\approx 200\;\mathrm{K}$, respectively (Fig.~\ref{fig:dists_gra01_wo_sca_SG}f-h). On the other hand, in the presence of metal cooling, the initial disk begins to collapse in a vertical direction at the start of the simulation and a very thin disk forms at $t\approx 6\;\kyr$ (Fig.~\ref{fig:dists_gra01_wo_sca_MTL_SG}c,d). After that, the system roughly keeps a quasi-steady state. The hydrogen number density in the thin disk is $10^{7\operatorname{-}9}\;\nden$. The gas temperature in the thin disk varies from $\approx 10^{4}\;\mathrm{K}$ at the inner part of the disk to $\approx 100\;\mathrm{K}$ at the outer part of the disk (e.g., Fig.~\ref{fig:dists_gra01_wo_sca_MTL_SG}g,h), where most of hydrogen takes the form of molecular hydrogen. The true scale height of the neutral disk should be intermediate between the scale heights of both models, because we use a overestimated metal cooling function as described in \S~\ref{subsec:ISM_dust_model}. The inner edge ($R\approx 0.38\operatorname{-}0.56\;\pc$) of the neutral thin disk fragments into a clumpy structure and cloudlets are occasionally blown off likely due to ram pressure and radiation pressure (e.g., Fig.~\ref{fig:dists_gra01_wo_sca_MTL_SG}d). These cloudlets are promptly destroyed by combination of ram pressure stripping and photoevaporation. In addition, chimney-like structures are frequently formed at the disk surface in the model with metal cooling (Fig.~\ref{fig:dists_gra01_wo_sca_MTL_SG}c). The time evolution of these inhomogeneous features depends on spatial resolution and we come back to this point in \S~\ref{subsec:resolution_effects}.

In order to examine the reasons why the neutral disk is geometrically thin and make clear the contribution of IR photons to the vertical support of the neutral disk, we perform additional five simulations (\texttt{gra01\_wo\_sca}, \texttt{gra01\_w\_sca}, \texttt{gra01\_wo\_sca\_noRP-IR}, \texttt{gra01\_wo\_sca\_noRP-drct}, and \texttt{gra01\_wo\_sca\_noRP-all}) and compare them each other. We first check if dust scattering of IR photons, which is neglected in the reference models, changes the structure of the neutral disk, by comparing the result of model \texttt{gra01\_wo\_sca} with that of model \texttt{gra01\_w\_sca}. Figs~\ref{fig:nH_dists_various_models}(a),(b) show the spatial distributions of hydrogen number density at $t=8.75\;\kyr$ in these two models. As is clear from the figure, there is no difference between them, indicating that dust scattering of IR photons does not affect the disk structure. Next, we compare the result of model \texttt{gra01\_wo\_sca} with that of model \texttt{gra01\_wo\_sca\_noRP-IR}, in which radiation pressure due to IR photons is switched off, to examine the contribution of IR photons to the vertical support of the disk. The density distributions of both models are shown in Figs~\ref{fig:nH_dists_various_models}(a),(c). The disk thickness is almost the same in these two models, suggesting that the vertical support due to IR photons is not effective. This is supported by the comparison between models \texttt{gra01\_wo\_sca} and \texttt{gra01\_wo\_sca\_noRP-drct}, in the latter of which only radiation pressure due to IR photons is taken into account. Fig.~\ref{fig:nH_dists_various_models}(d) shows the density distribution of model \texttt{gra01\_wo\_sca\_noRP-drct} when the total gas mass is $\approx 2.2\times 10^{4}\;\Msolar$. The disk thickness is virtually the same as that of model \texttt{gra01\_wo\_sca}. Thus, the disk thickness does not depend on the presence or the absence of radiation pressure due to IR photons. However, radiation pressure due to IR photons enhances the outflow rate as shown by the comparison between Figs.~\ref{fig:nH_dists_various_models}(a) and (e). The total gas mass at $t=8.75\;\kyr$ in model \texttt{gra01\_wo\_sca\_noRP-drct} is smaller than that in model \texttt{gra01\_wo\_sca}. Finally, we compare the result of model \texttt{gra01\_wo\_sca} with that of model \texttt{gra01\_wo\_sca\_noRP-all}, in which all of radiation pressure is switched off, and as a result, the disk thickness is determined only by thermal pressure. Fig.~\ref{fig:nH_dists_various_models}(f) show the density distribution at $t=8.75\;\kyr$ in model \texttt{gra01\_wo\_sca\_noRP-all}. The disk thickness in this model is much the same as that in model \texttt{gra01\_wo\_sca}. From these results, we conclude that the neutral disk is mainly supported by gas pressure, rather than radiation pressure of IR photons.

To make it more clear how radiation pressure due to IR photons acts, we plot the spatial distribution of the ratios of various types of radiative accelerations normalized by the acceleration due to the external gravity in Fig.~\ref{fig:acceleration_ratios}. Fig.~\ref{fig:acceleration_ratios}(a) shows that the outflowing gas is partially supported in the vertical direction by radiation pressure of IR photons reemitted by the neutral disk, while the neutral disk is \textit{compressed} by IR photons reemitted by the outflowing gas. Thus, IR photons is not effective in supporting the neutral disk. One of other reasons why the neutral disk does not expand vertically is that radiation pressure from direct photons and thermal pressure of the outflowing gas compress or confine the neutral thin disk. This can be confirmed by both Fig.~\ref{fig:nH_dists_various_models} and Figs.~\ref{fig:acceleration_ratios}(c),(d). Figs~\ref{fig:nH_dists_various_models}(a),(c), and (e) show that the presence of radiation pressure due to direct photons reduces the decreasing rate of gas mass in the system. Figs~\ref{fig:acceleration_ratios}(c),(d) show that radiation pressure due to direct photons near the surface of the neutral disk is as large as $a^{\mathrm{rad,drct}}_{\mathrm{R}}/|a^{\mathrm{grv,ext}}_{\mathrm{R}}|\approx 10$. Another reason is certainly that $\Tgr$ in the neutral disk does not become large enough due to both the anisotropic AD radiation and dust absorption of direct photons in the outflow region.

Figure~\ref{fig:gas_mass_evolution} shows the time evolution of gas mass in the computational box for each model. Mass outflow rate can be estimated from the time derivative. In the absence of metal cooling, the outflow rate are $0.05\operatorname{-}0.1\;\Msolar\;\yr^{-1}$ during $t=2\operatorname{-}6\;\kyr$ depending on the X-ray luminosity fraction. These values correspond to $\approx 20\%\operatorname{-}50\%$ of Eddington mass accretion rate $\dot{M}_{\mathrm{Edd}}$, which is given by
\begin{equation}
\dot{M}_{\mathrm{Edd}} = \frac{L_{\mathrm{Edd}}}{\eta c^{2}} \approx 0.23\;\Msolar\;\yr \left(\frac{\eta}{0.1}\right)^{-1}\left(\frac{\MBH}{10^{7}\;\Msolar}\right),
\end{equation}
where $\eta$ is the mass-to-radiation energy conversion factor. At later stage of the simulations ($t=6\operatorname{-}10\;\kyr$), the outflow rate increases up to $\approx 0.3\;\Msolar\;\yr^{-1}$. This is because dense neutral gas begins to go out of the computational box along the mid-plane due to radiation pressure. However, the gas does not have velocity large enough to escape from the galactic center region. Hence, large outflow rates in later stage of the simulations are superficial. The outflow rate of model \texttt{gra01\_wo\_sca\_noRP-IR} is slightly smaller than that of model \texttt{gra01\_wo\_sca}, indicating that IR photons enhance outflow rate although they have little influence on the structure of the neutral disk as described above. The models with metal cooling also show similar outflow rates after the formation of a very thin disk ($t\gtrsim 6\;\kyr$). 

Finally, we comment on the effects of self-gravity, which can be deduced from the comparison between model \texttt{gra01\_wo\_sca\_SG} and model \texttt{gra01\_wo\_sca}. Fig.~\ref{fig:dists_gra01_wo_sca_SG}(d) and Fig.~\ref{fig:nH_dists_various_models}(a) show the density distribution of these two models at $t=8.75\;\kyr$ and we find that both are very similar to each other. Thus, self-gravity does not affect the disk structure for the model considered in this study. Note that this is consistent with the fact that the initial value of the mid-plane number density ($\nHmid=10^{7}\;\nden$) is self-gravitationally stable.

\begingroup
\begin{figure*}
\centering
\begin{tabular}{cc}
\begin{minipage}{0.5\hsize}
\includegraphics[clip,width=\linewidth]{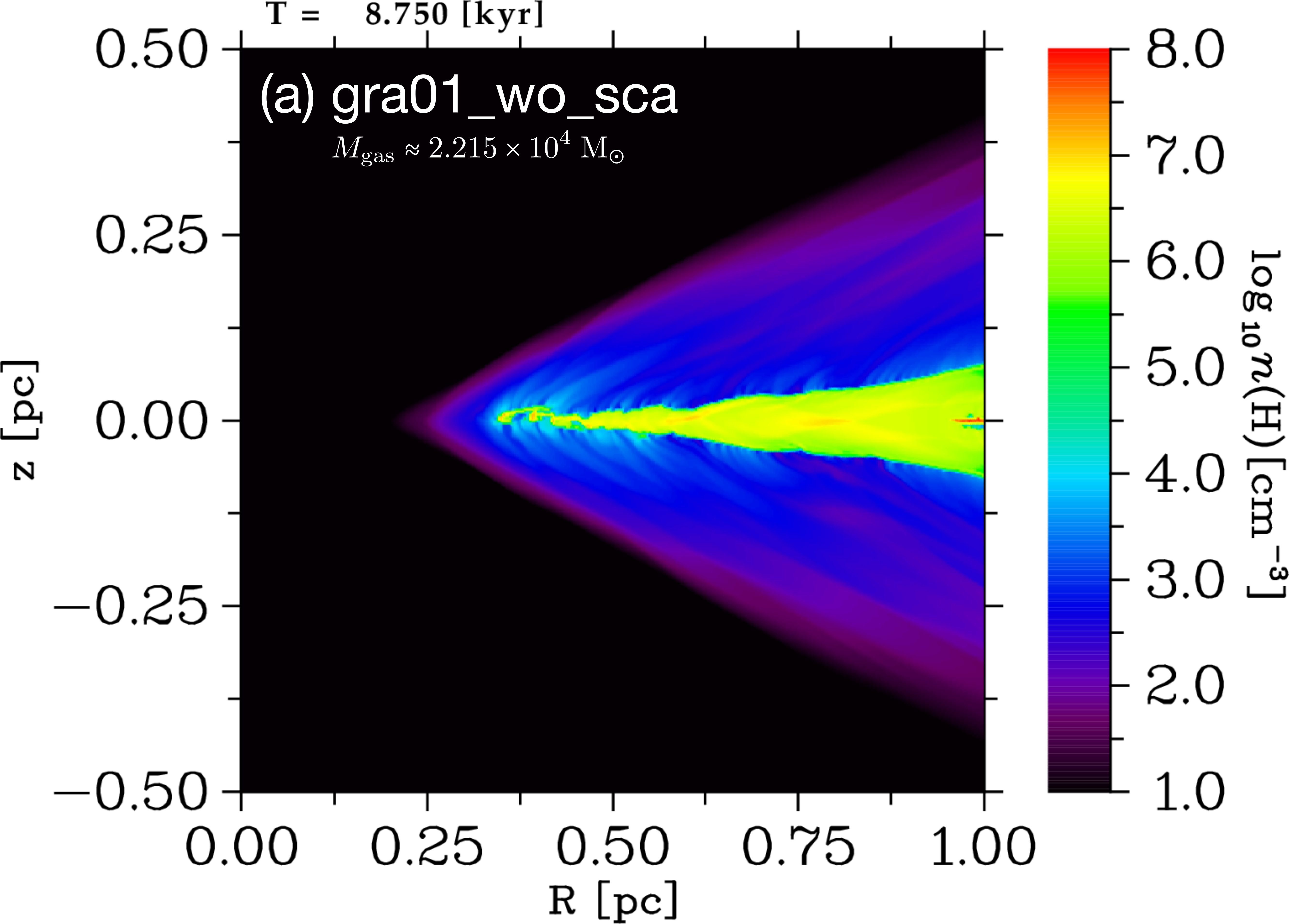}
\end{minipage}
\begin{minipage}{0.5\hsize}
\includegraphics[clip,width=\linewidth]{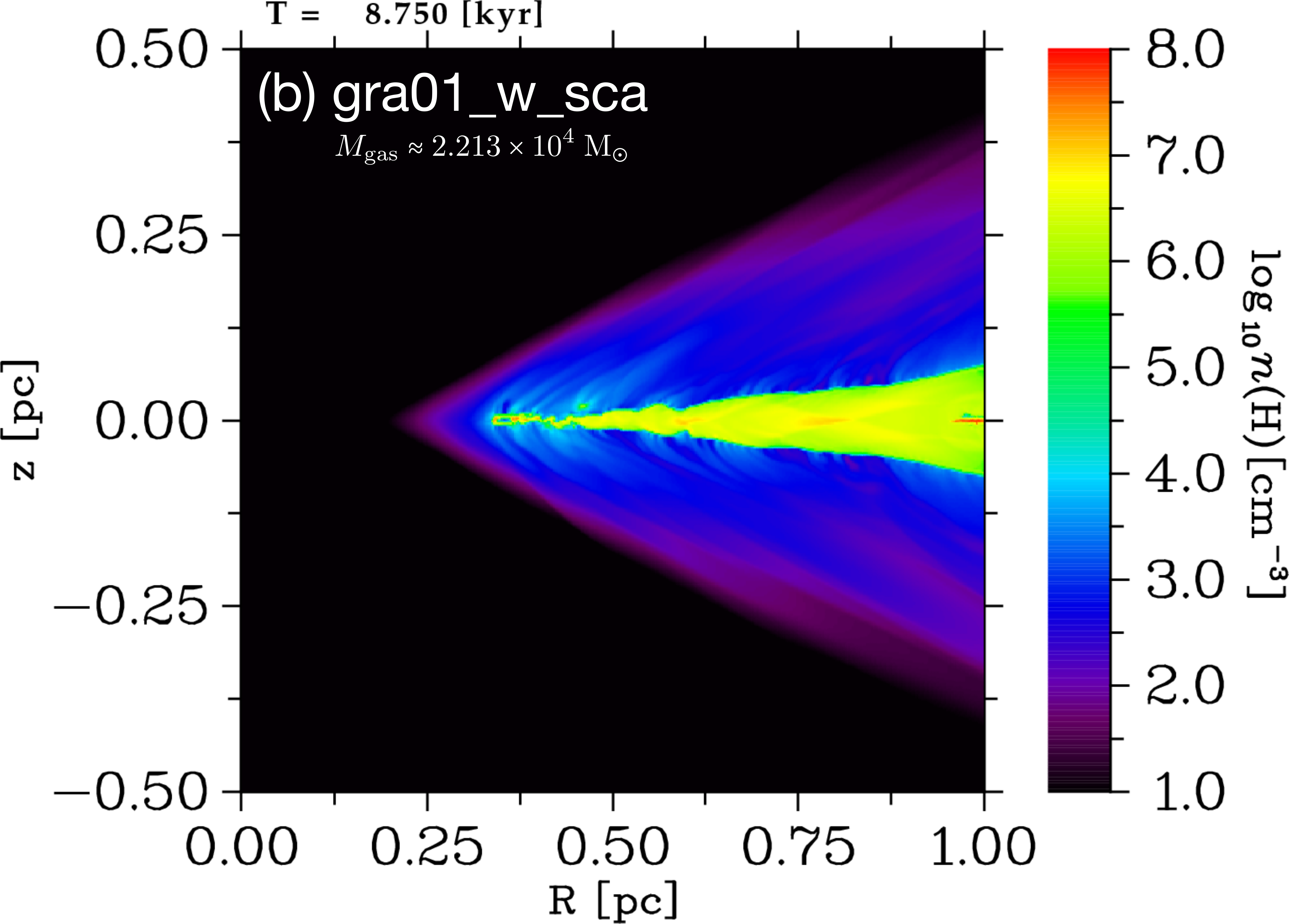}
\end{minipage}
\end{tabular}
\begin{tabular}{cc}
\begin{minipage}{0.5\hsize}
\includegraphics[clip,width=\linewidth]{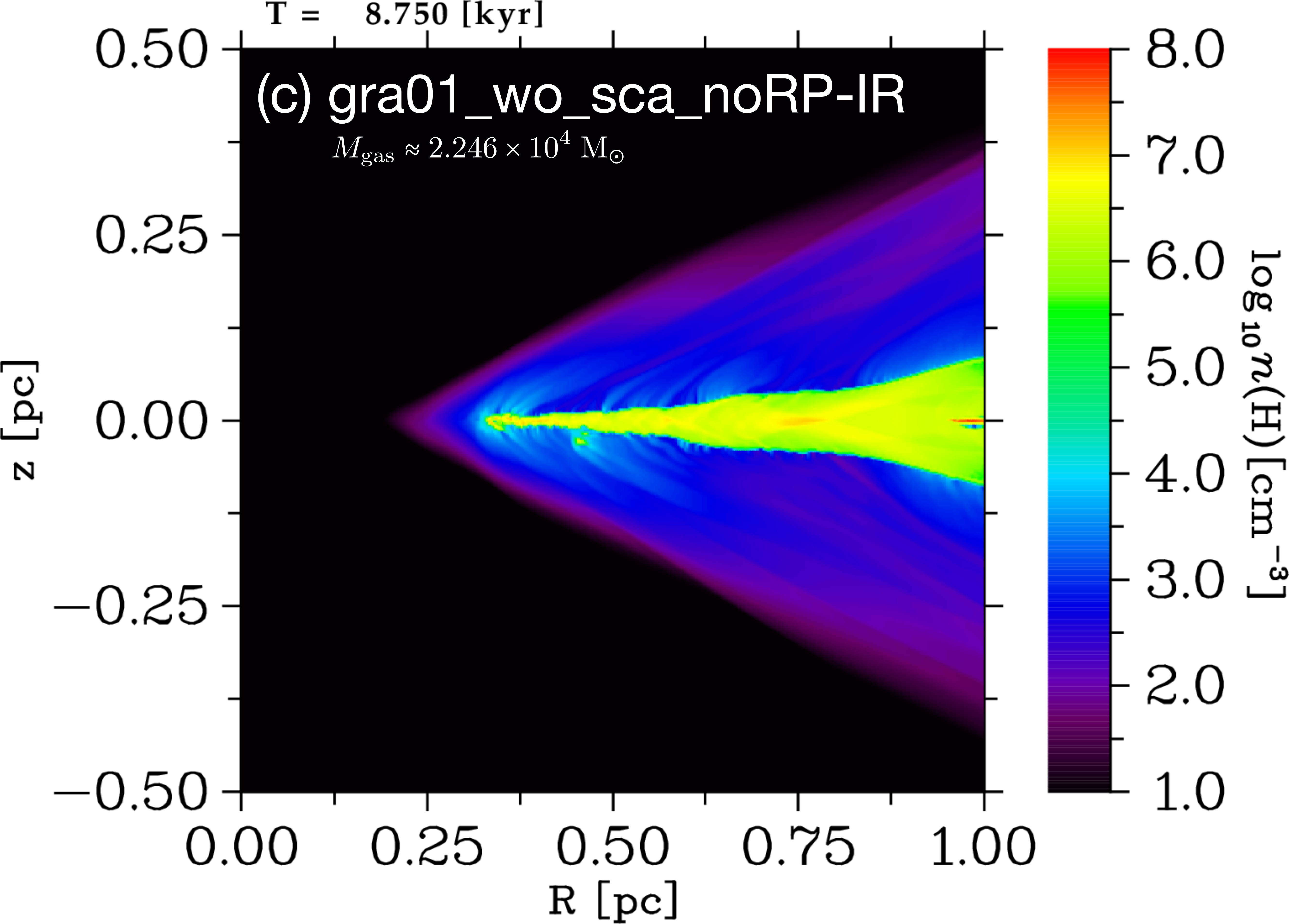}
\end{minipage}
\begin{minipage}{0.5\hsize}
\includegraphics[clip,width=\linewidth]{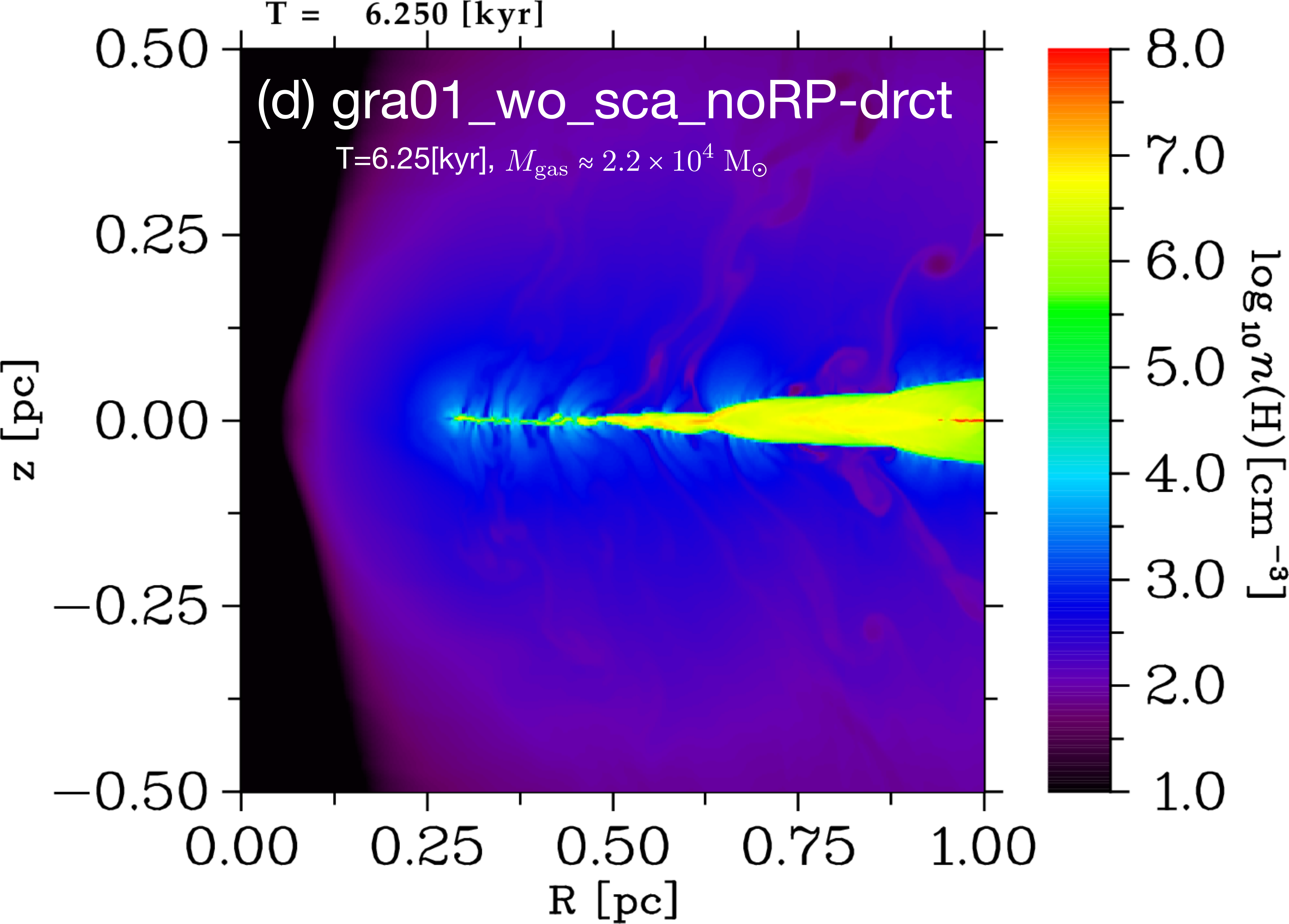}
\end{minipage}
\end{tabular}
\begin{tabular}{cc}
\begin{minipage}{0.5\hsize}
\includegraphics[clip,width=\linewidth]{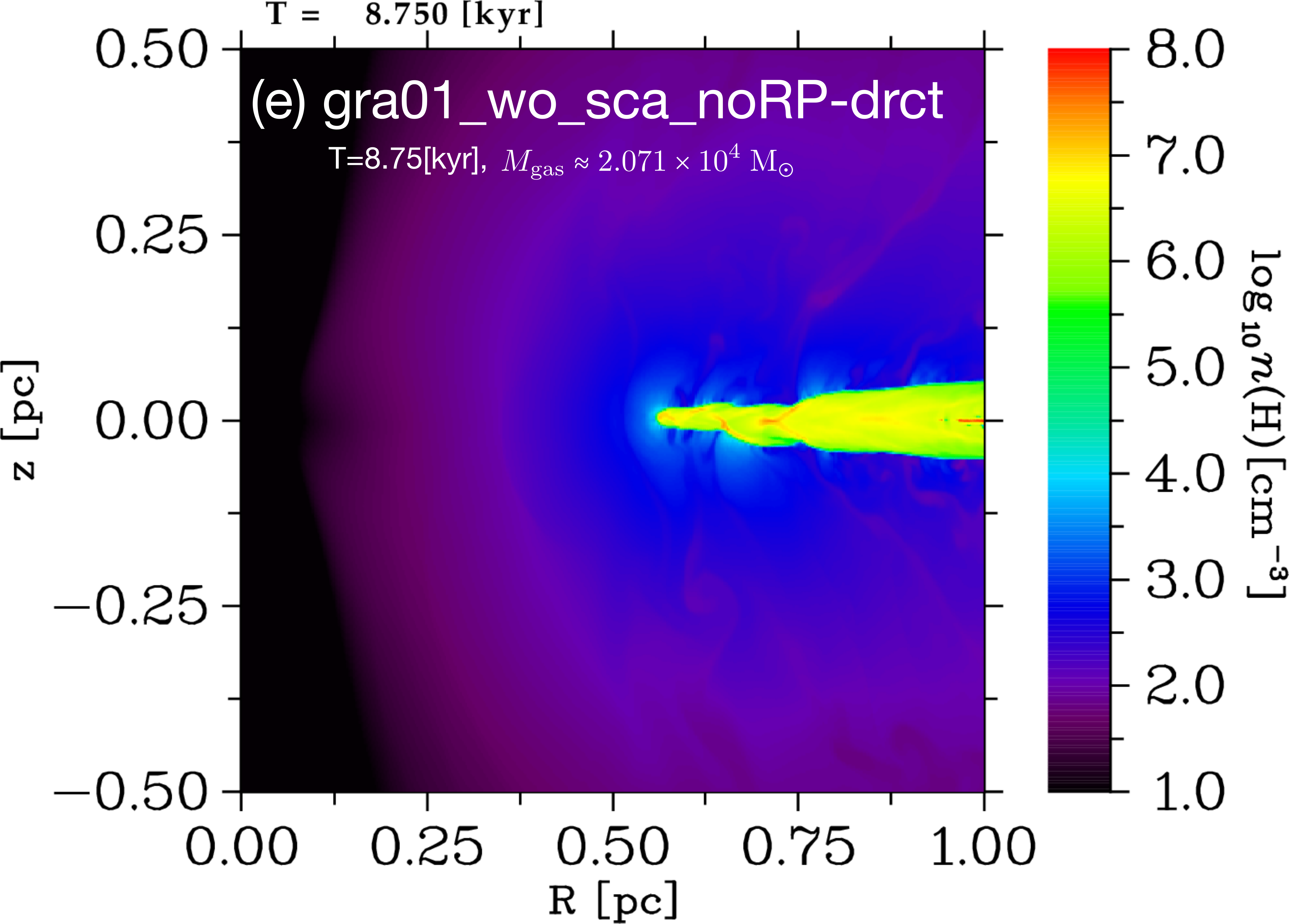}
\end{minipage}
\begin{minipage}{0.5\hsize}
\includegraphics[clip,width=\linewidth]{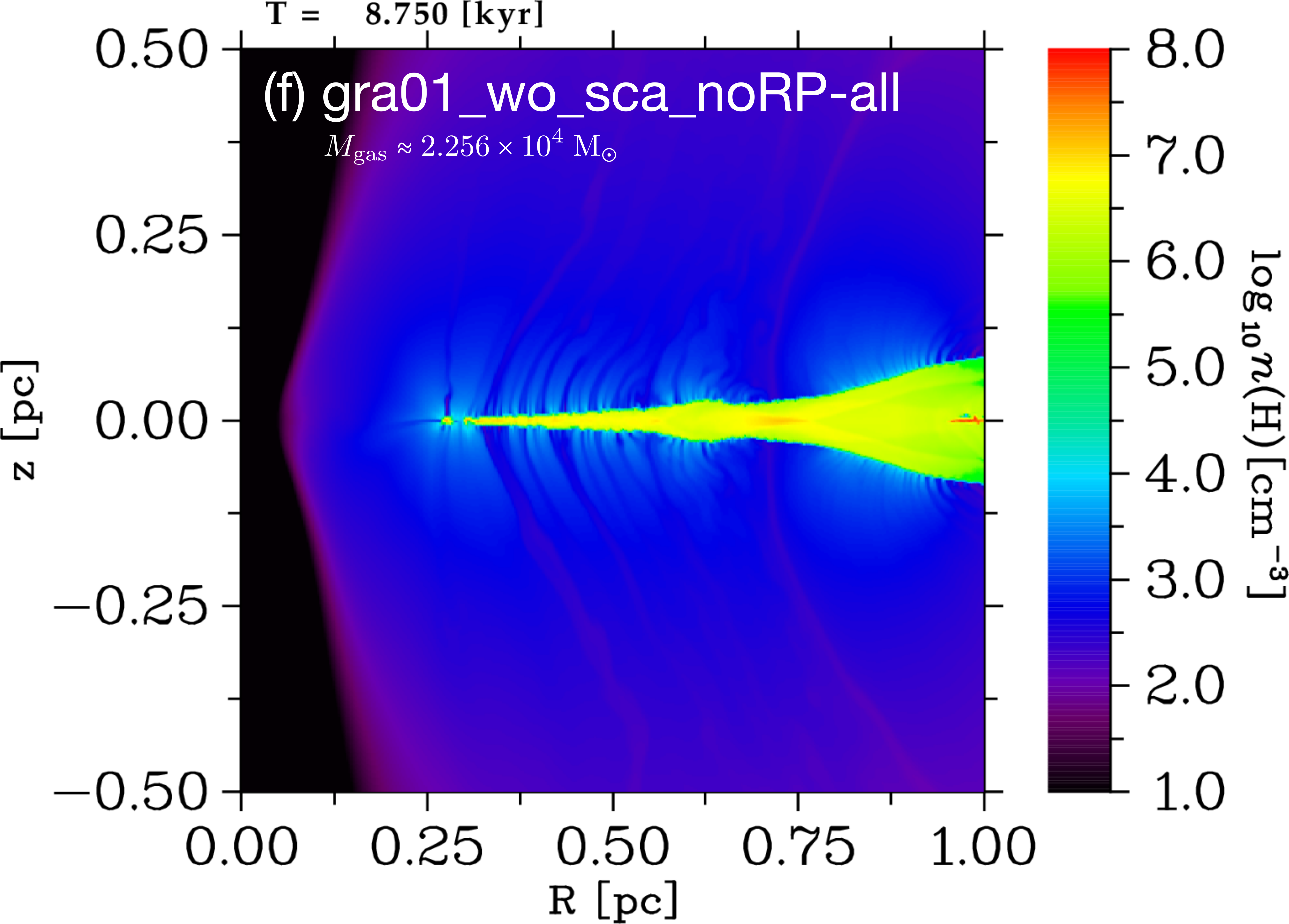}
\end{minipage}
\end{tabular}
\caption{Spatial distribution of hydrogen number density in various models described in \S~\ref{subsec:overview}, \S~\ref{subsec:grain_size_effects}, and \S~\ref{subsec:thermal_sputtering_effects}. Model name and the calculation time are described in the upper right portion of each panel. In panels (a)-(f), the total gas mass in the computational box is also shown for a fair comparison.}
\label{fig:nH_dists_various_models}
\end{figure*}
\endgroup

\begingroup
\begin{figure*}
\centering
\begin{tabular}{cc}
\begin{minipage}{0.5\hsize}
\includegraphics[clip,width=\linewidth]{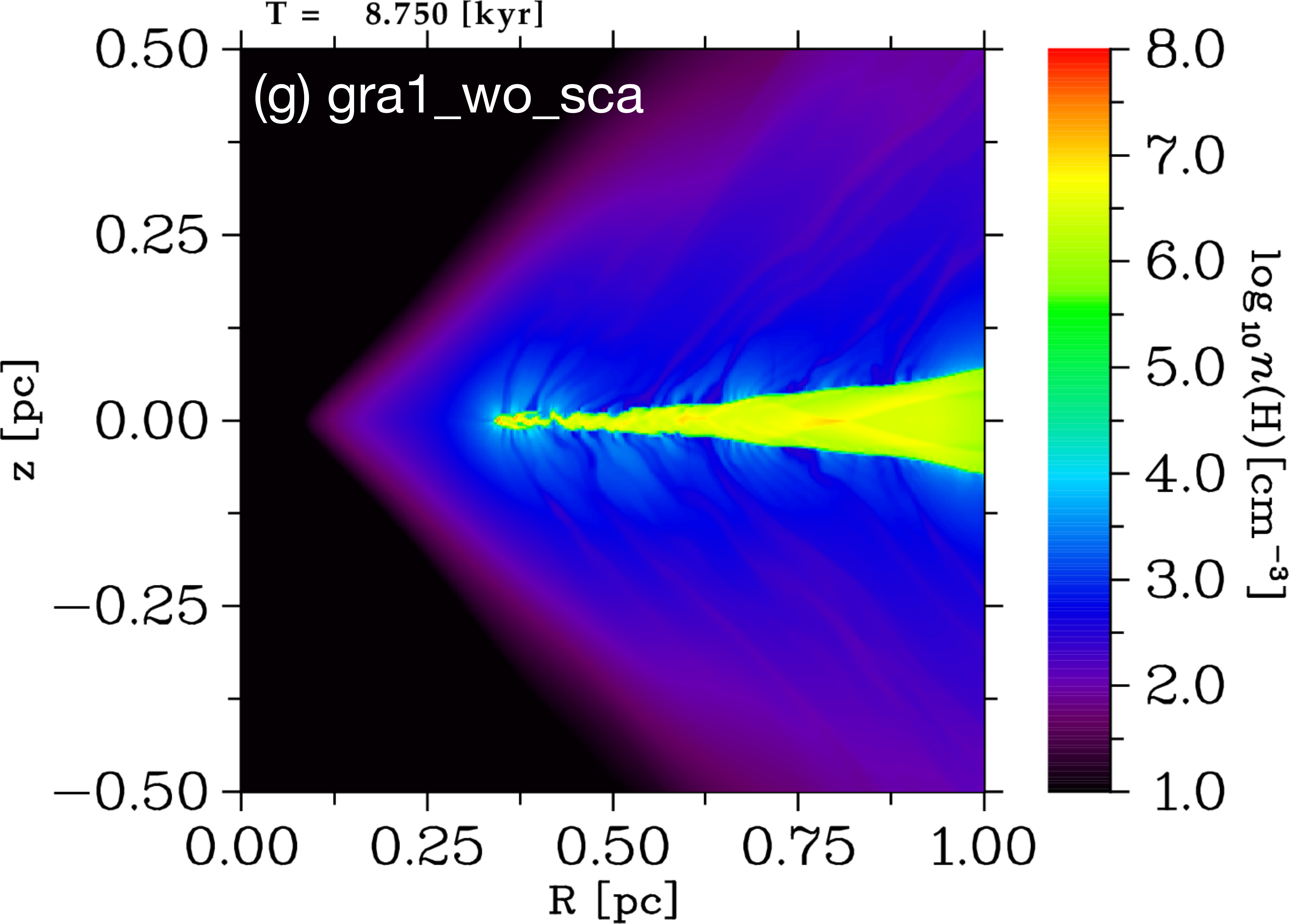}
\end{minipage}
\begin{minipage}{0.5\hsize}
\includegraphics[clip,width=\linewidth]{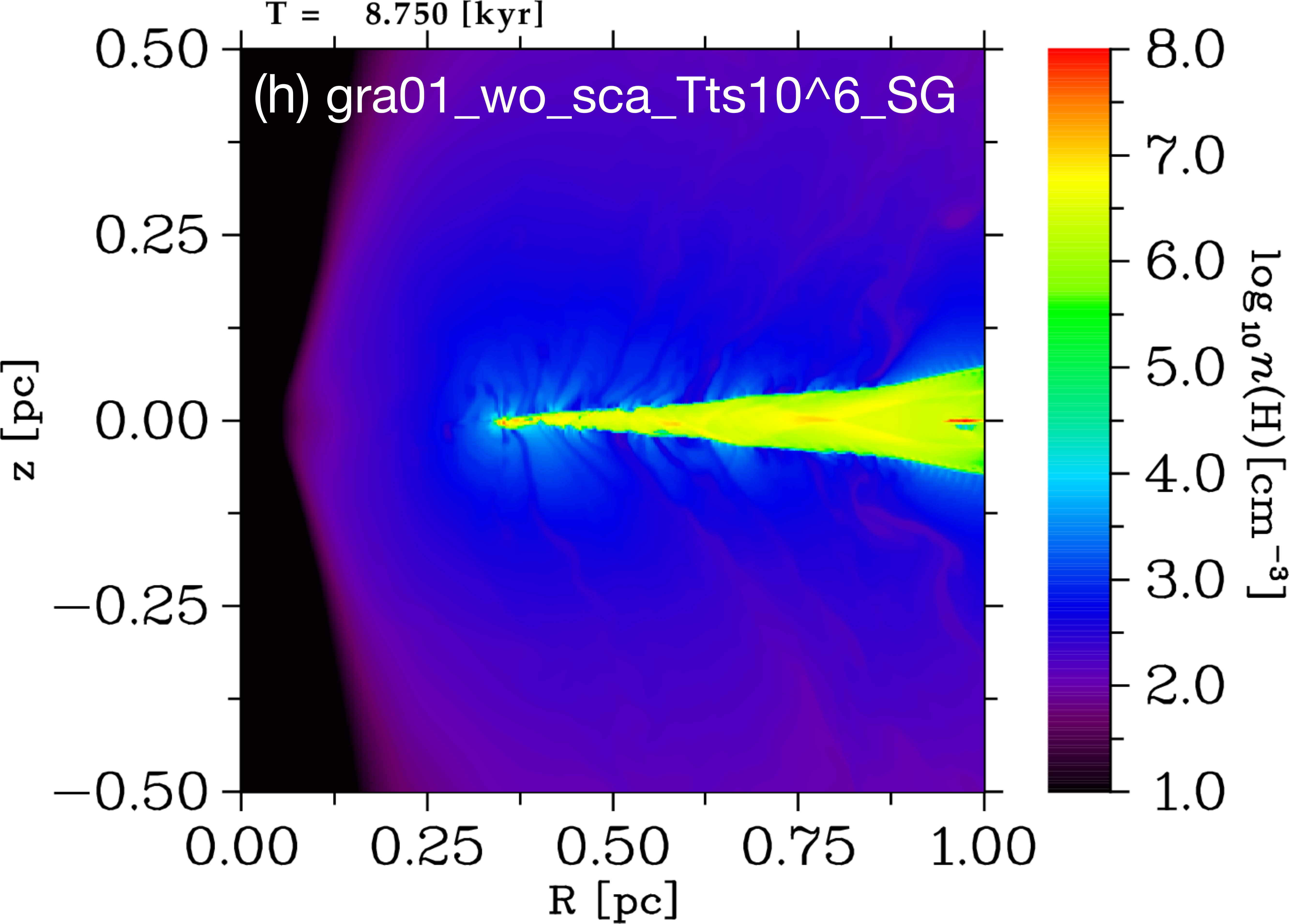}
\end{minipage}
\end{tabular}
\contcaption{}
\label{fig:nH_dists_various_models_cont}
\end{figure*}
\endgroup

\begin{figure*}
\centering
\begin{tabular}{cc}
\begin{minipage}{0.5\hsize}
\includegraphics[clip,width=\linewidth]{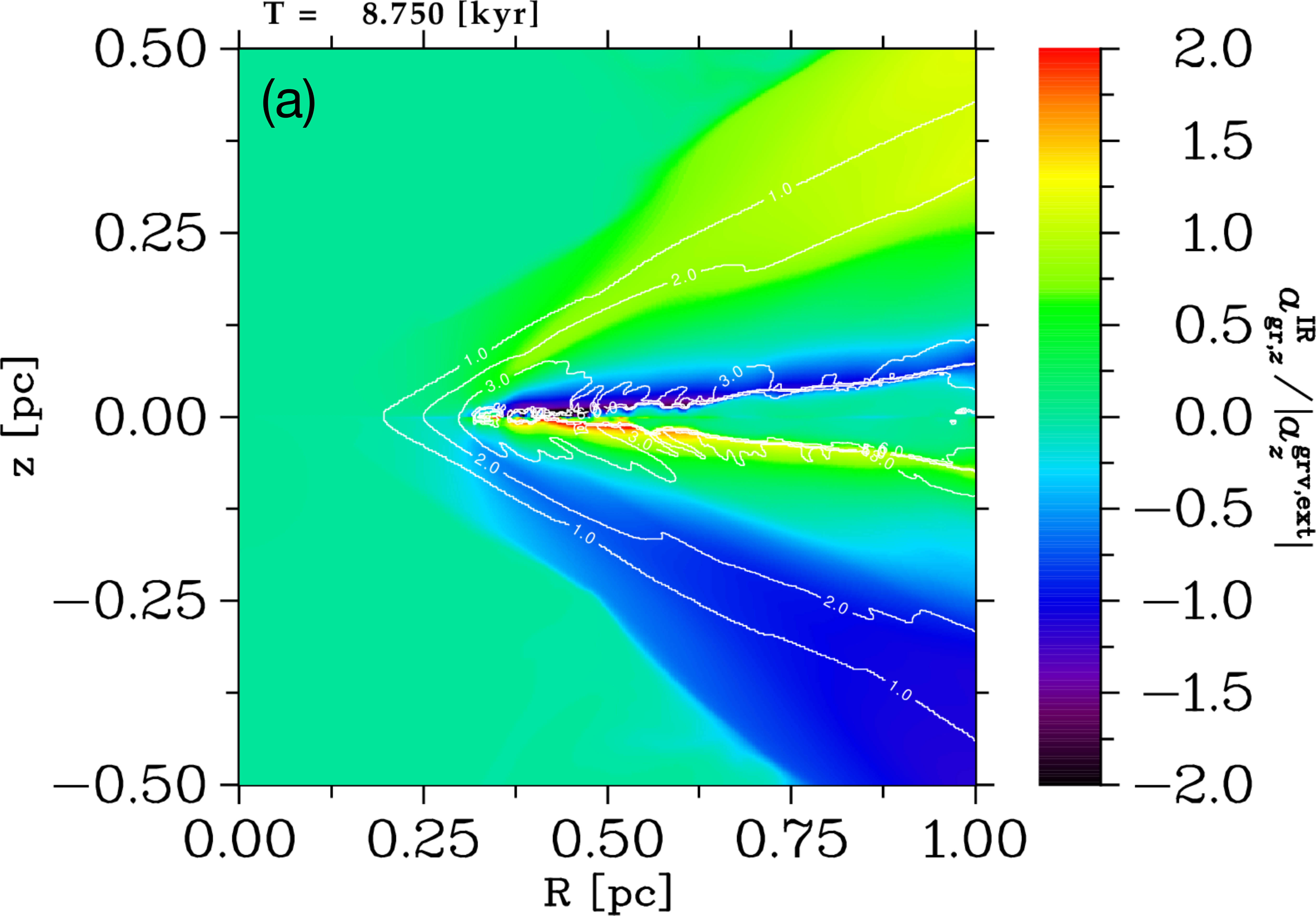}
\end{minipage}
\begin{minipage}{0.5\hsize}
\includegraphics[clip,width=\linewidth]{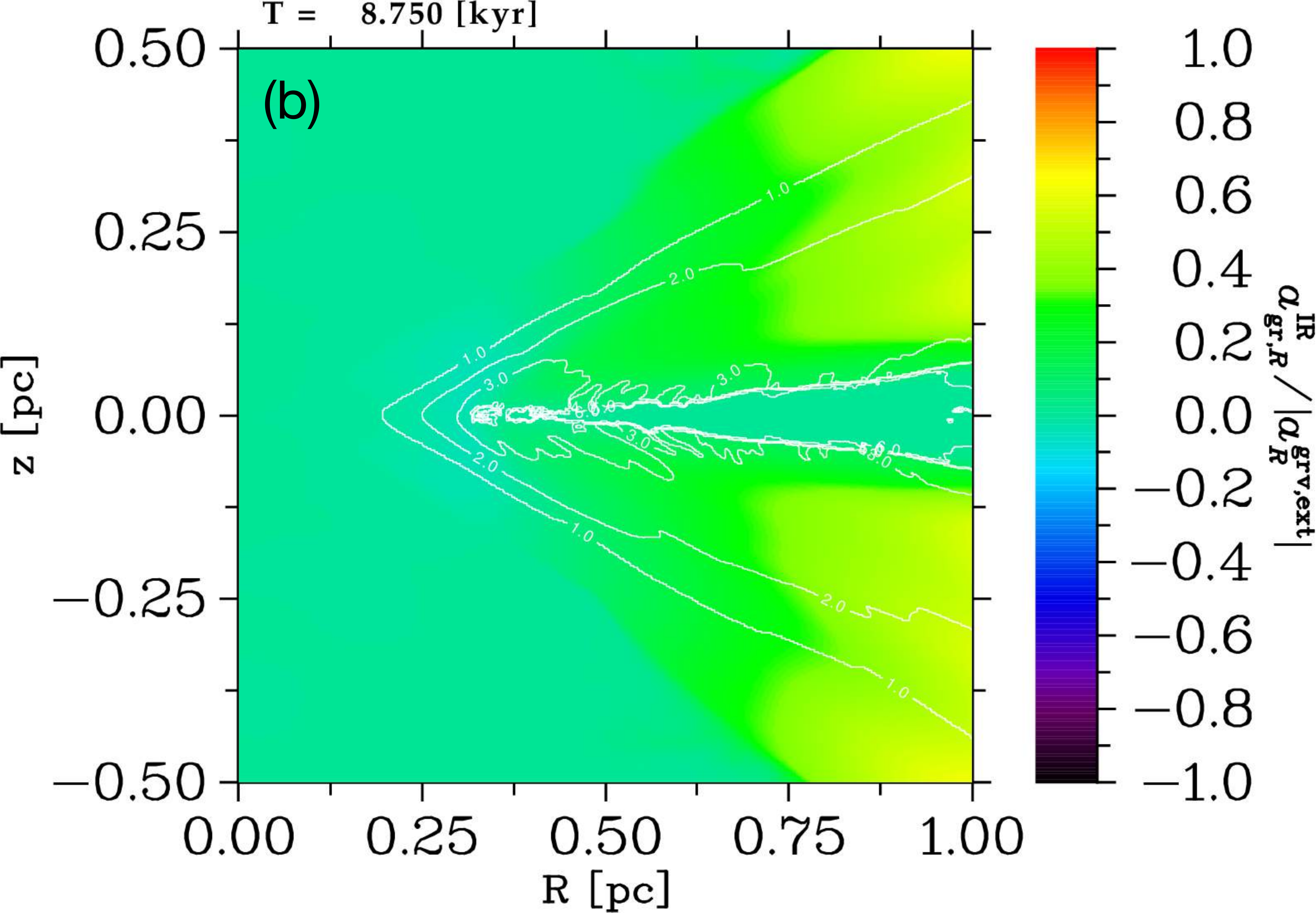}
\end{minipage}
\end{tabular}
\begin{tabular}{cc}
\begin{minipage}{0.5\hsize}
\includegraphics[clip,width=\linewidth]{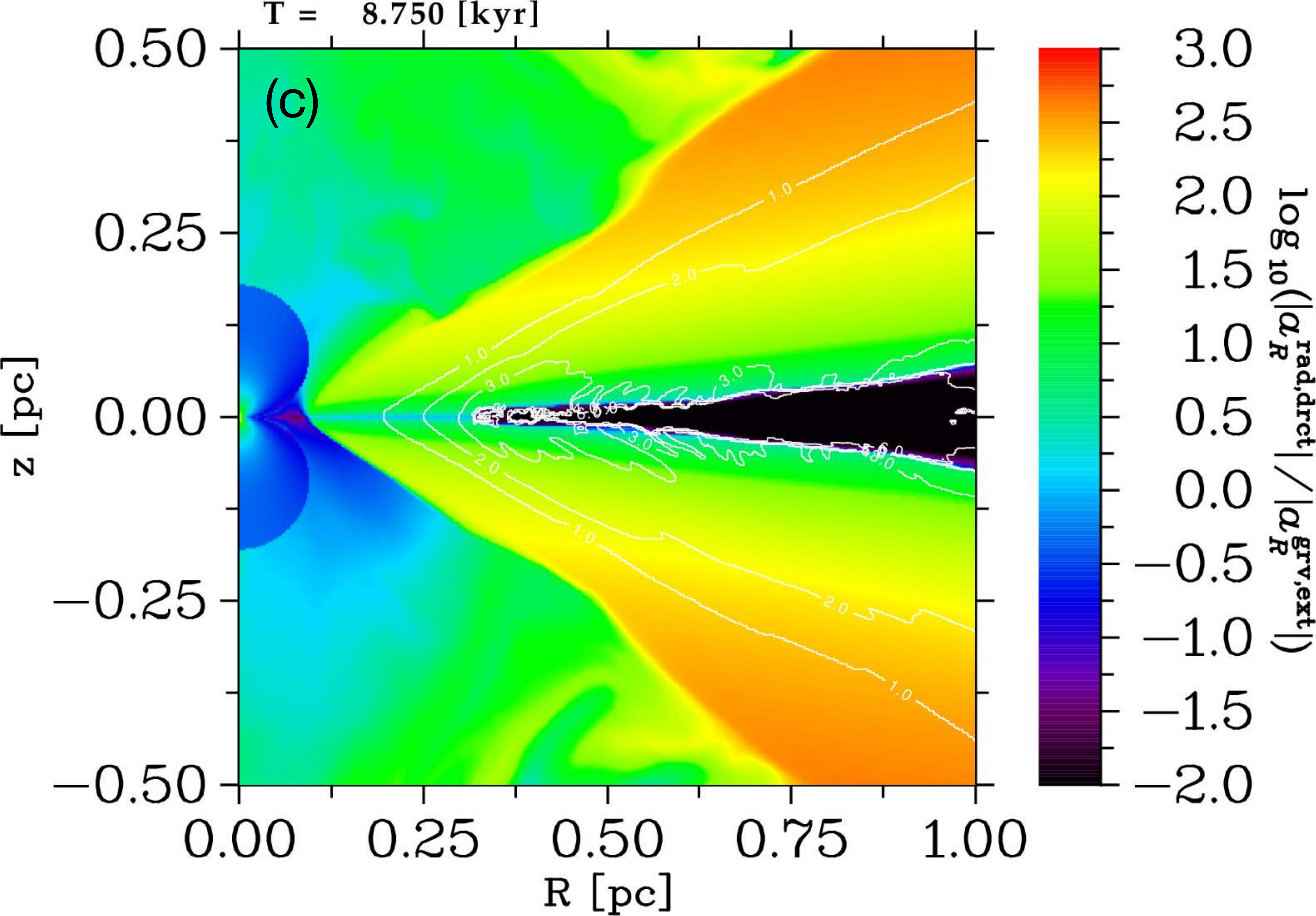}
\end{minipage}
\begin{minipage}{0.5\hsize}
\includegraphics[clip,width=\linewidth]{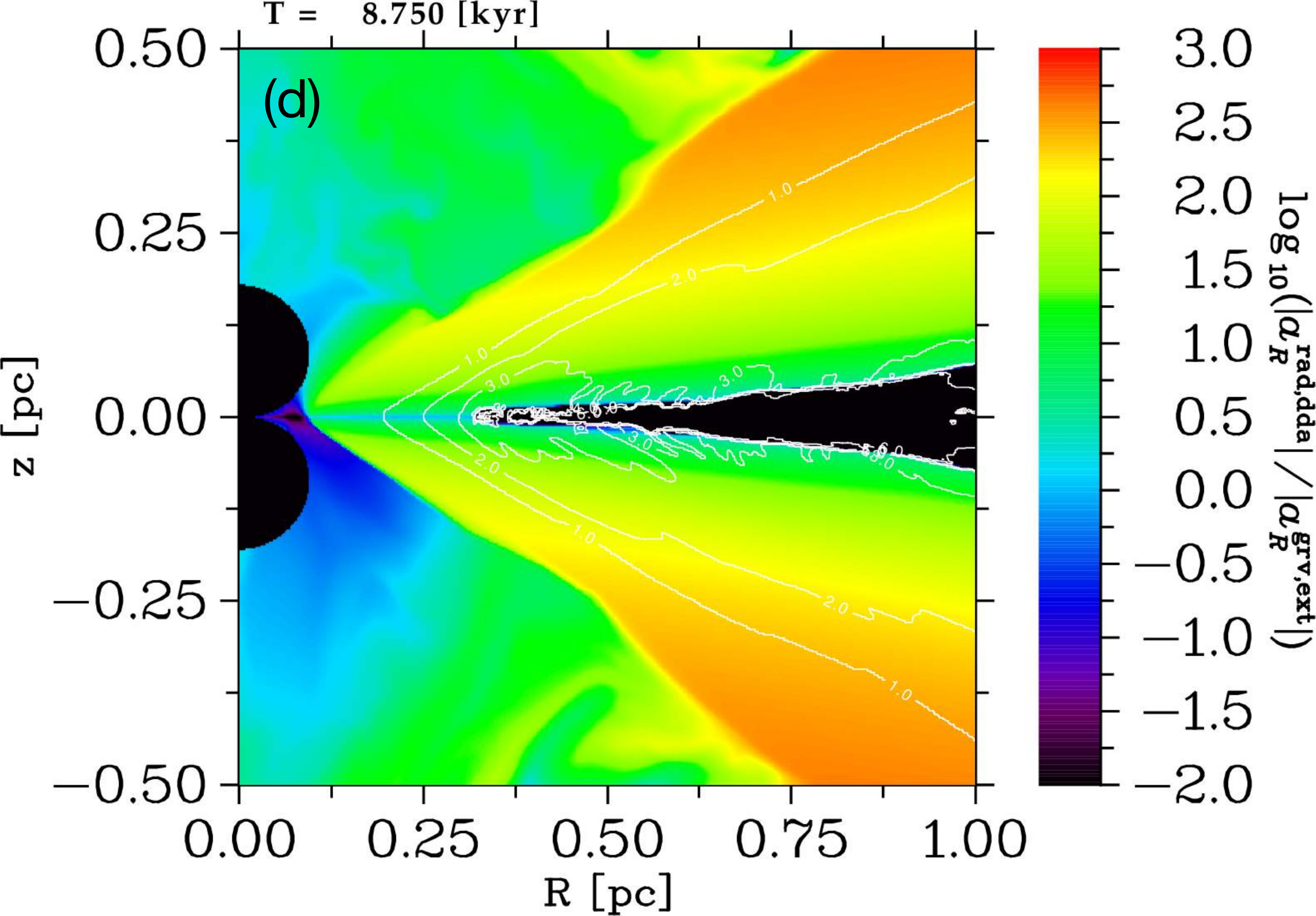}
\end{minipage}
\end{tabular}
\caption{Spatial distribution of the ratio of radiative acceleration due to particular processes to the acceleration of the external gravity in model \texttt{gra01\_wo\_sca\_SG} at $t=8.75\;\kyr$. The top panels show $a^{\mathrm{IR}}_{\mathrm{gr},z(R)}/|a^{\mathrm{grv,ext}}_{z(R)}|$ in linear scale and the lower panels show $\log_{10}(|a^{\mathrm{rad,drct}}_{R}|/|a^{\mathrm{grv,ext}}_{R}|)$ and $\log_{10}(|a^{\mathrm{rad,dda}}_{R}|/|a^{\mathrm{grv,ext}}_{R}|)$, where $a^{\mathrm{IR}}_{\mathrm{gr},z(R)}$ is the $z$($R$)-component of radiative acceleration due to IR photons, $a^{\mathrm{rad,drct}}_{R}$ is the $R$-component of radiative acceleration due to absorption of direct radiation, and $a^{\mathrm{rad,dda}}_{R}$ is the $R$-component of radiative acceleration due to \textit{dust} absorption of direct radiation. The white contours plot $\log_{10}\nH$ from $1$ to $7$ at intervals of $1$. Note that $\log_{10}(|a^{\mathrm{rad,drct}}_{z}|/|a^{\mathrm{grv,ext}}_{z}|)$ and $\log_{10}(|a^{\mathrm{rad,dda}}_{z}|/|a^{\mathrm{grv,ext}}_{z}|)$ are the same as $\log_{10}(|a^{\mathrm{rad,drct}}_{R}|/|a^{\mathrm{grv,ext}}_{R}|)$ and $\log_{10}(|a^{\mathrm{rad,dda}}_{R}|/|a^{\mathrm{grv,ext}}_{R}|)$, respectively, because all of $\bmath{a}^{\mathrm{rad,drct}}$, $\bmath{a}^{\mathrm{rad,dda}}$, and $\bmath{a}^{\mathrm{grv,ext}}$ are parallel to the radial unit vector $\hat{\bmath{e}}_{r}$.}
\label{fig:acceleration_ratios}
\end{figure*}

\begin{figure}
\centering
\includegraphics[clip,width=\linewidth]{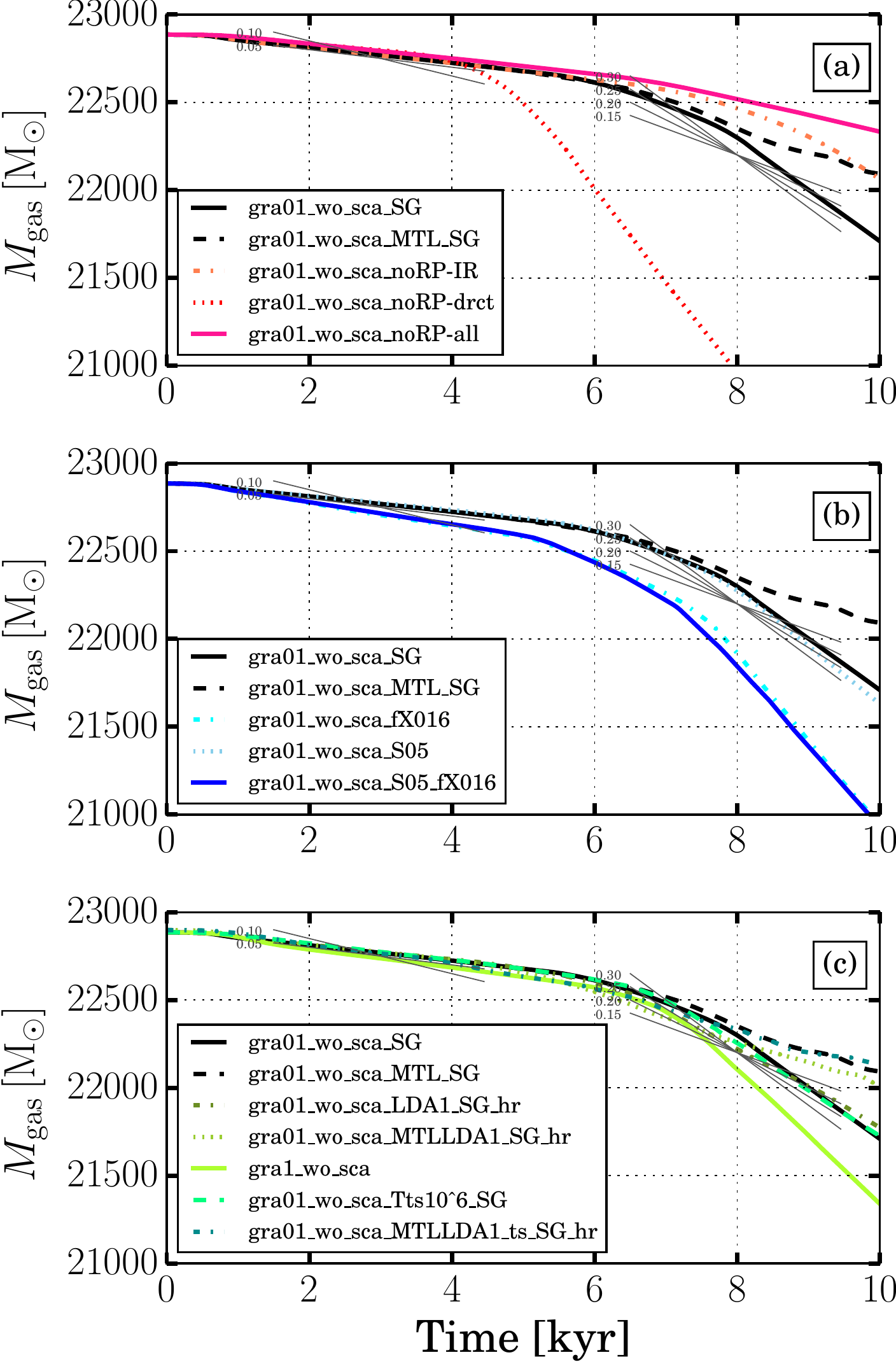}
\caption{Time evolution of gas mass in the computational box. The short black lines with numbers show constant values of $\diff M_{\mathrm{gas}}/\diff t$ in the unit of $\Msolar\;\yr^{-1}$ and are used to estimate the mass outflow rates.}
\label{fig:gas_mass_evolution}
\end{figure}

\subsection{The effects of cooling photons} \label{subsec:cooling_photon_effects}
\begin{figure}
\centering
\includegraphics[clip,width=\linewidth]{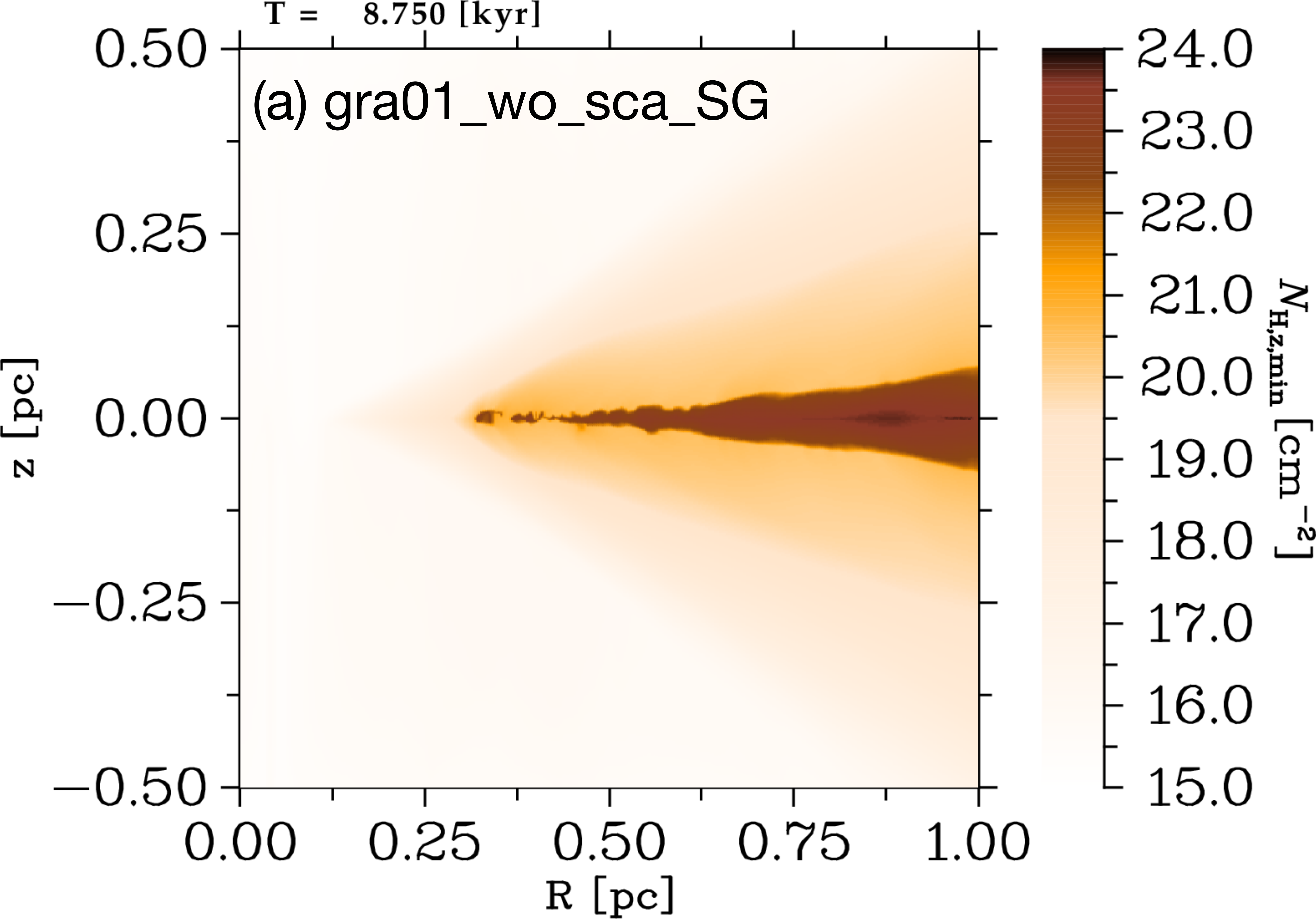}\\
\includegraphics[clip,width=\linewidth]{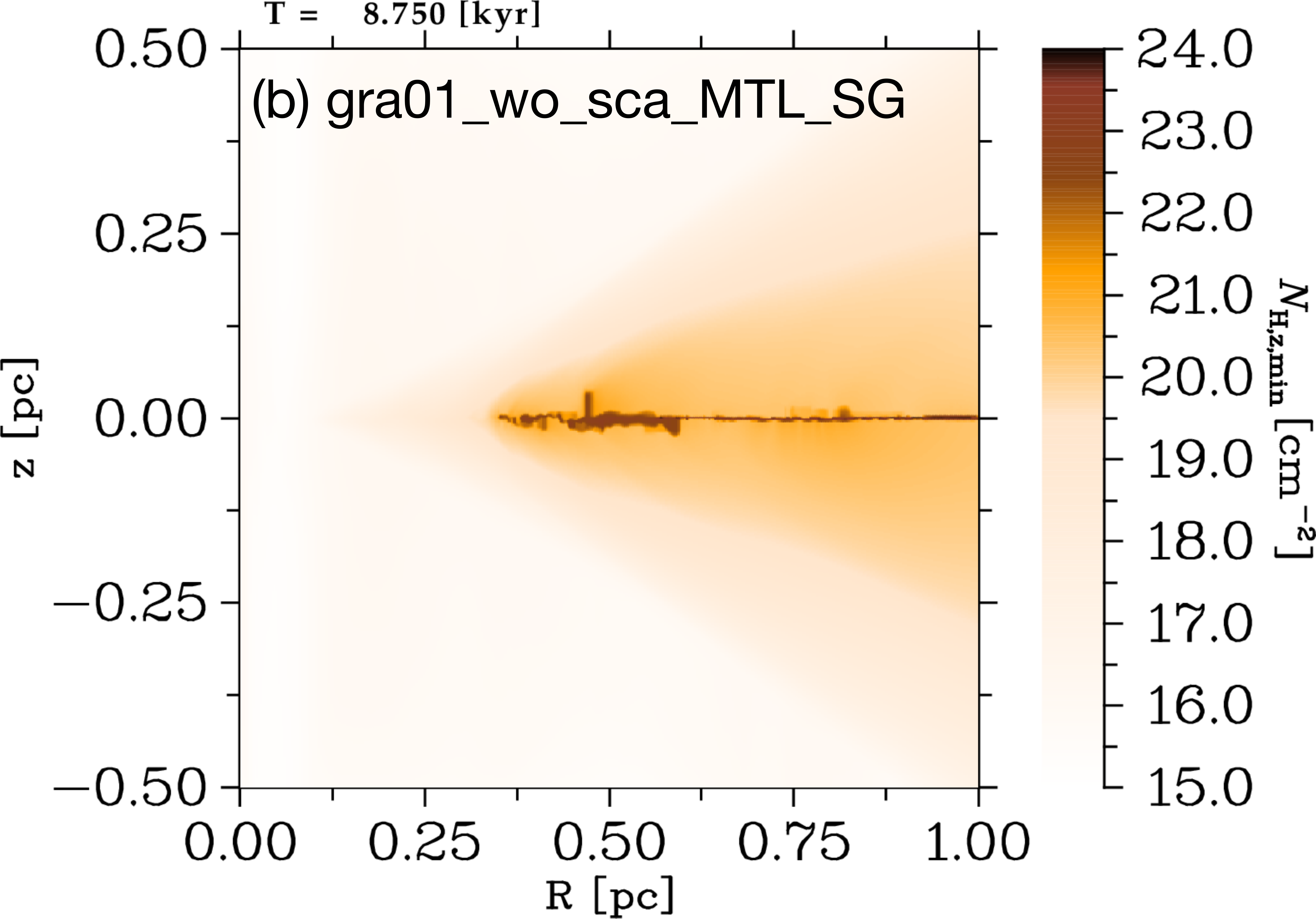}
\caption{Spatial distribution of minimum column density of hydrogen nuclei along the vertical direction at $t=8.75\;\kyr$ in the reference models (\textit{left}: \texttt{gra01\_wo\_sca\_SG}, \textit{right}: \texttt{gra01\_wo\_sca\_MTL\_SG}). $N_{\mathrm{H,z,min}}$ is defined as $\min[\int^{z_{\max}}_{z}\nH \diff z,\int^{z}_{z_{\min}}\nH \diff z]$, where $z_{\min(\max)}$ is the z coordinates at the lower (upper) boundaries of the computational box.}
\label{fig:NH_z_min}
\end{figure}

\begin{figure}
\centering
\includegraphics[clip,width=\linewidth]{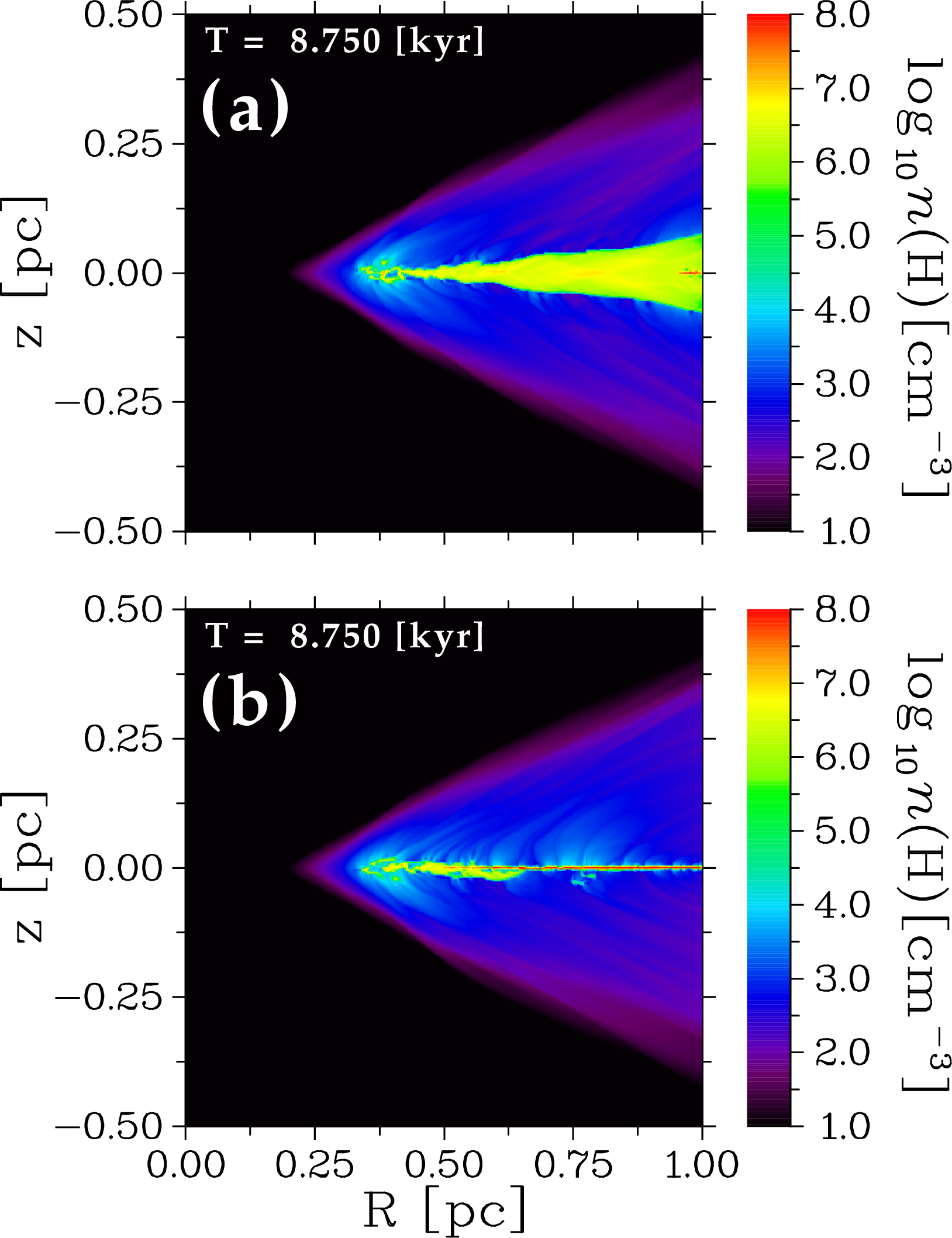}
\caption{Spatial distribution of hydrogen number density at $t=8.75\;\kyr$ in models \texttt{gra01\_wo\_sca\_LDA1\_SG} (\textit{left}) and \texttt{gra01\_wo\_sca\_MTLLDA1\_SG} (\textit{right}). In these models, the LDA approximation is used.}
\label{fig:nH_dists_LDA_models}
\end{figure}

As described in \S~\ref{subsec:radtr_chemistry}, we have assumed in the reference models that cooling photons from gas are optically-thin. However, hydrogen column density along the vertical direction is as high as $10^{23}\;\cden$ (Fig.~\ref{fig:NH_z_min}). Hence, the optically-thin approximation may not be appropriate. In order to check the effects of self-absorption of cooling photons, we perform simulations assuming that all of the cooling photons emitted by low temperature ($\Tgas\lesssim 3\times 10^{4}\;\mathrm{K}$), dense ($\nH\gtrsim 10^{5}\;\nden$) gas are \textit{locally} absorbed by dust grains\footnote{More specifically, we assume that a dust grain in each fluid cell is heated by absorption of cooling photons that arose in the cell at a rate $(\Lambda_{\mathrm{CP}}/\ngr) \exp(-\Tgas/T_{\mathrm{co}})\exp(-n_{\mathrm{co}}/\nH)$, where $\Lambda_{\mathrm{CP}}$ is the sum of cooling rates due to radiative processes that end with an emission of photons, $T_{\mathrm{co}}$ and $n_{\mathrm{co}}$ are the cutoff temperature and cutoff density, respectively. In this study, we assume $T_{\mathrm{co}}=3\times 10^{4}\;\mathrm{K}$ and $n_{\mathrm{co}}=10^{5}\;\nden$.}. By this, cooling photons are locally converted to IR photons and are transferred as IR photons. Thus, we can take into account radiation pressure due to cooling photons approximately. We call this approximation local dust absorption (LDA) approximation. Figure~\ref{fig:nH_dists_LDA_models} shows the density distributions at $t=8.75\;\kyr$ in the reference models \textit{with} the LDA approximation. By comparing this figure with Fig.~\ref{fig:dists_gra01_wo_sca_SG}(d) and Fig.~\ref{fig:dists_gra01_wo_sca_MTL_SG}(d), we find that the gas structure does not depend on the use of the LDA approximation, suggesting that the structure of the neutral thin disk does not change even if we perform the exact RT of cooling photons.

\subsection{The effects of spatial resolution} \label{subsec:resolution_effects}
\begin{figure*}
\centering
\includegraphics[clip,width=\linewidth]{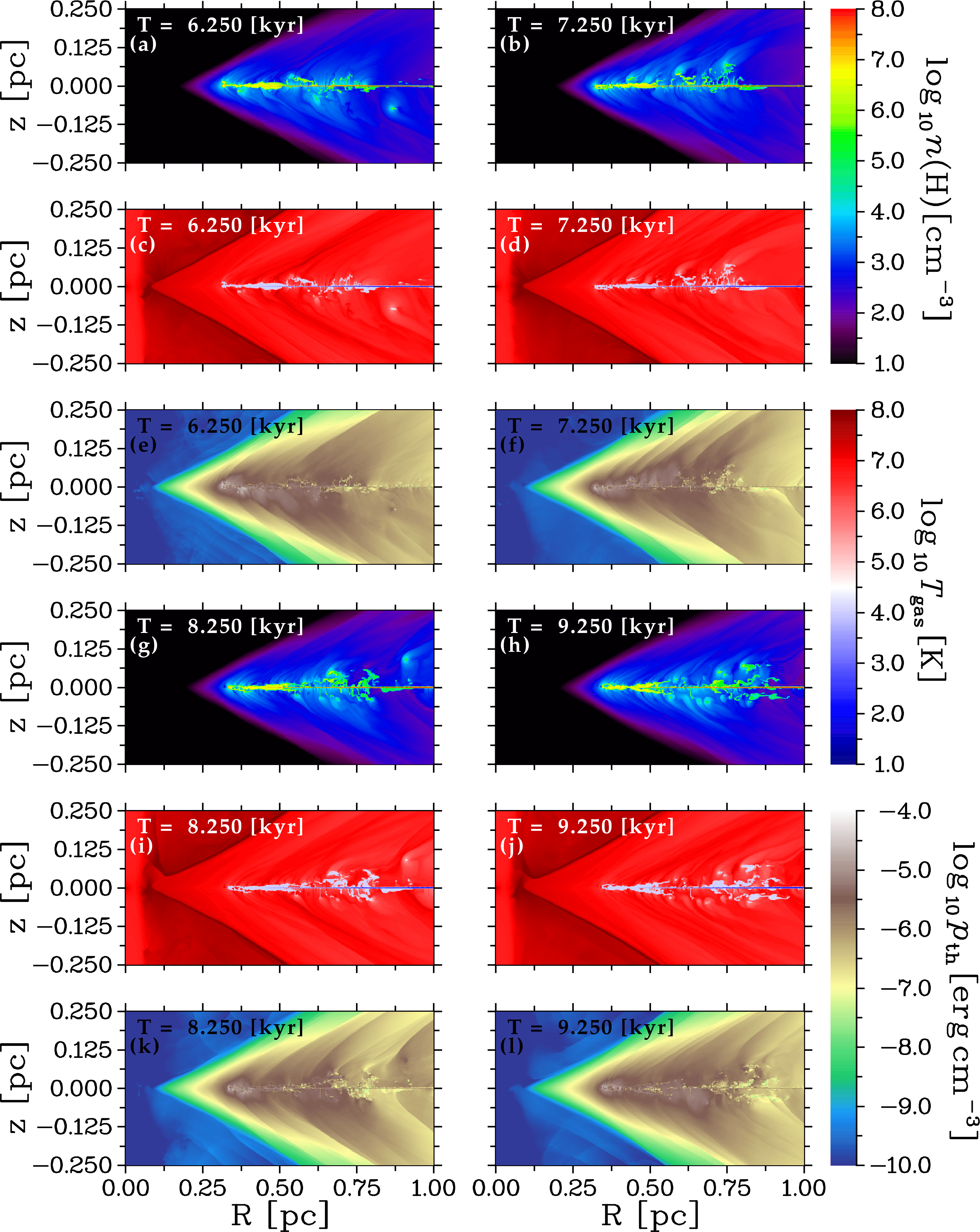}
\caption{Time evolution of the distribution of $\nH$, $\Tgas$, $\pth$ in model \texttt{gra01\_wo\_sca\_MTLLDA1\_SG\_hr}.}
\label{fig:dists_gra01_wo_sca_MTLLDA1_SG_hr}
\end{figure*}

\begin{figure}
\centering
\includegraphics[clip,width=\linewidth]{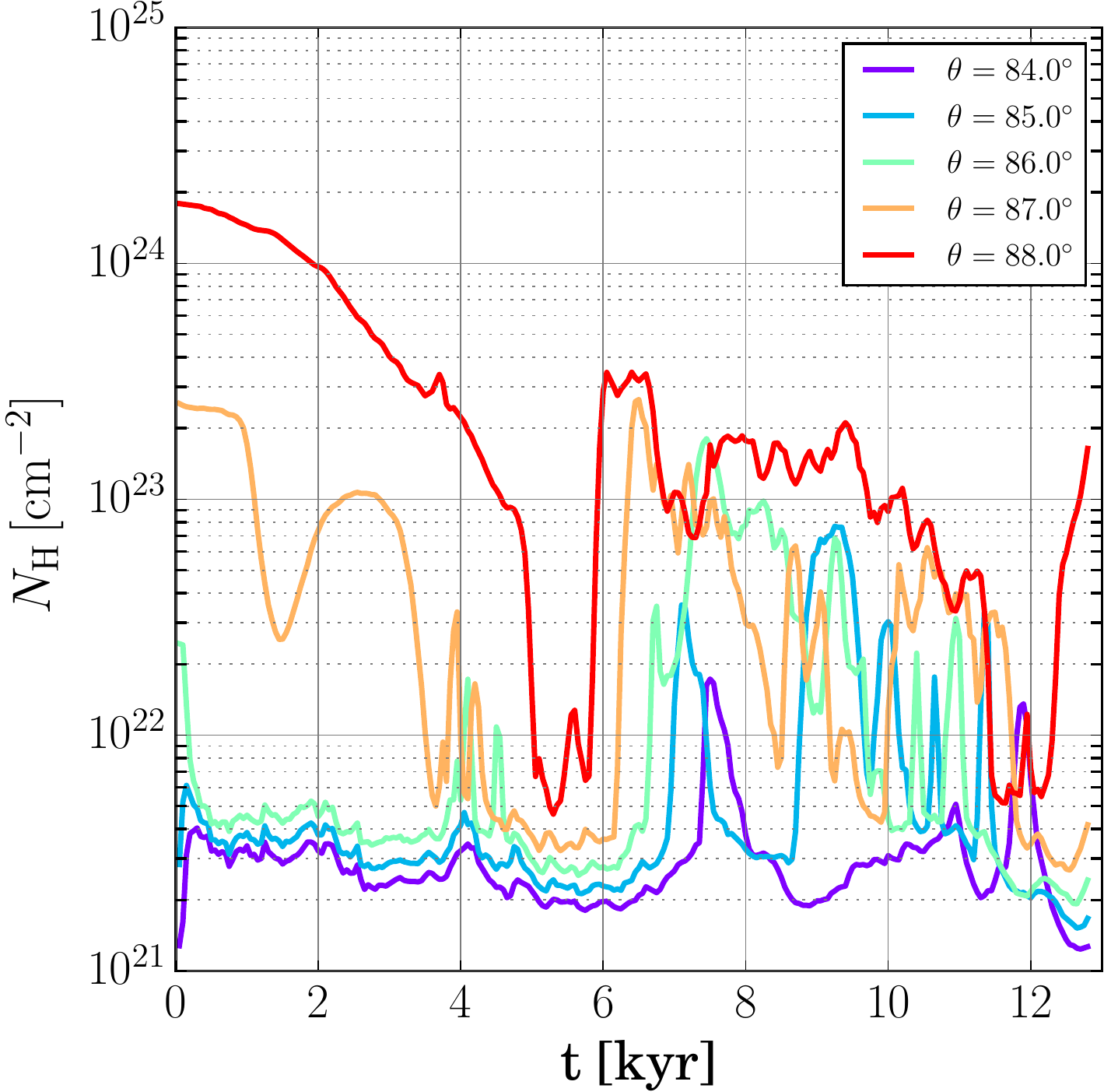}
\caption{Time evolution of hydrogen column density for different line-of-sights in model \texttt{gra\_wo\_sca\_MTLLDA1\_SG\_hr}. Note that a very thin disk forms at $t\approx 6\;\kyr$ and the formation of clumpy or chimney-like structures occurs after that.}
\label{fig:NH_evolution}
\end{figure}

In order to check the effects of spatial resolution on the gas structure and the gas flow, we perform simulations with high resolution (models \texttt{gra01\_wo\_sca\_LDA1\_SG\_hr} and \texttt{gra01\_wo\_sca\_MTLLDA1\_SG\_hr}). In the absence of metal cooling, the high resolution model shows time evolution very similar to that of the low resolution model, indicating that the numerical results converge. In contrast, there are differences in density structure between two different resolution models when metal cooling is taken into account. Figure~\ref{fig:dists_gra01_wo_sca_MTLLDA1_SG_hr} shows the time evolution of $\nH$, $\Tgas$, and $\pth$ during $t\approx 6\operatorname{-}10\;\kyr$ in model \texttt{gra01\_wo\_sca\_MTLLDA1\_SG\_hr}. After the formation of a very thin disk ($t\approx 6\;\kyr$), a number of clumpy or chimney-like structures repeatedly form immediately-above the disk surface until the end of the simulation (Fig.~\ref{fig:dists_gra01_wo_sca_MTLLDA1_SG_hr}a,b,g,h). This is not seen in the low resolution model. These clumpy structures have temperatures of $(1\operatorname{-}3)\times 10^{4}\;\mathrm{K}$ and are subject to strong thermal pressure of the outflowing gas (Fig.~\ref{fig:dists_gra01_wo_sca_MTLLDA1_SG_hr}e,f,k,l). These structures are not stable; they break up into smaller clumps probably because of the complex interplay of ram pressure, radiation pressure, and photoheating. Then, the clumps are blown off by both ram and radiation pressures. A part of the clumps is destroyed by both ram pressure stripping and photoevaporation in an early stage of acceleration. The other clumps gain altitude at first, but finaly become failed wind without having enough velocity. These failed winds do not perturb the structure of the neutral disk because the mass and the infall velocity of failed clumps are small. On one hand, these clumps increase the hydrogen column density in the outflow region (see the right panel of Fig.~\ref{fig:NH_dist}). As can be seen from Fig.~\ref{fig:NH_dist} and Fig.~\ref{fig:NH_evolution}, the column density of a group of clumps ranges from a few $\times 10^{22}\;\cden$ to a few $\times 10^{23}\;\cden$ and the column density for a particular light-of-sight show a rapid time variation with timescales of $\approx 10^{2\operatorname{-}3}\;\yr$.

A possible mechanism for the formation of clumpy structures may be coupled with thermal instability (e.g., \citealt{krolik81:_two,begelman90:_global,mckee90:_stead,rozanska99:_therm,calves07:_therm_x,czerny09:_therm_x,proga15:_cloud}), i.e., a transition from cold phase ($\Tgas\approx 10^{2\operatorname{-}3}\;\mathrm{K}$) to hot phase ($\Tgas\approx (1\operatorname{-}3)\times 10^{4}\;\mathrm{K}$), and the subsequent ejection of hot phase gas. The comparison between the high resolution model and the low resolution one suggests that the spatial resolution is important for such transition to occur. This can be understood by considering the effects of spatial resolution on the evolution of a moderately optically-thick gas parcel in the radially-perturbed neutral disk. If the gas parcel is not spatially resolved, the AGN just heats the whole of the gas parcel gently and the gas cannot become hot phase. On the other hand, if we are able to resolve the irradiated surface of the gas parcel and the shielded region behind it, the gas in the irradiated side could transition from cold phase to hot phase because of large photoheating rate. Thus, the spatial resolution is important to capture the correct radial variations of photoheating rate. To understand the detailed conditions of triggering active clump formation, a more detailed study is needed and we will address this in the future.

Outflow rate is nearly independent of spatial resolution regardless of the presence or absence of metal cooling (see Fig.~\ref{fig:gas_mass_evolution}), suggesting that the spatial resolution of $\approx 2\times 10^{-3}\;\pc$ is enough to measure accurately the outflow rate from a subparsec-scale disk.

\subsection{The effects of grain size} \label{subsec:grain_size_effects}
There are some observational suggestions that the average size of dust grains in AGNs is larger than that of normal ISM. For instance, \citet{maiolino01:_dust_i,maiolino01:_dust_ii} analyzed spectral properties of 19 Seyfert galaxies and found that their $E_{\mathrm{B-V}}/N_{\mathrm{H}}$ ratios are significantly lower than the Galactic value (by a factor of $3\operatorname{-}100$) and their spectra do not show significant carbon $2175\;\Angstrom$ absorption features expected from the Galactic extinction curve. They showed that these results can be understood if large grains dominate in AGN environment. \citet{lyu14:_dust} investigated the ratio of the visual extinction ($A_{\mathrm{V}}$) to $9.7\;\micron$ silicate absorption depth ($\Delta\tau_{9.7}$) for 110 type 2 AGNs and found that a mean value of $A_{\mathrm{V}}/\Delta\tau_{9.7}$ ($\lesssim 5.5$) is smaller than the Galactic value ($\approx 18.5$). They argued that this result could be explained by the predominance of larger grains in AGN tori. Other interesting discussion about dust properties in AGNs are found in \citet{lutz02:_infrar,maiolino02:_large,smith10:_anomal_liner_m81,xie14}.

Roughly speaking, dust absorption coefficient is inversely proportional to grain radius:
\begin{eqnarray}
\alpha_{\nu} & \equiv & \ngr\Qabs \pi \agr^{2} \\
& = & \frac{\rho Y(\mathrm{gr})}{\frac{4}{3}\pi \agr^{3}\rho_{\mathrm{gr}}} \times \Qabs \pi \agr^{2} \propto \agr^{-1},
\end{eqnarray}
where $Y(\mathrm{gr})$ is the dust-to-gas mass ratio, $\rho_{\mathrm{gr}}$ is the mass density of dust grains. Hence, radiation pressure on dust is smaller for larger grain. A different grain size could make a difference in the thickness of the neutral disk.

To examine the effects of grain size, we perform a simulation assuming $\agr=1\;\micron$ (model \texttt{gra1\_wo\_sca}) and its density distribution at $t=8.75\;\kyr$ is shown in Fig.~\ref{fig:nH_dists_various_models}(g). When compared with the result of model \texttt{gra01\_wo\_sca} (Fig.~\ref{fig:nH_dists_various_models}a), the outflowing gas attains a larger height (see also Fig.~\ref{fig:NH_dist}), but the thickness of the neutral disk is about the same. The outflow rate is a little bit larger than $\agr=0.1\;\micron$ cases (Fig.~\ref{fig:gas_mass_evolution}) because the confinement effect of photoevaporation flow by radiation pressure is reduced.

\subsection{The effects of thermal sputtering} \label{subsec:thermal_sputtering_effects}
We have noted in \S~\ref{subsec:overview} that small dust grains in the outflow region can be destroyed by thermal sputtering. Based on \citet{draine79}, the destruction timescale for graphite grain $\tau_{\mathrm{dest}}$ is given by
\begin{eqnarray}
\tau_{\mathrm{dest}} \approx 
\begin{cases}
700\;\kyr\left(\frac{\Tgas}{10^{5}\;\mathrm{K}}\right)^{-2.504}n^{-1}_{\mathrm{H},2}a_{\mathrm{gr},-1}, & \Tgas \lesssim 10^{6}\;\mathrm{K}, \\
3.09\;\kyr\left(\frac{\Tgas}{10^{6}\;\mathrm{K}}\right)^{-0.5}n^{-1}_{\mathrm{H},2}a_{\mathrm{gr},-1}, & \Tgas \approx 10^{6\operatorname{-}7}\;\mathrm{K}, \\
1.08\;\kyr\left(\frac{\Tgas}{10^{7}\;\mathrm{K}}\right)^{0.204}n^{-1}_{\mathrm{H},2}a_{\mathrm{gr},-1}, & \Tgas \gtrsim 10^{7}\;\mathrm{K}.
\end{cases}
\end{eqnarray}
where $n_{\mathrm{H},2}=\nH/10^{2}\;\nden$ and $a_{\mathrm{gr},-1}=\agr/0.1\;\micron$. Because the outflowing gas has a density of $\nH=10^{2\operatorname{-}3}\;\nden$, dust grains of radius $0.1\;\micron$ ($1\;\micron$) are destroyed after moving a distance of $0.1\;\pc$ ($1\;\pc$), if dust grains are embedded in a wind with velocity of $100\;\kms$. Thus, thermal sputtering should affect the dynamics of the outflowing gas at subparsec scale.

To examine the effects of thermal sputtering, we perform two simulations assuming that dust grains are instantaneously destroyed if $\Tgas>10^{6}\;\mathrm{K}$ (models \texttt{gra01\_wo\_sca\_Tts10\textasciicircum 6\_SG} and \texttt{gra01\_wo\_sca\_MTLDLA1\_ts\_SG\_hr}). The density distribution of the former model at $t=8.75\;\kyr$ is shown in Fig.~\ref{fig:nH_dists_various_models}(h). As shown in the figure, the outflowing gas extends largely in the vertical direction (see also Fig.~\ref{fig:NH_dist}). This is because there are no dust grains in the outflow region due to thermal sputtering, and as a result, the radiation pressure on the outflowing gas is significantly reduced. For the same reason, the outflow velocity is considerably reduced; $\approx 10^{3}\;\kms$ near the boundaries of the computational box. The density distribution is smoother than the cases where dust exists. By contrast, the thickness of the neutral disk is nearly unchanged. The outflow rate is also approximately same as that of the reference model \texttt{gra01\_wo\_sca\_SG} (Fig.~\ref{fig:gas_mass_evolution}), although the opacity of the outflowing gas is much smaller than that of model \texttt{gra01\_wo\_sca\_SG}. This is seemingly contradictory to the result found in model \texttt{gra1\_wo\_sca}, in which a lower opacity results in a higher outflow rate (\S~\ref{subsec:grain_size_effects}). These facts indicate that the confinement of photoevaporation or the reduction of photoevaporation rate due to radiation pressure are realized by absorption of direct photons in the thin surface layer of the neutral disk (the opacity in the neutral disk in model \texttt{gra01\_wo\_sca\_Tts10\textasciicircum 6\_SG} is higher than that in model \texttt{gra1\_wo\_sca}).

\begin{figure}
\centering
\includegraphics[clip,width=\linewidth]{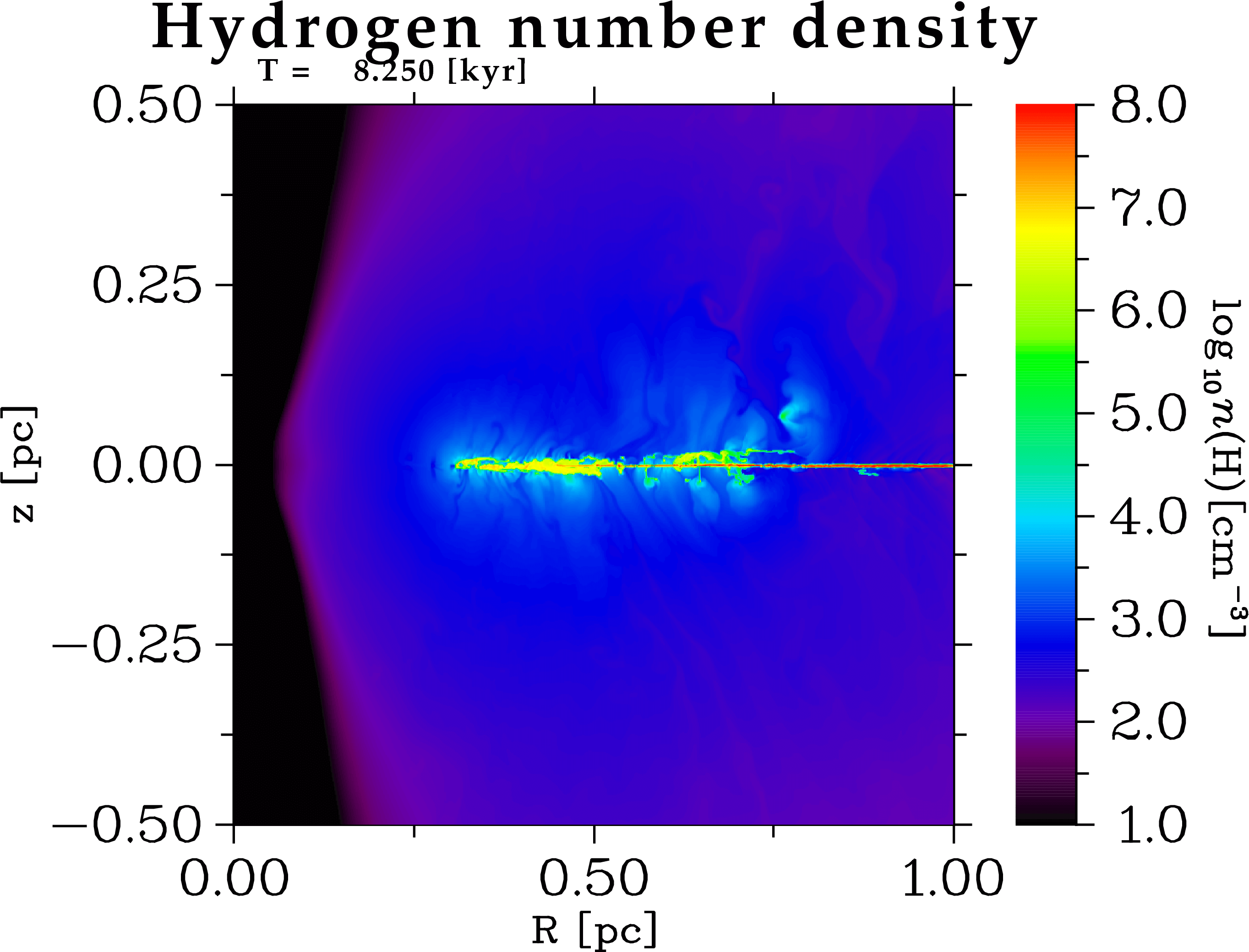}
\caption{Spatial distribution of $\nH$ at $t=8.25\;\kyr$ in model \texttt{gra01\_wo\_sca\_MTLLDA1\_ts\_SG\_hr}.}
\label{fig:nH_gra01_wo_sca_MTLLDA1_ts_SG_hr}
\end{figure}

Figure~\ref{fig:nH_gra01_wo_sca_MTLLDA1_ts_SG_hr} shows the number density distribution of model \texttt{gra01\_wo\_sca\_MTLLDA1\_ts\_SG\_hr} at $t=8.25\;\kyr$, which should be compared with Fig.\ref{fig:dists_gra01_wo_sca_MTLLDA1_SG_hr}(g). The distribution of the outflow is very similar to that of model \texttt{gra01\_wo\_sca\_Tts10\textasciicircum 6\_SG}. A number of clumps or chimneys form above the disk surface as in model \texttt{gra01\_wo\_sca\_MTLLDA1\_SG\_hr}, but it is less frequent. These newly-created clumps tend to pile up at the disk surface, probably because of the small ram pressure of the outflow. 

\subsection{The effects of Compton heating/cooling} \label{subsec:Compton_heatcool_effects}
\begin{figure}
\centering
\includegraphics[clip,width=\linewidth]{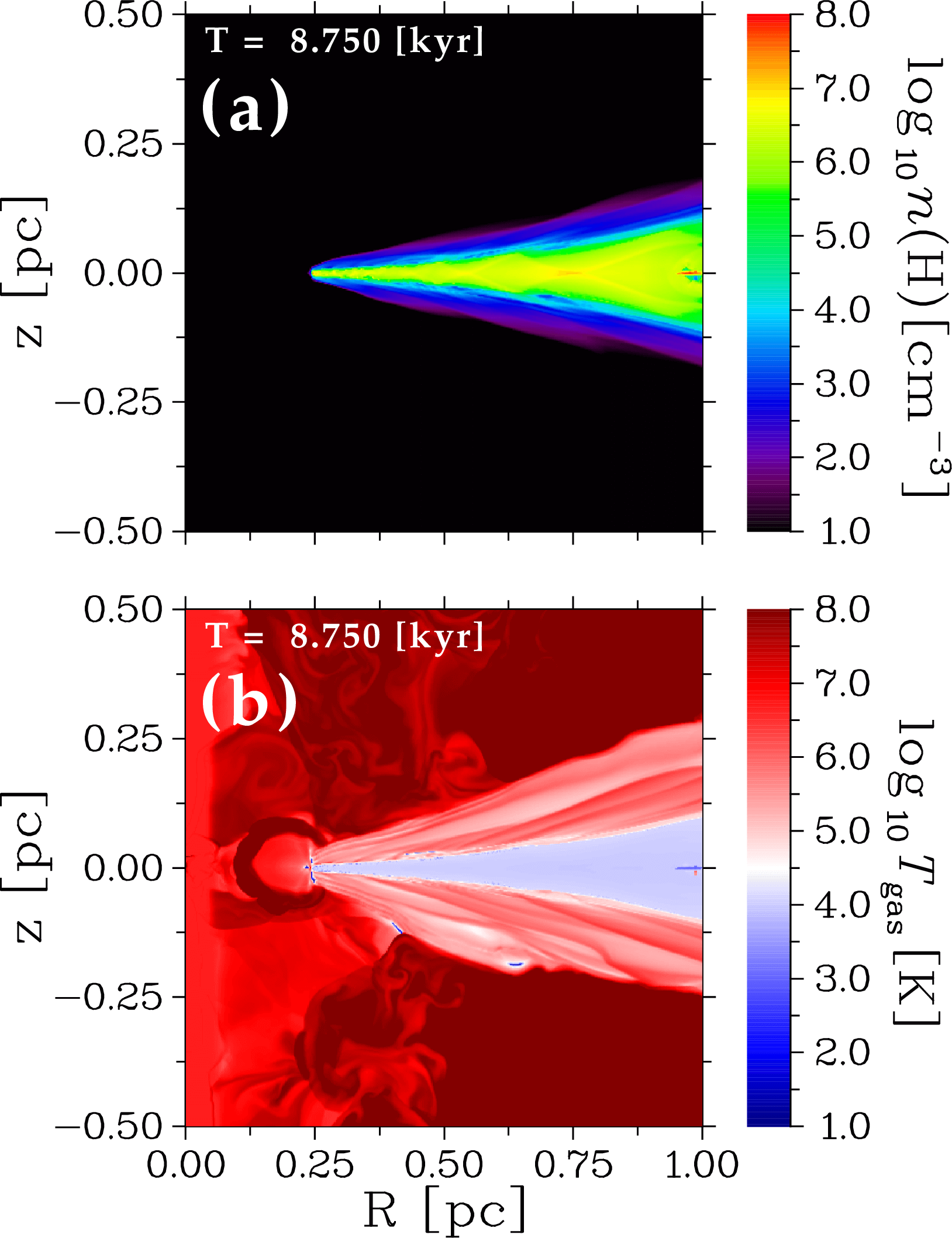}
\caption{Spatial distribution of $\nH$ and $\Tgas$ at $t=8.75\;\kyr$ in model \texttt{gra01\_wo\_sca\_noCompton\_SG}.}
\label{fig:dists_noCompton}
\end{figure}

It is useful to make clear the importance and the effects of X-ray heating, since some previous studies (e.g., \citealt{chan15:_radiat}) did not take into account X-ray heating. Here, we focus on the effects of Compton heating and Compton cooling only, because it is self-evident that switching off coronal emission leads to the reduction of scale height of the neutral disk. To this end, we perform a simulation in which Compton heating/cooling are switched off (model \texttt{gra01\_wo\_sca\_noCompton\_SG}). Figure~\ref{fig:dists_noCompton} shows the density and gas temperature distributions at $t=8.75\;\kyr$ in the model. Compared to the reference model \texttt{gra01\_wo\_sca\_SG}, the neutral disk is $\approx 1.6$ times thicker likely because the compressional effect of the neutral disk by thermal pressure of the outflowing gas is weaken due to lower gas temperature ($\Tgas\approx 10^{5\operatorname{-}6}\;\mathrm{K}$). On the other hand, the vertical extent of the outflow becomes small significantly. Thus, the outflow gas cannot reach a large height without Compton heating.

\section{Discussions} \label{sec:discussions}
In the present study, we have investigated the evolution of a dusty gas disk in regions near the dust sublimation radius for the case of $\MBH=10^{7}\;\Msolar$ and $\Lbol/L_{\mathrm{Edd}}\approx 0.77$ and have shown that the quasi-steady state of the dusty gas disk takes the form of a two-layer structure: a neutral, geometrically-thin, dense disk and a high-velocity outflow that is launched from the disk surface. The existence of metal cooling dramatically changes the disk thickness as shown in \S~\ref{subsec:overview}. However, at this time, it may be premature to conclude that the case with metal cooling is more realistic, because the photoionization of metal is not included in our simulations. But, we can safely say that the disk thickness is likely to become as small as $h/R < 0.06$, where $h$ is the vertical scale height of the disk. The dust content in the outflow strongly affects the covering factor of the outflow as shown in \S~\ref{subsec:overview},\ref{subsec:grain_size_effects}, and \ref{subsec:thermal_sputtering_effects} (see also Fig.~\ref{fig:NH_dist}). We have also shown that irradiation by an AGN and radiation pressure due to IR photons reemitted by dust grains \textit{alone} cannot form a geometrically thick, obscuring structure in the regions and that the outflowing gas does not provide a sufficient column density at least within $r\approx 1\;\pc$ (Fig.~\ref{fig:NH_dist}). These results do not depend largely on the various assumptions adopted in this study or the presence or the absence of a particular physical process as shown in \S~\ref{subsec:cooling_photon_effects}-\S~\ref{subsec:Compton_heatcool_effects}.

In the following, we first discuss uncertainties of the gas temperatures in both the outflow region and the neutral disk by comparing our results with the results of photoionization calculations by the \textsc{Cloudy} code (\S~\ref{subsec:uncertainties}). Next, we discuss the relations between our study and previous studies (\S~\ref{subsec:previous_studies}). Then, we examine other possible mechanisms that inflate the neutral disk and give implications for AGN torus (\S~\ref{subsec:implications}).

\subsection{Uncertainties} \label{subsec:uncertainties}
In this study, we have made several assumptions on the numerical treatments of radiative processes and RT in order to reduce computational cost. For instance, we have taken into account the effects of metals in the form of cooling function without solving ionization/recombination reactions of metals directly and we have applied the cooling function only for low temperature gas. Also, we have neglected multiple Compton scattering of direct photons and Compton scattering of IR photons. These simplifications may cause a large deviation of gas temperature from its true value. To check this possibility, we perform photoionization calculations using the \textsc{Cloudy} and compare the resultant profiles of gas temperature with that of our numerical results. In the calculations shown below, we make the following assumptions:
\begin{enumerate}[topsep=0pt,leftmargin=*]
\item The gas is assumed to be at the rest.
\item The hydrogen number density distribution is the same as that along $\theta=88.854^{\circ}$ at $t=8.75\;\kyr$ in the reference model \texttt{gra01\_wo\_sca\_SG} (see the magenta dotted line in Fig.~\ref{fig:dists_gra01_wo_sca_SG}d).
\item The incident AGN SED is also the same as that along $\theta=88.854^{\circ}$ in the reference model.
\item We assume that dust consists of graphite and has a radius of $0.1\;\micron$.
\item We assume the chemical composition shown in Table~\ref{tbl:chemical_composition}.
\end{enumerate}

\begingroup
\renewcommand{\arraystretch}{1.0}
\begin{table}
\centering
\begin{minipage}{\hsize}
\caption{Chemical composition assumed in the photoionization calculations described in \S~\ref{subsec:uncertainties}. All the abundance is given in $\log_{10}[n(X)/\nH]$, where $n(X)$ is the number density of element $X$. The abundance pattern is similar to the \textsc{Cloudy}'s abundance set \texttt{ISM} for the case of $Z=4\;\Zsolar$. The main difference between them is that most of carbon is assumed to deplete into graphite grains in this study. This is needed to make the dust-to-gas mass ratio be $0.01$. Note that beryllium and scandium are turned off in the photoionization calculations.}
\label{tbl:chemical_composition}
\begin{tabular}{@{}llll@{}}
\hline
Element & Abundance & Element & Abundance \\
\hline
\ding{110} \textbf{Gas-phase} &&& \\
H  &  0.0000 & S  & -3.8874 \\
He & -0.4067 & Cl & -6.3979 \\
Li & -9.6655 & Ar & -4.9477 \\
B  & -9.4485 & K  & -7.3565 \\
C  & -7.1761 & Ca & -8.7851 \\
N  & -3.4981 & Ti & -8.6345 \\
O  & -2.7445 & V  & -9.3979 \\
F  & -7.0969 & Cr & -7.3979 \\
Ne & -3.3080 & Mn & -7.0362 \\
Na & -5.8982 & Fe & -3.8738 \\
Mg & -3.7409 & Co & -7.6270 \\
Al & -6.4981 & Ni & -7.1378 \\
Si & -3.8421 & Cu & -8.2218 \\
P  & -6.1938 & Zn & -7.0969 \\
\hdashline[0.4pt/1pt]
\ding{110} \textbf{Grain} &&& \\
C & -2.6742 && \\
\hline
\end{tabular}
\end{minipage}
\end{table}
\endgroup

\begin{figure*}
\centering
\includegraphics[clip,width=\linewidth]{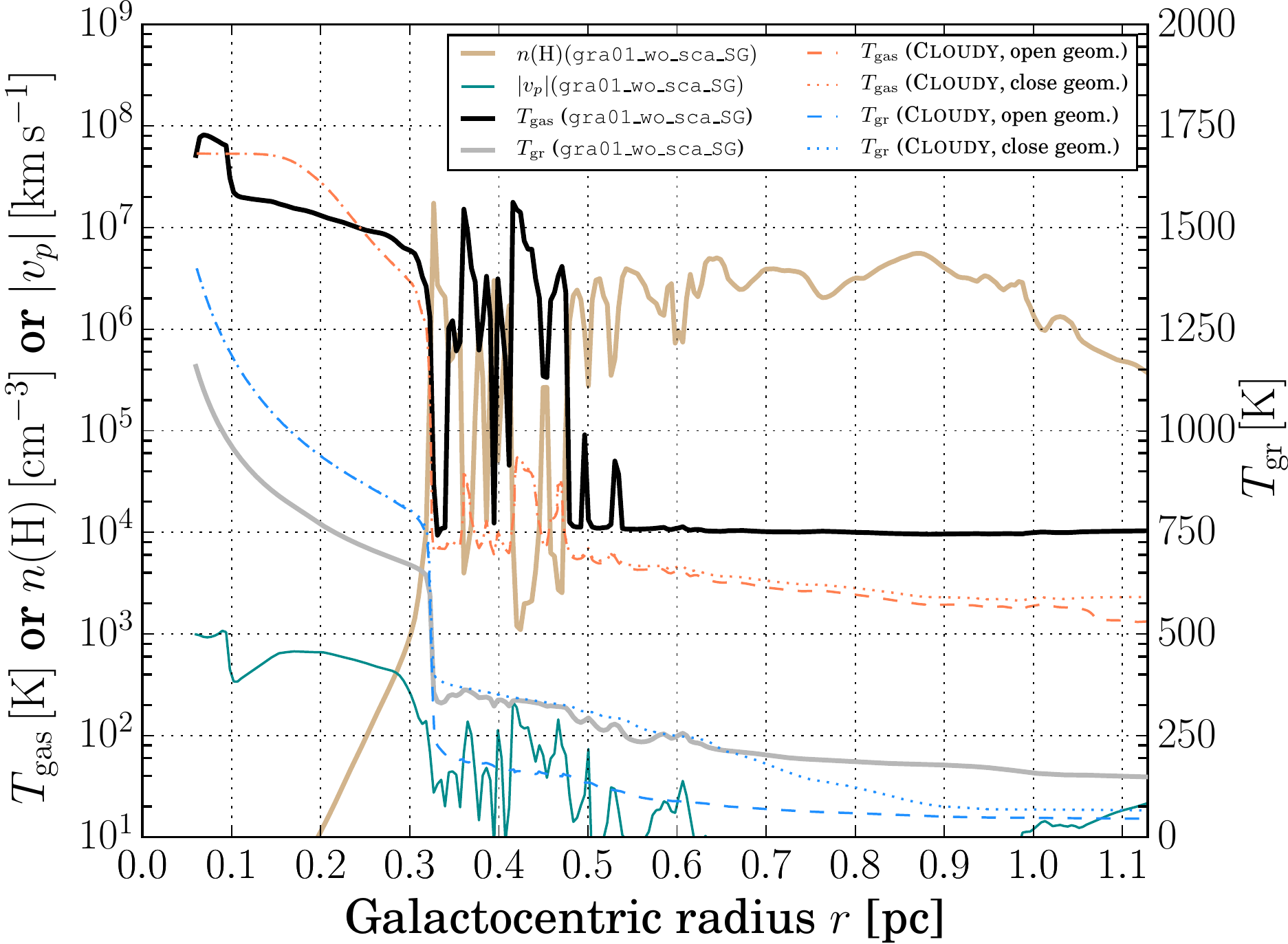}
\caption{A comparison of the spatial distributions of gas and dust temperatures along the line-of-sight $\theta=88.854^{\circ}$ in the reference model \texttt{gra01\_wo\_sca\_SG} at $t=8.75\;\kyr$ with those obtained by the photoionization calculations. The spatial distribution of $\nH$, $\Tgas$, $\Tgr$, and poloidal velocity $|v_{p}|$ in the reference model are shown by the brown line, the black line, the gray line, and the darkgreen line, respectively. $\Tgas$ and $\Tgr$ in the photoionization calculations are shown by the orange lines and the light blue lines, respectively.}
\label{fig:temperature_comparison}
\end{figure*}

We perform two photoionization calculations. One is with open geometry and another is with closed geometry. In the latter, it is assumed that an irradiated gas surrounds the AGN and photons emitted by the gas at one side interact with the gas at the opposite side of the AGN. The spatial distributions of gas and dust temperatures obtained by the calculations are shown in Fig.~\ref{fig:temperature_comparison}. The two photoionization calculations predict similar results except that dust temperature at $r\approx 0.3\operatorname{-}0.9\;\pc$ is larger in the closed geometry case than in the open geometry case. The difference arises because dust in the region of $r\approx 0.3\operatorname{-}0.9\;\pc$ is heated by absorbing photons reradiated by dust at the opposite side of the AGN. Note that the gas temperature is completely different from the dust temperature, suggesting that the assumption of $\Tgas=\Tgr$ used both in \citet{chan15:_radiat} and \citet{dorodnitsyn15:_parsec} does not hold.

In Fig.~\ref{fig:temperature_comparison}, we also show the gas and dust temperatures in the reference model. In a low density region near the AGN ($r\lesssim 0.3\;\pc$), the gas temperature in the reference model are generally consistent with the results of the photoionization calculations, indicating that gas temperatures in low density regions in our simulations are sufficiently accurate (i.e., the neglect of multiple Compton scattering of direct photons and Compton scattering of IR photons does not lead to a large deviation). By contrast, there is a systematic difference (of the order of $200\;\mathrm{K}$) in dust temperature in that region. This is perhaps due to the difference in treatment of gas-dust collision in a hot gas between our code and the \textsc{Cloudy} code\footnote{This interpretation is supported by the following circumstantial evidence: (i) Our code and the \textsc{Cloudy} code use the same dust optical constants, and hence heating rate due to direct radiation is also the same; (ii) The two photoionization calculations predict almost the same dust temperatures in this low density region, suggesting that photons from the opposite side do not play a role in heating of dust in the low density region; (iii) Heating by optical/UV photons backscattered at $R\approx 0.3\;\pc$ (our code neglects dust scattering of direct radiation) is not a factor because there is essentially no difference between the two photoionization calculations. In other words, if backscattered photons are one of dominant sources of heating, a difference must arise in dust temperature between the two photoionization calculations since the low density region is optically-thin ($N_{\mathrm{H}}\approx 10^{19\operatorname{-}20}\;\cden$), and therefore, optical/UV photons arisen at the opposite side should transmit the low density region at that side and then heat the dust at this side). The remaining possibility is that the systematic difference in dust temperature is due to the difference in the numerical treatment of gas-dust collisional energy transfer between our code and the \textsc{Cloudy} code. In fact, the \textsc{Cloudy} code calculates the rate of energy transfer in a more sophisticated way, in which various microscopic physical processes that are not taken into account in our study are self-consistently solved (see the function \texttt{GrainCollHeating} in the source code \texttt{grains.cpp}). Unfortunately, it is difficult to confirm the above possibility because the \textsc{Cloudy} code cannot output the heating rate of dust due to collision with gas particles.}.

In an intermediate region ($r\approx 0.3\operatorname{-}0.5\;\pc$), where $\nH$ changes rapidly, there are several peaks in gas temperature both in the reference model and the photoionization calculations and the temperature peaks in the reference model is much larger than those in the photoionization calculations. This is because the gas in the reference model is strongly heated by shock heating. This is confirmed by the facts that all the large temperature jumps associate with the discontinuities of $\nH$ and $|v_{p}|$ and that the sizes of the discontinuities in $|v_{p}|$ are of the order of several tens of $\kms$ to $200\;\kms$, which correspond to the postshock temperatures of $\approx 10^{5\operatorname{-}7}\;\mathrm{K}$. The local minimum values of gas temperature in the reference model are $\approx 10^{4}\;\mathrm{K}$, which are in good agreement with those in the photoionization calculations ($\Tgas\approx 8000\;\mathrm{K}$). On the other hand, the dust temperature in the reference model agrees well with that of the photoionization calculation with the closed geometry.

In the outer region ($r\gtrsim 0.5\;\pc$), the gas temperature in the reference model is virtually constant at $\Tgas\approx 10^{4}\;\mathrm{K}$, while the photoionization calculations predict that it decreases from $\Tgas\approx 8000\;\mathrm{K}$ at $r\approx 0.5\;\pc$ to $\Tgas\approx (1\operatorname{-}2)\times 10^{3}\;\mathrm{K}$ at $r=1.1\;\pc$, indicating the importance of metal cooling. Note that this is an expected deviation because metal cooling is not taken into account in the reference model \texttt{gra01\_wo\_sca\_SG}. The dust temperature in the region of $r\gtrsim 0.7\;\pc$ is larger than that in the photoionization calculation with the closed geometry. This is likely because the dust in the reference model is heated by IR photons reemitted by the outflowing gas.

In summary, this comparison tells us (i) that only unexpected deviation is the systematic difference in dust temperature in the low density region ($r\lesssim 0.5\;\pc$) and (ii) that metal cooling does lower gas temperature in a moderately obscured ($N_{\mathrm{H}}\gtrsim 10^{22}\;\cden$), high density ($\nH\approx 10^{6}\;\nden$) gas at the regions of $r\gtrsim 0.5\;\pc$, indicating that we must include the effects of metal cooling in some form to obtain the true structure of the neutral disk. This result confirm that the true scale height of the neutral disk is certainly intermediate between those of the two reference models \texttt{gra01\_wo\_sca\_SG} and \texttt{gra01\_wo\_sca\_MTL\_SG}, which give upper and lower limits of the true scale height.

\subsection{Relations to previous studies} \label{subsec:previous_studies}
In this section, we discuss relations to previous studies on RHD modeling of AGN tori. A list of physical processes taken into account in previous studies is shown in Table~\ref{tbl:comparison_of_recent_studies}.

The formation of two-layer structure was also reported in the RHD simulations by \citet{dorodnitsyn12:_activ_galac_nucleus_obscur_winds}, although they considered an extremely dense gas disk. As shown their Figs.~1, 3, and 5, the dense disk in their simulations is actually geometrically-thin. Considering together with our results, it may suggest that a thin dense disk forms in the regions near the dust sublimation radius for a wide range of conditions. This result is contrary to the results of \citet{krolik07:_agn} and \citet{shi08:_radiat_x}, in which they showed that a geometrically-thick, radiation-supported, hydrostatic structure can be formed in a region near an AGN. The averaged outflow rate found in our study ($0.05\operatorname{-}0.1\;\Msolar\;\yr^{-1}$) is in good agreement with that reported by \citet{dorodnitsyn12:_activ_galac_nucleus_obscur_winds} ($0.1\operatorname{-}0.2\;\Msolar\;\yr^{-1}$). Contrary to our study, they argued that AGN obscuration is realized at parsec-scales by a dust wind and that the critical angle at which the optical depth for Thomson scattering is unity is $\theta=72^{\circ}\operatorname{-}75^{\circ}$. However, the outflowing gas in their simulations has an extremely high density of the order of $\nH\approx 10^{10\operatorname{-}12}\;\nden$, which is much higher than the critical density of self-gravitational instability. Thus, the outflow structure shown in their simulations is totally unstable and it it not clear that AGN obscuration at parsec-scales is real.

\begin{figure}
\centering
\includegraphics[clip,width=\linewidth]{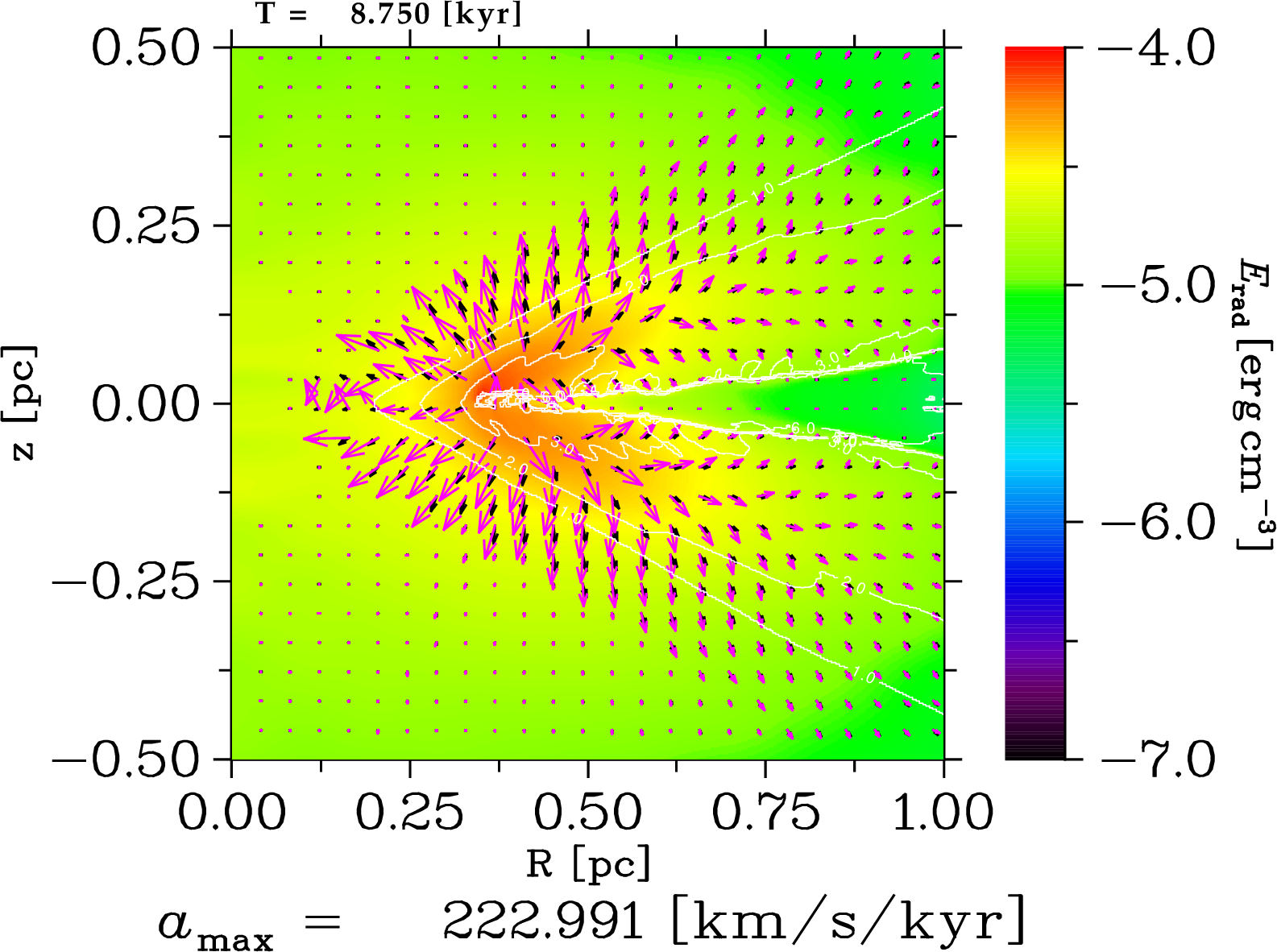}
\caption{Comparison between radiative accelerations computed by the finite-volume method (FVM) ($\bmath{a}^{\mathrm{FVM}}_{\mathrm{IR}}$; black arrows) and by the flux-limited diffusion (FLD) approximation ($\bmath{a}^{\mathrm{FLD}}_{\mathrm{IR}}$; magenta arrows). The colors and the white contours show the radiation energy density and the hydrogen number density of the reference model (\texttt{gra01\_wo\_sca\_SG}) at $t=8.75\;\kyr$, respectively. The largest arrow corresponds to $223\;\kms\;\kyr^{-1}$. We calculate $\bmath{a}^{\mathrm{FLD}}_{\mathrm{IR}}$ by $\bmath{a}^{\mathrm{FLD}}_{\mathrm{IR}}=-\rho^{-1}\Lambda(R)\nabla E_{\mathrm{rad}}$, where $\Lambda(R)=(2+R)/(6+3R+R^{2})$ (\citealt{levermore81:_flux_diffus_theor}), $R=|\nabla E_{\mathrm{rad}}|/(\chi^{\mathrm{abs}}_{\mathrm{R}} E_{\mathrm{rad}})$, and $\chi^{\mathrm{abs}}_{\mathrm{R}}$ is the Rosseland-mean absorption coefficient.}
\label{fig:FVMRT_vs_FLDapprox}
\end{figure}

Just recently, gas structure around an AGN with lower Eddington ratios ($\Lbol/L_{\mathrm{Edd}}\lesssim 0.1$) were investigated by \citet{chan15:_radiat} and \citet{dorodnitsyn15:_parsec} using RHD simulations and both studies showed that a geometrically-thick, obscuring structure can be formed in a region near the dust sublimation radius. We cannot directly compare our results with their results because of different Eddington ratio. Nevertheless, it is useful to go through the influences of the assumptions adopted in these two studies. As described in the introduction, \citet{chan15:_radiat} assumed that the AGN radiates isotropically and they did not take into account the effects of X-ray heating. In addition, they used artificially-reduced dust opacity and assumed $\Tgas=\Tgr$. The isotropic AGN radiation allows for a torus to receive more radiation energy than the anisotropic AGN radiation does. The lack of X-ray heating tends to result in a thicker torus, because the confinement effect due to hot gas does not work (see \S~\ref{subsec:Compton_heatcool_effects}). The use of reduced opacity also weakens the confinement effect due to direct radiation (cf. \S~\ref{subsec:overview}) and allows the interior of the disk to be heated \footnote{In \S~4.3 in their paper, \citet{chan15:_radiat} explored the effects of reduced opacity by performing a simulation assuming a UV-to-IR opacity ratio twice larger than their fiducial value $\kappa_{\mathrm{UV}}/\kappa_{\mathrm{IR}}=4$. However, the ``twice'' is insufficient because the true ratio is much larger as they described in their paper.}. The assumption $\Tgas=\Tgr$ forces gas to cool at a rate proportional to $\Tgas^{4}$. As a result, gas pressure is underestimated and radiation pressure is overestimated. As shown in \S~\ref{sec:numerical_results} and \ref{subsec:uncertainties}, $\Tgas\neq \Tgr$ in general. Thus, all the assumptions seem to lead to a thicker torus. \citet{dorodnitsyn15:_parsec} assumed the isotropic AGN radiation and a large X-ray luminosity fraction ($0.5$). The latter allows for deeper parts of the torus to be heated strongly, and thus, it is possible that this assumption operates in favor of the formation of a thicker torus. In their simulations, the flux-limited diffusion (FLD) approximation is used. In order to examine the influence of the FLD approximation, we compare the radiative acceleration computed by the finite-volume method (FVM) with that computed by the FLD approximation. The result is shown in Fig.~\ref{fig:FVMRT_vs_FLDapprox}. This figure shows that the radiative acceleration computed by the FLD approximation deviates from that computed by the FVM in regions with large $|\nabla E_{\mathrm{rad}}|/E_{\mathrm{rad}}$. This deviation may affect the covering factor of the outflowing gas, and possibly, the structure of the neutral disk. It is unclear what gas structure forms in the regions near the dust sublimation radius in lower Eddington ratio cases ($\Lbol/L_{\mathrm{Edd}}\lesssim 0.1$) when all the assumptions discussed above are removed. We will investigate this in the next paper.

\begingroup
\renewcommand{\arraystretch}{1.0}
\begin{table*}
\centering
\begin{minipage}{\hsize}
\caption{A comparison of physical processes and assumptions adopted in recent studies}
\label{tbl:comparison_of_recent_studies}
\begin{tabularx}{\hsize}{XXXXX}
\hline
 Physical effect  & This work & DK12$^{\dagger}$ & CK15$^{\dagger}$ & DKP15$^{\dagger}$ \\
\hline
Direct radiation
& Yes
\begin{itemize}[nosep,leftmargin=*]
\item Ray-trace
\item A realistic AGN SED and its frequency dependency
\end{itemize}
& Yes
\begin{itemize}[nosep,leftmargin=*]
\item Ray-trace
\end{itemize}
& Yes
\begin{itemize}[nosep,leftmargin=*]
\item Ray-trace
\end{itemize}
& Yes
\begin{itemize}[nosep,leftmargin=*]
\item Ray-trace
\end{itemize} \\
\hdashline[0.4pt/1pt]
IR photons
& Yes
\begin{itemize}[nosep,leftmargin=*]
\item FVM w/ $c=\infty$ approx. (direct solver for time-independent RT Eq.)
\item Frequency dependency
\end{itemize}
& Yes
\begin{itemize}[nosep,leftmargin=*]
\item FLD approx.
\end{itemize}
& Yes
\begin{itemize}[nosep,leftmargin=*]
\item Direct solver for time-dependent RT Eq. (\citealt{jiang14})
\end{itemize}
& Yes
\begin{itemize}[nosep,leftmargin=*]
\item FLD approx.
\end{itemize} \\
\hdashline[0.4pt/1pt]
Anisotropy of AD radiation  &  Yes                     & No                     & No & No \\
\hdashline[0.4pt/1pt]
X-ray from AD corona        &  Yes ($\fX \approx 0.1$)  & Yes ($\LX/\Lbol=0.5$)$^{\ddagger}$ & No & Yes ($\LX/\Lbol=0.5$)$^{\ddagger}$ \\
\hdashline[0.4pt/1pt]
Separate handling of $\Tgas$ and $\Tgr$
& Yes
& $\triangle$
& No
\begin{itemize}[nosep,leftmargin=*]
\item $\Tgas=\Tgr$ at all times
\end{itemize}
& $\triangle$ \\
\hdashline[0.4pt/1pt]
Self-gravity
& Yes
& No
& No
& No \\
\hdashline[0.4pt/1pt]
Others
&
\begin{itemize}[nosep,leftmargin=*]
\item Rotation-supported disk
\end{itemize}
&
\begin{itemize}[nosep,leftmargin=*]
\item Extremely high gas density
\end{itemize}
&
\begin{itemize}[nosep,leftmargin=*]
\item Sub-Kepler rotation disk
\end{itemize}
& \\
\hline
\end{tabularx}
\end{minipage}
\begin{flushleft}
$^{\dagger}$ DK12 --- \citet{dorodnitsyn12:_activ_galac_nucleus_obscur_winds}; CK15 --- \citet{chan15:_radiat}; DKP15 --- \citet{dorodnitsyn15:_parsec}. \\
$^{\ddagger}$ They have not given the exact definition of X-ray luminosity $\LX$.
\end{flushleft}
\end{table*}
\endgroup

\subsection{Other possible effects that inflate the neutral disk and implications for AGN torus} \label{subsec:implications}
\begin{figure}
\centering
\includegraphics[clip,width=\linewidth]{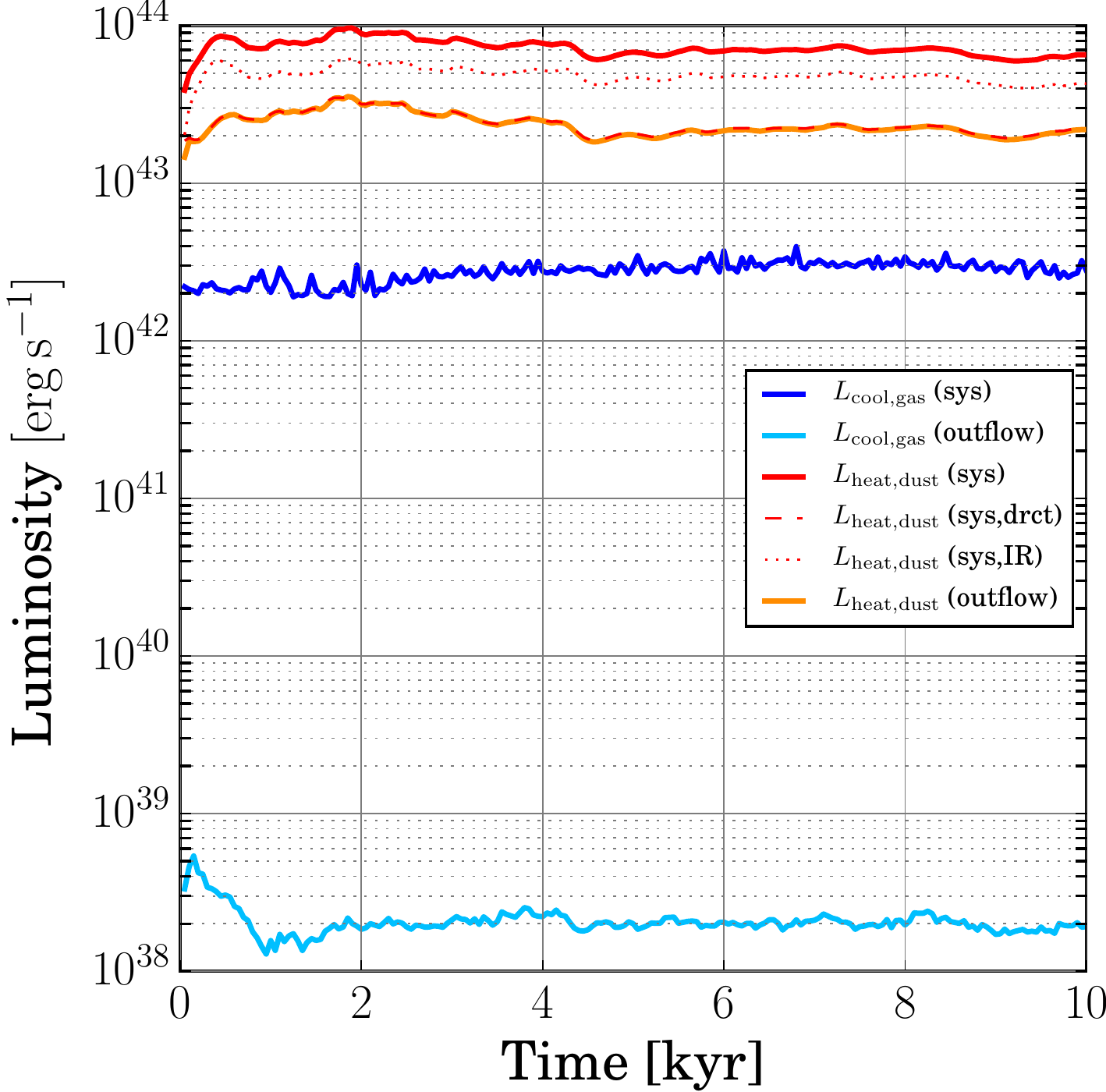}
\caption{Cooling and heating luminosities in various parts of the system in the reference model \texttt{gra01\_wo\_sca\_SG}. The blue and lightblue line show the cooling luminosities from gas in the whole of the system and in the outflow region, respectively. Here, we define the outflow region as regions with $\nH<10^{4}\;\nden$. The red and orange solid lines show the heating luminosities from dust in the whole of the system and in the outflow region, respectively. The red dashed and dotted line show the heating luminosities due to direct radiation and IR radiation, respectively.}
\label{fig:Lcool_Lheat}
\end{figure}

Are there other possible effects that help the neutral disk inflate vertically? One of the effects neglected in this study is photoheating of dust grains in the neutral disk by photons scattered by dust in the outflow region or reemitted by gas in the outflow region, which might enhance radiation pressure due to IR photons in the neutral disk. In order to check this possibility, we compare heating luminosity of dust grains in the neutral disk with both cooling luminosity of gas in the outflow region and heating luminosity of dust grains in the outflow region. Figure~\ref{fig:Lcool_Lheat} shows the time evolution of heating and cooling luminosities in various parts of the system. The heating luminosity of dust in the neutral disk can be estimated as $L_{\mathrm{heat,dust}}(\mathrm{sys})-L_{\mathrm{heat,dust}}(\mathrm{outflow})$ and is $\approx 4\times 10^{43}\;\mathrm{erg\;s^{-1}}$ at $t=8.75\;\kyr$. Compared to this, the cooling luminosity of gas in the outflow region is only $\approx 2\times 10^{38}\;\mathrm{erg\;s^{-1}}$, showing that cooling photons from gas in the outflow region cannot heat dust grains in the neutral disk sufficiently. Because $\sigma^{\mathrm{sca}}_{\mathrm{gr}}\sim \sigma^{\mathrm{abs}}_{\mathrm{gr}}$ at UV/optical wavelengths, luminosity of photons scattered by dust in the outflow must be the same order as the heating luminosity of dust in the outflow region (i.e., $L_{\mathrm{heat,dust}}(\mathrm{outflow})$) and is at most $\approx 2\times 10^{43}\;\mathrm{erg\;s^{-1}}$, which is smaller than heating luminosity of dust in the neutral disk. Thus, the structure of the neutral disk does not change if we take into account the above effects. From this result and results in \S~\ref{sec:numerical_results}, we conclude that it is difficult to form a geometrically-thick, radiation-supported, hydrostatic structure in regions near the dust sublimation radius only by radiation pressure of dust reemission and a thin disk will be formed there without additional heat or energy sources.

One of possible heating mechanisms is heating by the stellar radiation of bulge stars. However, we can rule out this possibility by a simple consideration. Assuming that the stellar mass of bulge within $1\;\pc$ is $\sim 10\MBH$ and that each star has a mass of $1\;\Msolar$, the luminosity of inner bulge is $\sim 3.85\times 10^{41}\;\mathrm{erg\;s^{-1}}$, which is much smaller than the heating luminosity of dust in the neutral disk discussed above. The stellar radiation also has an effect to confine the dusty gas disk by its radiation pressure. Thus, it can work against the formation of thick disk. Related to this, \citet{ohsuga01:_format} demonstrated that radiation force and gravity of a circumnuclear starburst allow stable orbits of gas motion around an AGN. As a result, a geometrically-thin, optically-thick, vertical wall can form with a large covering factor. This mechanism can create an optically-thick structure at sub-parsec or parsec scales for a lower luminosity AGN. We will examine this interesting possibility in the future study.

Another possibility is that a geometrically-thick structure is formed by stellar feedbacks at sub-parsec scales. To examine this possibility, we estimate a required star formation rate (SFR) in this region based on the theoretical framework developed by \citet{wada02:_obscur_mater_aroun_seyfer_nuclei_with_starb} and \citet{kawakatu08:_coevol_of_super_black_holes}. Assuming that the gas disk is vertically supported by turbulent pressure, that turbulent energy dissipation is balanced by energy input due to stellar feedbacks, and $h=\alpha r$ ($h$ is the disk scale height), we can obtain a depletion timescale of gas,
\begin{eqnarray}
t_{\mathrm{dep}} \equiv \frac{\Sigma}{\dot{\Sigma}_{\ast}} & = & 17.43\;\kyr\; \alpha^{-2} \eta \tilde{E}_{\ast,51} M^{-3/2}_{\mathrm{BH},7} r^{5/2}_{-1},
\end{eqnarray}
where $\Sigma$ is the gaseous surface density, $\dot{\Sigma}_{\ast}$ is the surface star formation rate, $\tilde{E}_{\ast,51}$ is the total energy injected by stars in any form per unit mass in the unit of $10^{51}\;\mathrm{erg\;\Msolar^{-1}}$, $\eta$ is the efficiency that represents what fraction of energy from stars is converted to turbulent energy, $M_{\mathrm{BH},7}=\MBH/10^{7}\;\Msolar$, and $r_{-1}=r/0.1\;\pc$. Stars will form if the disk is gravitationally unstable, and therefore, a star-forming disk should has a surface density
\begin{equation}
\Sigma = \frac{v_{t}\Omega}{\pi G}Q^{-1},
\end{equation}
where $v_{t}$ is the velocity dispersion of turbulence, $\Omega$ is the orbital frequency, $Q$ is the Toomre parameter ($\lesssim 1$). Using the assumptions introduced above, we can eliminate $v_{t}$ and the surface density can be rewritten into $\Sigma=\MBH/(Q \pi r^{2})$. In this case, the surface star formation rate is
\begin{eqnarray}
\dot{\Sigma}_{\ast} = 1.83\times 10^{4}\;\Msolar\;\pc^{-2}\;\yr^{-1}\; \alpha^{2} Q^{-1} \eta^{-1} \tilde{E}^{-1}_{\ast,51} M^{5/2}_{\mathrm{BH,7}} r^{-9/2}_{-1}.
\end{eqnarray}
The SFR in the region of $r=0.1\operatorname{-}1\;\pc$ is calculated as
\begin{eqnarray}
\mathrm{SFR} = 458\;\Msolar\;\yr^{-1}\; \alpha^{2} Q^{-1} \eta^{-1} \tilde{E}^{-1}_{\ast,51} M^{5/2}_{\mathrm{BH},7}.
\end{eqnarray}
Thus, an extremely large SFR is needed to realize $\alpha\sim 1$ at sub-parsec scale even if a high conversion efficiency is assumed ($\eta\approx 1$). We conclude that it is difficult to form a long-lived, geometrically-thick, obscuring structure at sub-parsec scales by stellar feedbacks.

Lastly, we give some implications for AGN tori. Two important unanswered questions about obscuring tori are (i) physical mechanisms that maintain their thicknesses, and (ii) their typical sizes. According to recent numerical studies (\citealt{dorodnitsyn12:_activ_galac_nucleus_obscur_winds,wada12:_radiat,chan15:_radiat}) and our study, a fully radiation-supported, hydrostatic, obscuring structure is probably ruled out at least in regions near the dust sublimation radius. The studies cited suggest obscuration by a dusty wind (\citealt{dorodnitsyn12:_activ_galac_nucleus_obscur_winds,chan15:_radiat}) or a (dusty) failed wind-driven circulation flow (\citealt{wada12:_radiat}). As for typical sizes of tori, a clear answer has not yet been obtained. The sizes of tori are closely related with the obscuring hydrogen column density $N_{\mathrm{H}}$ and a value of $10^{22}\;\cden$ may correspond to typical sizes. As shown in Fig.~\ref{fig:NH_dist}, $N_{\mathrm{H}} \lesssim (1\operatorname{-}3)\times 10^{21}\;\cden$ in the most part of the outflow region and the critical angle at which $N_{\mathrm{H}}$ exceeds $10^{22}\;\cden$ is $\theta\approx 85^{\circ}$ in the reference model (\texttt{gra01\_wo\_sca\_SG}). In contrast, X-ray observation suggests that the critical angle should be $66^{\circ}$ for X-ray luminosity of $10^{44}\;\mathrm{erg\;s^{-1}}$ (\citealt{hasinger08:_absor}). This means that the outflowing gas in the region of $r<1\;\pc$ in our simulations does not provide a sufficient column density and more outflowing gas is needed to explain observations. There are two possibilities to explain observations. Obvious first one is to take into account the outflowing gas at $r>1\;\pc$. In this case, $N_{\mathrm{H}}$ will exceed $10^{22}\;\cden$ at larger scales. This is naturally expected because the outflow in our simulations has a velocity enough to reach larger radii. In this case, the outflowing gas itself can be regard as a ``torus''. In addition, a failed wind must help obscuration as shown by \citet{wada12:_radiat}. Thus, one possible picture is a \textit{large-scale torus} ($r\approx 10\operatorname{-}100\;\pc$). Second interesting possibility is that a denser dusty wind launched from the outer part of AD ($R\sim 10^{4}\;R_{g}\approx 10^{-2}\;\pc\;M_{\mathrm{BH,7}}$; $R_{g}$ is the Schwarzschild radius) provides obscuration at sub-parsec scales. In the outer part of AD, the disk temperature drops below $1800\;\mathrm{K}$, and as a result, dust grains certainly form (\citealt{czerny11,czerny15:_broad_line_region_agn}). Also, the $\alpha$ viscosity acts as an additional heat source and it may help a denser wind being launched or it could create a radiation-supported, hydrostatic structure. Thus, another possible picture is a \textit{small-scale torus} ($r\lesssim 0.1\operatorname{-}$ a few $\pc$). These two pictures can coexist if a dusty wind is the main obscuration mechanism and it extends from $\approx 10^{4}\;R_{g}$ to $\sim 10\;\pc$. In order to deepen our understanding of AGN tori as well as the true gas structure and gas dynamics in the torus-AD transition zone, we need to perform RHD simulations of the outer part of AD. As shown in \S~\ref{subsec:grain_size_effects}, \ref{subsec:thermal_sputtering_effects}, and \ref{subsec:uncertainties}, the properties of dust can affect the dynamics of the outflow. Hence, it is very important to take into account the details of dust physics such as formation, growth, and destruction of dust grains. We will address this problem in the future.

\section{Summary} \label{sec:summary}
In this paper, we have performed axisymmetric RHD simulations of a dusty gas disk of radius $\approx 1\;\pc$ around an AGN of black hole mass $10^{7}\;\Msolar$ and Eddington ratio $0.77$ in order to investigate typical density and temperature distributions realized near the dust sublimation radius and to measure the mass outflow rate. Our simulations are the first RHD simulations that take into account the following important effects: \ding{202} anisotropy of AD radiation, \ding{203} X-ray heating by corona with typical X-ray luminosity fraction, \ding{204} RT of IR photons reemitted by dust grains, \ding{205} frequency dependency of direct radiation and IR photons, and \ding{206} separate temperatures for gas and dust ($\Tgas$, $\Tgr$).

The present study is summarized as follows:
\begin{enumerate}[leftmargin=*]
\item In the quasi-steady state, a nearly-neutral, dense ($\approx 10^{6\operatorname{-}8}\;\nden$), geometrically-thin ($h/r < 0.06$; $h$ the height of the disk surface) disk forms near the dust sublimation radius and a high velocity ($\approx 200\operatorname{-}3000\;\kms$) outflow is launched from the disk surface.
\begin{itemize}[nosep,leftmargin=*]
\item[-] The disk scale height strongly depends on the presence or absence of metal cooling; $h/r \ll 0.06$ if metal cooling is included (see Figs.~\ref{fig:dists_gra01_wo_sca_SG} and \ref{fig:dists_gra01_wo_sca_MTL_SG}). At this time, it is difficult to conclude which case is more realistic, because the photoionization of metal is not taken into account in our simulations. But, it would be safer to say that the true scale height is within a range of $0 < h/r < 0.06$. 
\item[-] The covering factor of the outflow strongly depends on the presence or absence of dust in the outflow and on the grain size. In the absence of dust (e.g., due to thermal sputtering), the outflow attains a large height of $|h_{\mathrm{out}}|/R \gtrsim 1\operatorname{-}2$, where $h_{\mathrm{out}}$ is the height of the outflow surface (see Fig.~\ref{fig:nH_dists_various_models}h). $h_{\mathrm{out}}/R$ gets smaller as the opacity of the outflow increases: $h_{\mathrm{out}}/R\approx 1$ and $0.3$ for $\agr=1\;\micron$ and $0.1\;\micron$, respectively (see Figs.~\ref{fig:nH_dists_various_models}a,g). This is because $h_{\mathrm{out}}/R$ is determined by the competition between vertical thermal expansion and radial acceleration due to radiation pressure.
\end{itemize}
\item Contrary to the results of \citet{krolik07:_agn} and \citet{shi08:_radiat_x}, the radiation pressure by IR photons is not effective to thicken the disk, but rather compresses it (see Fig.~\ref{fig:acceleration_ratios}a). The disk is almost totally supported by thermal pressure (the gas temperature of the disk is determined by the balance between X-ray heating and various cooling [mainly, dust cooling and metal cooling]). Thus, it seems difficult for a radiation-supported, geometrically-thick, obscuring torus to form in the region near the dust sublimation radius.
\item The mass outflow rate is $0.05\operatorname{-}0.1\;\Msolar/\yr$ depending the fraction of X-ray (>1 keV) luminosity (see Fig.~\ref{fig:gas_mass_evolution}). These values correspond to $20\operatorname{-}40$\% of the Eddington mass accretion rate for the mass-to-radiation energy conversion factor 0.1. The column density of the outflow is $N_{\mathrm{H}}\lesssim 10^{21}\;\cden$ in the region of $r<1\;\pc$, which is insensitive to the X-ray luminosity fraction within a range of $0.08\operatorname{-}0.16$.
\item To check the accuracy of our simulations, we have performed photoionization calculations for a density slice taken from one of our simulations using the \textsc{Cloudy} code and compared the resultant temperature distributions, $\Tgas$ and $\Tgr$, with those of our simulation. As a result, we have found that there is no large (unexpected) mismatch between them (see \S~\ref{subsec:uncertainties}). 
\item Based on the results above, we have discussed the typical sizes of AGN tori. In order to explain observed type-II AGN fraction (e.g., \citealt{hasinger08:_absor,toba13:_microm_lumin_funct_various_types_galax_akari,toba14:_lumin_soloan_digit_sky_survey}), it is required that outflow gas is extended to larger radii ($r \gtrsim 10\;\pc$) or that a denser dusty wind is launched from smaller radii ($r\sim 10^{4}\;R_{g}$) (see \S~\ref{subsec:implications}).
\end{enumerate}

\section*{Acknowledgments}
We thank the anonymous referee for helpful suggestions or comments that improved the paper. We thank Professor Matthias Steinmetz for providing us the 1D PPM data of the Evrard test and Dr. J{\'e}r{\^o}me Loreau for giving detailed information about the charge transfer  in $\proton$-$\HeI$ collisions. We also thank the developers of the softwares \texttt{DISLIN}, \texttt{matplotlib}, and \texttt{Asymptote}, with which we visualized our numerical results. The numerical simulations were carried out on Cray XC30 at Center for Computational Astrophysics (CfCA) of National Astronomical Observatory of Japan (NAOJ) and Cray XC30 at Institute for Information Management and Communication in Kyoto University. This work was supported by the Ministry of Education, Culture, Sports, Science and Technology (MEXT) Grant-in-Aid for Young Scientists (B)25800100 (DN) and by MEXT SPIRE Field 5 and JICFuS. This research was also supported in part by Interdisciplinary Computational Science Program in Center for Computational Sciences, University of Tsukuba, and Grant-in-Aid for Scientific Research (B) by JSPS (15H03638).

\bibliographystyle{mnras}
\bibliography{ms,bibDB_AGN_obs,bibDB_AGN_theory,bibDB_ISM_physics,bibDB_GalForm,bibDB_Galaxy,bibDB_RT,bibDB_SPH,bibDB_Nbody}

\appendix

\section{Notes on reaction rate coefficients} \label{appendix:sec:chemical_reactions}
The chemical reactions adopted in this study and their rate coefficients are summarized in Table~\ref{tbl:chemical_reactions}. In the following, we give more detailed descriptions for some of reactions. 

\begingroup
\renewcommand{\arraystretch}{1.0}
\begin{table*}
\centering
\begin{minipage}{\hsize}
\caption{Chemical reactions.}
\label{tbl:chemical_reactions}
\begin{tabular}{@{}lllp{3cm}@{}}
\hline
Number$^{\dag}$ & Reaction & Rate Coefficient or Cross Section$^{\ddag}$ & References$^{\sharp}$ \\
\hline
R1  & $\HI + \electron \rightarrow \proton + 2\electron$  &
$\kR{1} = \exp(-32.71396786375$ & \uwave{(16)},(34) \\
&& \hspace{4.5em} $+x(13.53655609057$ & \\
&& \hspace{4.5em} $+x(-5.739328757388$ & \\
&& \hspace{4.5em} $+x(1.563154982022$ & \\
&& \hspace{4.5em} $+x(-2.877056004391 \times 10^{-1}$ & \\
&& \hspace{4.5em} $+x(3.482559773736999 \times 10^{-2}$ & \\
&& \hspace{4.5em} $+x(-2.63197617559 \times 10^{-3}$ & \\
&& \hspace{4.5em} $+x(1.119543953861 \times 10^{-4}$ & \\
&& \hspace{5.5em} $  -2.039149852002 \times 10^{-6}x))))))))$, & \\
&& where $x =\ln\Tgas(\eV)$. & \smallskip\\

R2  & $\proton  + \electron \rightarrow \HI + \gamma$ &
$\kR{2} = 2.753\times 10^{-14}\dfrac{\lambda^{1.500}_{\HI}}{[1+(\lambda_{\HI}/2.740)^{0.407}]^{2.242}}$, & \uwave{(24)},(35) \\
&& where $\lambda _{\HI}=2(157807/\Tgas)$. & \smallskip\\

R3  & $\Hmol + \electron \rightarrow 2\HI + \electron$  &
$\kR{3} = 4.38\times 10^{-10}\Tgas^{0.35}\exp(-102000/\Tgas)$ & \uwave{(34)} \smallskip\\

\textbf{R4}  & $\HI + \gamma \rightarrow \proton + \electron$  &
$\sigma_{\HI}(\nu) =
\begin{cases}
6.3\times 10^{-18}\left(\dfrac{\nu}{\nu_{1}}\right)^{-4}\dfrac{\exp[4-4\tan^{-1}(\varepsilon)/\varepsilon]}{1-\exp(-2\pi/\varepsilon)}, & \nu > \nu_{1}, \\
0, & \text{otherwise},
\end{cases}$, & \uwave{(46)} \\
&& where $\varepsilon=\sqrt{\dfrac{\nu}{\nu_{1}}-1}$,
         $h\nu_{1}=13.6\;\eV$. & \smallskip\\

R5  & $\HI + \electron \rightarrow \Hminus + \gamma$ &
$ \kR{5} = 
\begin{cases}
1.429\times 10^{-18} \Tgas^{p_{1}}, & \Tgas \leq 6000\;\mathrm{K}, \\
3.802\times 10^{-17} \Tgas^{0.1998x} 10^{p_{2}}, & 6000\;\mathrm{K} < \Tgas \leq 10^{7}\;\mathrm{K}, \\
10^{-0.6(x-7) - 14.95325055247}, & 10^{7}\;\mathrm{K} < \Tgas,
\end{cases} $
& \uwave{(11)},(34), a linear extrapolation in the log-log space for $\Tgas>10^{7}\;\mathrm{K}$ \\
&& where $p_{1} = 0.762 + 0.1523x - 3.274\times 10^{-2}x^{2}$, & \\
&& \hspace{2.5em} $p_{2} = 4.0415\times 10^{-5}x^{6} - 5.447\times 10^{-3}x^{4}$, & \\
&& \hspace{2.5em} $x = \log_{10}\Tgas$. & \smallskip\\

R6   & $\Hminus + \HI \rightarrow \Hmol + \electron$ & 
$ \kR{6} = 
\begin{cases}
1.5\times 10^{-9}, & \Tgas < 300\;\mathrm{K}, \\
4\times 10^{-9} \Tgas^{-0.17}, & \Tgas \geq 300\;\mathrm{K}.
\end{cases}$
& \uwave{(23)},(41) \smallskip\\

R7   & $\Hminus + \electron \rightarrow \HI + 2\electron$ &
$\kR{7} = \exp(-18.01849334$ & \uwave{(16)},(34) \\
&& \hspace{4.5em} $ +x(2.3608522                $  & \\
&& \hspace{4.5em} $ +x(-2.8274430\times 10^{-1}  $  & \\
&& \hspace{4.5em} $ +x(1.62331664\times 10^{-2}  $  & \\
&& \hspace{4.5em} $ +x(-3.36501203\times 10^{-2} $  & \\
&& \hspace{4.5em} $ +x(1.17832978\times 10^{-2}  $  & \\
&& \hspace{4.5em} $ +x(-1.65619470\times 10^{-3} $  & \\
&& \hspace{4.5em} $ +x(1.06827520\times 10^{-4}  $  & \\
&& \hspace{5.5em} $   -2.63128581\times 10^{-6}x)))))))) $, \\
&& where $x=\ln\Tgas(\eV)$. & \smallskip\\

R8   & $\Hminus + \HI \rightarrow 2\HI + \electron$ &
$ \kR{8} = 
\begin{cases}
2.5634\times 10^{-9} \Tgas(\eV)^{1.78186}, & \Tgas(\eV) < 0.1 \\
\exp(-20.37260896 \\[-2pt]
\hspace{1.5em} +x(1.13944933  \\[-2pt]
\hspace{1.5em} +x(-1.4210135 \times 10^{-1} \\[-2pt]
\hspace{1.5em} +x(8.4644554 \times 10^{-3}  \\[-2pt]
\hspace{1.5em} +x(-1.4327641 \times 10^{-3} \\[-2pt]
\hspace{1.5em} +x(2.0122503 \times 10^{-4}  \\[-2pt]
\hspace{1.5em} +x(8.6639632 \times 10^{-5}  \\[-2pt]
\hspace{1.5em} +x(-2.5850097 \times 10^{-5} \\[-2pt]
\hspace{1.5em} +x(2.4555012 \times 10^{-6}  \\[-2pt]
\hspace{2.5em}   -8.0683825 \times 10^{-8}x))))))))),
& \Tgas(\eV) \geq 0.1,
\end{cases}$
& \uwave{(16)},(34) \\
&& where $x = \ln\Tgas(\eV)$ & \smallskip\\

R9   & $\Hminus + \proton \rightarrow 2\HI$ &
$ \kR{9} = 2.4\times 10^{-6} \Tgas^{-0.5} (1 + 5\times 10^{-5}\Tgas)$ & \uwave{(38)},(48) \smallskip\\

R10  & $\Hminus + \proton \rightarrow \Hmolplus + \electron$ &
$ \kR{10} =
\begin{cases}
2.291 \times 10^{-10} \Tgas(\eV)^{-0.4}, & \Tgas(\eV) < 1.719, \\
8.4258 \times 10^{-10} \Tgas(\eV)^{-1.4} \exp(-1.301/\Tgas(\eV)), & \Tgas(\eV) \geq 1.719.
\end{cases}$
& (18),(34) \smallskip\\

R11  & $\Hminus + \HeI \rightarrow \HI + \HeI + \electron$ &
$\kR{11} = 4.1\times 10^{-17} \Tgas^{2} \exp(-19870/\Tgas)$ & \uwave{(13)},(48) \smallskip\\

R12  & $\Hminus + \Hmolplus \rightarrow \HI + \Hmol$ &
$\kR{12} = 5\times 10^{-7} (100/\Tgas)^{0.5} $ &  \uwave{(17)},(34) \smallskip\\

\hline
\end{tabular}
\end{minipage}
\end{table*}
\endgroup

\begingroup
\renewcommand{\arraystretch}{1.0}
\begin{table*}
\centering
\begin{minipage}{\hsize}
\contcaption{}
\label{tbl:chemical_reactions_cont}
\begin{tabular}{@{}lllp{3cm}@{}}
\hline
Number$^{\dag}$ & Reaction & Rate Coefficient or Cross Section$^{\ddag}$ & References$^{\sharp}$ \\
\hline

\textbf{R13}  & $\Hminus + \gamma \rightarrow \HI + \electron$ &
$\sigma_{\Hminus}(\lambda) = 
\begin{cases}
10^{c_{1}(\log_{10}\lambda+c_{2})+c_{3}}, & x<x_{\min}, \\
10^{-18}x^{3}y^{1.5} \\[-2pt]
\hspace{1.0em} \times (1.52519 \times 10^{2} \\[-2pt]
\hspace{1.5em} +4.9534 \times 10^{1} y^{0.5} \\[-2pt]
\hspace{1.5em} -1.18858 \times 10^{2} y \\[-2pt]
\hspace{1.5em} +9.2536 \times 10^{1} y^{1.5} \\[-2pt]
\hspace{1.5em} -3.4194 \times 10^{1} y^{2} \\[-2pt]
\hspace{1.5em} +4.982 y^{2.5}), & x_{\min} \leq x \leq x_{\max}, \\
0, & x>x_{\max},
\end{cases}$ 
& \uwave{(11)}, (21), a linear extrapolation in the log-log space for $x<x_{\min}$ \\
&& where $c_{1}=0.99562198151188$, $c_{2}=4.90308998699194$, & \\
&& \hspace{2.5em} $c_{3}=-17.2651201972074$, & \\
&& \hspace{2.5em} $x=\lambda/\micron$, $x_{\min}=0.125$, $x_{\max}=1.6419$, & \\
&& \hspace{2.5em} $y = x^{-1}- x^{-1}_{\max}$. & \smallskip\\

R14  & $\Hmol + \HI \rightarrow 3\HI$         &
$ \kR{14} = d\left(\frac{8E}{\pi \mu}\right)^{0.5}\dfrac{aE^{b-1}\Gamma(b+1)\exp(-E_{0}/E)}{(1+cE)^{b+1}}$, & (37) \\
&& where $\displaystyle E=\dfrac{\kBoltz \Tgas}{27.21\;\eV}$, $E_{0}=0.168$, $\mu=2\mH/3$, & \\
&& \hspace{2.5em} $a=54.1263$, $b=2.5726$, $c=3.4500$, $d=1.849\times 10^{-22}$. & \smallskip\\

R15  & $\Hmol + \Hmol \rightarrow \Hmol + 2\HI$ &
$\kR{15}$ is obtained by the same formula used in $\kR{14}$, & (37) \\
&& but with the following parameters: & \\
&& \hspace{2.5em} $E_{0}=0.1731$, $\mu=\mH$, & \\
&& \hspace{2.5em} $a=40.1008$, $b=4.6881$, $c=2.1347$.  & \smallskip\\

R16  & $2\HI + \dust \rightarrow \Hmol + \dust$ &
$\kR{16} = \sqrt{\dfrac{8k_{\mathrm{B}}\Tgas}{\pi \mH}}S_{\mathrm{H}}f_{a}\sigmagr$, & (10),(40),(42),(43),(49) \\
&& where $\displaystyle S_{\mathrm{H}}=\dfrac{1}{[1+0.04(\Tgas+\Tgr)^{0.5}+0.002\Tgas + 8\times 10^{-6}\Tgas^{2}]}$, & \\
&& \hspace{2.5em}
$ f_{a} =
\begin{cases}
1,   & 5\;\mathrm{K} \leq \Tgr \leq 20\;\mathrm{K}, \\
0.2, & 20\;\mathrm{K}< \Tgr \leq 500\;\mathrm{K}, \\
0,   & \mathrm{otherwise}.
\end{cases}$
& \smallskip\\

R17  & $3\HI \rightarrow \Hmol + \HI$ &
$\kR{17} = 5.5\times 10^{-29}/\Tgas$ & (14) \smallskip\\

R18 & $2\HI + \Hmol \rightarrow 2\Hmol$ &
$\kR{18} = 6.875\times 10^{-30}/\Tgas$  & (14) \\ 

\textbf{R19}  & $\Hmol + \gamma \rightarrow 2\HI$ & \rdelim\}{4}{4ex}[see Appendix~\ref{appendix:subsec:R19_R22}] & \multirow{4}{3cm}{\uwave{(9)},\uwave{(19)},\uwave{(25)},\uwave{(26)},\\\uwave{(28)},(36),(39)}  \\

\textbf{R20}  & $\Hmol + \gamma \rightarrow \Hmolplus + \electron$ &  &  \\

\textbf{R21}  & $\Hmol + \gamma \rightarrow \HI + \proton + \electron$ &  &  \\

\textbf{R22}  & $\Hmol + \gamma \rightarrow 2\proton + 2\electron$ &  &  \smallskip\\

R23  & $\HI + \proton \rightarrow \Hmolplus + \gamma$ &
$\kR{23} = 10^{-19.38 - 1.523x + 1.118x^{2} - 0.1269x^{3}}$, & \uwave{(8)},(47) \\
&& where $x=\log_{10}\Tgas$. & \smallskip\\

R24$^{\spadesuit}$  & $\Hmolplus + \HI \rightarrow \Hmol^{*} + \proton$ &
$\kR{24} = 6.4\times 10^{-10}$ & \uwave{(12)},(34) \smallskip\\  

R25  & $\Hmol + \proton \rightarrow \Hmolplus + \HI$ &
$\kR{25} = \exp(-21237.15/\Tgas)$ & \uwave{(44),(45)},(47) \\
&& \hspace{2.25em} $\times (-3.3232183 \times 10^{-7}$ & \\
&& \hspace{3.25em} $+x(3.3735382 \times 10^{-7}$ & \\
&& \hspace{3.25em} $+x(-1.4491368 \times 10^{-7}$ & \\
&& \hspace{3.25em} $+x(3.4172805 \times 10^{-8}$ & \\
&& \hspace{3.25em} $+x(-4.7813720 \times 10^{-9}$ & \\
&& \hspace{3.25em} $+x(3.9731542 \times 10^{-10}$ & \\
&& \hspace{3.25em} $+x(-1.8171411 \times 10^{-11}$ & \\
&& \hspace{5.00em} $   +3.5311932 \times 10^{-13}x)))))))$, & \\ 
&& where $x=\ln\Tgas$. & \\

R26  & $\Hmolplus + \electron \rightarrow 2\HI$ &
$\kR{26} = 
\begin{cases}
10^{-8}, & \Tgas < 617, \\
1.32 \times 10^{-6} \Tgas^{-0.76}, & \Tgas \geq 617.
\end{cases}$
& \uwave{(27)},(34) \smallskip\\

\textbf{R27}  & $\Hmolplus + \gamma \rightarrow \HI + \proton$ & 
$\sigma_{\Hmolplus,R27}(\nu) = 
\begin{cases}
10^{-40.97 + E(6.03 + E(-0.504 + 1.387 \times 10^{-2}E))}, & 2.65 < E < 11.27, \\
10^{-30.26 + E(2.79 + E(-0.184 + 3.535 \times 10^{-3}E))}, & 11.27 \leq E < 21.0 \\
0, & \mathrm{otherwise},
\end{cases}$
& \uwave{(3)},(18) \\
&& where $E = h\nu/\eV$. \smallskip\\

\textbf{R28}  & $\Hmolplus + \gamma \rightarrow 2\proton + \electron$ &
$\sigma_{\Hmolplus,R28}(\nu) =
\begin{cases}
10^{-16.926 + E(-4.528\times 10^{-2} + E(2.238\times 10^{-4} + 4.245\times 10^{-7}E))}, & 30 < E < 90, \\
0, & \text{otherwise},
\end{cases}$
& \uwave{(4)},(18) \\
&& where $E = h\nu/\eV$. \smallskip\\

\hline
\end{tabular}
\end{minipage}
\end{table*}
\endgroup

\begingroup
\renewcommand{\arraystretch}{1.0}
\begin{table*}
\centering
\begin{minipage}{\hsize}
\contcaption{}
\label{tbl:chemical_reactions_cont2}
\begin{tabular}{@{}llll@{}}
\hline
Number$^{\dag}$ & Reaction & Rate Coefficient or Cross Section$^{\ddag}$ & References$^{\sharp}$ \\
\hline

R29 & $\HeI + \electron \rightarrow \HeII + 2\electron$ &
$\kR{29} = \exp(-44.09864886$ & \uwave{(16)},(34) \\
&& \hspace{4.5em} $+x(23.91596563$ & \\
&& \hspace{4.5em} $+x(-10.7532302$ & \\
&& \hspace{4.5em} $+x(3.05803875$ & \\
&& \hspace{4.5em} $+x(-5.6851189 \times 10^{-1}$ & \\
&& \hspace{4.5em} $+x(6.79539123 \times 10^{-2}$ & \\
&& \hspace{4.5em} $+x(-5.00905610 \times 10^{-3} $ & \\
&& \hspace{4.5em} $+x(2.06723616 \times 10^{-4}  $ & \\
&& \hspace{5.5em} $ -3.64916141 \times 10^{-6}x))))))))$, \\ 
&& where $x = \ln\Tgas(\eV)$. & \smallskip\\

R30  & $\HeII + \electron \rightarrow \HeIII + 2\electron$ &
$\kR{30} = \exp(-68.71040990 $ & (34) \\
&& \hspace{4.5em} $+x(43.93347633 $ & \\
&& \hspace{4.5em} $+x(-18.4806699 $ & \\
&& \hspace{4.5em} $+x(4.70162649  $ & \\
&& \hspace{4.5em} $+x(-0.76924663 $ & \\
&& \hspace{4.5em} $+x(8.113042 \times 10^{-2} $ & \\
&& \hspace{4.5em} $+x(-5.32402063 \times 10^{-3} $ & \\
&& \hspace{4.5em} $+x(1.97570531 \times 10^{-4} $ & \\
&& \hspace{5.5em} $  -3.16558106 \times 10^{-6}x))))))))$, & \\
&& where $\ln\Tgas(\eV)$. & \smallskip\\

R31  & $\HeII + \electron \rightarrow \HeI + \gamma$ &
$\kR{31,r} = 1.26\times 10^{-14}\lambda_{\HeI}^{0.75}$ (Case B), & \uwave{(1),(5)},(35) \\
&& $\kR{31,d} = 1.9 \times 10^{-3}\Tgas^{-1.5}\exp(-473421/\Tgas)(1 + 0.3\exp(-94684/\Tgas))$ (dielectric), &  \\
&& where $\lambda_{\HeI}=2(285335/\Tgas)$. & \smallskip\\

R32  & $\HeIII + \electron \rightarrow \HeII + \gamma$ &
$\kR{32} = 5.506 \times 10^{-14} \dfrac{\lambda_{\HeII}^{1.5}}{[1+(\lambda_{\HeII}/2.74)^{0.407}]^{2.242}}$, & \uwave{(24)},(35) \\
&& where $\lambda_{\HeII} = 2(631515/\Tgas)$. & \smallskip\\

R33  & $\HeII + \HI \rightarrow \HeI + \proton + \gamma$ &
$\kR{33} = 1.2 \times 10^{-15} (\Tgas/300)^{0.25}$ & \uwave{(22)},(47) \smallskip\\

R34  & $\HeI + \proton \rightarrow \HeII + \HI$ & see Appendix~\ref{appendix:subsec:R34} & \uwave{(50)} \smallskip\\

\textbf{R35}  & $\HeI + \gamma \rightarrow \HeII + \electron$ &
$\sigma_{\HeI}(\nu) = 
\begin{cases}
7.42 \times 10^{-18}(1.66x^{-2.05}-0.66x^{-3.05}), & x > 1, \\
0, & \text{otherwise}, 
\end{cases}$
& \uwave{(6)},(34) \\
&& where $x = h\nu/(24.6\;\eV)$. & \smallskip\\

\textbf{R36}  & $\HeII + \gamma \rightarrow \HeIII + \electron$ &
$\sigma_{\HeII}(\nu) = 
\begin{cases} 
1.58 \times 10^{-18} \left(\dfrac{\nu}{\nu_{1}}\right)^{-4} \dfrac{\exp[4-4\tan^{-1}(\varepsilon)/\varepsilon]}{1-\exp(-2\pi/\varepsilon)}, & \nu>\nu_{1}, \\
0, & \text{otherwise},
\end{cases}$
& \uwave{(46)} \\
&& where $\varepsilon=\sqrt{\dfrac{\nu}{\nu_{1}}-1}$, $h\nu_{1}=54.4\;\eV$. & \smallskip\\

R37  & $\Hmol + \HeI \rightarrow 2\HI + \HeI$ &
$\kR{37} = 
\begin{cases}
10^{\log k_{H} - (\log k_{H}-\log k_{L})/[1+(n_{\mathrm{He}}/n_{\mathrm{crit}})^{1.09}]}, & 2000\;\mathrm{K} \leq \Tgas \leq 10^{4}\;\mathrm{K}, \\
0, & \text{otherwise}
\end{cases}$
& \uwave{(20)},(48) \\
&& where $n_{\mathrm{crit}}=10^{5.0792(1-1.23\times 10^{-5}(\Tgas-2000))}$, & \\
&& \hspace{2.5em} $\log k_{H} = -1.75x - 2.729  - 23474/\Tgas$, \\
&& \hspace{2.5em} $\log k_{L} = 3.801x - 27.029 - 29487/\Tgas$, \\
&& \hspace{2.5em} $x = \log_{10}\Tgas$. & \smallskip\\

R38 & $\Hminus + \Hmolplus \rightarrow 3\HI$ &
$\kR{38} = 1.4 \times 10^{-7}(\Tgas/300)^{-0.5}$ & \uwave{(17)},(48) \smallskip\\  

R39 & $\Hmol + \electron \rightarrow \Hminus + \HI$ &
$\kR{39} = 2.7 \times 10^{-8} \Tgas^{-1.27} \exp(-4.3\times 10^{4}/\Tgas)$ & \uwave{(2)},(48) \smallskip\\

\hline
\end{tabular}
\end{minipage}
\end{table*}
\endgroup

\begingroup
\renewcommand{\arraystretch}{1.0}
\begin{table*}
\centering
\begin{minipage}{\hsize}
\contcaption{}
\label{tbl:chemical_reactions_cont3}
\begin{tabular}{@{}lllp{3cm}@{}}
\hline
Number$^{\dag}$ & Reaction & Rate Coefficient or Cross Section$^{\ddag}$ & References$^{\sharp}$ \\
\hline

R40 & $\Hmol + \HeII \rightarrow \HeI + \HI + \proton$ &
$\kR{40} = 3.7 \times 10^{-14} \exp(35/\Tgas)$ & \uwave{(15)},(48) \smallskip\\

R41 & $\Hmol + \HeII \rightarrow \Hmolplus + \HeI$ &
$\kR{41} = 7.2 \times 10^{-15}$ & \uwave{(15)},(48) \smallskip\\

R42 & $\HeII + \Hminus \rightarrow \HeI + \HI$ & see Appendix~\ref{appendix:subsec:R42} & \uwave{(29),(32)} \smallskip\\ 

R43 & $2\HI + \HeI \rightarrow \Hmol + \HeI$ &
$\kR{43} = 6.9 \times 10^{-32} \Tgas^{-0.4}$ & \uwave{(7)},(48) \smallskip\\

\hline
\end{tabular}
\begin{flushleft}
{\footnotesize REFERENCES.}---
(1) \citet{burgess60:_radiat_he};
(2) \citet{schulz67:_isotop_effec_dissoc_attac_low_energ};
(3) \citet{dunn68:_photod};
(4) \citet{bates68:_undul};
(5) \citet{aldrovandi73:_radiat_dielec_recom_coeff_compl_ions};
(6) \citet{osterbrock74:_astrop_gaseous_nebul};
(7) \citet{walkauskas75:_gas_phase_hydrog_atom_recom};
(8) \citet{ramaker76:_molec};
(9) \citet{oneil78:_photoion};
(10) \citet{hollenbach79:_molec_format_and_infrar_emiss};
(11) \citet{wishart79};
(12) \citet{karpas79};
(13) \citet{huq82:_total_elect_detac_cross_section};
(14) \citet{palla83:_primor_star_format};
(15) \citet{barlow84:_dynam_stored_ions_low_temper};
(16) \citet{janev87:_elemen_proces_hydrog_helium_plasm};
(17) \citet{dalgarno87:_chemis};
(18) \citet{shapiro87:_hydrog_molec_radiat_coolin_pregal_shock};
(19) \citet{dujardin87:_doubl};
(20) \citet{dove87:_excit_dissoc_molec_hydrog_shock};
(21) \citet{john88:_contin};
(22) \citet{zygelman89:_radiat};
(23) \citet{launay91};
(24) \citet{ferland92:_anisot_line_emiss_geomet_broad};
(25) \citet{sadeghpour93:_doubl};
(26) \citet{chung93:_dissoc};
(27) \citet{schneider94:_dissoc_recom_molec_ions_hydrog};
(28) \citet{samson94:_total};
(29) \citet{peart94:_merged_h_d};
(30) \citet{stancil94:_contin_absor_cool_white_dwarf};
(31) \citet{draine96:_struc_of_station_photod_front};
(32) \citet{olamba96:_improv};
(33) \citet{chibisov97:_one};
(34) \citet{abel97:_model}$^{\clubsuit}$;
(35) \citet{hui97:_equat};
(36) \citet{yan98:_photoion_cross_section_he};
(37) \citet{martin98:_collis_induc_dissoc_of_molec};
(38) \citet{croft99:_rate_h_h};
(39) \citet{wilms00:_x};
(40) \citet{hirashita02:_effec};
(41) \citet{glover03:_compar_gas_phase_grain_catal_format};
(42) \citet{cazaux04:_format_on_grain_surfac};
(43) \citet{cazaux04:_molec_hydrog_format_on_dust};
(44) \citet{savin04:_rate_x_sigma_j};
(45) \citet{savin04:_errat};
(46) \citet{osterbrock06:_astrop_gaseous_nebul_activ_galac_nuclei};
(47) \citet{yoshida06:_format_of_primor_stars_in};
(48) \citet{glover08:_uncer_hd};
(49) \citet{cazaux10:_errat};
(50) \citet{loreau14:_charg}. \\
$^{\dag}$ For convenience, the reaction number for photoionization or photodissociation process is shown in boldface. \\
$^{\ddag}$ The rate coefficient is in $\mathrm{cm^{3}\;s^{-1}}$ except for the reactions R17, R18, and R43, which are in $\mathrm{cm^{6}\;s^{-1}}$. The temperatures are in $\mathrm{K}$ unless otherwise stated. The cross section is in $\mathrm{cm}^{2}$. The definition of the formation efficiency $f_{a}$ in the reaction R16 is based on the results of \citet{cazaux04:_molec_hydrog_format_on_dust} and \citet{cazaux10:_errat}. \\
$^{\sharp}$ Reference for the original experimental data or theoretical calculation is indicated by wavy line whenever possible. \\
$^{\clubsuit}$ We actually use the rate coefficients adopted in the T0D code, which is published in \url{http://www.slac.stanford.edu/\textasciitilde tabel/PGas/codes.html}. \\
$^{\spadesuit}$ $\Hmol^{*}$ indicates an excited molecular hydrogen. In the simulations, we treat it as the ground-state molecular hydrogen for simplicity. 

\end{flushleft}
\end{minipage}
\end{table*}
\endgroup

\subsection{Photodissociation and photoionization of $\Hmol$ (R19-R22)} \label{appendix:subsec:R19_R22}
\subsubsection{Photodissociation}
According to \citet{abel97:_model}, molecular hydrogen can be photodissociated through the following processes:
\begin{equation}
\Hmol + \gamma \rightarrow \Hmol^{*} \rightarrow  2\HI \quad \text{(the two-step Solomon process)} \label{eq:Solomon_process}
\end{equation}
and
\begin{equation}
\Hmol + \gamma \rightarrow 2\HI \quad \text{(the direct photodissociation)}.
\end{equation}
For the former process, we employ the same approximation used in \citet{namekata14:_agn} to avoid (numerically expensive) radiative transfer calculation of the Lyman-Werner band photons required for an accurate treatment of the process (\ref{eq:Solomon_process}) (for detail, see \S 3.2 in \citealt{namekata14:_agn}). As for the direct photodissociation, \citet{abel97:_model} gives a fit for cross section data of \citet{allision69:_photod_lyman_werner}. In this study, we use it:
\begin{equation}
\sigma_{\mathrm{R19}} = \dfrac{1}{y+1}(\sigma_{L0}+\sigma_{W0}) + \left(1-\dfrac{1}{1+y}\right)(\sigma_{L1}+\sigma_{W1}),
\end{equation}
where $y$ is the ortho-to-para ratio of $\Hmol$, 
\begin{eqnarray}
\scalebox{0.85}{$\displaystyle
\sigma_{L0} =
\begin{cases}
\dex(-2.8711 - 1.05139E), & 14.675 < E < 16.82, \\
\begin{aligned}[c]
\dex(&-49.41  \\
     &+1.8042\times 10^{-2} E^{3} \\
     &-4.2339\times 10^{-5}E^{5}),
\end{aligned} & 16.82 \leq E \leq 17.6 \\
0, & \text{otherwise}, 
\end{cases}
$}
\end{eqnarray}
\begin{eqnarray}
\scalebox{0.85}{$\displaystyle
\sigma_{W0} = 
\begin{cases}
\dex(-4.4689 + 0.9182618E), & 14.675 < E < 17.7, \\
0, & \text{otherwise},
\end{cases}
$}
\end{eqnarray}
\begin{eqnarray}
\scalebox{0.85}{$\displaystyle
\sigma_{L1} =
\begin{cases}
\dex(-5.9781594 - 0.819429E), & 14.159 < E < 15.302, \\
\dex(-1.95356   - 1.082438E), & 15.302 < E < 17.2, \\
0, & \text{otherwise},
\end{cases}
$}
\end{eqnarray}
\begin{eqnarray}
\scalebox{0.85}{$\displaystyle
\sigma_{W1} =
\begin{cases}
\dex(-5.12633 - 0.85088597E), & 14.159 < E < 17.2, \\
0, & \text{otherwise},
\end{cases}
$}
\end{eqnarray}
and $E=h\nu/\eV$ and $\mathrm{dex}(x)\equiv 10^{x}$. In this study, we assume $y=3$ for simplicity.

\subsubsection{Photoionization}
\begin{figure}
\centering
\includegraphics[clip,width=\linewidth]{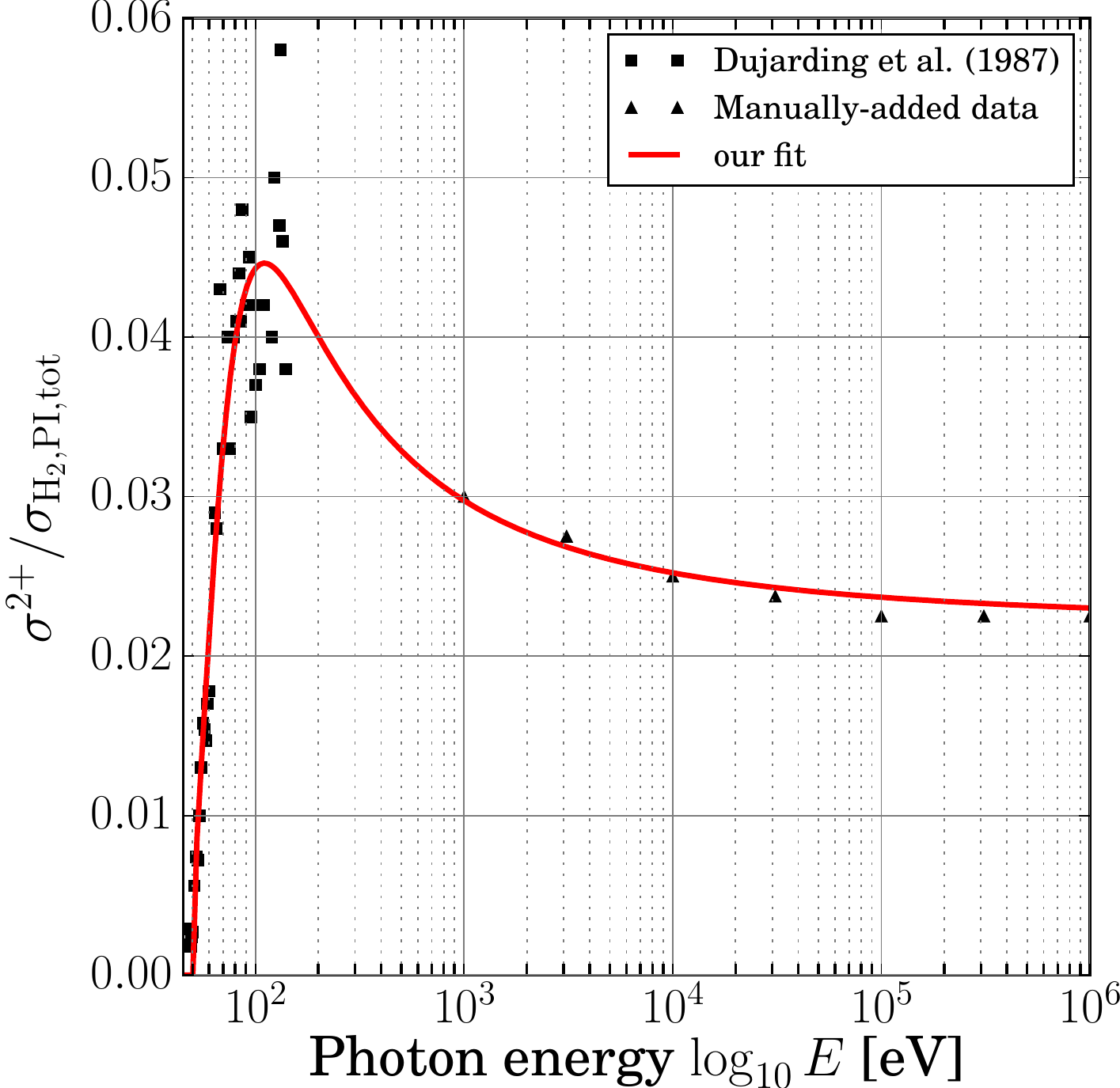}
\caption{Comparison of the branch ratio of the DPI adopted in this study (the red solid line) with data by \citet{dujardin87:_doubl} (the filled squares). The filled triangles indicate manually-added data which are introduced for the branch ratio to smoothly approaches the asymptotic value $0.0225$.}
\label{fig:BR_Dujardin1987}
\end{figure}

Previous studies have shown that molecular hydrogen can be photoionized via the following three reations:
\begin{eqnarray}
&& \Hmol + \gamma \rightarrow  \Hmolplus + \electron \label{eq:H2_simple_PI} \\
&& \Hmol + \gamma \rightarrow  \HI + \proton + \electron \qquad \text{(dissociative photoionization)} \label{eq:H2_dissociative_PI} \\
&& \Hmol + \gamma \rightarrow  2\proton + 2\electron \qquad \text{(double photoionization[DPI])} \label{eq:H2_DPI}
\end{eqnarray}
\citet{yan98:_photoion_cross_section_he} and \citet{wilms00:_x} give a fit for \textit{total} photoionization cross section of $\Hmol$, $\sigmaHmolPItot$, based on the results of early theoretical and experimental studies (e.g., \citealt{oneil78:_photoion,samson94:_total}). The ratios of the cross sections of the processes (\ref{eq:H2_simple_PI}-\ref{eq:H2_DPI}) to the total photoionization cross section are generally called the \textit{branch(ing) ratios}. Unfortunately, the branch ratios are not precisely measured for the whole range of photon energies observed in AGNs (i.e., $h\nu \approx 10\;\eV$-$100\;\keV$) to the best of our knowledge. Hence, we are forced to use very crude approximations. In this study, the branch ratio of the DPI is computed based on the results of \citet{dujardin87:_doubl} and \citet{sadeghpour93:_doubl}. \citet{dujardin87:_doubl} experimentally investigated the cross section of DPI, $\sigma^{2+}$, for photon energies ranging from $47.5\;\eV$ to $140\;\eV$ and the result is summarized in their Table 2 where the ratios of $\sigma^{2+}$ to the total \textit{photoabsorption} cross section, $\sigmaHmolPAtot$ ($\sigma_{\mathrm{abs}}$ in the original paper), are tabulated. \citet{sadeghpour93:_doubl} predicted that the ratio of the cross section of DPI to that for single photoionization is $0.0225$ in the limit of high photon energy. To obtain a rough approximation of the branch ratio of the DPI, we extend the data by \citet{dujardin87:_doubl} to high energy so that its value asymtotically approaches the one predicted by \citet{sadeghpour93:_doubl} with the following assumptions: 
\begin{itemize}[leftmargin=*]
\item $\sigmaHmolPAtot \approx \sigmaHmolPItot$ for $E>51.4\;\eV$. 
\item Single photoionization is dominant ionization process for $E>51.4\;\eV$; $\sigma^{2+}/\sigmaHmolPItot$ in the high energy limit is approximated by $0.0225$.
\end{itemize}
The branch ratio of the DPI adopted in this study is given by
\begin{eqnarray}
\BRD & = &
\begin{cases}
0.0225 + \dfrac{N_{\mathrm{DPI}}(E)}{D_{\mathrm{DPI}}(E)}, & E \geq 51.4 \\
0, & E < 51.4
\end{cases} \\
N_{\mathrm{DPI}}(E) & = & -1.32750962930545   \nonumber\\
                & + & x(0.911485133374493 \nonumber\\
                &   &\hspace{.75em}-0.0937820997652705x), \\
D_{\mathrm{DPI}}(E) & = & 29.0918253433397 \nonumber\\
                & + & x(3.5939984406517 \nonumber\\
                & + & x(-31.4630245382174 \nonumber\\
                &   &\hspace{.75em}+11.8863104773797x)),
\end{eqnarray}
where $E=h\nu/\eV$ and $x=\log_{10}E$. In Fig.~\ref{fig:BR_Dujardin1987}, we compare $\BRD$ with data by \citet{dujardin87:_doubl}. Note that the shape of $\BRD$ for $E>140\;\eV$ is highly uncertain because there are no theoretical and experimental restrictions.

\begin{figure}
\centering
\includegraphics[clip,width=\linewidth]{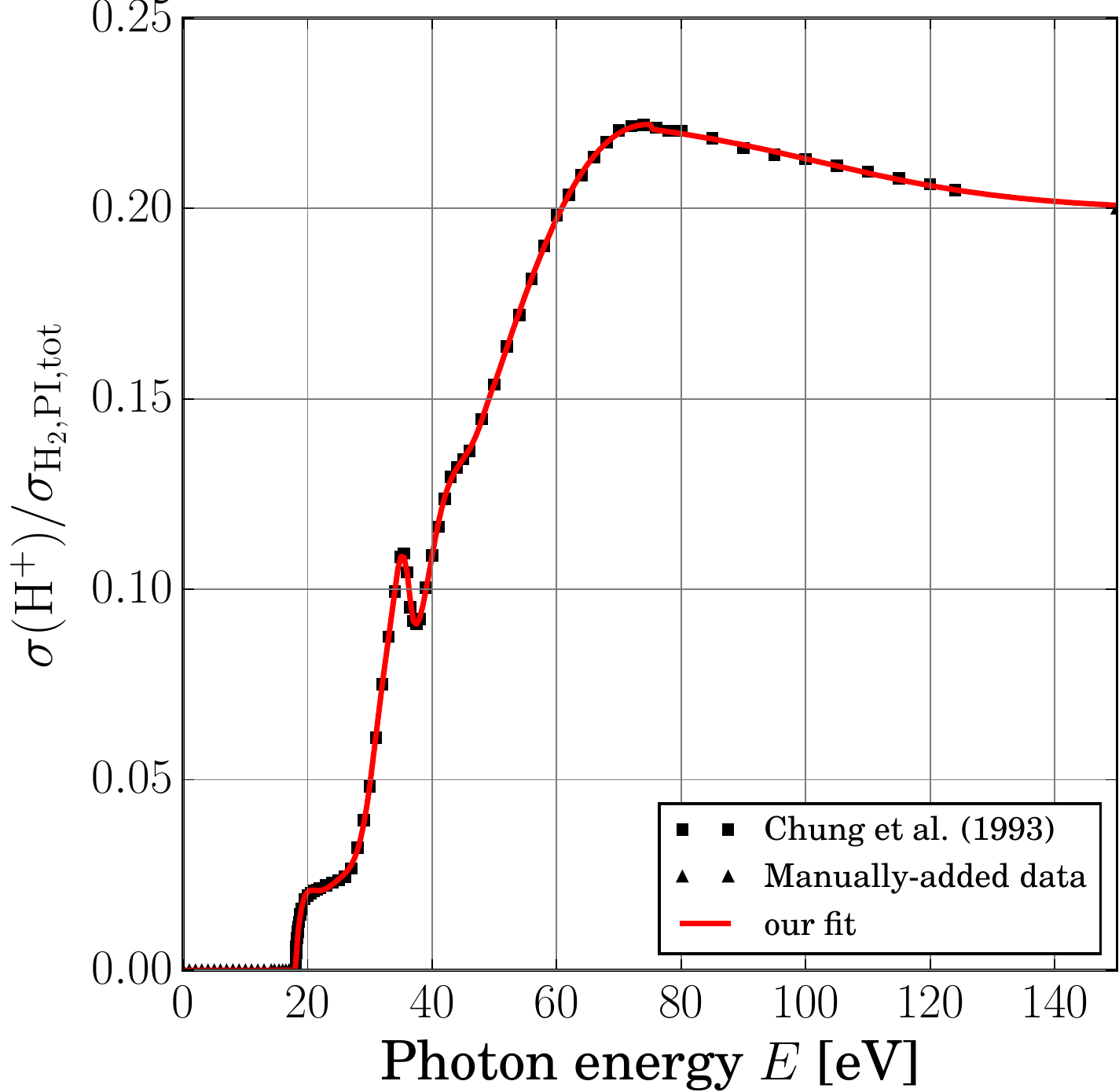}
\caption{Comparison of the branch ratio for the processes (\ref{eq:H2_dissociative_PI}+\ref{eq:H2_DPI}) (the red solid line) with data by \citet{chung93:_dissoc} (the filled square). The filled triangles indicate manually-added data points which is introduced to make the fitting process numerically stable.}
\label{fig:BR_Chung1993}
\end{figure}

As for the remaining processes (\ref{eq:H2_simple_PI}) and (\ref{eq:H2_dissociative_PI}), \citet{chung93:_dissoc} experimentally measured the branch ratios for photon energies ranging from $\approx 18\;\eV$ to $124\;\eV$. The results are summarized in Table 2 in their paper. $\sigma(\mathrm{abs})$ and $\sigma(\mathrm{H^{+}})$ in their table correspond to $\sigmaHmolPAtot$ and the sum of the cross sections for the processes (\ref{eq:H2_dissociative_PI}) and (\ref{eq:H2_DPI}), respectively. In this study, we assume $\sigmaHmolPItot=\sigmaHmolPAtot$ for simplicity since the ratio $\sigmaHmolPItot/\sigmaHmolPAtot$ is $\approx 1$ for photon energies examined in their study\footnote{As noted in \citet{chung93:_dissoc}, fluorescence cross sections have been measured by \citet{glass-maujean86:_photod} and \citet{glass-maujean88:_photod} for photon energies $E\approx 30\operatorname{-}40\;\eV$. However, its cross section is at most 8\% of $\sigmaHmolPItot$ as shown in Table 1 of \citet{chung93:_dissoc}. Therefore, we ignore it here.}. Then, we can calculate the branch ratio of processes (\ref{eq:H2_dissociative_PI}+\ref{eq:H2_DPI}) from $\sigma(\mathrm{H^{+}})/\sigma(\mathrm{abs})$. For $E>124\;\eV$, we simply assume a constant value of $0.2$, the branch ratio at $E=124\;\eV$. The adopted branch ratio for the processes (\ref{eq:H2_dissociative_PI}+\ref{eq:H2_DPI}) is given by
\begin{eqnarray}
\BRC = 
\begin{cases}
0, & E \leq E_{0}, \\
f_{01}(E), & E_{0} < E \leq E_{1}, \\
f_{12}(E), & E_{1} < E \leq E_{2}, \\
f_{23}(E), & E_{2} < E \leq E_{3}, \\
f_{34}(E), & E_{3} < E \leq E_{4}, \\
f_{4\infty}(E), & \text{otherwise},
\end{cases}
\end{eqnarray}
where $E=h\nu/\eV$, $E_{0}=18.08$, $E_{1}=31.5$, $E_{2}=36.5$, $E_{3}=50$, $E_{4}=75$, 
\begin{eqnarray}
\begin{aligned}[c]
f_{01}(E) & = -101.458630090864 \\[-2pt]
         & + E(24.6904573956157 \\[-2pt]
         & + E(-2.48962385790305 \\[-2pt]
         & + E(0.133146904967389 \\[-2pt]
         & + E(-0.00398253789826133 \\[-2pt]
         & + E(6.31510327097242 \times 10^{-5} \\[-2pt]
         &\hspace{1.5em}-4.14575959485418 \times 10^{-7}E))))),
\end{aligned}
\end{eqnarray}
\begin{eqnarray}
\begin{aligned}[c]
f_{12}(E) & = 27012.7669603569 \\[-2pt]
         & + E(-4889.93779623088 \\[-2pt]
         & + E(368.405573161181 \\[-2pt]
         & + E(-14.7856344026732 \\[-2pt]
         & + E(0.333394792938545 \\[-2pt]
         & + E(-4.00448693372017 \times 10^{-3} \\[-2pt]
         &\hspace{1.5em}+2.00161570302515 \times 10^{-5}E))))),
\end{aligned}
\end{eqnarray}
\begin{eqnarray}
\begin{aligned}[c]
f_{23}(E) & = 746.973740382314 \\[-2pt]
         & + E(-94.0242442135252 \\[-2pt]
         & + E(4.87764010685915 \\[-2pt]
         & + E(-0.133304546185048 \\[-2pt]
         & + E(2.02085520657692 \times 10^{-3} \\[-2pt]
         & + E(-1.60733315675007 \times 10^{-5} \\[-2pt]
         &\hspace{1.5em}+5.22196051715457 \times 10^{-8}E))))),
\end{aligned}
\end{eqnarray}
\begin{eqnarray}
&& f_{34}(E) = N_{34}(E)/D_{34}(E), \\
&& \begin{aligned}[c]
N_{34}(E) & = 1.16610687888582 \\[-2pt]
         & + E(3.84246976428773 \\[-2pt]
         & + E(-0.218714768189657 \\[-2pt]
         &\hspace{1.5em}+6.79610772742364 \times 10^{-3}E)),
\end{aligned} \\
&& \begin{aligned}
D_{34}(E) & = 1.00180058229909 \\[-2pt]
         & + E(1.13734973913559 \\[-2pt]
         & + E(3.42145181343494 \\[-2pt]
         & + E(-7.86343556225973 \times 10^{-2} \\[-2pt]
         &\hspace{1.5em}+7.11642133202197 \times 10^{-4}E))),
\end{aligned}
\end{eqnarray}
\begin{eqnarray}
&& f_{4\infty}(E) = 0.2 + N_{4\infty}(E)/D_{4\infty}(E), \\
&& \begin{aligned}[c]
N_{4\infty}(E) & = 0.910389337740106 \\[-2pt]
            & + E(16.6678477014188 \\[-2pt]
            & + E(-0.140616352265613 \\[-2pt]
            &\hspace{1.5em}+2.3935834710775\times 10^{-4}E)),
\end{aligned} \\
&& \begin{aligned}[c]
D_{4\infty}(E) & = 1.01378383859614 \\[-2pt]
            & + E(1.55801362381919 \\[-2pt]
            & + E(19.6430609488683 \\[-2pt]
            & + E(-0.319084566564548 \\[-2pt]
            &\hspace{1.5em}+1.60889570527996\times 10^{-3}E))).
\end{aligned}
\end{eqnarray}
Figure~\ref{fig:BR_Chung1993} compares $\BRC$ with data by \citet{chung93:_dissoc}. Note that $\BRC$ is highly uncertain for $E>124\;\eV$ by the same reason as the one for $\BRD$.

In summary, the cross sections of the processes (\ref{eq:H2_simple_PI})-(\ref{eq:H2_DPI}) are computed by
\begin{eqnarray}
\sigma_{\mathrm{R20}}(E) & = & \sigmaHmolPItot\; (1-\BRC), \\
\sigma_{\mathrm{R21}}(E) & = & \sigmaHmolPItot\; (\BRC-\BRD), \\
\sigma_{\mathrm{R22}}(E) & = & \sigmaHmolPItot\; \BRD.
\end{eqnarray}
For $\sigmaHmolPItot$, we refer to \citet{yan98:_photoion_cross_section_he} and \citet{wilms00:_x}.

\subsection{Charge transfer in $\proton$-$\HeI$ collision (R34)} \label{appendix:subsec:R34}
Previous studies (e.g., \citealt{galli98:_univer,glover03:_radiat_x,yoshida06:_format_of_primor_stars_in,glover08:_uncer_hd}) have used the following formula 
\begin{equation}
k_{\rm R34,K93} = 
\begin{cases}
1.26\times 10^{-9}\Tgas^{-0.75}\exp\left(-\dfrac{127500}{\Tgas}\right), & \Tgas < 10^{4}\;\mathrm{K}, \\
4\times 10^{-37}\Tgas^{4.74}, & \Tgas > 10^{4}\;\mathrm{K}. 
\end{cases}
\end{equation}
This is based on the result of \citet{kimura93:_rate}. However, the rate coefficient shown in Table 3 in \citet{kimura93:_rate} is likely incorrect because they cannot be reproduced from the cross section data given by Table 1 in \citet{kimura93:_rate}. Recently, \citet{loreau14:_charg} computes the cross section for this reaction in the wide range of collision energy from $10\;\mathrm{eV/u}$ to $10\;\mathrm{MeV/u}$ and gives fits for the cross section. Therefore, in this study, we compute the reaction rate coefficient using their fit for total cross section. The following assumptions are made in our calculation:
\begin{itemize}[leftmargin=*]
\item The total cross section is linearly extrapolated in the $\log_{10} E$-$\log_{10}\sigma(E)$ space for $E/\mu > 10\;\mathrm{MeV/u}$, where $E$ is the collision energy, $\mu$ the reduced mass of a $\proton$-$\HeI$ pair, and $u$ the unified atomic mass unit.
\item The threshold collision energy of the reaction is $0.403\;\mathrm{hartree}$.
\item The velocity distribution is Maxwell-Boltzmann.
\end{itemize}
Our fit to the calculated rate coefficient is given by 
\begin{eqnarray}
\scalebox{0.85}{$\displaystyle
\kR{34} = 
\begin{cases}
0, & x < x_{1} \smallskip\\
\begin{aligned}[c]
\dex(&-1.76287220621074\times 10^{4}     \\
    +& x(2.1445848497045 \times 10^{4}    \\
    +& x(-1.06074374926831 \times 10^{4}  \\
    +& x(2.65427074189074 \times 10^{3}   \\
    +& x(-3.34973736941601 \times 10^{2}  \\
    &\hspace{.75em}+17.0141958222459x))))),
\end{aligned} & x_{1} \leq x < x_{2} \smallskip\\
\begin{aligned}[c]
\dex(& 1.09553146475757 \times 10^{5}     \\
    +& x(-1.72975474377577 \times 10^{5} \\
    +& x(1.1819752981393 \times 10^{5}   \\
    +& x(-4.57115774717229 \times 10^{4} \\
    +& x(1.09520685214731 \times 10^{4}  \\
    +& x(-1.6655912987892 \times 10^{3}  \\
    +& x(1.5708504189844 \times 10^{2}   \\
    +& x(-8.40288638997883 \\
    &\hspace{.75em}+0.19524950645106x)))))))),
\end{aligned} & x_{2} \leq x < x_{3} \smallskip\\
\begin{aligned}[c]
\dex(& 1.42783030693601 \times 10^{3}     \\
    +& x(-1.11398874929773 \times 10^{3}  \\
    +& x(3.59957914543843 \times 10^{2}   \\
    +& x(-63.966577618147   \\
    +& x(6.89280662474058   \\
    +& x(-0.463872626389121 \\
    +& x(1.91125347347231 \times 10^{-2}  \\
    +& x(-4.41955097831939 \times 10^{-4} \\
    &\hspace{.75em}+4.39952423140937 \times 10^{-6}x)))))))),
\end{aligned} & x_{3} \leq x < x_{4} \smallskip\\
\begin{aligned}[c]
\dex(& 5.93253991498311  \\
    -& 1.49993832690037x),
\end{aligned} & x \geq x_{4},
\end{cases}
$}
\end{eqnarray}
where $x=\log_{10}\Tgas$, $x_{1}=2.477121255$, $x_{2}=4.25$, $x_{3}=6.5$, and $x_{4}=12.5$.

\subsection{Mutual neutralization in $\HeII$-$\Hminus$ collision (R42)} \label{appendix:subsec:R42}
\citet{glover08:_uncer_hd} have used the reaction rate coefficient given by
\begin{equation}
k_{\rm R42,GA08} = 2.32\times 10^{-7} \left(\frac{\Tgas}{300}\right)^{-0.52}\exp\left(\frac{\Tgas}{22400}\right).
\end{equation}
However, the value of this formula gets large significantly in high temperature regime and this causes incorrect chemical abundance in our simulations. Therefore, we recalculate the rate coefficient from the original cross section data of \citet{peart94:_merged_h_d} and \citet{olamba96:_improv}, the latter of which agrees well with a recent theoretical calculation by \citet{chibisov97:_one}. We first read out the cross section data from Fig.7\footnote{Data of \citet{peart94:_merged_h_d} is also shown in the figure.} in \citet{olamba96:_improv} using a digitizer software because the data is not given in tabular form. Next, we derive a fit for the cross section data. Finally, we compute the rate coefficient using the analytic function fitted for the cross section data and assuming the Maxwell-Boltzmann distribution. Our fit to the rate coefficient is given by
\begin{eqnarray}
\scalebox{0.85}{$\displaystyle 
\kR{42} =
\begin{cases}
\begin{aligned}[c]
\dex(&-7.36259040688347     \\
    +& x(5.73551820505366 \times 10^{-2} \\
    +& x(-6.40028343922151 \times 10^{-2} \\
    +& x(4.21597448292023 \times 10^{-2} \\
    +& x(-4.92314886406864 \times 10^{-2} \\
    +& x(4.17244612509607 \times 10^{-2} \\
    +& x(-1.70596424267789 \times 10^{-2} \\
    +& x(3.22894205021298 \times 10^{-3} \\
     &\hspace{.75em}-2.28791697512786 \times 10^{-4}x)))))))),
\end{aligned} & x < 3 \smallskip\\
\begin{aligned}[c]
\dex(& -92.7574528385805      \\
    +& x(1.45918325797873 \times 10^{2} \\
    +& x(-1.05864511112237 \times 10^{2} \\
    +& x(42.5804596528042 \\
    +& x(-10.381328443496 \\
    +& x(1.56990347212684 \\
    +& x(-1.43638981364979 \times 10^{-1} \\
    +& x(7.26694756199783 \times 10^{-3} \\
     &\hspace{.75em}-1.55781479113395 \times 10^{-4}x)))))))).
\end{aligned} & 3 \leq x < 8.5 \smallskip\\
\begin{aligned}[c]
\dex(& 5.42380317714043  \\
    -& 1.48800503180287x),
\end{aligned} & x \geq 8.5,
\end{cases}
$}
\end{eqnarray}
where $x=\log_{10}\Tgas$.

\section{Self-gravity calculation for an axisymmetric system} \label{appendix:sec:self_gravity}
Here, we explain the computational method of self-gravity in the axisymmetric cylindrical coordinate.
In the cylindrical coordinate, the $z$- and $R$-components of the acceleration due to self-gravity are written as
\begin{eqnarray}
&& a^{\mathrm{SG}}_{z}(R,z) \nonumber\\
&& \qquad = -\iiint R' \diff R' \diff \phi' \diff z' \frac{G\rho(R',z')(z-z')}{D}, \\
&& a^{\mathrm{SG}}_{R}(R,z) \nonumber\\
&& \qquad = -\iiint R' \diff R' \diff \phi' \diff z' \frac{G\rho(R',z')(R-R'\cos\phi')}{D}, \\
&& D = \{R^{2} + R'^{2} - 2RR'\cos\phi' + (z-z')^{2}\}^{3/2}, 
\end{eqnarray}
where $G$ is the gravitational constant. By introducing variables $a \equiv R^{2}+R'^{2}+(z-z')^{2}$, $b \equiv 2RR'$, and $s \equiv b/a$, we can rewrite the equations above into
\begin{eqnarray}
&& a^{\mathrm{SG}}_{z}(R,z) = -\iint R' \diff R' \diff z' \frac{G\rho(R',z')(z-z')F_{1}(s)}{a^{3/2}}, \label{eq:agz} \\
&& a^{\mathrm{SG}}_{R}(R,z) \nonumber\\ 
&& \qquad = -\iint R' \diff R' \diff z' \frac{G\rho(R',z')(RF_{1}(s)-R'F_{2}(s))}{a^{3/2}}, \label{eq:agR}
\end{eqnarray}
where
\begin{eqnarray}
F_{1}(s) & \equiv & \int^{2\pi}_{0} \diff \phi' \frac{1}{\{1-s\cos\phi'\}^{3/2}}, \label{eq:f1} \\
F_{2}(s) & \equiv & \int^{2\pi}_{0} \diff \phi' \frac{\cos\phi'}{\{1-s\cos\phi'\}^{3/2}}. \label{eq:f2}
\end{eqnarray}
By definition, $s \in (0,1]$.  
With the usual discrete representation, equations (\ref{eq:agz}) and (\ref{eq:agR}) can be discretized as
\begin{eqnarray}
a^{\mathrm{SG}}_{z,ij} & = & -\sum_{i',j'} \frac{G m_{i'j'}(z_{j}-z_{j'})F_{1}(s_{i'j'ij})}{a^{3/2}_{i'j'ij}}, \label{eq:agz_discretized} \\
a^{\mathrm{SG}}_{R,ij} & = & -\sum_{i',j'} \frac{G m_{i'j'}(R_{i}F_{1}(s_{i'j'ij})-R_{i'}F_{2}(s_{i'j'ij}))}{a^{3/2}_{i'j'ij}}, \label{eq:agR_discretized} \\
m_{i'j'} & = & R_{i'} \Delta R_{i'} \Delta z_{j'} \rho_{i'j'}.
\end{eqnarray}
Thus, if we prepare tables for $F_{1}(s)$ and $F_{2}(s)$ in advance\footnote{There are a few points to note: (1) The integrals in the right-hand sides of equations (\ref{eq:f1}) and (\ref{eq:f2}) should be performed \textit{simultaneously}. In other words, the same stepsize $\Delta \phi'$ should be used for the computation of both integrals. Without this, numerical errors in $F_{1}(s)$ and $F_{2}(s)$ are different each other and the ratio $F_{1}(s)/F_{2}(s)$ can have a large error. As the result, an error in the acceleration can be large. In this study, the integrals are simultaneously performed by the fourth order Runge-Kutta method with adaptive stepsize control described in \citet{press92:_numer_recip_in_fortr} (see their \S 16.2). (2) the tables for $F_{1}(s)$ and $F_{2}(s)$ need to have sufficient resolution near $s\approx 0$ and $\approx 1$, because $F_{1}(s)$ and $F_{2}(s)$ change rapidly there. In this study, we use $2^{14}$ grid points, a quarter of which is used to cover the ranges $(0,0.1]$ and $[0.9,1]$ logarithmically (the grids become finer toward $s=0$ or $1$). The range $[0.1,0.9]$ is uniformly covered by the remaining three-quarter grid points.}, the acceleration due to self-gravity for cell $(i,j)$ can be calculated by simply summing contributions from all other cells. However, such a straightforward approach is very time-consuming and a faster computational method is needed. In this study, we apply the tree method (\citealt{barnes86:_nlog_n}), which is widely used in $N$-body simulations, to accelerate the calculation.

\begin{figure}
\centering
\includegraphics[clip,width=\linewidth]{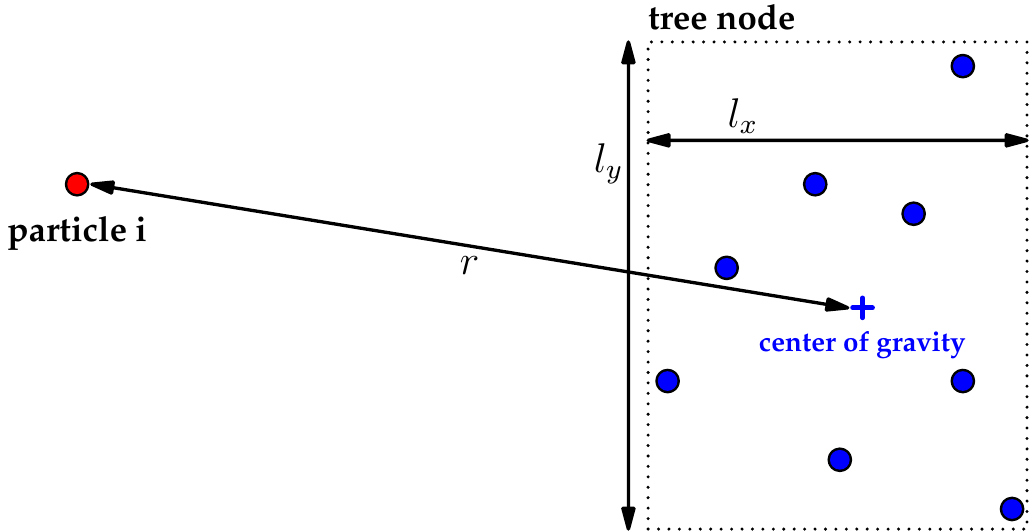}
\caption{A schematic illustration of a gravitational interaction between a particle and a tree node in $N$-body simulation. For simplicity, two dimensional case is shown. The red point labeled "particle $i$" is the particle which we're currently computing gravitational force acting on.}
\label{fig:tree_method_nbody}
\end{figure}

\begin{figure}
\centering
\includegraphics[clip,width=\linewidth]{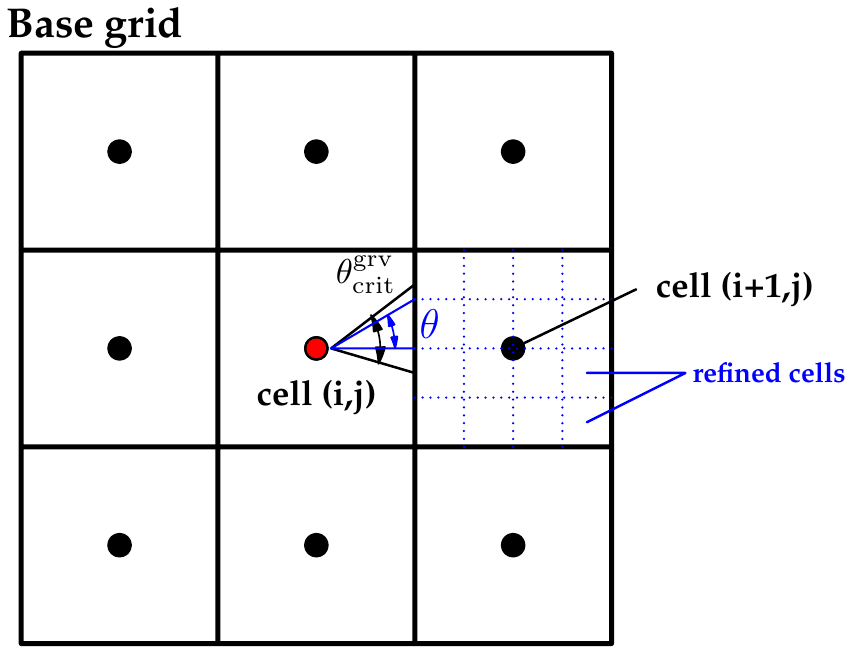}
\caption{A schematic illustration of mesh refinement of cell $(i+1,j)$ required by a given opening criterion $\theta^{\mathrm{grv}}_{\mathrm{crit}}$. In the above example, cell $(i+1,j)$ is refined to $4\times 4$ subcells in order for each subcell $(k,l)$ to satisfy the opening angle condition $\theta_{k,l}<\theta^{\mathrm{grv}}_{\mathrm{crit}}$ and the subcells can be replaced by `rings` passing through the centers of the subcells. }
\label{fig:tree_method_MR}
\end{figure}

  Figure~\ref{fig:tree_method_nbody} illustrates a gravitational interaction between particle $i$ and a tree node in $N$-body simulations schematically. In the tree method, particles in the node are replaced by one virtual particle if the following condition is satisfied,
\begin{equation}
\theta \equiv \frac{\max(l_{x},l_{y})}{r} < \theta^{\mathrm{grv}}_{\mathrm{crit}}, \label{eq:tree_condition}
\end{equation}
where $r$ is the distance between particle $i$ and the center of gravity of particles in the node, $l_{x}$ and $l_{y}$ are the sizes of the node along $x$- and $y$-axes, respectively. In this way, the computational cost is reduced. The simplest way to apply the tree method is to treat cells as particles in $N$-body simulations. This approach, however, requires mesh refinement because, for a target cell $(i,j)$, even its adjacent cells do not satisfy the condition (\ref{eq:tree_condition}) for a typical value of $\theta^{\mathrm{grv}}_{\mathrm{crit}}$ (e.g., $0.5$) due to finite sizes of cells. This situation is schematically shown in Fig.~\ref{fig:tree_method_MR}, where cell $(i+1,j)$ is refined to $4\times 4$ subcells because the opening angle of cell $(i+1,j)$ seen from the center of cell $(i,j)$ is larger than a given criterion. The subcells shown in Fig.~\ref{fig:tree_method_MR} satisfy the condition (\ref{eq:tree_condition}) and can be replaced by 'rings' passing through the centers of the subcells. The refinement level $l_{\mathrm{MR}}$, which is defined such that the number of subcells is $2^{l_{\mathrm{MR}}}\times 2^{l_{\mathrm{MR}}}$,  of each cell is purely determined by $\theta^{\mathrm{grv}}_{\mathrm{crit}}$ and the aspect ratios of the adjacent cells $\Delta R_{i'}/\Delta z_{j'}$, where $(i',j')=(i\pm 1,j),\;(i,j\pm 1)$. For instance, $l_{\mathrm{MR}}$ should be $\geq 2$ for $\theta^{\mathrm{grv}}_{\mathrm{crit}}=0.5$ and an uniform grid ($\Delta R_{i}/\Delta z_{j}=1$ for all $(i,j)$). Thus, the total number of the tree nodes is much larger than that of the cells in this approach and this is not desirable in the light of computational efficiency.

\begin{figure}
\centering
\includegraphics[clip,width=\linewidth]{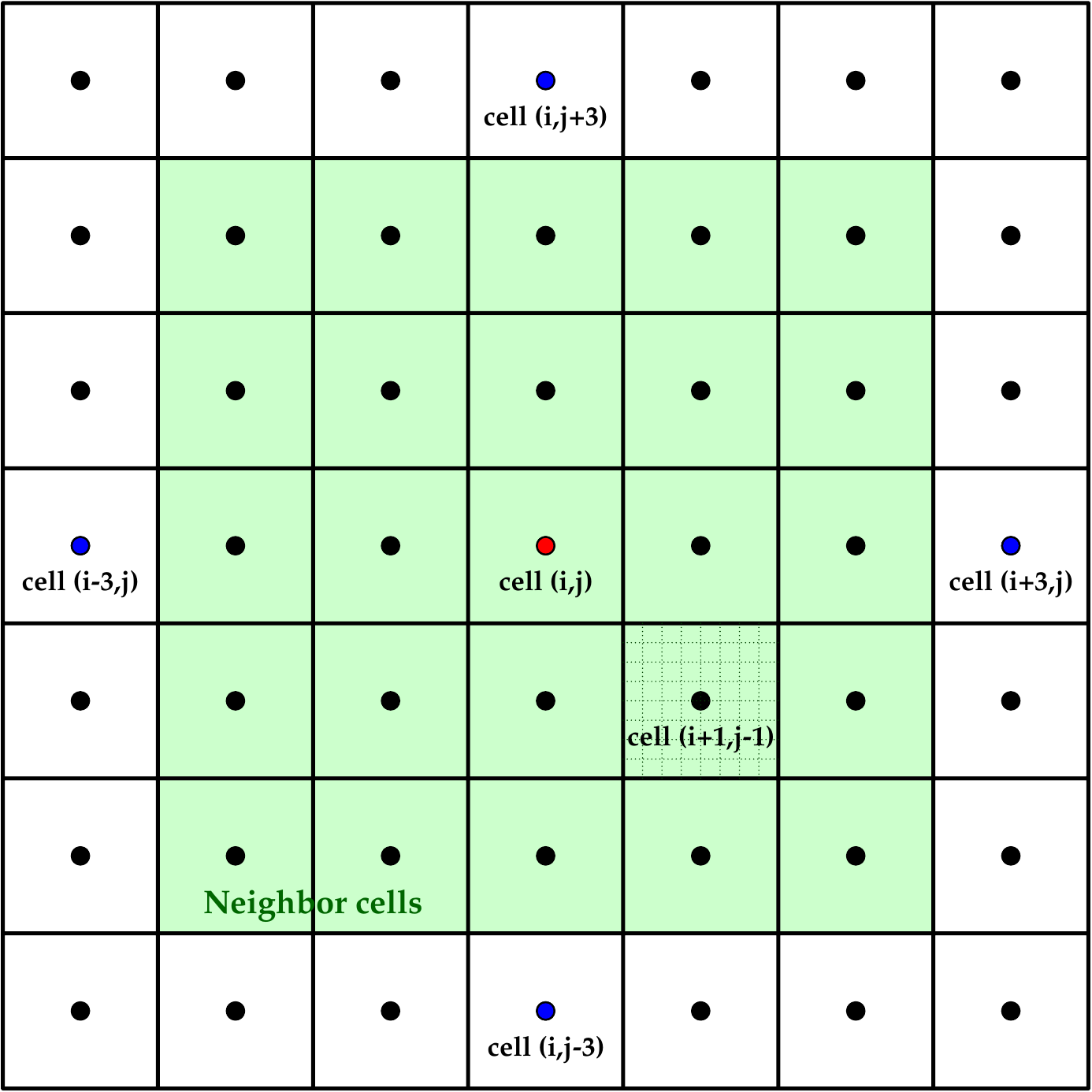}
\caption{A schematic illustration of neighbor cells. The neighbor cells of cell $(i,j)$ are shown by lightgreen cells. Each neighbor cell is refined to a level specified by $l_{\mathrm{MR,near}}$ like cell $(i+1,j-1)$. The cells with blue points are used to determine the refinement level of cell $(i,j)$.}
\label{fig:tree_method_NFdiv}
\end{figure}

The refinement level can be reduced by splitting gravitational force into contributions from neighbor cells and the others. Here, we call the former \textit{near-field force} and the latter \textit{far-field force}. An example of the split is shown in Fig.~\ref{fig:tree_method_NFdiv}, where the neighbor cells of cell $(i,j)$ is defined as $5\times 5$ cells around cell $(i,j)$. In this case, the refinement level of cell $(i,j)$ is computed by $\theta^{\mathrm{grv}}_{\mathrm{crit}}$ and the coordinates of cells $(i\pm 3,j),\; (i,j\pm 3)$ (the blue points in the figure). Thus, the refinement level becomes small compared the case shown in Fig.~\ref{fig:tree_method_MR}. In this study, we control the number of neighbor cells by the parameter $N_{\mathrm{nb}}$ and $N_{\mathrm{nb}}=25$ is assumed. Note that this reduction method is effective only on uniform grid.

The acceleration by the near-field force is computed summing contributions from \textit{refined} neighbor cells as follows:
\begin{eqnarray}
a^{\mathrm{near}}_{z,ij} & = & - \sum_{i',j'}\sum_{k'l'} \frac{G m_{k'l'}(z_{j}-z_{l'})F_{1}(s)}{a^{3/2}}, \label{eq:agz_near} \\
a^{\mathrm{near}}_{R,ij} & = & - \sum_{i',j'}\sum_{k'l'} \frac{G m_{k'l'}(R_{i}F_{1}(s)-R_{k'}F_{2}(s))}{a^{3/2}}, \label{eq:agR_near}
\end{eqnarray}
\begin{eqnarray}
a & = & R^{2}_{i} + \tilde{R}^{2}_{k'} + (z_{j}-z_{l'})^{2} \\
b & = & 2R_{i}\tilde{R}_{k'} \\
s & = & b/a \\
m_{k'l'} & = & \rho_{k'l'} R_{k'}\delta R_{i'}\delta z_{j'}
\end{eqnarray}
\begin{eqnarray}
z_{l'} & = & 
\begin{cases}
z_{j'} - \delta z_{j'}\left(\frac{n}{2}-l'+\frac{1}{2}\right), & 1 \leq l' \leq \frac{n}{2} \\
z_{j'} + \delta z_{j'}\left(l'-\frac{n}{2}-\frac{1}{2}\right), & \frac{n}{2} < l' \leq n  
\end{cases} \\
\tilde{R}_{k'} & = & R_{k'} + \frac{\delta R^{2}_{i'}}{12 R_{k'}} \\
R_{k'} & = & 
\begin{cases}
R_{i'} - \delta R_{i'}\left(\frac{n}{2}-k'+\frac{1}{2}\right), & 1 \leq k' \leq \frac{n}{2} \\
R_{i'} + \delta R_{i'}\left(k'-\frac{n}{2}-\frac{1}{2}\right), & \frac{n}{2} < k' \leq n  
\end{cases}
\end{eqnarray}
\begin{eqnarray}
\delta R_{i'} & = & \frac{\Delta R_{i'}}{n} \\
\delta z_{j'} & = & \frac{\Delta z_{j'}}{n} \\
n & = & 2^{l_{\mathrm{MR,near}}} 
\end{eqnarray}
where $(i',j')$ indicates cell $(i,j)$ or its neighbor cells, $(k',l')$ the indices for subcells of cell $(i',j')$, $l_{\mathrm{MR,near}}$ the refinement level of the neighbor cells. Assuming that density is uniform within a cell (i.e., $\rho_{k'l'}=\rho_{i'j'}$), we can perform the summation over $k',l'$ for each combination of $(i'j',ij)$. As the result, the equations above can be rewritten in the form $a^{\mathrm{near}}_{R[z],ij}=\sum_{i'j'}\rho_{i'j'}F_{R[z],i'j'ij}$, where $F_{R[z],i'j'ij}$ stores the result of the summation and depends only on the geometry of the grid. Thus, the computational cost is effectively independent of $l_{\mathrm{MR,near}}$. In this study, we assume $l_{\mathrm{MR,near}}=6$. The accelerations by the far-field force is calculated by the tree method in the usual manner.

\begin{figure}
\centering
\includegraphics[clip,width=\linewidth]{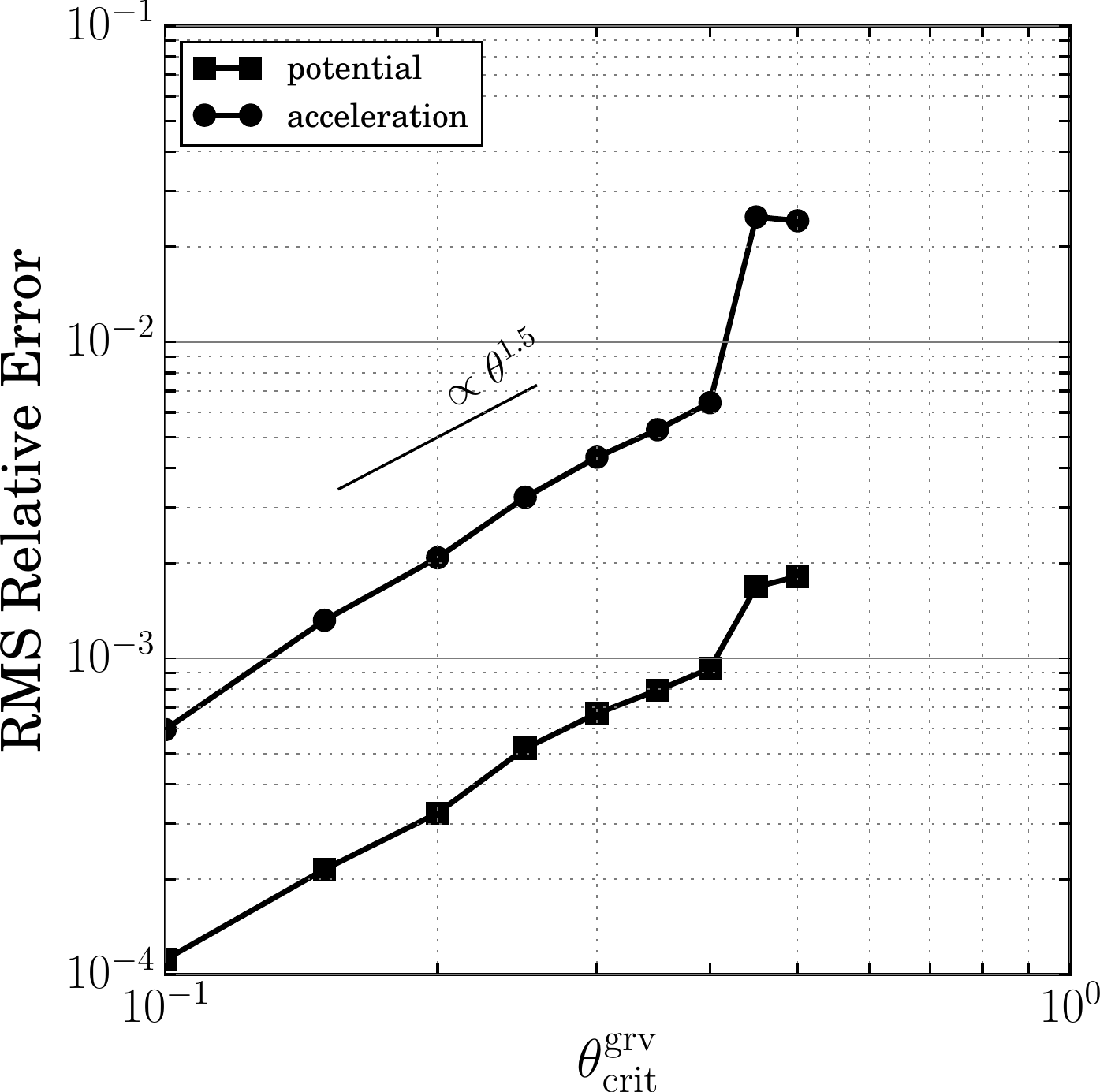}
\caption{RMS relative errors of gravitational potential and acceleration for different values of $\theta^{\mathrm{grv}}_{\mathrm{crit}}$. In the measurements, the same density distribution as that used for the simulations in \S~\ref{sec:numerical_results} is assumed and the number of grid points is $512^{2}$. The RMS relative error of acceleration is calculated by the formula $\sqrt{\frac{1}{N}\sum_{ij}\left(\frac{|\bmath{a}_{ij,\mathrm{tree}}-\bmath{a}_{ij,\mathrm{exact}}|}{|\bmath{a}_{ij,\mathrm{exact}}|}\right)^{2}}$, where $N$ is the number of grid points. A similar formula is used for the computation of the RMS relative error of potential.}
\label{fig:RMS_relative_errors}
\end{figure}

In order to check the accuracy of our method, we compare the gravitational acceleration computed by the Tree method with an `exact` solution for different values of $\theta^{\mathrm{grv}}_{\mathrm{crit}}$ and examine the dependency of a root-mean-square (RMS) relative error on the opening angle criterion $\theta^{\mathrm{grv}}_{\mathrm{crit}}$. The `exact` solution is numerically obtained by applying Eqs.(\ref{eq:agz_near}-\ref{eq:agR_near}) directly to \textit{all the cells} assuming they are neighbor cells. Figure~\ref{fig:RMS_relative_errors} shows the RMS relative error as a function of $\theta^{\mathrm{grv}}_{\mathrm{crit}}$ for the same density distribution as the initial condition used in the simulations in \S\ref{sec:numerical_results}. We can see from the figure that the RMS relative error is approximately proportional to $(\theta^{\mathrm{grv}}_{\mathrm{crit}})^{1.5}$ below $\theta^{\mathrm{grv}}_{\mathrm{crit}}=0.4$ and it abruptly increases for $\theta^{\mathrm{grv}}_{\mathrm{crit}}>0.4$. To examine the reason of the abrupt increase of the RMS relative errors, we make a comparison of the relative error distributions between $\theta^{\mathrm{grv}}_{\mathrm{crit}}=0.35$ and $0.5$ and find that locations of large relative errors in the $\theta^{\mathrm{grv}}_{\mathrm{crit}}=0.5$ case are concentrated near the mid-plane, while such concentration is not found for the $\theta^{\mathrm{grv}}_{\mathrm{crit}}=0.35$ case. However, unfortunately, the exact cause and mechanism of the abrupt increase are not identified. Because the distribution of relative error are relatively smooth (i.e., locations of large relative errors are not concentrated at small portions of the computational region) and its magnitude is acceptably small in the $\theta^{\mathrm{grv}}_{\mathrm{crit}}=0.35$ case, we decide to use $\theta^{\mathrm{grv}}_{\mathrm{crit}}=0.35$ in this study.

\begin{figure}
\centering
\includegraphics[clip,width=\linewidth]{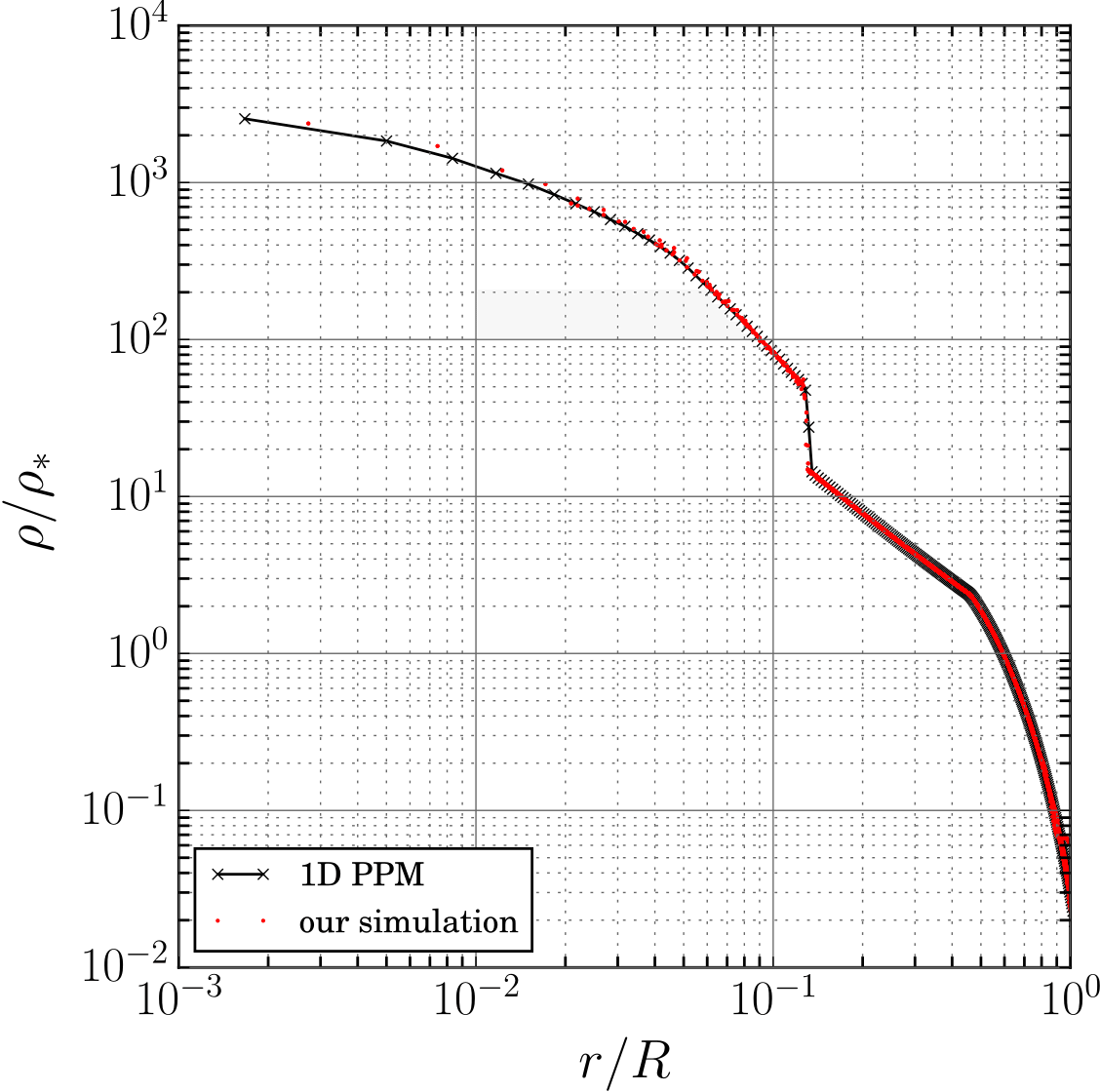}
\caption{Comparison of spatial distribution of the normalized density $\rho/\rho_{\ast}$ in the Evrard test at $t/t_{\ast}=0.7$. The red dots show the result of our simulation with $\theta^{\mathrm{grv}}_{\mathrm{crit}}=0.35$, while the black line with crosses shows that of a (spherically-symmetric) one-dimensional piece-wise parabolic method (PPM) calculation (\citealt{steinmetz93}). For clarity, a tenth of our result is plotted.}
\label{fig:1D_rho_Evrard_test}
\end{figure}

\begin{figure}
\centering
\includegraphics[clip,width=\linewidth]{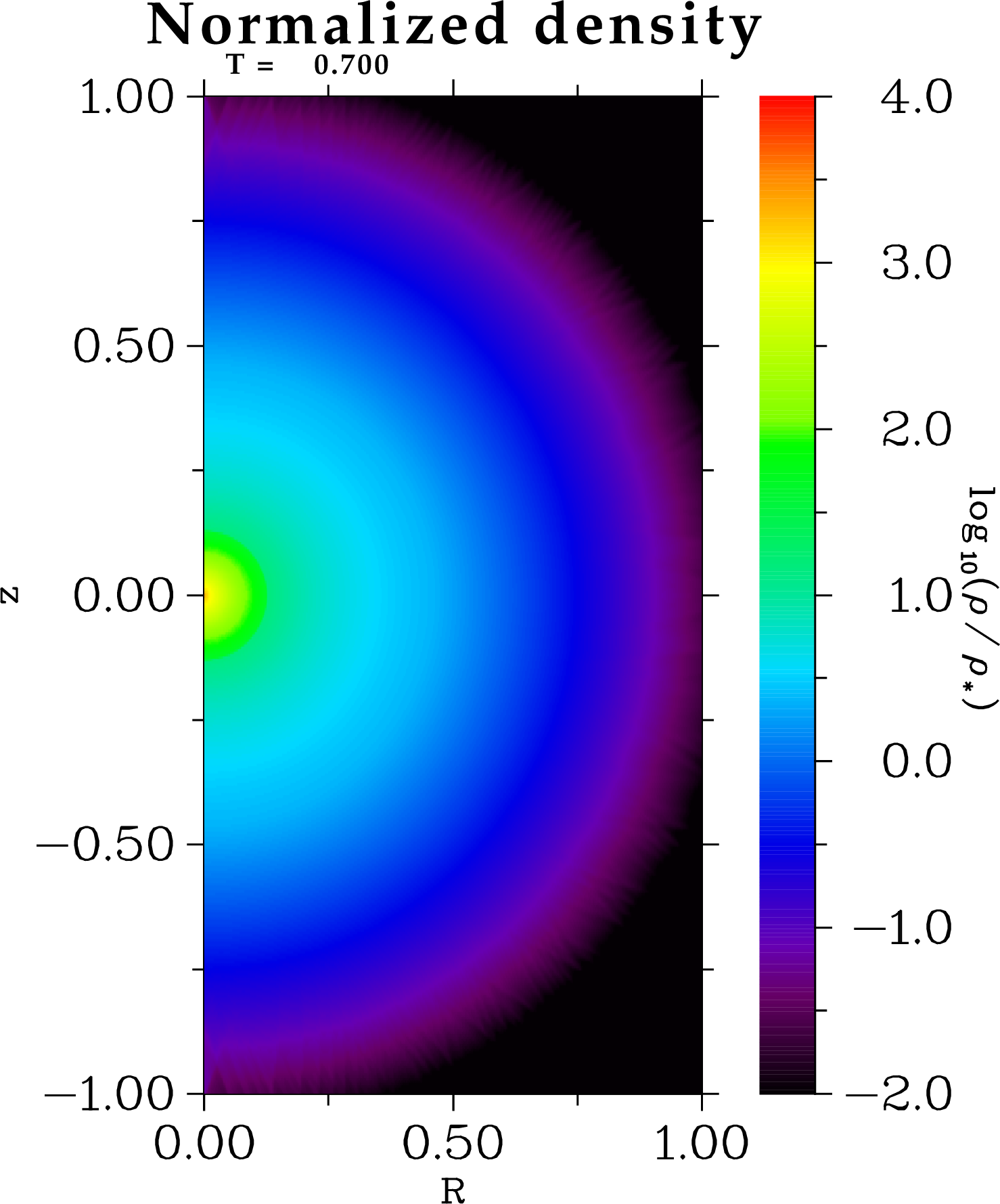}
\caption{Two-dimensional (normalized) density distribution in the Evrard test at $t/t_{\ast}=0.7$.}
\label{fig:2D_rho_Evrard_test}
\end{figure}

In order to check if the decision above is appropriate and for further investigation on the reliability of the Tree method, we performed two test simulations. One is the Evrard test (\citealt{evrard88:_beyon_n}) which follows self-gravitational adiabatic collapse of a cold gas sphere. This test is often used to check the validity and accuracy of simulation codes (e.g., \citealt{steinmetz93,springel05:_gadget,cullen10:_invis}) and its initial setup is detailed in \citet{evrard88:_beyon_n}. In the same unit system as \citet{evrard88:_beyon_n}, a computational box of size $(R_{\max},z_{\min},z_{\max})=(2.5,-2.5,2.5)$ is used in this study. We uniformly cover the box with $1024^{2}$ grid points. Figure~\ref{fig:1D_rho_Evrard_test} compares the result obtained by the Tree method with that of a piece-wise parabolic method (PPM) calculation (\citealt{steinmetz93}). Our result agrees with that of the PPM calculation. In addition, there is no prominent features breaking the spherical symmetry as shown in Fig.~\ref{fig:2D_rho_Evrard_test}, although there are very fine structures around $r/R\approx 1$.

As an another test, we re-perform a simulation for model \texttt{gra01\_wo\_sca\_MTL\_SG} with $\theta^{\mathrm{grv}}_{\mathrm{crit}}=0.2$ and compare the result with that with $\theta^{\mathrm{grv}}_{\mathrm{crit}}=0.35$. We choose this model because a very thin dense ($>10^{8}\;\nden$) disk forms along the mid-plane where self-gravity is strong. We find that there is no essential difference between them. From these results, we conclude that the tree method with $\theta^{\mathrm{grv}}_{\mathrm{crit}}=0.35$ is sufficiently accurate to treat the physical system considered in this study.

\section{Boundary zone} \label{appendix:sec:boundary_zone}
As described in \S~\ref{subsec:numerical_setup}, density or thermal pressure sometimes becomes negative at exceptionally low density ($\nH \ll 10^{-20}\;\nden$) regions that are formed near the symmetric axis owing to strong radiation pressure from AGN. To prevent such unphysical behaviors, we place an boundary zone around the symmetric axis, in which physical quantities $\rho$, $\bmath{v}$, $\Tgas$ and $p_{\mathrm{th}}$ are artificially modified. In this section, we describe the detail of the algorithm adopted in this study.

At the end of each time step, we first reduce a deviation of the physical quantities $q=(\rho, \Tgas, \bmath{v})$ of low density ($\rho < \rho_{0}$) cells in the zone from user-specified values $q_{0}=(\rho_{0}, T_{\mathrm{gas},0}, \bmath{v}_{0})$ at a rate $a_{d}$ as follows (\citealt{machida09:_format_galac_center_magnet_loops}) :
\begin{eqnarray}
q^{\mathrm{new}} & = & q - a_{d}(q-q_{0}), \\
a_{d} & = & 0.1[1-\tanh(6\tilde{R} - 3)],
\end{eqnarray}
where $\tilde{R}\equiv R/R_{\mathrm{BZ}}$, $R_{\mathrm{BZ}}=6.25\times 10^{-2}\;\pc$ is the radius of the boundary zone. The user-specified values are chosen as follows: $\rho_{0}=10^{-16}\mH\;\mathrm{g\;cm^{-3}}$, $T_{\mathrm{gas,0}}=5\times 10^{6}\;\mathrm{K}$, and $\bmath{v}_{0}=(\mathrm{sign}(v_{R})v_{\mathrm{MP}}(|v_{R}|/|\bmath{v}|),\; v_{\mathrm{c}},\; \mathrm{sign}(v_{z})v_{\mathrm{MP}}(|v_{z}|/|\bmath{v}|))$, where $v_{\mathrm{MP}}=4\times 10^{3}\;\kms$ and $v_{\mathrm{c}}$ is the circular velocity. The $R$ and $z$ components of the velocity are updated only if $|v_{i}|>v_{\mathrm{MP}}(|v_{i}|/|\bmath{v}|)$ ($i=R,z$). In addition, if $|v_{i}|> v_{i,\max}\equiv 5\times 10^{3}(|v_{i}|/|\bmath{v}|) \;\kms$ ($i=R,z$), we immediately set $|v_{i}|=v_{i,\max}$. The operation above makes unphysical behaviors harder to occur by artificially increasing $e_{\mathrm{th}}$ and decreasing $e_{\mathrm{kin}}$ $(\equiv \frac{1}{2}\rho \bmath{v}^{2}$). An another cause of unphysical behaviors is a large pressure difference between a low density cell and its neighbor cell(s) with normal density. To reduce it, we update $p_{\mathrm{th}}$ of low density cells and its adjacent cells as follows:
\begin{equation}
p^{\mathrm{new}}_{\mathrm{th}} = p_{\mathrm{th}} + \kappa \nabla^{2} p_{\mathrm{th}},
\end{equation}
where $\kappa = 0.1/\max_{(i,j)\in \mathcal{T}}(|\nabla^{2}p_{\mathrm{th}}|/p_{\mathrm{th}})$, $(i,j)$ is the cell index, and $\mathcal{T}$ is the set of cell indices of low density cells and their neighbor cells. We repeat the update until the number of the iteration reaches 16 or the following inequality is satisfied:
\begin{equation}
\max_{(i,j)\in \mathcal{L}, i-1\leq i'\leq i+1,j-1\leq j'\leq j+1} \frac{|p_{\mathrm{th},i'j'}-p_{\mathrm{th},ij}|}{p_{\mathrm{th},ij}} < 10^{3},
\end{equation}
where $\mathcal{L}$ is the set of cell indices of low density cells.

\section{Finite volume method for radiative transfer} \label{appendix:sec:FVM_RT}
In this section, we describe the detail of the finite volume method for radiative transfer (FVMRT) and present test calculations including a comparison between the FVMRT and the short-characteristics method (SCM), which is used in astrophysical applications.

\subsection{FVMRT} \label{appendix:subsec:FVMRT}
The FVMRT has been developed in the field of engineering application (e.g., \citealt{raithby90:_finit_volum_method_predic_radian,chui92:_predic_radiat_trans_cylin_enclos,chai94:_finit_volum_method_radiat_heat_trans,murthy98:_radiat_heat_trans_axisy_geomet,salah04:_numer,kim05:_model,tian05:_two_schem_axisy_radiat_heat,kim08:_asses}) and it has the significant advantage that radiation energy is fully conserved. This important property does not reside in the SCM on curvilinear coordinates (e.g., \citealt{dullemond00,busche00,noort02:_multid_non_lte_radiat_trans,milic13:_trans}). Another advantage of using the FVMRT is that it does not require a closure relation unlike the flux-limited diffusion (FLD) approximation method and the M-1 closure method, because we solve the radiative transfer equation (RTE, Eq.(\ref{eq:RTE})) directly. The scheme adopted in this study is almost the same as that used in \citet{kim08:_asses}, but some of coefficients in the original scheme are modified. Here, we explain the detail of our scheme. 

First of all, we summarize the assumptions made here: (A1') emissivity is isotropic, (A2') scattering is isotropic and coherent, (A3') gas is at rest. Hence, all the terms proportional to $\frac{v}{c}$ are ignored. Thus, the scheme shown below is valid only for $\frac{v}{c} \ll 1$. For simplicity, we omit the frequency dependency of intensity.

Next, we give the definition of coordinate grids that are used in the following explanation:
\begin{description}[topsep=1pt,leftmargin=*,font=\normalfont\itshape\textbullet\space]
\item[Axisymmetric cylindrical grid] This is the same grid as that is used in hydrodynamic calculation. The positions of cell centers, the cell widths, and the unit vectors are denoted by the symbols $R_{i}$, $z_{j}$, $\Delta R_{i}$, $\Delta z_{j}$, $\hat{\bmath{e}}_{R}$, and $\hat{\bmath{e}}_{z}$, respectively. The indices in $R_{i}$ and $z_{j}$ can be half integers when they represent cell faces (e.g., $R_{i+\frac{1}{2}}$). We solve the RTE on this grid.
\item[Local Cartesian coordinate] This is defined at every spacial point. Its origin coincides with the spatial point we choose and their $x$ and $z$ axes point toward the directions of $\hat{\bmath{e}}_{R}$ and $\hat{\bmath{e}}_{z}$, respectively. This coordinate is used to define the directions of discretized rays.
\item[Three-dimensional (3D) cylindrical grid] The positions of cell centers are denoted by ($R_{i}$, $\Phi_{m}$, $z_{j}$), where $\Phi_{m}$ is the azimuth. The corresponding 3D Cartesian coordinate is denoted by ($X$, $Y$, $z$). These coordinates are used to link the intensity at cell face to the intensity at cell center.
\end{description}

The basic approach to obtain a discretized equation is almost same as the finite-volume method in computational fluid dynamics (hereafter, FVMCFD). As in the FVMCFD, we first derive the integral form of the RTE by integrating the differential form of the RTE (\ref{eq:RTE}) over a control volume $\Delta V_{ij}$ and a control solid angle $\Delta\Omega^{lm}$. Assuming that intensity is constant over both the control volume and the control angle at a given time and using the Gauss theorem, the integral form of the equation can be written as
\begin{eqnarray}
& & \dfrac{1}{c}\frac{I^{lm}_{ij}-I^{lm}_{0,ij}}{\Delta t}\Delta V_{ij}\Delta\Omega^{lm} + \sum_{f}\int_{\Delta\Omega^{lm}}\int_{A_{f}} I^{lm}_{f} (\bmath{s}^{lm}\cdot \diff \bmath{A})\diff\Omega  \nonumber\\
&=& (-\chi_{ij} I^{lm}_{ij} + j_{\mathrm{eff},ij})\Delta V_{ij}\Delta\Omega^{lm}, \label{eq:RTE_integral1}
\end{eqnarray}
where $I^{lm}_{0,ij}$ and $I^{lm}_{ij}$ are the intensity at the initial and final states, respectively (corresponding to times $t$ and $t+\Delta t$). $I^{lm}_{f}$ is the intensity at cell face $f$ (generally, there are 6 cell faces in the cylindrical grid and we denote them by the symbols $R^{\pm}$, $\Phi^{\pm}$, $z^{\pm}$ as shown in the upper right portion of Fig.~\ref{fig:FVMRT_overview}), $A_{f}$ is the surface area of cell face $f$, and $\bmath{s}^{lm}$ is the unit vector along ray direction $(l,m)$. The effective emissivity $j_{\mathrm{eff},ij}$ is defined as follows:
\begin{equation}
j_{\mathrm{eff},ij} = j_{\mathrm{em},ij} + \dfrac{\alpha^{\mathrm{sca}}_{ij}}{2\pi}\sum_{l',m'}I^{l'm'}_{ij}\Delta\Omega^{l'm'}, \label{eq:jeff}
\end{equation}
where $j_{\mathrm{em},ij}$ is the emissivity and $\alpha^{\mathrm{sca}}_{ij}$ is the scattering coefficient. The summation in Eq.(\ref{eq:jeff}) covers the half of the entire range of solid angle due to the axisymmetry. To solve Eq.(\ref{eq:RTE_integral1}), we must determine the way of solid angle discretization (\S~\ref{subsubsec:solid_angle_discretization}), evaluate the double integrals in Eq.(\ref{eq:RTE_integral1}) (\S~\ref{subsubsec:double_integral}), and write the face intensity $I^{lm}_{f}$ as a function of $I^{lm}_{ij}$ to obtain the final form of discretized equation (\S~\ref{subsubsec:discretized_RTE}).

\subsubsection{Solid angle discretization}\label{subsubsec:solid_angle_discretization}
\begin{figure*}
\centering
\includegraphics[clip,width=\linewidth]{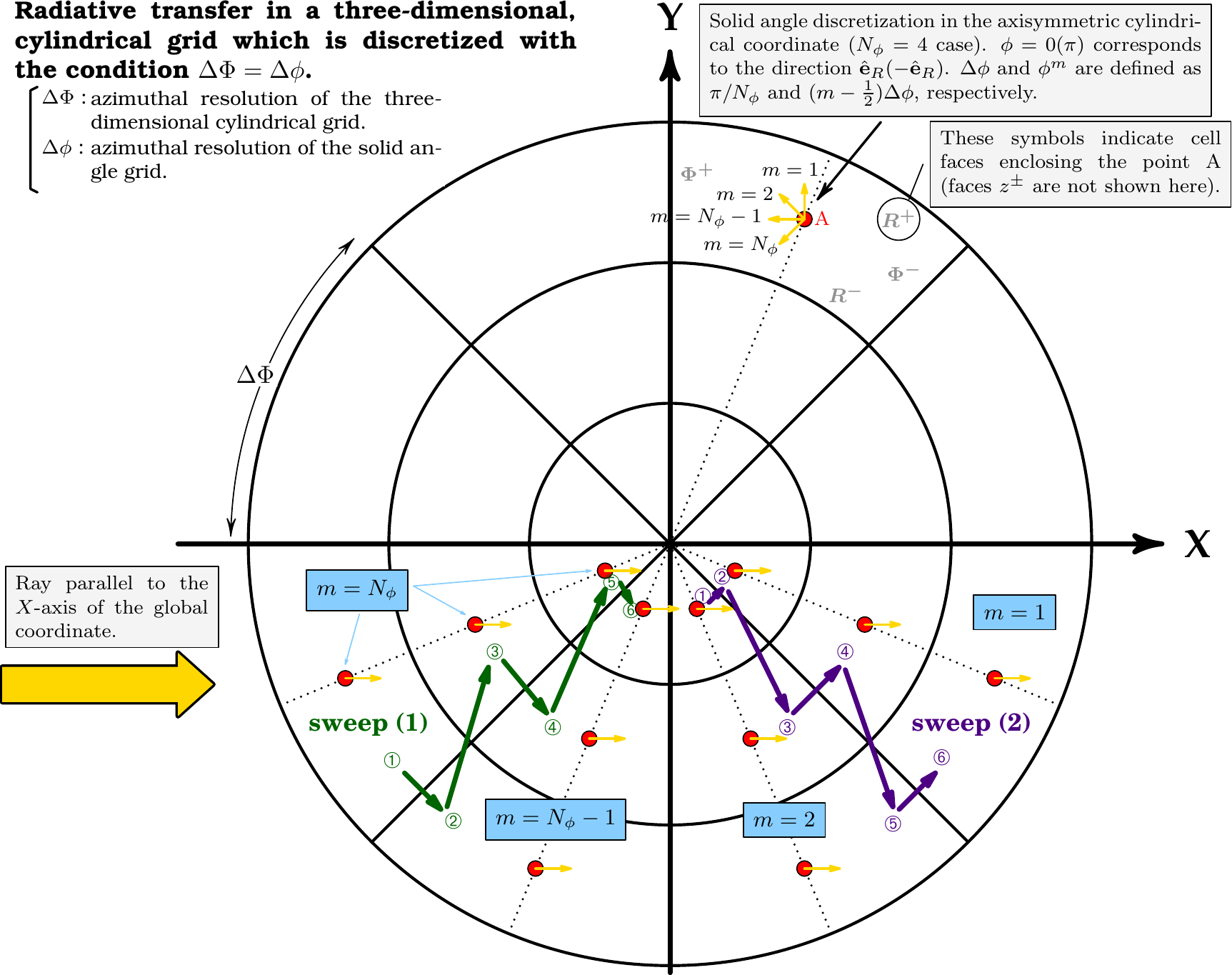}
\caption{A correspondence relationship between RT in the axisymmetric cylindrical grid and RT in the 3D cylindrical grid. The upper right portion of the figure schematically illustrates the way of discretization of solid angle in the axisymmetric cylindrical grid. As shown in the lower portion of the figure, RT in the axisymmetric cylindrical grid becomes equivalent to RT in the 3D cylindrical grid if its azimuth $\Phi$ is discretized so that $\Delta\Phi=\Delta\phi$, where $\Delta\Phi$ is the azimuthal resolution of the 3D cylindrical grid and $\Delta\phi$ is the azimuthal resolution of discretized solid angle. Owing to the condition $\Delta\Phi=\Delta\phi$, the set $\{\phi^{m}|m=1\operatorname{-}N_{\phi}\}$ is in bijection with the set $\{\Phi_{m}|m=1\operatorname{-}N_{\phi}\}$. All we need to do is to perform a single RT calculation for a ray which is parallel to the $X$ axis of the 3D Cartesian grid. In the RT calculation, the grid must be swept in the order indicated by green and purple arrows.}
\label{fig:FVMRT_overview}
\end{figure*}

We describe solid angle at each point in space by two angles $\phi$ and $\theta$, which are defined as the angles measured from the $x$ and $z$ axes of \textit{local Cartesian coordinate}. In this definition, $(\phi,\theta)=(0,\pi/2)$ corresponds to the direction $\hat{\bmath{e}}_{R}$ (see Fig.~\ref{fig:FVMRT_overview}). We discretize $\phi$ and $\theta$ by using $N_{\phi}$ and $N_{\theta}$ grid points and denote discretized angles by $\phi^{m}$ and $\theta^{l}$. Then, we define $\Delta\Omega^{lm}$ as follows:
\begin{equation}
\Delta\Omega^{lm} = \left[\cos\left(\theta^{l-\frac{1}{2}}\right)-\cos\left(\theta^{l+\frac{1}{2}}\right)\right]\left(\phi^{m+\frac{1}{2}}-\phi^{m-\frac{1}{2}}\right),
\end{equation}
where $\phi^{m\pm\frac{1}{2}}$ and $\theta^{l\pm\frac{1}{2}}$ are the boundaries of the control solid angle and they satisfy the relations: $\phi^{m}=(\phi^{m+\frac{1}{2}}+\phi^{m-\frac{1}{2}})/2$ and $\theta^{l}=(\theta^{l+\frac{1}{2}}+\theta^{l-\frac{1}{2}})/2$. For simplicity, we assume (i) uniform grids in both angles (i.e., $\Delta\theta^{l}\equiv \theta^{l+\frac{1}{2}}-\theta^{l-\frac{1}{2}}=\Delta\theta=\pi/N_{\theta}$ and $\Delta\phi^{m} \equiv \phi^{m+\frac{1}{2}}-\phi^{m-\frac{1}{2}}=\Delta\phi=\pi/N_{\phi}$)\footnote{Note that we only have to solve the RTE for $\phi=0\operatorname{-}\pi$ because of the axisymmetry,}, and (ii) that both $N_{\phi}$ and $N_{\theta}$ are even numbers.

Adopting the above discretization of solid angle, we can replace a RT problem in the axisymmetric cylindrical grid by a RT problem in the 3D cylindrical grid where its azimuth $\Phi$ is discretized with the condition $\Delta\Phi = \Delta\phi$. An specific example for $N_{\phi}=4$ is schematically shown in Fig.~\ref{fig:FVMRT_overview}, the upper right portion of which illustrates the discretization of solid angle and the lower portion of which shows that solving the RTE in the 3D cylindrical grid is equivalent to solving the RTE in the axisymmetric cylindrical grid (there is a one-to-one correspondence between $\phi^{m}$ and $\Phi_{m}$ owing to the condition $\Delta\Phi=\Delta\phi$). This viewpoint is used both to calculate the double integrals in Eq.(\ref{eq:RTE_integral1}) and to consider the relation between $I^{lm}_{f}$ and $I^{lm}_{ij}$.

\subsubsection{Double integrals}\label{subsubsec:double_integral}
We calculate the double integrals in Eq.(\ref{eq:RTE_integral1}) on the 3D cylindrical grid assuming that the face intensity $I^{lm}_{f}$ is constant over both the cell face $f$ and the control solid angle $\Delta\Omega^{lm}$. For example, the double integral for face $R^{+}$ is calculated as follows:
\begin{eqnarray}
\mathrm{DI}_{R^{+}} & \equiv & \int_{\Delta\Omega^{lm}}\int_{A_{R^{+}}} (\bmath{s}^{lm}\cdot \diff\bmath{A})\diff\Omega \nonumber\\ 
&=& \int_{\Delta\Omega^{lm}}\int_{A_{R^{+}}} \sin\theta \cos(\phi-\Phi)\diff A \diff\Omega \nonumber\\
&=& \int^{\theta^{l+\frac{1}{2}}}_{\theta^{l-\frac{1}{2}}}\sin\theta^{2} \diff\theta \int^{\phi^{m+\frac{1}{2}}}_{\phi^{m-\frac{1}{2}}}\diff\phi \nonumber\\
& & \qquad\times \int^{z_{j+\frac{1}{2}}}_{z_{j-\frac{1}{2}}}\diff z \int^{\frac{\Delta\Phi}{2}}_{-\frac{\Delta\Phi}{2}}\diff \Phi R_{i+\frac{1}{2}}\cos(\phi-\Phi) \nonumber\\
&=& R_{i+\frac{1}{2}}\Delta z_{j} \Delta\phi\; \Theta^{l}\dfrac{\sin\frac{\Delta\phi}{2}}{\frac{\Delta\phi}{2}}\left(\sin\phi^{m+\frac{1}{2}}-\sin\phi^{m-\frac{1}{2}}\right), \label{eq:DIRp}
\end{eqnarray}
where $\mathrm{DI}_{R^{+}}$ is computed for the face of a cell whose azimuthal extent is $[-\frac{\Delta\Phi}{2},\frac{\Delta\Phi}{2}]$ (we can choose such a cell without the loss of generality because of the axisymmetry) and we used the relation $\Delta\Phi=\Delta\phi$ to derive the last line. The function $\Theta^{l}$ is defined as follows:
\begin{equation}
\Theta^{l} = \dfrac{1}{2}\left(\Delta\theta + \cos\theta^{l-\frac{1}{2}}\sin\theta^{l-\frac{1}{2}} - \cos\theta^{l+\frac{1}{2}}\sin\theta^{l+\frac{1}{2}}\right).
\end{equation}

The double integrals for the other faces are similarly calculated and the results are summarized as follows:
\begin{align}
\mathrm{DI}_{R^{-}} & = & - & R_{i-\frac{1}{2}}\Delta z_{j} \Delta\phi\; \Theta^{l}\dfrac{\sin\frac{\Delta\phi}{2}}{\frac{\Delta\phi}{2}}\left(\sin\phi^{m+\frac{1}{2}}-\sin\phi^{m-\frac{1}{2}}\right), \\
\mathrm{DI}_{\Phi^{+}} & = &   & \Delta R_{i}\Delta z_{j} \Delta\phi\; \Theta^{l}\dfrac{\cos\phi^{m-1}-\cos\phi^{m}}{\Delta\phi}, \\
\mathrm{DI}_{\Phi^{-}} & = & - & \Delta R_{i}\Delta z_{j} \Delta\phi\; \Theta^{l}\dfrac{\cos\phi^{m}-\cos\phi^{m+1}}{\Delta\phi}, \\
\mathrm{DI}_{z^{\pm}}  & = & \pm & R_{i}\Delta R_{i}\; \Delta\phi \dfrac{1}{2}\left(\sin^{2}\theta^{l+\frac{1}{2}}-\sin^{2}\theta^{l-\frac{1}{2}}\right).
\end{align}
Note that $\mathrm{DI}_{R^{\pm}}$ and $\mathrm{DI}_{\Phi^{\pm}}$ are different from those in \citet{kim08:_asses} because they ignore the dependency of $(\bmath{s}^{lm}\cdot\diff\bmath{A})$ on $\Phi$ (e.g., $\cos(\phi-\Phi)$ in Eq.(\ref{eq:DIRp})).

\subsubsection{Algebraic equations to be solved}\label{subsubsec:discretized_RTE}
We adopt a first-order upwind scheme to relate $I^{lm}_{f}$ to $I^{lm}_{ij}$ in this study. Namely, $I^{lm}_{f}$ is approximated by an intensity at the center of a cell that is nearest to face $f$ and is located upstream of face $f$. This scheme is called the STEP scheme in the field of engineering application and is known to a numerically stable scheme\footnote{It may be possible to use high-order schemes instead of the STEP scheme since there are several studies for applications of high-order interpolation techniques developed in the CFD such as the total variation diminishing (TVD) interpolation to the discrete ordinates method, which is similar to the FVMRT (e.g., \citealt{balsara01:_fast,coelho08,godoy10,coelho14:_applic_nvd_tvd}). }. The nearest cell upstream of a given cell face $f$ can be easily identified by using a figure similar to Fig.~\ref{fig:FVMRT_overview}. For example, face intensity $I^{lm}_{R^{+}}$ is approximated as follows: $I^{lm}_{R^{+}}=I^{lm}_{i+1,j}$ for $\frac{N_{\phi}}{2}+1\leq m \leq N_{\phi}$ ($\frac{\pi}{2} < \phi^{m} < \pi$) and $I^{lm}_{R^{+}}=I^{lm}_{ij}$ for $1\leq m \leq \frac{N_{\phi}}{2}$ ($0 < \phi^{m} < \frac{\pi}{2}$). Before deriving the algebraic equations to be solved, we introduce the following auxiliary coefficients:
\begin{eqnarray}
D^{lm}_{z} & \equiv & \dfrac{1}{2}\left(\sin^{2}\theta^{l+\frac{1}{2}}-\sin^{2}\theta^{l-\frac{1}{2}}\right)\Delta\phi
\end{eqnarray}
\begin{eqnarray}
K^{lm}_{0} & \equiv & \Theta^{l} \dfrac{\sin\frac{\Delta\phi}{2}}{\frac{\Delta\phi}{2}}\left(\sin\phi^{m+\frac{1}{2}}-\sin\phi^{m-\frac{1}{2}}\right),
\end{eqnarray}
\begin{eqnarray}
C^{lm}_{+} & \equiv & 
\begin{cases}
\Theta^{l} \dfrac{\cos\phi^{m-1}-\cos\phi^{m}}{\Delta\phi}, & m > 1, \\
0, & m=1,
\end{cases}
\end{eqnarray}
\begin{eqnarray}
C^{lm}_{-} & \equiv &
\begin{cases}
\Theta^{l}\dfrac{\cos\phi^{m+1}-\cos\phi^{m}}{\Delta\phi}, & m < N_{\phi}, \\
0, & m=N_{\phi},
\end{cases}
\end{eqnarray}
where $C^{l1}_{+}=0$ and $C^{lN_{\phi}}_{-}=0$ is due to the axisymmetry.

Using these auxiliary coefficients and assuming the STEP scheme, Eq.(\ref{eq:RTE_integral1}) can be rewritten into algebraic equations
\begin{align}
& \left[\dfrac{D^{lm}_{z}}{\Delta z_{j}} + \dfrac{R_{i+\frac{1}{2}}}{R_{i}\Delta R_{i}}C^{lm}_{+} - \dfrac{R_{i-\frac{1}{2}}}{R_{i}\Delta R_{i}}(K^{lm}_{0}+C^{lm}_{+}) \right.\nonumber\\
& \quad + \left.\left(\chi_{ij}+\dfrac{1}{c\Delta t}\right)\Delta\Omega^{lm} \right]I^{lm}_{ij} \nonumber\\
& = \left(j_{\mathrm{eff},ij}+\dfrac{I^{lm}_{0,ij}}{c\Delta t}\right)\Delta\Omega^{lm} + \dfrac{D^{lm}_{z}}{\Delta z_{j}}I^{lm}_{i,j-1} \nonumber\\
& \quad - \dfrac{R_{i+\frac{1}{2}}}{R_{i}\Delta R_{i}}\left(K^{lm}_{0}I^{lm}_{i+1,j} + C^{lm}_{-}I^{l,m+1}_{ij}\right) + \dfrac{R_{i-\frac{1}{2}}}{R_{i}\Delta R_{i}} C^{lm}_{-}I^{l,m+1}_{ij}, \nonumber\\
& \qquad \text{for}\; \mu^{l}\equiv \cos\theta^{l} > 0, \frac{N_{\phi}}{2}+1 \leq m \leq N_{\phi},
\end{align}
\begin{align}
& \left[\dfrac{D^{lm}_{z}}{\Delta z_{j}} + \dfrac{R_{i+\frac{1}{2}}}{R_{i}\Delta R_{i}}(K^{lm}_{0}+C^{lm}_{+}) - \dfrac{R_{i-\frac{1}{2}}}{R_{i}\Delta R_{i}}C^{lm}_{+} \right.\nonumber\\
& \quad + \left.\left(\chi_{ij}+\dfrac{1}{c\Delta t}\right)\Omega^{lm} \right]I^{lm}_{ij} \nonumber\\
& = \left(j_{\mathrm{eff},ij} + \dfrac{I^{lm}_{0,ij}}{c\Delta t}\right)\Delta\Omega^{lm} + \dfrac{D^{lm}_{z}}{\Delta z_{j}}I^{lm}_{i,j-1} \nonumber\\
& \quad - \dfrac{R_{i+\frac{1}{2}}}{R_{i}\Delta R_{i}}C^{lm}_{-}I^{l,m+1}_{ij} + \dfrac{R_{i-\frac{1}{2}}}{R_{i}\Delta R_{i}}\left(K^{lm}_{0}I^{lm}_{i-1,j} + C^{lm}_{-}I^{l,m+1}_{ij}\right), \nonumber\\
& \qquad \text{for}\; \mu^{l} > 0, 1 \leq m \leq \frac{N_{\phi}}{2}, 
\end{align}
where we only show the equations for $\cos\theta^{l}>0$ because of space limitations. We can obtain the solution for $I^{lm}_{ij}$ by solving the equations above in the order shown in Fig.~\ref{fig:FVMRT_overview} (see green and purple arrows labeled by `sweep (1)` and `sweep (2)`, respectively). The algebraic equations for the limit of $c\rightarrow\infty$ can be obtained simply by dropping the terms with $c^{-1}$. The number of grid sweeps is unity in the absence of scattering and time derivative term; otherwise, we need to solve the equations iteratively (we repeat grid sweep and update $I^{lm}_{ij}$ until a converged solution is obtained).

\subsection{Tests} \label{appendix:subsec:FVMRT_Tests}
\begin{figure}
\centering
\includegraphics[clip,width=\linewidth]{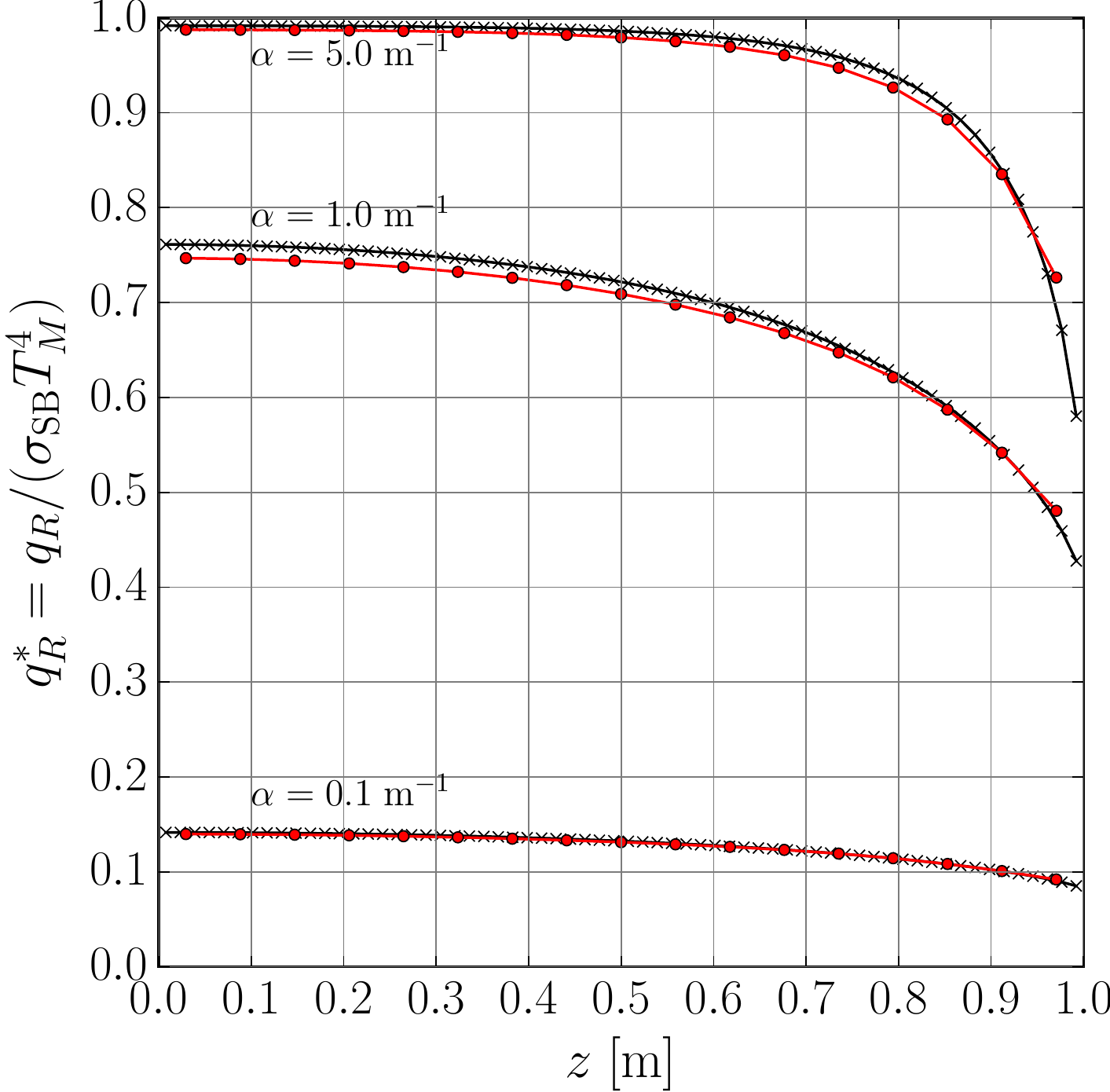}
\caption{Comparison between $q_{R}$ calculated by the FVMRT and a quasi-exact solution. The black line with crosses and the red line with points indicate a quasi-exact solution and the result of the FVMRT calculation, respectively. To obtain a quasi-exact solution, we divide solid angle $\Omega_{0} \equiv \{(\phi,\theta)| \phi \in [0,\frac{\pi}{2}],\; \theta\in[0,\pi]\}$ into $128 \times 256$ elements ($I \neq 0$ only on $\Omega_{0}$ at $R=R_{c}$) and perform a numerical integration of Eq.(\ref{eq:qR}) on $\Omega_{0}$ using the rectangle rule. As for the FVMRT, we assume $N_{R}=18$, $N_{z}=34$, and $N_{\phi}=N_{\theta}=12$.}
\label{fig:FVMRT_Test1}
\end{figure}

To check the validity of our implementation, we perform a test calculation described in \citet{chui92:_predic_radiat_trans_cylin_enclos}, where we compute a radiation flux $q_{R}$ at the surface of a finite cylindrical enclosure containing an absorbing-emitting but nonscattering medium by the FVMRT and compare the result with the exact solution. The numerical setup is detailed in \citet{chui92:_predic_radiat_trans_cylin_enclos}. Briefly, the enclosure has a radius $R_{c}=1\;\mathrm{m}$ and a height $h_{c}=2\;\mathrm{m}$. Its wall is infinitesimally thin and is assumed to be cold ($T_{w}=0\;\mathrm{K}$) and black (no reflection at the wall). The medium has a constant temperature $T_{\mathrm{M}}=100\;\mathrm{K}$ and a constant absorption coefficient ($\alpha=0.1$, $1$, $5\;\mathrm{m^{-1}}$). In this case, the net radiation flux at the wall ($R=R_{c}$) is described by
\begin{equation}
q_{R}(z) = \int I(R_{c},z,\phi,\theta) (\bmath{s} \cdot \hat{\bmath{e}}_{R})\diff \Omega, \label{eq:qR}
\end{equation}
where $\bmath{s}$ is an unit vector in the direction of $(\phi,\theta)$, $I(R_{c},z,\phi,\theta) = B(T_{\mathrm{M}})(1-e^{-\tau})$, $\tau$ is the optical depth at the wall, $B(T_{\mathrm{M}})=(\sigma_{\mathrm{SB}}T^{4}_{\mathrm{M}})/\pi$, $\sigma_{\mathrm{SB}}$ is the Stefan-Boltzmann constant. A quasi-exact solution can be obtained by numerically integrating the equation above. Figure~\ref{fig:FVMRT_Test1} compares the result obtained by the FVMRT with the quasi-exact solution.  We find that the numerical result agrees with the quasi-exact solution in a satisfactory level.

As an another test, we make a comparison between the FVMRT and the SCM. First, we give a short summary of our implementation of the SCM. The SCM on the axisymmetric cylindrical coordinate is described in detail by \citet{noort02:_multid_non_lte_radiat_trans} and we basically follow it. We define the intensities at cell corners in keeping with \citet{voegler05:_simul} and \citet{hayek10:_radiat_mhd}. The spatial and angular interpolations required to obtain an upwind intensity and source functions in a short-characteristics are performed by quadratic or cubic B{\'e}zier interpolations (e.g., \citealt{hayek10:_radiat_mhd,rodriguez13:_delo_bezier}). Also, a quadratic B{\'e}zier interpolant is used to integrate the RTE (e.g, \citealt{auer03:_formal_solut,hayek10:_radiat_mhd}). Solid angle is discretized by the Carlson's Set A (\citealt{carlson63:_numer_theor_neutr_trans,carlson68:_trans_theor}; see also Appendix in \citealt{bruls99:_comput}), but we extend the Set A by adding grid points at $\phi=0$, $\frac{\pi}{2}$, and $\pi$ for each $\mu$ ($\equiv \cos\theta$) to avoid the use of extended short-characteristics method (ESCM) (\citealt{dullemond00}). The point weights for these newly-added points is set to 0.

\begin{figure*}
\centering
\begin{tabular}{ccc}
\begin{minipage}{0.333\hsize}
\includegraphics[clip,width=\linewidth]{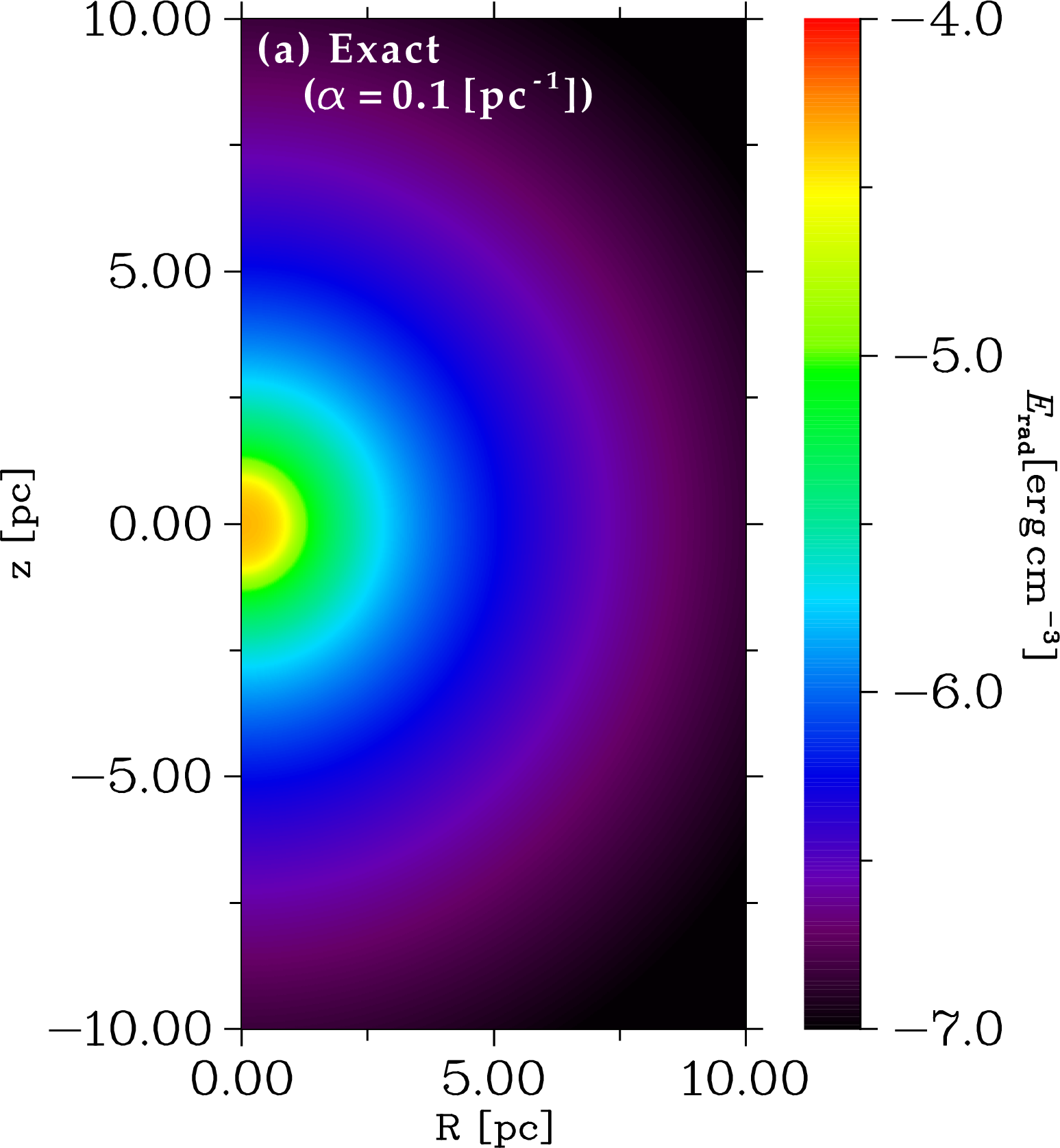}
\end{minipage}
\begin{minipage}{0.333\hsize}
\includegraphics[clip,width=\linewidth]{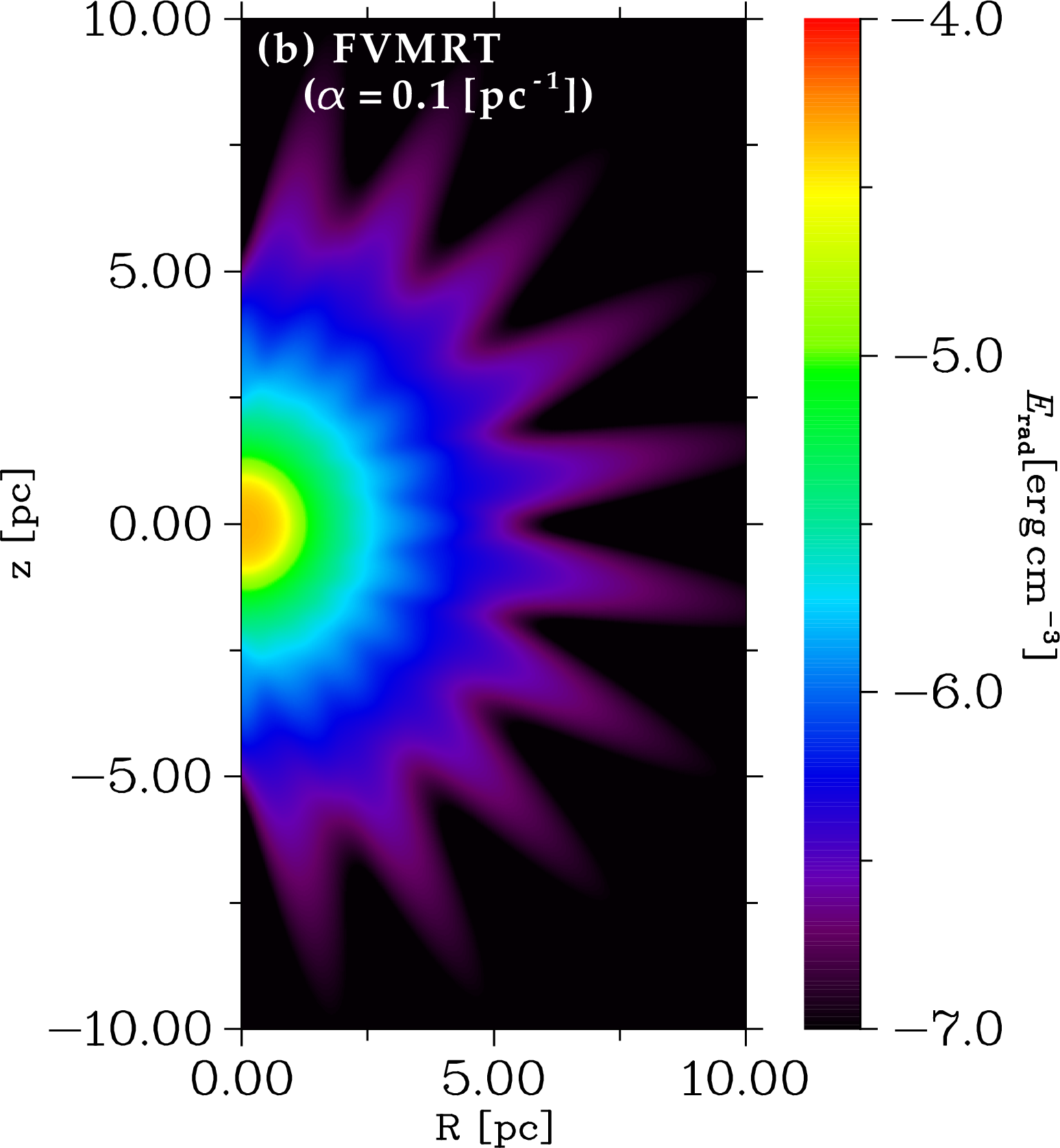}
\end{minipage}
\begin{minipage}{0.333\hsize}
\includegraphics[clip,width=\linewidth]{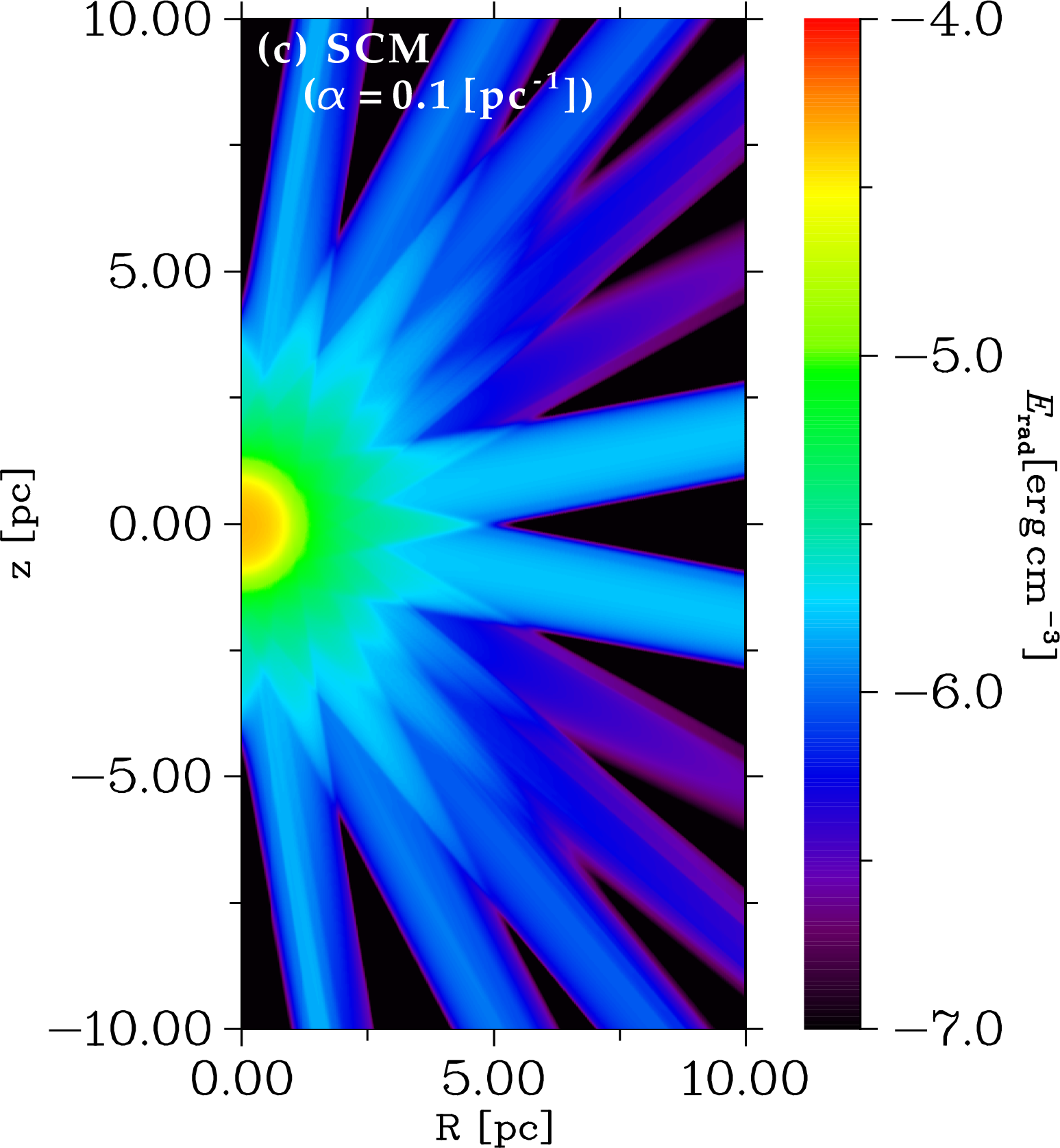}
\end{minipage}
\end{tabular}
\caption{Spatial distribution of radiation energy density in the comparison test described in \S~\ref{appendix:subsec:FVMRT_Tests} (from the left to the right, $E^{\mathrm{exact}}_{\mathrm{rad}}$, $E^{\mathrm{FVM}}_{\mathrm{rad}}$, and $E^{\mathrm{SCM}}_{\mathrm{rad}}$ are plotted). $E^{\mathrm{exact}}_{\mathrm{rad}}$ can be easily calculated using the fact that the point $(R,z)$ is equivalent to the point $(0,r[\equiv \sqrt{R^{2}+z^{2}}])$ and we can perform $\phi$-integration immediately because of the spherical symmetry. For $r < r_{c}$, we divide the polar angle $\theta$ ($\in [0,\pi]$) into $16384$ elements and perform a numerical integration of Eq.(\ref{eq:Erad}) with the rectangle rule. For $r > r_{c}$, $I\neq 0$ only for $\theta \in [0,\sin^{-1}(r_{c}/r)]$. Therefore, we consider that $\theta$ range and divide it into $16384$ elements. $E^{\mathrm{FVM}}_{\mathrm{rad}}$ is computed assuming $N_{\phi}=N_{\theta}=10$. $E^{\mathrm{SCM}}_{\mathrm{rad}}$ is computed assuming the \textit{maximum} order of the quadrature ($n=12$; the number of discrete ordinates with non-zero point weight is 84). We use an uniform spatial grid ($N_{R}=512$, $N_{z}=1024$) for all the calculations.}
\label{fig:Comparison_of_RT_solvers}
\end{figure*}

The numerical setup used is very simple. We consider an uniform sphere of gas with a radius $r_{c}=1\;\pc$, which is located at the center of a cylindrical computational box with a radius $R_{\max}=10\;\pc$ and a height $h=z_{\max}-z_{\min}=20\;\pc$. The gas consists of an absorbing-emitting but nonscattering medium with a temperature $T_{\mathrm{M}}=500\;\mathrm{K}$ and is assumed to be a black body. Its absorption coefficient is $\alpha=0.1\;\pc^{-1}$. We compute the distribution of steady-state radiation energy density by both methods (denoted by $E^{\mathrm{FVM}}_{\mathrm{rad}}$ and $E^{\mathrm{SCM}}_{\mathrm{rad}}$) and compare them with a quasi-exact solution ($E^{\mathrm{exact}}_{\mathrm{rad}}$), which can be obtained by numerically integrating the following equation:
\begin{equation}
E_{\mathrm{rad}}(R,z) = \dfrac{1}{c}\int I(R,z,\phi,\theta)\diff\Omega. \label{eq:Erad}
\end{equation}
The results are shown in Fig.~\ref{fig:Comparison_of_RT_solvers}. As expected, $E^{\mathrm{exact}}_{\mathrm{rad}}$ is roughly constant inside the sphere and it decreases with distance from the sphere ($E^{\mathrm{exact}}_{\mathrm{rad}} \propto r^{-2}$ at large distances). Compared to this, $E^{\mathrm{SCM}}_{\mathrm{rad}}$ shows significant deviations from $E^{\mathrm{exact}}_{\mathrm{rad}}$ for $r \gtrsim 2 r_{c}$. Its distribution is a superposition of a finite number of radial beams, which is far from the spherical symmetry. Most problematic is that $E^{\mathrm{SCM}}_{\mathrm{rad}}$ does not decrease with distance along a beam. This results in a substantial overestimation of radiation pressure at large distances. Thus, the SCM significantly violates the conservation of radiation energy (see Table.~\ref{tbl:energy_balance}). On the other hand, $E^{\mathrm{FVM}}_{\mathrm{rad}}$ decreases with distance at a rate similar to $E^{\mathrm{exact}}_{\mathrm{rad}}$, although its distribution is not spherical symmetric at $r \gtrsim 2r_{c}$. Most important is that the FVMRT accurately conserves the radiation energy (see Table~\ref{tbl:energy_balance}). The amplitude of the spherically-asymmetric feature at $r>2r_{c}$ can be reduced in the FVMRT by increasing $N_{\phi}$ and $N_{\theta}$ at the cost of computational time. By contract, there is a limit to such reduction in the SCM because there exists the maximum number of discrete ordinates (see \citealt{carlson63:_numer_theor_neutr_trans}), which is already assumed in the calculation of $E^{\mathrm{SCM}}_{\mathrm{rad}}$ in Fig.~\ref{fig:Comparison_of_RT_solvers}.

\begingroup
\renewcommand{\arraystretch}{1.0}
\begin{table}
\centering
\begin{minipage}{\hsize}
\caption{Energy balances in the FVMRT and the SCM}
\label{tbl:energy_balance}
\begin{tabular}{@{}lll@{}}
\hline
Rate of energy $[\mathrm{erg\;s^{-1}}]$ $^{\dag}$ & FVMRT & SCM$^{\ddag}$ \\
\hline
\ding{192} $\equiv \dot{E}_{\mathrm{em}}$ & $5.6435004\times 10^{43}$ & $5.6435004\times 10^{43}$ \\
\ding{193} $\equiv \dot{E}_{\mathrm{out}}$ & $1.7771012\times 10^{43}$ & $1.1497229\times 10^{45}$ \\
\ding{194} $\equiv \dot{E}_{\mathrm{abs}}$ & $3.8663993\times 10^{43}$ & $4.0017076\times 10^{42}$ \\
\ding{195} $\equiv$ |\ding{192}-(\ding{193}+\ding{194})| & $2.9710561\times 10^{29}$ & $1.0972896\times 10^{45}$ \\
\ding{195} $\div$ \ding{192} & $\mathbf{5.2645625\times 10^{-15}}$ & $\mathbf{1.9443422\times 10^{1}}$ \\
\hline
\end{tabular}
\begin{flushleft}
$^{\dag}$ $\dot{E}_{\mathrm{em}}$ is the total luminosity of the medium, $\dot{E}_{\mathrm{out}}$ is the net amount of radiation energy flowing out from the computational box per unit time, and $\dot{E}_{\mathrm{abs}}$ is the amount of radiation energy absorbed by the medium per unit time. The conservation of radiation energy requires the equality $\dot{E}_{\mathrm{em}}=\dot{E}_{\mathrm{out}}+\dot{E}_{\mathrm{abs}}$.\\
$^{\ddag}$ In the SCM, an intensity at the center of a cell is approximated by an arithmetic average of intensities at the four corners of the cell.
\end{flushleft}
\end{minipage}
\end{table}
\endgroup

\bsp	
\label{lastpage}
\end{document}